\documentclass[aip,jcp,amsmath,amssymb,reprint]{revtex4-2}

\usepackage[utf8]{inputenc}
\usepackage[T1]{fontenc}
\usepackage{microtype}

\usepackage{array,booktabs}
\usepackage{graphicx}
\usepackage{dcolumn}
\usepackage{multirow}
\usepackage{bm}
\usepackage{setspace}

\usepackage{longtable}

\usepackage[breaklinks=true,colorlinks=true,linkcolor=blue,urlcolor=blue,citecolor=blue,bookmarks=true,bookmarksopenlevel=2]{hyperref}
\usepackage[referable]{threeparttablex}
\usepackage[table]{xcolor}
\definecolor{navyblue}{HTML}{000080}
\usepackage[normalem]{ulem}

\newcommand{\changed}[1]{#1}
\usepackage{xspace}
\usepackage{textcomp}
\usepackage[textsize=footnotesize]{todonotes}
\usepackage{rotating}
\usepackage{braket}
\makeatletter
\newcommand*{\Cite}[2][]{%
  \begingroup
  \let\NAT@mbox=\mbox
  \let\@cite\NAT@citenum
  \let\NAT@space\NAT@spacechar
  \let\NAT@super@kern\relax
  \renewcommand\NAT@open{}%
  \renewcommand\NAT@close{}%
  \cite[#1]{#2}%
  \endgroup
}
\makeatother

\usepackage{siunitx}
\newcommand{\sandwich}[3]{\left\langle #1 \vphantom{#2 #3} \right\vert #2 \left\vert \vphantom{#1 #2} #3 \right\rangle}

\begin{document}

\title{
Relativistic and quantum electrodynamics effects on NMR shielding tensors of Tl$X$ ($X$ = H, F, Cl, Br, I, At) molecules
}

\author{Karol Kozio{\l}}
\email[Electronic mail: ]{karol.koziol@ncbj.gov.pl}
\affiliation{\NCBJ}

\author{I. Agust\'{\i}n Aucar}
\affiliation{\IMIT}
\affiliation{\RUG}

\author{Konstantin Gaul}
\author{Robert Berger}
\email[Electronic mail: ]{robert.berger@uni-marburg.de}
\affiliation{\Philipps}

\author{Gustavo A. Aucar}
\email[Electronic mail: ]{gaaucar@conicet.gov.ar}
\affiliation{\IMIT}

\newcommand{\NCBJ}{
Narodowe Centrum Bada\'{n} J\k{a}drowych (NCBJ), Andrzeja So{\l}tana 7, 05-400 Otwock-\'{S}wierk, Poland
\vspace*{0.1cm}
}

\newcommand{\IMIT}{
Instituto de Modelado e Innovaci\'on Tecnol\'ogica (UNNE-CONICET),
Facultad de Ciencias Exactas y Naturales y Agrimensura,
Universidad Nacional del Nordeste, Avda.~Libertad 5460, Corrientes, Argentina
\vspace*{0.1cm}
}

\newcommand{\Philipps}{
Fachbereich Chemie, Philipps–Universit\"at Marburg, Hans-Meerwein-Stra{\ss}e 4, 35032 Marburg, Germany
\vspace*{0.1cm}
}

\newcommand{\RUG}{
Van Swinderen Institute for Particle Physics and Gravity,
University of Groningen, 9747 AG Groningen, The Netherlands\\
\vspace*{0.1cm}
}

\date{\today}

\begin{abstract}
Results of relativistic calculations of nuclear magnetic resonance shielding tensors (${\bm \sigma}$) for the thallium monocation (Tl$^+$), thallium hydride (TlH) and thallium halides (TlF, TlCl, TlBr, TlI, and TlAt) are presented as obtained within a four-component polarization propagator formalism and a two-component linear response approach within the zeroth-order regular approximation. \changed{In addition to a detailed analysis of relativistic effects performed in this work,} some quantum electrodynamical (QED) effects on those NMR shieldings \changed{as well as other small contributions} are estimated. A strong dependence of ${\bm \sigma}$(Tl) on the bonding partner is found, together with a very weak dependence of QED effects with them. In order to explain the trends observed, the excitation patterns associated with relativistic $ee$ (or paramagnetic-like) and $pp$ (or diamagnetic-like) contributions to ${\bm \sigma}$ are analyzed. For this purpose, also the electronic spin-free and spin-dependent contributions are separated within the two-component zeroth-order regular approximation, and the influence of spin-orbit coupling on involved molecular orbitals is studied, which allows for a thorough understanding of the underlying mechanisms.
\end{abstract}

\maketitle

\section{Introduction}
\label{sec:introduction}
%
Reliable theoretical predictions of the two main NMR spectroscopic parameters, namely the nuclear magnetic resonance shielding tensor ${\bm \sigma}$ and the indirect nuclear spin-spin coupling tensor ${\bm J}$ for heavy-atom-containing molecules require to account for relativistic effects. To do this one typically works currently within four-component (4C) or two-component (2C) frameworks, and applies state-of-the-art electronic structure methods.\cite{ILRusakova_Magnetochem2023}

The relativistic polarization propagator theory (RelPPT) is among the most powerful ones for obtaining an accurate reproduction of measurements of response properties, together with information about the electronic origin of their absolute values and trends.\cite{Aucar1993,Aucar2010} Its approximate schemes have the capacity to improve its results in a systematic manner though at the moment only 4C calculations at the first-order level of perturbation, known as random-phase approximation (RPA), can be performed.\cite{JOD_CPR1984} What one usually does is to run response calculations on top of either relativistic Dirac--Hartree--Fock (DHF) or density-functional theory (DFT). Furthermore, RelPPT allows for quantifying relativistic effects by working only within the 4C framework and letting the velocity of light $c$ go to infinity. It also permits to identify the contribution of the most involved virtual excitations among molecular orbitals (MOs) in each kind of contributions, meaning $ee$ (or paramagnetic-like) and $pp$ (or diamagnetic-like), as detailed in Sec.~\ref{sec:theory} of this work.\cite{Aucar1993}

Additionally, the quasi-relativistic zeroth-order regular approximation (ZORA) is used in the present work, although it is commonly considered to display significant deficiencies in the description of \emph{absolute} NMR shieldings. This approach has, however, the advantage that spin-free and spin-dependent contributions from individual operators can be easily analysed separately. Moreover, as we demonstrate in the present work, reasonable agreement with 4C calculations can be achieved with a simple procedure to re-scale the ZORA molecular orbital energies, which are identified herein to be the main source of error in ZORA calculations of NMR shielding constants.

Searching for a way to include smaller and more intricate effects on the NMR spectroscopic parameters, some of the present authors developed and published first results concerning the estimation of quantum electrodynamics (QED) effects on ${\bm \sigma}$ and ${\bm J}$.\cite{Gimenez2016,Koziol2019,ColomboJofre2022} 
They started working on a model for including QED corrections to the NMR shielding tensor, and applied it to He-like and Be-like atomic systems with atomic number $Z$ in the range $10 \le Z \le 86$.\cite{Gimenez2016} That model scaled previous results obtained by Yerokhin \textit{et al.} for H-like atoms to the aforementioned ionic systems.\cite{Yerokhin2011,Yerokhin2012} Such a procedure is similar to the way QED effects are usually introduced in many-electron atomic systems.\cite{Koziol2018a} 
In a subsequent paper, they estimated QED effects on the shielding of neutral and ionic atoms with $10 \le Z \le 86$, and diatomic halogen molecules by extending the previous approach.\cite{Koziol2019} QED effects on shieldings were found to have a negative sign, being in magnitude greater than 1\,\% of the relativistic effects for high-$Z$ atoms such as Hg and Rn, and up to 0.6\,\% of its total 4C value for neutral Rn. In the most recent paper of this series a model was developed to estimate the QED corrections to the indirect nuclear spin--spin coupling constant $J$.\cite{ColomboJofre2022} This new model was applied to estimate QED effects on $J$-couplings of the $X_2^{2+}$ and $X_3^{2+}$ ($X$ = Zn, Cd, Hg) linear molecular ions. Those effects were found to be within the interval $(-0.4~;~-0.2)$\% of the total isotropic coupling constant for Zn-containing ions, and they increase until the range $(-1.2~;~-0.8)$\% for Hg-containing ions. \changed{
QED corrections of hyperfine interactions were computed previously for hydrogen-like atoms (see e.g. Refs.~\citenum{schneider:1994,sunnegren:1998,Jentschura:06}), few electron atoms (see Refs.~\citenum{shabaev:1998,Artemyev:01,Shabaev:01,Sapirstein2001} and references therein), and atoms with one valence electron (see Refs.~\citenum{Sapirstein2003,Ginges2017}). To the best of our knowledge, QED effects on the NMR shielding were computed previously only for light, few-electron atoms,\cite{KPachucki_JCP2009, Yerokhin2011, KPachucki_PRL2021, KPachucki_PRA2022, Yerokhin2024} but} no other assessment has been made regarding the magnitude of the QED effects on the NMR shielding \changed{in molecules}. Our findings show that QED effects hold considerable significance and must be taken into account in order to achieve theoretical values that pursue to align closely with experimental results. At the present time, precise absolute measurements of NMR shielding in molecules can be achieved through gas-phase experiments. The uncertainty associated with these measurements may be smaller than the magnitudes of QED corrections observed in molecules containing heavy atoms.\cite{KJackowski_JMolStruct2005, KJackowski_JPCA2010, KJackowski_Prog2012, KJackowski_PCCP2016}

In order to extend our studies on relativistic and QED effects, we have selected the Tl$X$ ($X$ = H, F, Cl, Br, I, At) molecules. Few of them are of great interest presently as they feature pronounced relativistic effects and will be studied in high-precision experiments.\cite{Grasdijk_2021,Blanchard_PRR2023} In fact there are only few calculations of NMR shielding or ${\bm \sigma}$(Tl) reported like the paper by Jaszu{\'n}ski et al.,\cite{Jaszunski2014} where computational values of ${\bm \sigma}$(Tl) in TlF were presented. These authors applied coupled-cluster methods and 4C relativistic DFT. Another set of calculations for the whole series of molecules was published by Bashi et al.,\cite{Bashi2020} but they focused on solid state NMR. Then there is only one previous study of ${\bm \sigma}$(Tl) in gas phase of TlF with 4C methods.

This work pursues three aims: i) to obtain accurate relativistic and QED contributions to the NMR shielding tensor of Tl nucleus (${\bm \sigma}$(Tl)) and ligand atoms (${\bm \sigma}$($X$)) in the series of thallium hydride and thallium halide diatomic molecules: TlH, TlF, TlCl, TlBr, TlI, and TlAt; ii) to learn about the electronic origin of relativistic effects in that tensor, meaning the trends of relativistic effects on the paramagnetic-like and diamagnetic-like contributions to the shieldings and their influence on the tensor components perpendicular and parallel to the interatomic axis, and iii) to apply an effective model to estimate QED effects to NMR magnetic shielding constant, described in Ref.~\citenum{Koziol2019}, to the calculations for those series of thallium halide molecules. 

In Sec.~\ref{sec:theory} we shall give the theoretical framework used for getting accurate theoretical results and the way we analyze the contribution of each pattern of excitation associated with relativistic $ee$ (or paramagnetic-like) and $pp$ (or diamagnetic-like) terms of ${\bm \sigma}$ in order to explain the origin of the trends observed for ${\bm \sigma}$(Tl) in the Tl$X$ ($X$ = H, F, Cl, Br, I, At) series. We also explain the model used for estimating QED effects. In Sec.~\ref{sec:results} we shall show how large relativistic effects are and for which terms they are most important. We will demonstrate that in these molecules relativity mainly affects the perpendicular contribution to ${\bm \sigma}$(Tl) and that the trend of shieldings observed is explained from that contributing term. Furthermore, we shall show that there is a well-defined pattern of excitation for the $ee$ and $pp$ terms, and finally, that QED effects on ${\bm \sigma}$(Tl) are not vanishingly small and depend on the halogen bonded to the thallium atom.

\medskip

\section{Theoretical models and computational details}\label{sec:theory}

\subsection{Shieldings calculated with relativistic polarization propagators}
\label{sec:shield-rpp}
As mentioned above we worked with the RelPPT theory which has been frequently used in recent years (see, e.g., Refs.~\citenum{Aucar1999, Saue2003, Aucar2010, Aucar2014}).
Within such a theory it has been shown that both, the well known paramagnetic and diamagnetic contributions to the NMR spectroscopic parameters arise naturally when the non-relativistic (NR) limit is applied. On the other hand, one can work with relativistic generalizations of those contributions known as $ee$ and $pp$, respectively. Furthermore, the response properties we are interested in are static and so they must be calculated at zero frequency. For this reason we shall not include the frequency dependence explicitly in equations. 

Within the RelPPT the NMR shielding tensor has the following expression (the SI system of units is used throughout this work):\cite{Aucar2010}
%
\begin{eqnarray}
\label{eq:sigma_tensor}
{\bm \sigma}(K) &=& 
\langle \langle \hat{\bf P}_K; \hat{\bf Q}_B \rangle\rangle \nonumber  \\
&=&
\frac{\mu_0\,e^2}{8\pi} \left\langle \left\langle \dfrac{c{\bm \alpha} \times {\bf r}_K}{r_K^3}; c{\bm \alpha} \times {\bf r}_G \right\rangle \right\rangle ,
\end{eqnarray}
%
where $\hat{\bf P}_K=-(\mu_0/(4\pi)) e c {\bm \alpha} \times ({\bf r}_K/r_K^3)$ and $\hat{\bf Q}_B=-(e/2) c{\bm \alpha} \times {\bf r}_G$ are the operators arising from the perturbative Hamiltonians that account for electromagnetic interactions between the electrons and both the nuclear spin of nucleus $K$ and an external uniform magnetic field, respectively. Besides, $\mu_0$ is the permeability of free space, $e$ is the elementary charge, $c$ is the speed of light in vacuum, $\bm{\alpha}$ stands for the Dirac matrices in standard representation, and ${\bf r}_K$ and ${\bf r}_G$ are the position operators of the electrons relative to the nucleus $K$ and to the fixed gauge origin (GO) for the external magnetic potential, respectively.
At the RPA (or first-order) level of approach, Eq.~\eqref{eq:sigma_tensor} can also be expressed as
\begin{widetext}
\begin{equation}
\label{eq:RPP}
{\bm \sigma}(K) = 
\begin{pmatrix}
 \tilde{\bf P}_K^{ee} & \tilde{\bf P}_K^{ep} & \tilde{\bf P}_K^{pe}   
\end{pmatrix}
\begin{pmatrix}
{\bf M}^{ee, ee} & {\bf M}^{ee, ep} &  {\bf M}^{ee, pe} \vspace{0.1cm}\\ 
{\bf M}^{ep, ee} & {\bf M}^{ep, ep} &  {\bf M}^{ep, pe} \vspace{0.1cm}\\ 
{\bf M}^{pe, ee} & {\bf M}^{pe, ep} &  {\bf M}^{pe, pe}
\end{pmatrix}^{-1}
\begin{pmatrix}
{\bf Q}_B^{ee} \vspace{0.1cm} \\
{\bf Q}_B^{ep} \vspace{0.1cm} \\
{\bf Q}_B^{pe}
\end{pmatrix}.
\end{equation}
\end{widetext}

In Eq.~\eqref{eq:RPP}, each sub-block of the principal propagator (${\bf M}^{-1}$) and the perturbators (${\bf P}_K$ and ${\bf Q}_B$) is written in such a way that the excitation manifold is grouped into two different types of virtual excitations: one between occupied electronic states ($i$) and positive-energy unoccupied electronic states ($a$), and another one between occupied electronic states and negative-energy unoccupied electronic states ($\tilde a$). The former are named as $ee$ excitations, whereas the latter give rise to the so-called $ep$ and $pe$ excitations. It must be highlighted that while the $ep$ terms involve the creation of virtual electron-positron pairs, the $pe$ ones consider their annihilation.\cite{Aucar2010} 
Therefore, Eq.~\eqref{eq:RPP} can be rewritten as
\begin{eqnarray}
\label{eq:sigma-ee+pp}
{\bm \sigma}(K) &=&
\begin{pmatrix}
 \tilde{\bf P}_K^{ee} & \tilde{\bf P}_K^{ep} & \tilde{\bf P}_K^{pe}   
\end{pmatrix}
\begin{pmatrix}
{\bf X}_B^{ee} \vspace{0.1cm} \\
{\bf X}_B^{ep} \vspace{0.1cm} \\
{\bf X}_B^{pe} 
\end{pmatrix}
\nonumber \\
\nonumber \\
&=& {\bm \sigma}^{ee}(K) + {\bm \sigma}^{ep}(K) + {\bm \sigma}^{pe}(K), 
%
\end{eqnarray}
\noindent where the solution vector $\textbf{X}_B$ of the response equation $\textbf{M} \; \textbf{X}_B = \textbf{Q}_B$ is first expanded into a linear combination of trial vectors and subsequently contracted with the property matrix $\textbf{P}_K$, according to Eq.~\eqref{eq:RPP}.

The first of the three terms on the last line of Eq.~\eqref{eq:sigma-ee+pp} includes all those contributions to the NMR shielding tensor involving virtual excitations between the $i$-th occupied MO and the $a$-th positive-energy unoccupied MO. Besides, the second and third terms enclose all the remaining contributions: those implying virtual excitations from occupied MOs to negative-energy unoccupied MOs ($\tilde a$), and those including virtual de-excitations, meaning virtual excitations from negative-energy unoccupied MOs to occupied MOs. The sum of these two last terms can then be grouped into ${\bm \sigma}^{pp}(K)$. In short, this implies that
\begin{eqnarray}
\label{eq:sigma-ee+pp2}
{\bm \sigma}^{ee}(K) &=& \sum_{ia} {\bm \sigma}_{ia}(K) \nonumber \\
{\bm \sigma}^{ep}(K) = \sum_{i\tilde a} {\bm \sigma}_{i\tilde a}(K) &;&
{\bm \sigma}^{pe}(K) = \sum_{i\tilde a} {\bm \sigma}_{\tilde a i}(K) \nonumber \\
{\bm \sigma}^{pp}(K) &=& {\bm \sigma}^{ep}(K) + {\bm \sigma}^{pe}(K).
\end{eqnarray}

\subsection{Estimating QED effects\label{sec:qed_theory}}
%
The model used to estimate QED effects on the NMR shielding constants of nuclei in atoms and molecules was first proposed in Ref.~\citenum{Gimenez2016}, and additional details were given later in Ref.~\citenum{Koziol2019}.

We assume here that the inner MOs are similar to equivalent inner atomic orbitals (AOs). Based on this assumption, we proposed a model that allows estimating the QED effects on the nuclear magnetic shieldings of nuclei in molecules, starting from the atomic calculations for H-like systems reported by Yerokhin \textit{et al.}\cite{Yerokhin2011,Yerokhin2012} Assuming that the pattern of the ratio between the self-energy (SE) and vacuum polarization (VP) effects is similar for each $ns$ subshell, as occurs for orbital energies (see the Supplementary Material of Ref.~\citenum{Koziol2018a}), we have been able to extend our model by considering excitations starting from $ns$ subshells, with $n>1$.

Employing this model, the leading-order QED effects can consistently be taken into account in some specific contributions to the NMR shielding tensors. In fact, these effects are included in the calculation of this property in the following manner:
\begin{eqnarray}\label{eq:DC+QED}
{\bm \sigma}^\text{DC+QED}(K) &=&
\sum_{\substack{ia}} {\bm \sigma}_{ia}^\text{DC}(K) \left( 1 + \delta_{iq} \delta_{ar} C_{qr}^\text{QED/DC}(K)  \right) 
\nonumber \\
&& + \sum_{\substack{i\tilde a}} \left( {\bm \sigma}_{i\tilde a}^\text{DC}(K) + {\bm \sigma}_{\tilde a i}^\text{DC}(K) \right),
\end{eqnarray}
where the indices $q$ represent the inner (occupied) s-type MOs, whereas $r$ are the virtual (unoccupied) positive-energy s-type MOs. Besides, $\delta$ stands for the Kronecker delta and the superscript DC implies that the linear response functions associated with the NMR shielding tensors (see Eqs.~\eqref{eq:RPP} and \eqref{eq:sigma-ee+pp}) are based on the solution of the Dirac--Coulomb Hamiltonian. 

In Eq.~\eqref{eq:DC+QED} it can be seen that some contributions to the NMR shielding tensor arising from excitations between specific sets of MOs (i.e., s-type MOs) are scaled by the coefficients $C_{qr}^\text{QED/DC}(K)$. As shown in Ref.~\citenum{Koziol2019}, these molecular coefficients can be reliably replaced by their atomic analogues. In other words, the factors $C_{qr}^\text{QED/DC}(K)$ are assumed to be equal to $C_{QR}^\text{QED/DC}(K)$, where $Q$ and $R$ are s-type AOs analogous to the $q$ and $r$ MOs. These last coefficients are expressed as
\begin{widetext}
\begin{equation}
\label{eq:C_QED}
C_{QR}^\text{QED/DC}(K) =
\left( \frac{C_{QR}^\text{VP/DC}(K)-\left(\Delta\varepsilon_{QR}^\text{VP}(K)/\Delta\varepsilon_{QR}^\text{DC}(K)\right)}
{1+\left(\Delta\varepsilon_{QR}^\text{VP}(K)/\Delta\varepsilon_{QR}^\text{DC}(K)\right)} \right) 
\left( \frac{D_\text{VP}(Z\alpha)+D_\text{SE}(Z\alpha)}{D_\text{VP,po}(Z\alpha)} \right),
\end{equation}
\end{widetext}
\noindent where $D_\text{VP}(Z\alpha)$, $D_\text{SE}(Z\alpha)$, and $D_\text{VP,po}(Z\alpha)$ (with \text{VP,po} being the perturbed-orbital contributions to the vacuum polarization influence on ${\bm \sigma}(K)$) are functions that are parameterized to include the QED effects on the NMR shielding of hydrogen-like ions whose atomic number $Z$ refers to the nucleus $K$. These coefficients were taken from recent works by Yerokhin and co-workers.\cite{Yerokhin2011,Yerokhin2012} 
Besides, $\Delta \varepsilon_{QR}^\text{DC}(K)$ are the energy differences between the $Q$-th and $R$-th AOs obtained at the Dirac-Coulomb (DC) level of theory, $\Delta \varepsilon_{QR}^\text{VP}(K)$ are the differences between the VP corrections to the orbital energies of the same AOs, and the coefficients $C_{QR}^\text{VP/DC}(K)$ are expressed, for a given pair of s-type AOs, as
\begin{widetext}
\begin{equation}\label{eq:C-atom}
C_{QR}^\text{VP/DC}(K) = 
\left(C_{QR}^\text{VP/DC}(K)\right)_{uu} =
\frac{\sandwich{Q}{
\left(\dfrac{c{\bm \alpha} \times {\bf r}_K}{r_K^3}\right)_u
}{R}^\text{DC+VP}}{\sandwich{Q}{
\left(\dfrac{c{\bm \alpha} \times {\bf r}_K}{r_K^3}\right)_u
}{R}^\text{DC}}
\;
\frac{\sandwich{Q}{
\left(c{\bm \alpha} \times {\bf r}_G\right)_u
}{R}^\text{DC+VP}}{\sandwich{Q}{
\left(c{\bm \alpha} \times {\bf r}_G\right)_u
}{R}^\text{DC}}
\;-\; 1,
\end{equation}
\end{widetext}
\noindent with $u$ being any of the $x$, $y$, or $z$ Cartesian coordinates, and $\left(C_{QR}^\text{VP/DC}(K)\right)_{xx}=\left(C_{QR}^\text{VP/DC}(K)\right)_{yy}=\left(C_{QR}^\text{VP/DC}(K)\right)_{zz}$. Additionally, in the calculations where Eq.~\eqref{eq:C-atom} is used, it was assumed that ${\bf r}_G={\bf r}_K$. These coefficients account for the perturbed-orbital VP influence on the perturbators by comparing calculations including (labeled as DC+VP) and not including (DC) the Uehling potential in the self-consistent field process.
\changed{
Because $D_\text{VP}(Z\alpha)$, $D_\text{SE}(Z\alpha)$, and $D_\text{VP,po}(Z\alpha)$ coefficients were used to describe corrections to matrix elements involving only s-type AOs,\cite{Yerokhin2011,Yerokhin2012} our method is also limited to s-to-s virtual excitations. 
In addition to that, we are unable to calculate the factors $\Delta\varepsilon_{QR}^\text{VP}$, $\sandwich{Q}{
\left(\frac{c{\bm \alpha} \times {\bf r}_K}{r_K^3}\right)_u
}{R}^\text{DC+VP}$, and $\sandwich{Q}{
\left(c{\bm \alpha} \times {\bf r}_G\right)_u
}{R}^\text{DC+VP}$ for negative energy atomic states and so, our method is limited at the moment to consider only positive energy $Q$ and $R$ AOs. 
} 

\subsection{Computational details}\label{sec:comp-det}
All 4C calculations have been performed using the \textsc{Dirac} code,\cite{DIRAC22,dirac-paper} and the solutions of the linear response equations were obtained there at the RPA level of approach using DHF wave functions built from the DC Hamiltonian. In order to investigate the influence of electron correlation effects, calculations at the DFT/PBE0 and DFT/LDA\cite{LDA,LDA2} Dirac--Kohn--Sham levels of theory were also performed, which are only reported in the Supplementary Material.

In all calculations, experimental internuclear distances were used for Tl$X$ ($X =$ H, F, Cl, Br, and I), taken from Ref.~\citenum{Huber1979}, whereas the internuclear distance for TlAt has been determined by performing a structure optimization with the \textsc{Dirac} code, employing Dirac--Kohn--Sham DFT with the NR exchange-correlation hybrid PBE0 functional.\cite{PBE0} The distances used are 1.87~\AA{} for TlH, 2.084438~\AA{} for TlF, 2.484826~\AA{} for TlCl, 2.61819~\AA{} for TlBr, 2.81367~\AA{} for TlI, and 2.907656~\AA{} for TlAt.

For all elements under consideration both in 4C and 2C calculations, the uncontracted Dyall's relativistic all-electron with additional diffuse functions and quadruple-$\zeta$ quality basis sets (dyall.aae4z) were employed.\cite{KD02,KD06,KD12,KD16} 
\changed{The nuclei were modeled using normalized spherical Gaussian nuclear density distributions $\varrho_K \left( \vec{r} \right) = \frac{\zeta_K^{3/2}}{\pi ^{3/2}} \text{e}^{-\zeta_K \left| \vec{r} -
\vec{r}_K \right| ^2}$ with $\zeta_K = \frac{3}{2 r^2
_\text{nuc},K}$. The root-mean-square radius $r_{\text{nuc},K}$ was chosen as suggested by Visscher and Dyall.\cite{visscher:1997} The effect of the nuclear charge density distribution $\rho_K \left( \vec{r} \right) = Z_K e \varrho_K \left( \vec{r} \right)$ on the hyperfine interaction is known as the Breit--Rosenthal (BR) effect \cite{rosenthal:1932} and will be shortly discussed in Section \ref{small_contributions}. In 4C calculations, we assumed a finite scalar magnetization density distribution $M_K \left( \vec{r} \right) = \mu_K \varrho_K \left( \vec{r} \right)$, where $\mu_K$ is the magnitude of the magnetic dipole moment of nucleus $K$. The effect of a finite magnetization density distribution is called the Bohr--Weisskopf (BW) effect \cite{bohr:1950} and is usually small. In 2C calculations, the nuclei were assumed to be point-like magnetic dipoles. The BW effect at the 2C and 4C levels is explicitly discussed for Tl$^+$ in Section \ref{small_contributions}.} Furthermore, in all computations the common-gauge-origin (CGO) approach was used, and the GO for the external magnetic potential was placed at the nuclear center of mass (CM). All values of fundamental constants were taken from the CODATA2018 database.\cite{codata2018}

In calculations carried out in \textsc{Dirac}, contributions from two-electron small component integrals of ($SS$$\mid$$SS$) type were included, and their magnitude is analyzed in the Supplementary Material. Besides, the unrestricted kinetic balance prescription (UKB) was employed to generate the small component basis sets. Additional 4C calculations of NMR shielding tensors were performed employing the gauge-independent atomic orbital (GIAO) scheme and the results are also given in the Supplementary Material. When this approach is used instead of the CGO one, it is well-known that the GO dependence for the NMR shieldings disappears. Comparing the results obtained with the two approaches, CGO and GIAO (and excluding in both cases the two-electron integrals of ($SS$$\mid$$SS$) type), it is observed that the results obtained using the CGO approximation are well converged.

In \textsc{Dirac}, the linear response functions associated with each of the NMR shielding tensor elements in Eqs.~\eqref{eq:sigma_tensor}, \eqref{eq:RPP} and \eqref{eq:sigma-ee+pp} were calculated by contracting each vector element of the perturbator ${\bf P}_K$ with the solution vector ${\bf X}_B$ (which is linked to ${\bf Q}_B$) in each direction. In other words, the property gradients associated with the external uniform magnetic field were taken into account to solve the response equations given in Eqs.~\eqref{eq:RPP} and \eqref{eq:sigma-ee+pp}.\cite{Saue2003}

The summations in Eq.~\eqref{eq:DC+QED} run over all indices $i$, $a$ and $\tilde{a}$. To calculate the QED contributions to the NMR shielding tensors based on this expression, we limit the number of ($q$ and $r$) MOs involved in the summation associated with the second of the four terms given on the right-hand-side of Eq.~\eqref{eq:DC+QED}. We have established a threshold that allowed us to retain there only those terms that contribute at least (in absolute value) 0.01\,\% of the total $ee$ term. To do this we employed the \textsc{Dirac} \texttt{ANALYZE} keyword from the \texttt{LINEAR RESPONSE} module, and we set \texttt{ANATHR} equal to $0.01$. Furthermore, we characterized each occupied and virtual MO by performing a Mulliken population analysis.

It is worth noting that the NR limit of Eq.~\eqref{eq:sigma_tensor} is recovered in 4C calculations by scaling $c$ to infinity.\cite{Aucar1993} To obtain the NR values shown in this work, we have set the speed of light in vacuum to 30 times the value of $c$ reported in CODATA.

Quasi-relativistic 2C calculations were performed using a modified version
\cite{berger:2005,nahrwold:09,isaev:2012,gaul:2020,bruck:2023,ColomboJofre2022,zulch:2022} of a two-component program \cite{wullen:2010} based on Turbomole.\cite{ahlrichs:1989} The ZORA approach was used and calculations were performed within the complex generalized Hartree-Fock (cGHF) framework.
Picture-change transformed NMR shielding tensors were computed with our toolbox approach detailed in Ref.~\citenum{gaul:2020} using CGO in the nuclear CM for all molecules. Besides, for optimization of the response functions, we used the approach detailed in Refs.~\citenum{ColomboJofre2022} and \citenum{bruck:2023}. In order to improve ZORA results of NMR shieldings, we renormalized the ZORA wave function and orbital energies by considering the overlap of the approximate small components as described by Eq.~(34) in Ref.~\citenum{gaul:2020}.

In a second approach, we rescaled the polarization propagator by replacing the ZORA orbital energy differences with those from a so-called X2C (exact two-component transformation) calculation. The latter orbital energies were computed using a model-potential approach to account for two-electron spin-orbit couplings.\cite{wullen:2005} For the replacement of orbital energies, it was assumed that the energetic order of orbitals of ZORA and X2C is unaltered. After the orbital energy replacement we applied the renormalization procedure described above. This combined re-scaling and renormalization results in a reasonably good agreement between the absolute shielding tensors obtained with (2C) ZORA and (4C) DHF.

For the sake of comparison between (2C) ZORA and (4C) DHF results, it is worth to mention here that the (2C) ZORA calculations do not include 
the contributions from integrals of ($SS$$\mid$$SS$) type.

\begin{figure}[!htb]
\centering
\includegraphics[width=\linewidth]{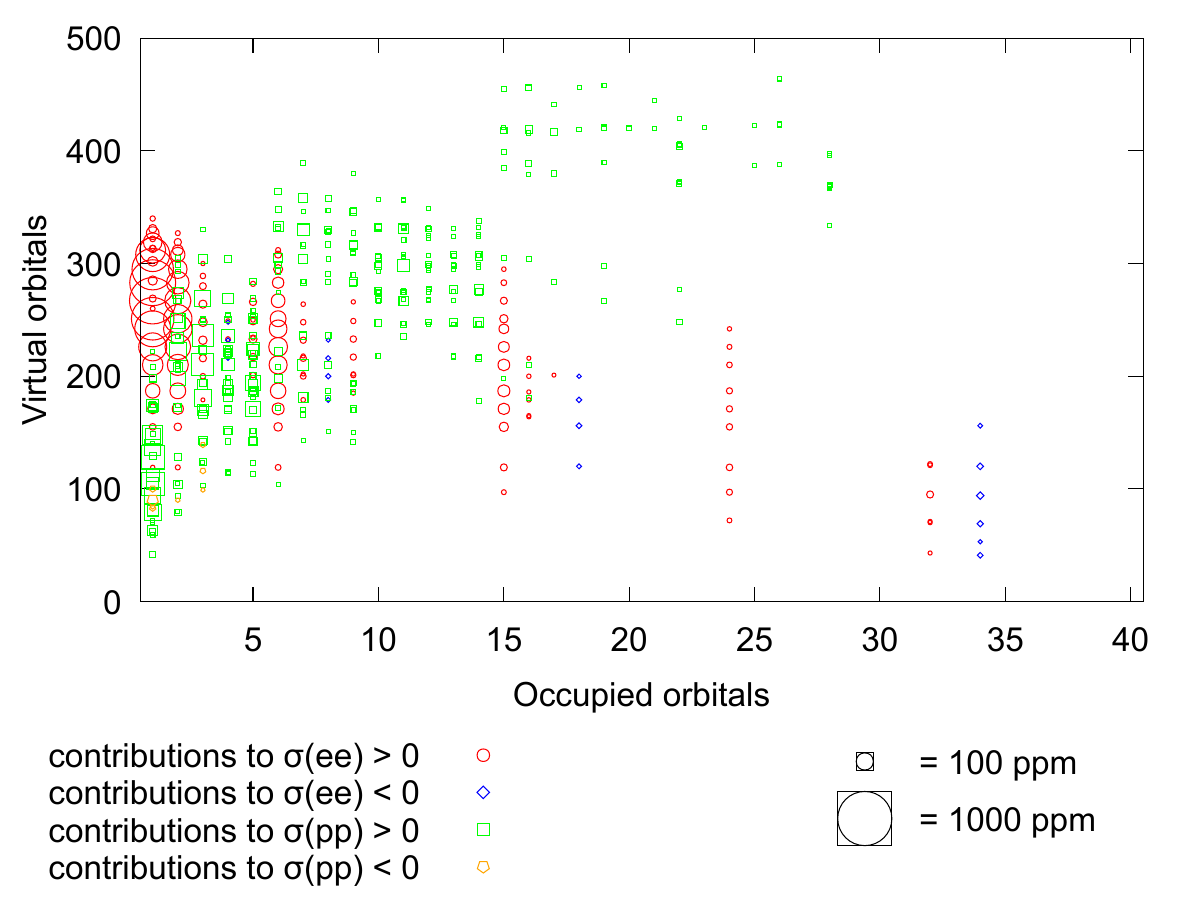}
\caption{Pattern of contributing excitations to $\sigma$(Tl) in Tl$^+$ at the DHF level of approach. The magnitude of each contribution is proportional to the marked area. The numbers that label the occupied and virtual Kramer's orbitals are such that: 1 refers to the lowest energy occupied pair and 40 corresponds to the highest energy occupied pair.}
\label{fig:ex_plot_Tl}
\end{figure}

\begin{figure*}[!htb]
\centering
\includegraphics[width=0.48\linewidth]{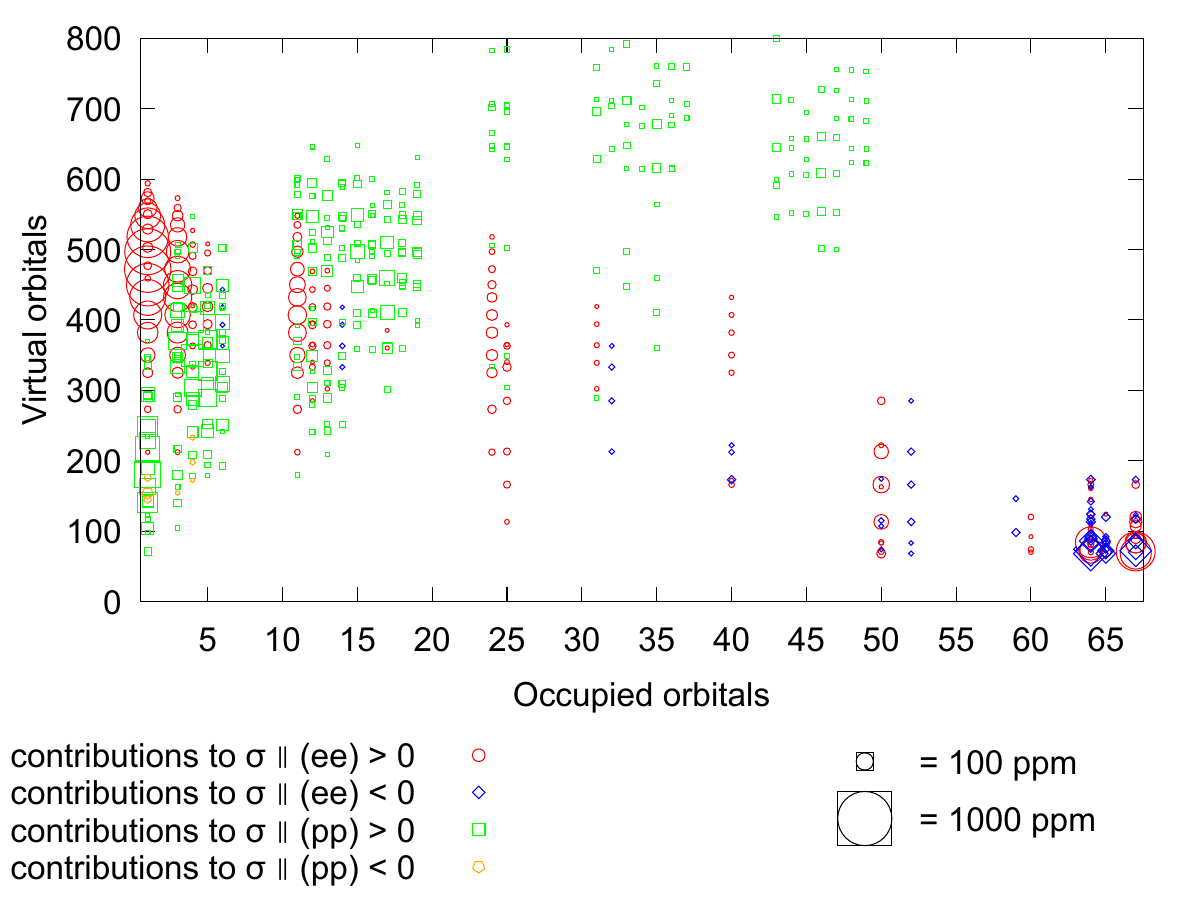}
\hfill
\includegraphics[width=0.48\linewidth]{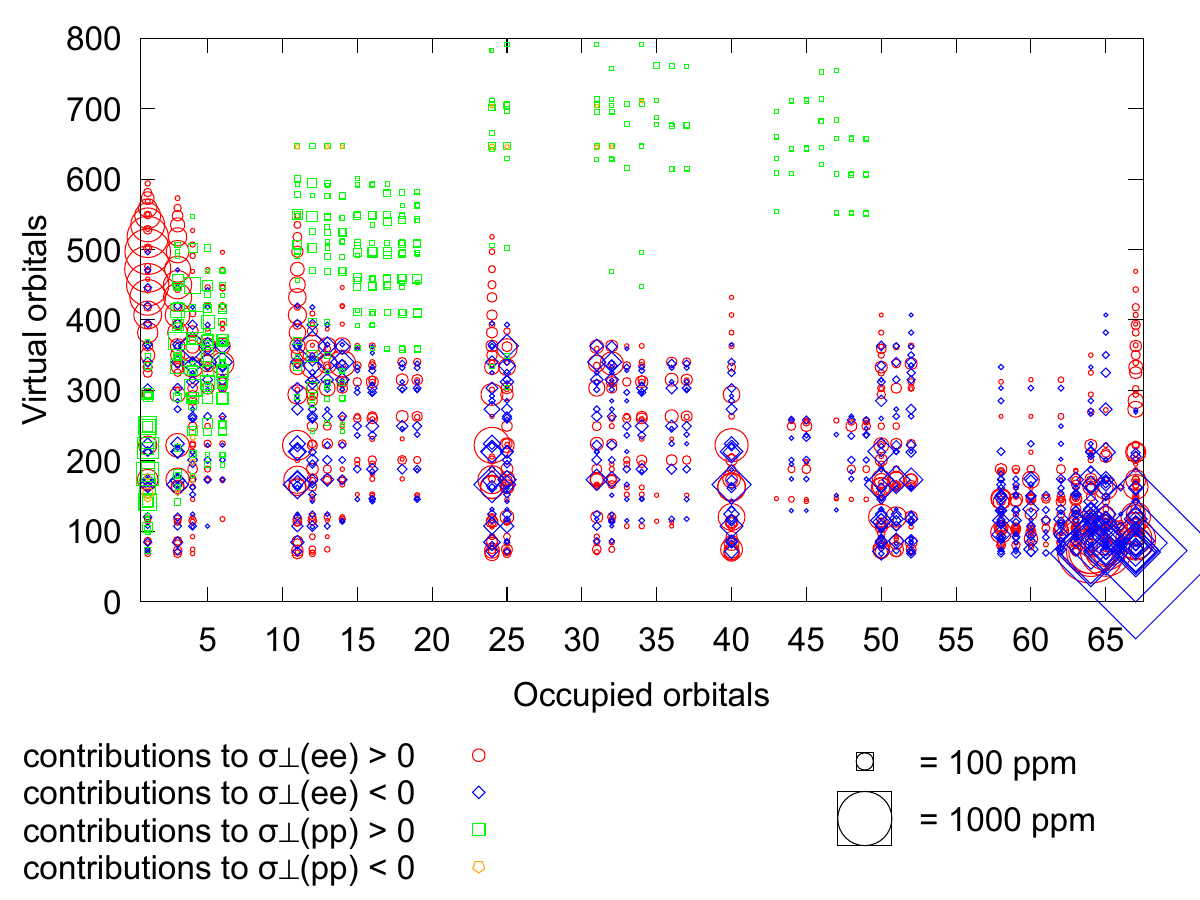}
\caption{Pattern of contributing excitations to $\sigma_\parallel$(Tl) (left) and $\sigma_\perp$(Tl) (right) for TlI molecule at the DHF level of approach. The magnitude of each contribution is proportional to the marked area. The numbers that label the occupied and virtual Kramer's orbitals are such that: 1 refers to the lowest occupied pair and 67 corresponds to the highest occupied pair.}
\label{fig:ex_plot_TlI}
\end{figure*}

\begin{figure*}[!htb]
\centering
\includegraphics[width=0.48\linewidth]{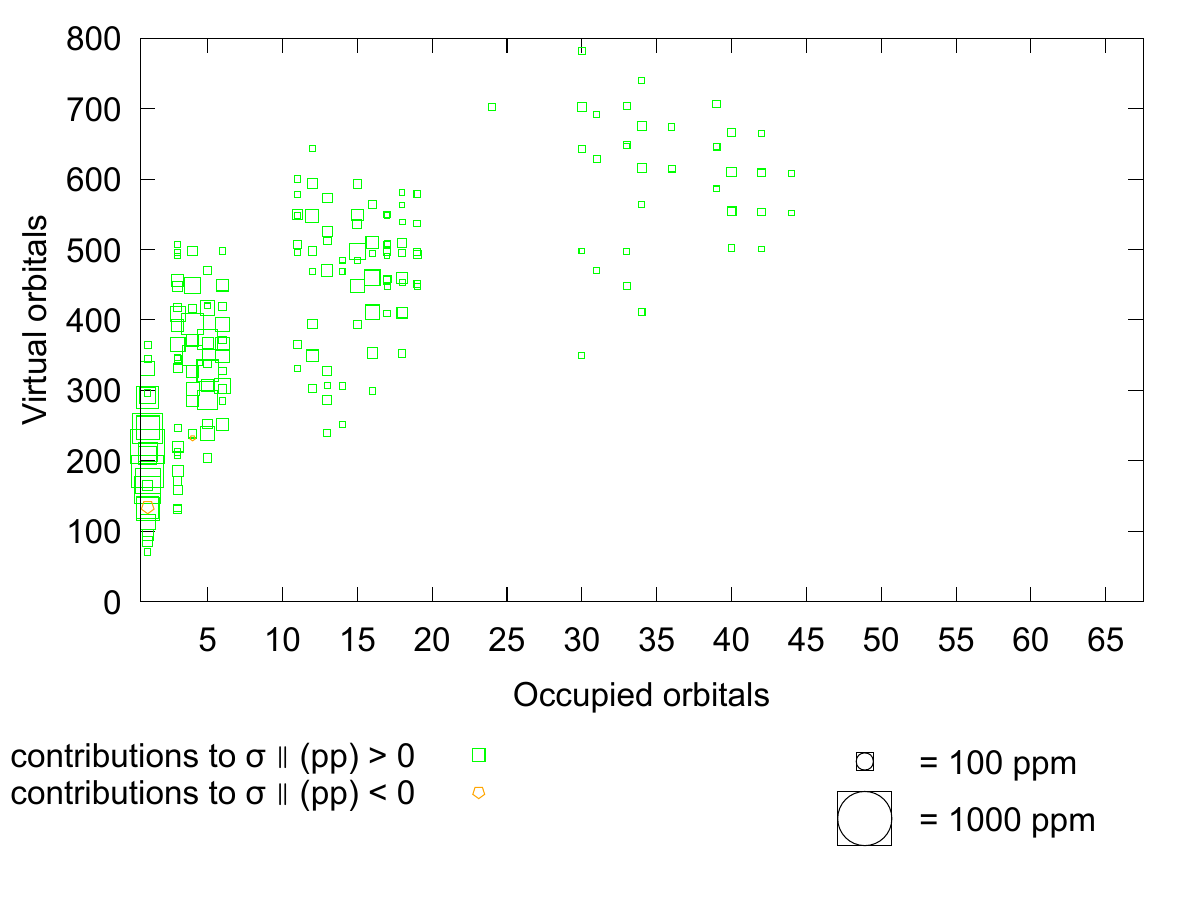}
\hfill
\includegraphics[width=0.48\linewidth]{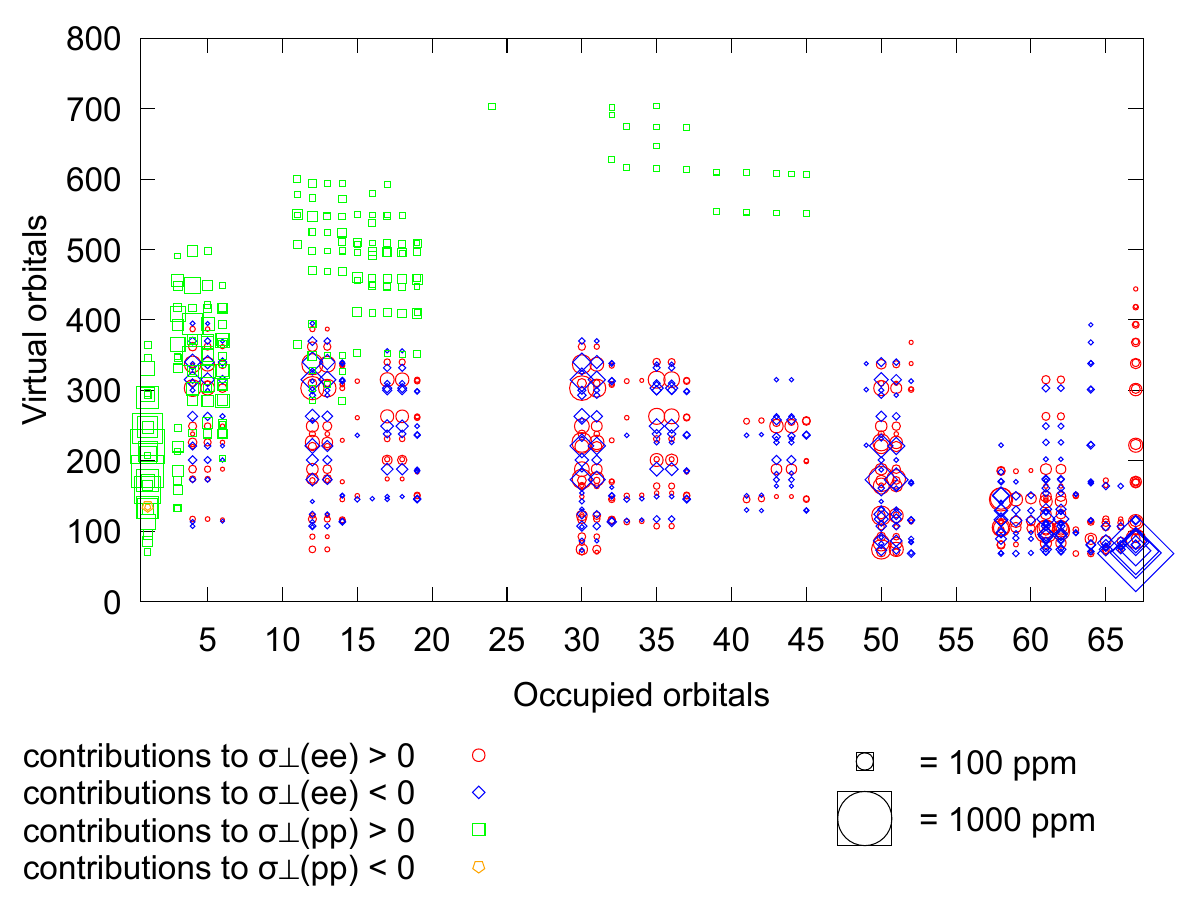}
\caption{Pattern of contributing excitations to $\sigma_\parallel$(Tl) (left) and $\sigma_\perp$(Tl) (right) for TlI molecule within the NR framework. The magnitude of each contribution is proportional to the marked area. The numbers that label the occupied and virtual Kramer's orbitals are such that: 1 refers to the lowest occupied pair and 67 corresponds to the highest occupied pair.}
\label{fig:ex_plot_TlI_nr}
\end{figure*}

In Figs.~\ref{fig:ex_plot_Tl} to \ref{fig:propagator_rZORA+eX2C-HF_TlI} we analyze the contributions that each individual virtual excitation makes to the 4C values of $\sigma_\perp^{ee}$, $\sigma_\parallel^{ee}$, $\sigma_\perp^{pp}$, and $\sigma_\parallel^{pp}$, as well as for the 2C calculations of $\sigma_\perp^{\mathrm{p}-\text{SD}}$, $\sigma_\perp^{\mathrm{p}-\text{SF}}$, $\sigma_\perp^{\mathrm{p}-\text{SFSD}}$, $\sigma_\parallel^{\mathrm{p}-\text{SD}}$, $\sigma_\parallel^{\mathrm{p}-\text{SF}}$, and $\sigma_\parallel^{\mathrm{p}-\text{SFSD}}$. In such figures we show only excitations contributing at least 0.01\,\% of the analyzed contribution to each shielding tensor element.

\begin{figure}[!htb]
\centering
\includegraphics[width=\linewidth]{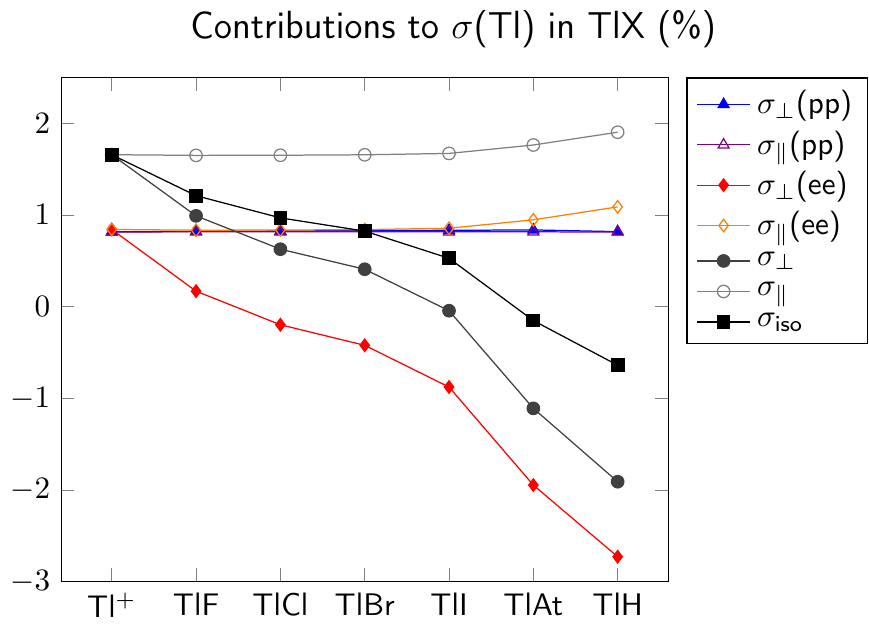}
\caption{Contributions to ${\bm \sigma}$(Tl) in Tl$^+$ ion and Tl$X$ ($X$ = H, F, Cl, Br, I, At) molecules, calculated at the DHF/RPA level of approach. }
\label{fig:sigma_Tl}
\end{figure}

\begin{figure}[!htb]
\centering
\includegraphics[width=\linewidth]{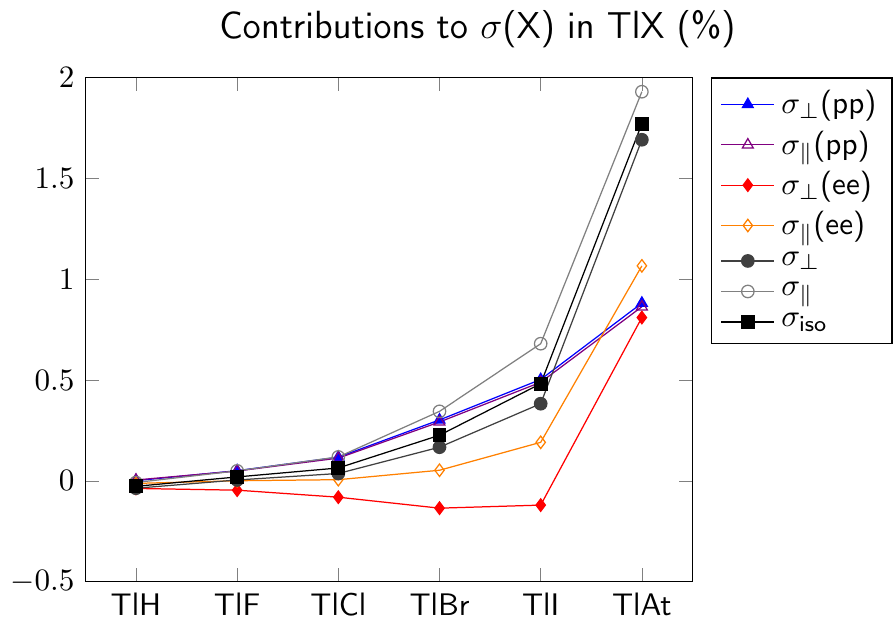}
\caption{Contributions to ${\bm \sigma}$($X$) in Tl$^+$ ion and Tl$X$ ($X$ = H, F, Cl, Br, I, At) molecules, calculated at the DHF/RPA level of approach. }
\label{fig:sigma_X}
\end{figure}


\begin{table*}[!htb]
\caption{Contributions to ${\bm \sigma}$(Tl) in Tl$^+$ and Tl$X$ ($X$ = H, F, Cl, Br, I, At), calculated at DHF/RPA level of approach and including 
two-electron integrals of the ($SS$$\mid$$SS$) type. Results in parentheses correspond to the NR limit. Values in ppm.}
\label{tab:sigma_Tl}
\begin{tabular*}{\linewidth}{@{}l @{\extracolsep{\fill}} *{7}{S[parse-numbers=false]}@{}}
\toprule
 &  \multicolumn{2}{c}{$pp$} & \multicolumn{2}{c}{$ee$} & \multicolumn{3}{c}{Total} \\
\cmidrule{2-3}\cmidrule{4-5}\cmidrule{6-8}
System & {$\sigma_\perp$} & {$\sigma_\parallel$} &  {$\sigma_\perp$} &  {$\sigma_\parallel$} &  {$\sigma_\perp$} &  {$\sigma_\parallel$} &  {$\sigma_\text{iso}$} \\
\midrule
Tl$^+$ & \multicolumn{2}{S[parse-numbers=false]}{8158.45} & \multicolumn{2}{S[parse-numbers=false]}{8431.69} & 16590.14 & 16590.14 & 16590.14 \\
& \multicolumn{2}{S[parse-numbers=false]}{(9884.56)} & \multicolumn{2}{S[parse-numbers=false]}{(8.64)} &&& (9893.20) \\[1.5ex]
TlH & 8172.91 & 8161.55 & -27966.46 & 10974.16 & -19793.55 & 19135.71 & -6817.13 \\
& (9893.50) & (9882.78) & (-6932.15) & (0.34) & (2961.36) & (9883.12) & (5268.61) \\ [1.5ex]
TlF & 8219.58 & 8162.91 & 1601.53 & 8348.26 & 9821.11 & 16511.17 & 12051.13 \\
& (9940.42) & (9884.45) & (-1964.67) & (9.13) & (7975.75) & (9893.57) & (8615.03) \\[1.5ex]
TlCl & 8245.43 & 8164.98 & -2118.37 & 8362.98 & 6127.06 & 16527.96 & 9594.03 \\
& (9965.69) & (9886.56) & (-3040.67) & (9.10) & (6925.02) & (9895.66) & (7915.23) \\[1.5ex]
TlBr & 8298.72 & 8166.25 & -4399.59 & 8412.06 & 3899.13 & 16578.31 & 8125.52 \\
& (10018.61) & (9887.85) & (-3500.92) & (9.34) & (6517.69) & (9897.19) & (7644.19) \\[1.5ex]
TlI & 8325.77 & 8167.84 & -9033.20 & 8565.86 & -707.43 & 16733.69 & 5106.28 \\
& (10045.65) & (9889.50) & (-4246.27) & (9.43) & (5799.38) & (9898.93) & (7165.90) \\[1.5ex]
TlAt & 8363.14 & 8168.94 & -20008.30 & 9509.61 & -11645.15 & 17678.55 & -1870.59 \\
& (10084.51) & (9890.79) & (-4702.16) & (10.26) & (5382.36) & (9901.05) & (6888.59) \\ 
%
\bottomrule
\end{tabular*}
\end{table*}

\begin{table*}[!htb]
\caption{Contributions to ${\bm \sigma}$($X$) in Tl$X$ ($X$ = H, F, Cl, Br, I, At), calculated at DHF/RPA level of approach and including 
two-electron integrals of the ($SS$$\mid$$SS$) type. Results in parentheses correspond to the NR limit. Values in ppm.}
\label{tab:sigma_X}
\begin{tabular*}{\linewidth}{@{}l @{\extracolsep{\fill}} *{7}{S[parse-numbers=false]}@{}}
\toprule
 &  \multicolumn{2}{c}{$pp$} & \multicolumn{2}{c}{$ee$} & \multicolumn{3}{c}{Total} \\
\cmidrule{2-3}\cmidrule{4-5}\cmidrule{6-8}
Molecule & {$\sigma_\perp$} & {$\sigma_\parallel$} &  {$\sigma_\perp$} &  {$\sigma_\parallel$} &  {$\sigma_\perp$} &  {$\sigma_\parallel$} &  {$\sigma_\text{iso}$} \\
\midrule
TlH & 10.91 & 45.08 & -382.19 & -129.67 & -371.28 & -84.59 & -275.72 \\
& (11.70) & (46.14) & (-0.12) & (-0.01) & (11.58) & (46.13) & (23.10) \\[1.5ex]
TlF & 505.80 & 489.78 & -463.06 & 7.29 & 42.74 & 497.07 & 194.18 \\
& (500.13) & (484.22) & (-327.48) & (0.04) & (172.65) & (484.26) & (276.52) \\[1.5ex]
TlCl & 1175.88 & 1128.62 & -813.12 & 59.21 & 362.76 & 1187.84 & 637.78 \\
& (1196.21) & (1148.83) & (-652.02) & (0.13) & (544.20) & (1148.96) & (745.79) \\[1.5ex]
TlBr & 3024.02 & 2920.37 & -1360.88 & 528.40 & 1663.14 & 3448.77 & 2258.35 \\
& (3225.47) & (3121.52) & (-1437.09) & (0.80) & (1788.38) & (3122.31) & (2233.02) \\[1.5ex]
TlI & 5032.08 & 4891.38 & -1216.56 & 1914.76 & 3815.52 & 6806.13 & 4812.39 \\
& (5648.10) & (5507.13) & (-2329.44) & (2.63) & (3318.66) & (5509.76) & (4049.03) \\[1.5ex]
TlAt & 8825.07 & 8637.83 & 8132.71 & 10678.13 & 16957.78 & 19315.96 & 17743.84 \\
& (10742.62) & (10556.81) & (-3927.91) & (10.98) & (6814.71) & (10567.79) & (8065.73) \\ 
%
\bottomrule
\end{tabular*}
\end{table*}

\begin{table*}[!htb]
\begin{threeparttable}
\caption{Deviation (dev.) of 2C ZORA (ZORA), ZORA with renormalized orbitals (rZORA) and ZORA with renormalized orbitals and re-scaled with X2C orbital energies (rZORA+X2Ce) results for the tensor elements of ${\bm \sigma}$(Tl) in Tl$^+$ ion and Tl$X$ ($X$ = H, F, Cl, Br, I, At) molecules from 4C DC including integrals of $(SS|SS)$ type (devDC) calculations. All results were obtained at the Hartree-Fock level. Contributions of the diamagnetic-like expectation values (Expec) and the paramagnetic-like linear responses (LR) are shown separately. Values in ppm.}
\label{tab:sigma_Tl_ZORA}
\begin{tabular*}{\linewidth}{@{}l @{\extracolsep{\fill}} r @{\hspace{1.0em}} *{4}{S[parse-numbers=false, table-format=7.1]} r *{4}{S[parse-numbers=false, table-format=7.1]} @{}}
\toprule
Molecule & Method
 &   \multicolumn{4}{c}{$\sigma_\perp$}
 &&  \multicolumn{4}{c}{$\sigma_\parallel$}  \\
 \cmidrule{3-6}\cmidrule{7-11}
 &&  {Expec} & {LR} & {Total} & {devDC/\%} & ~ &  {Expec} & {LR} & {Total} & {devDC/\%}  \\
\midrule
          \multirow{3}{*}{Tl+} &       ZORA &     10134.7 &     4878.1 &    15012.8 &    9.5 &&    10134.7 &     4878.1 &    15012.8 &    9.5\\
                               &      rZORA &      9786.8 &     5557.1 &    15343.9 &    7.5 &&     9786.8 &     5557.1 &    15343.9 &    7.5\\
                               & rZORA+X2Ce &      9786.8 &     7630.8 &    17417.6 &   -5.0 &&     9786.8 &     7630.8 &    17417.6 &   -5.0\\
 \\
          \multirow{3}{*}{TlH} &       ZORA &     10147.3 &   -31869.3 &   -21722.0 &   -9.7 &&    10137.7 &     7487.1 &    17624.9 &    7.9\\
                               &      rZORA &      9799.3 &   -31264.7 &   -21465.4 &   -8.4 &&     9789.8 &     8163.7 &    17953.5 &    6.2\\
                               & rZORA+X2Ce &      9799.3 &   -29222.6 &   -19423.2 &    1.9 &&     9789.8 &    10229.2 &    20019.0 &   -4.6\\                               
\\
          \multirow{3}{*}{TlF} &       ZORA &     10195.0 &    -1998.2 &     8196.8 &   16.5 &&    10139.2 &     4792.9 &    14932.1 &    9.6\\
                               &      rZORA &      9847.0 &    -1331.9 &     8515.1 &   13.3 &&     9791.2 &     5470.9 &    15262.1 &    7.6\\
                               & rZORA+X2Ce &      9847.0 &      741.7 &    10588.7 &   -7.8 &&     9791.2 &     7544.6 &    17335.8 &   -5.0\\
\\
         \multirow{3}{*}{TlCl} &       ZORA &     10220.1 &    -5741.9 &     4478.1 &   26.9 &&    10141.3 &     4808.1 &    14949.4 &    9.6\\
                               &      rZORA &      9872.0 &    -5082.0 &     4789.9 &   21.8 &&     9793.3 &     5485.6 &    15278.9 &    7.6\\
                               & rZORA+X2Ce &      9872.0 &    -3003.1 &     6868.9 &  -12.1 &&     9793.3 &     7572.8 &    17366.1 &   -5.1\\
\\
         \multirow{3}{*}{TlBr} &       ZORA &     10272.2 &    -8042.9 &     2229.3 &   42.8 &&    10142.5 &     4858.5 &    15001.0 &    9.5\\
                               &      rZORA &      9924.0 &    -7386.7 &     2537.3 &   34.9 &&     9794.5 &     5535.8 &    15330.4 &    7.5\\
                               & rZORA+X2Ce &      9924.0 &    -5284.3 &     4639.7 &  -19.0 &&     9794.5 &     7650.3 &    17444.9 &   -5.2\\
\\
          \multirow{3}{*}{TlI} &       ZORA &     10299.2 &   -12720.0 &    -2420.9 & -242.2\tnote{*} &&    10144.2 &     5016.2 &    15160.3 &    9.4\\
                               &      rZORA &      9950.6 &   -12073.0 &    -2122.4 & -200.0\tnote{*} &&     9796.1 &     5693.2 &    15489.3 &    7.4\\
                               & rZORA+X2Ce &      9950.6 &   -10024.6 &      -74.0 &   89.5\tnote{*} &&     9796.1 &     7765.5 &    17561.6 &   -4.9\\
\\
         \multirow{3}{*}{TlAt} &       ZORA &     10334.6 &   -24263.6 &   -13929.0 &  -19.6 &&    10145.0 &     5934.7 &    16079.7 &    9.0\\
                               &      rZORA &      9985.3 &   -23644.8 &   -13659.5 &  -17.3 &&     9797.0 &     6612.9 &    16409.9 &    7.2\\
                               & rZORA+X2Ce &      9985.3 &   -21657.4 &   -11672.1 &   -0.2 &&     9797.0 &     8687.6 &    18484.6 &   -4.6\\
\bottomrule
\end{tabular*}
\begin{tablenotes}
\item[*] Huge relative deviations stem from partial cancellations of Expec and LR contributions that lead to much smaller total values. Relative deviations are therefore not meaningful for the individual contributions, but only for the total isotropic value. Nevertheless, still the considerable improvement of X2C orbital energy corrected results can be seen.
\end{tablenotes}
\end{threeparttable}
\end{table*}

\section{Results}
\label{sec:results}

\subsection{Relativistic and non-relativistic shielding tensor elements}
%
Trends for the total values of ${\bm \sigma}$(Tl) and ${\bm \sigma}$($X$) in the molecules of Tl$X$ ($X$ = H, F, Cl, Br, I, At) and also for ${\bm \sigma}$(Tl) in the Tl$^+$ ion are given in Tables \ref{tab:sigma_Tl} and \ref{tab:sigma_X}, together with the contributions arising from the $ee$ and $pp$ parts of the linear response functions involved as well as the contributions from the perpendicular and parallel components of each ${\bm \sigma}$.

The behavior of each contributing term ($ee$ and $pp$) together with their components are shown in Figs.~\ref{fig:sigma_Tl} and \ref{fig:sigma_X}. It is easily seen that the behavior of most of the contributing terms to ${\bm \sigma}$(Tl) are opposite to the similar terms of ${\bm \sigma}$($X$).
Starting from the singly ionized Tl atom and then moving to TlAt, it can be seen that as Tl is bonded to heavier halogen atoms $\sigma_\perp$(Tl) and $\sigma_\text{iso}$(Tl) become more and more paramagnetic on the one hand. In other words, $\sigma_\text{iso}$(Tl) is being deshielded as $X$ becomes heavier, with the perpendicular contributions to the $ee$ terms being responsible for such a behavior. On the other hand, the behavior of $\sigma_\text{iso}$($X$) is opposite, being in this case both, the perpendicular and parallel contributions responsible for such a behavior.

As can be seen in Table~\ref{tab:sigma_Tl} and in Fig.~\ref{fig:sigma_Tl}, the contribution of the $pp$ part of $\sigma$(Tl) is only weakly dependent on the neighbour atom in the molecule. A similar behavior is observed in the case of $\sigma_\parallel^{ee}$(Tl). In contrast, $\sigma_\perp^{ee}$(Tl) strongly depends on the bonding partner. This pronounced dependence determines the value of $\sigma_\text{iso}$(Tl), which varies considerably among molecules. 
It is also worth to mention that, though $\sigma_\perp^{pp}$, $\sigma_\parallel^{pp}$, and $\sigma_\parallel^{ee}$ are positive, $\sigma_\perp^{ee}$ change from a large positive value (\SI{8431.69}{ppm}) to a large negative value (\SI{-20008.30}{ppm}) in the Tl$^+$, TlF, TlCl, TlBr, TlI, TlAt series. As a result, $\sigma_\text{iso}$ varies from \SI{16590.14}{ppm} for Tl$^+$ to \SI{-1870.59}{ppm} for TlAt, changing the character of the shielding of the thallium nucleus from diamagnetic to paramagnetic. 

In the case of ${\bm \sigma}$($X$) in Tl$X$ molecules ($X$ = F, Cl, Br, I, At), the $\sigma_\perp^{pp}$, $\sigma_\parallel^{pp}$, and $\sigma_\parallel^{ee}$ contributions are positive and they increase monotonously with increasing $Z$: with a dependence of about $Z^{1.2}$ for $\sigma_\text{iso}^{pp}$ and about $Z^{3.4}$ for $\sigma_\parallel^{ee}$. The contributions to $\sigma_\perp^{ee}($X$)$ are negative when $X$ = F, Cl, Br, and I, but they are positive for TlAt.

The TlH molecule is special. In this case, the presence of the very light hydrogen atom produces an effect on $\sigma_\text{iso}$(Tl) that is greater than the effects of the heavy At atom. In fact, in TlH the relativistic effect on the Tl nucleus is known as ``Heavy-Atom effect on the Heavy-Atom itself'' (HAHA effect), which has been defined long time ago.\cite{Edlund1987} The effect of large decreasing $\sigma_\text{iso}$(Tl) in TlH in comparison to the sole Tl$^+$ ion is in agreement with similar results obtained in compounds of hydrogen and 14- and 15-group elements.\cite{Lantto2006}

In order to estimate relativistic effects on ${\bm \sigma}$, calculations with $c=30\,c_0$ have been performed, mimicking NR calculations. 
As it can be seen in Tables~\ref{tab:sigma_Tl} and~\ref{tab:sigma_X}, the NR values of the perpendicular and parallel tensor elements of ${\bm \sigma}^{pp}$(Tl) and ${\bm \sigma}^{pp}$($X$) are greater than the relativistic ones. The relativistic effects on $\sigma^{pp}$(Tl$^+$) represent \SI{-21}{\percent} of its value, while the relativistic effects on $\sigma_\perp^{pp}$($X$) and $\sigma_\parallel^{pp}$($X$) for the Tl$X$ molecules ($X$ = H, F, Cl, Br, I, At) vary from \SI{+1}{\percent} in the case of $\sigma_\perp^{pp}$(F) and $\sigma_\parallel^{pp}$(F), to \SI{-22}{\percent} for $\sigma_\perp^{pp}$(At) and $\sigma_\parallel^{pp}$(At). It is worth highlighting that in the NR limit ($c \to \infty$), $\sigma_\parallel^{ee}$(Tl) and $\sigma_\parallel^{ee}$($X$) are known to be exactly zero. However, in our calculations that aim to approach the NR limit they are only close to zero because the value of $c$ was not chosen high enough. The NR values of the $\sigma_\perp^{ee}$ contributions vary in a complex way. For $\sigma_\perp^{ee}$(Tl) in TlF and TlCl the relativistic effect is positive, but for TlH, TlBr, TlI, and TlAt it is negative. For $\sigma_\perp^{ee}$($X$) in TlH, TlF, and TlCl the relativistic effect is negative, whereas for TlBr, TlI, and TlAt it becomes positive.

The 4C values of $\sigma_\parallel^{ee}$(Tl) in all the molecular systems studied here are relatively close to the computed value of $\sigma^{ee}$(Tl) in the Tl$^+$ ion, with the largest deviation of about 30\,\% being found for TlH at the DHF/RPA level of approach (see Table \ref{tab:sigma_Tl}). Considering that $\sigma_\parallel^{ee}$(Tl) must be exactly zero for all our diatomic systems in the NR limit (as it is observed in our calculations), and also that the NMR shielding is only diamagnetic in nature for closed-shell atoms within the NR regime, it can be concluded i) that $\sigma_\parallel^{ee}$(Tl) is purely relativistic, but ii) that it is closely related to the value of $\sigma^{ee}$(Tl) in the Tl$^+$ ion as well as to other contributions, some of which depend on the electronic environment.\cite{Aucar2016}
Furthermore, the 4C-DHF/RPA calculations of $\sigma_\perp^{pp}$(Tl) and $\sigma_\parallel^{pp}$(Tl) in all the molecules studied here are close to the value of $\sigma^{pp}$(Tl) in the Tl$^+$ ion (the deviations are less than 2.5\,\%).
From all these observations we can state (and this is confirmed by observing Fig.~\ref{fig:sigma_Tl}) that the $Z_X$-dependence of $\sigma_\text{iso}$(Tl) in Tl$X$ can be mainly attributed to the dependence of $\sigma_\perp^{ee}$(Tl) (whose relativistic effects are also very relevant) with $Z_X$, but also to a slight $Z_X$-dependence of $\sigma_\parallel^{ee}$(Tl). This latter effect was previously suggested to be related to one of the linear response functions involved in the nuclear spin-rotation constant (for details, please see Refs.~\citenum{Aucar2016} and \citenum{aucar2019}).

In Tables I to VI of the Supplementary Material we show a set of 4C values for the $ee$ and $pp$ contributions to the ${\bm \sigma}$(Tl) and ${\bm \sigma}$($X$) tensors, both in the perpendicular and parallel projections on the internuclear axis ($xx$ and $zz$ tensor elements, respectively, for molecules oriented along the $z$ axis). We show results obtained on the DHF and DKS levels of theory, and the effects arising from the inclusion of integrals of ($SS$$\mid$$SS$) type 
are also displayed in these tables. In order to get an idea of the order of magnitude of these latter contributions, we highlight here two cases specifically:
From Table V of the Supplementary Material, it is seen that at DHF level of theory, the contributions arising from the integrals of ($SS$$\mid$$SS$) type are about 0.2\,\% of the total value of $\sigma_\text{iso}^{ee}$(Tl) in TlF. On the other hand, these interactions represent about $-0.9\,\%$ of $\sigma_\text{iso}^{ee}$(Tl) in TlI at the same level of theory.

\begin{figure}[h!]
\includegraphics[width=.5\textwidth]{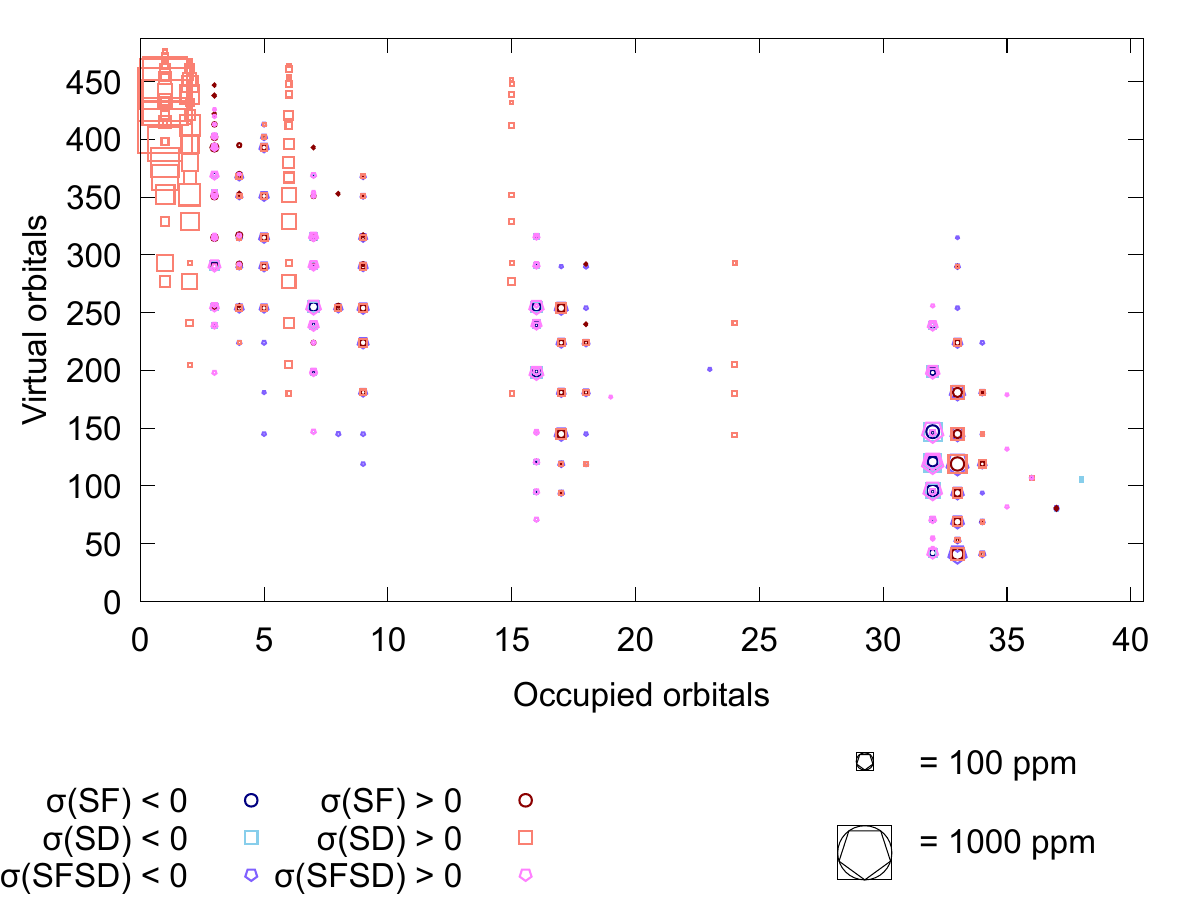}
\caption{Pattern of excitations for $\sigma_\parallel$(Tl) (left) and $\sigma_\perp$(Tl) (right) for Tl$^+$ shown separately for the SD and SF parts of the NMR shielding propagator at the level of rZORA+eX2C-HF. The magnitude of the amplitude of each type is proportional to marked area. The numbers that label the occupied and virtual orbitals correspond to the Kramers pairs, such that 1 refers to the lowest energy occupied pair and 40 corresponds to the highest energy occupied pair.}
\label{fig:propagator_rZORA+eX2C-HF_Tl+}
\end{figure}
    
      \begin{figure*}[htb!]
      \begin{minipage}{.48\textwidth}
      \includegraphics[width=\textwidth]{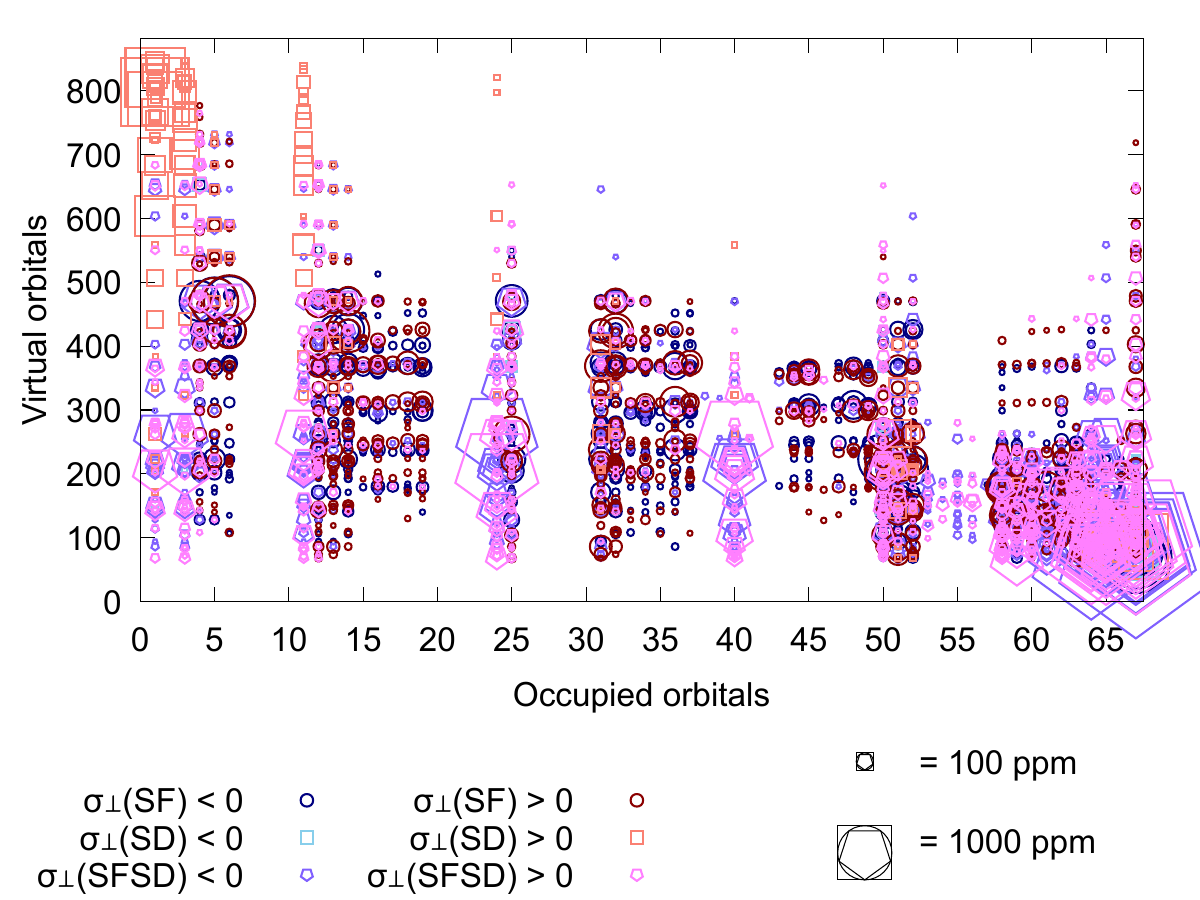}
      \end{minipage}
      \hfill
      \begin{minipage}{.48\textwidth}
      \includegraphics[width=\textwidth]{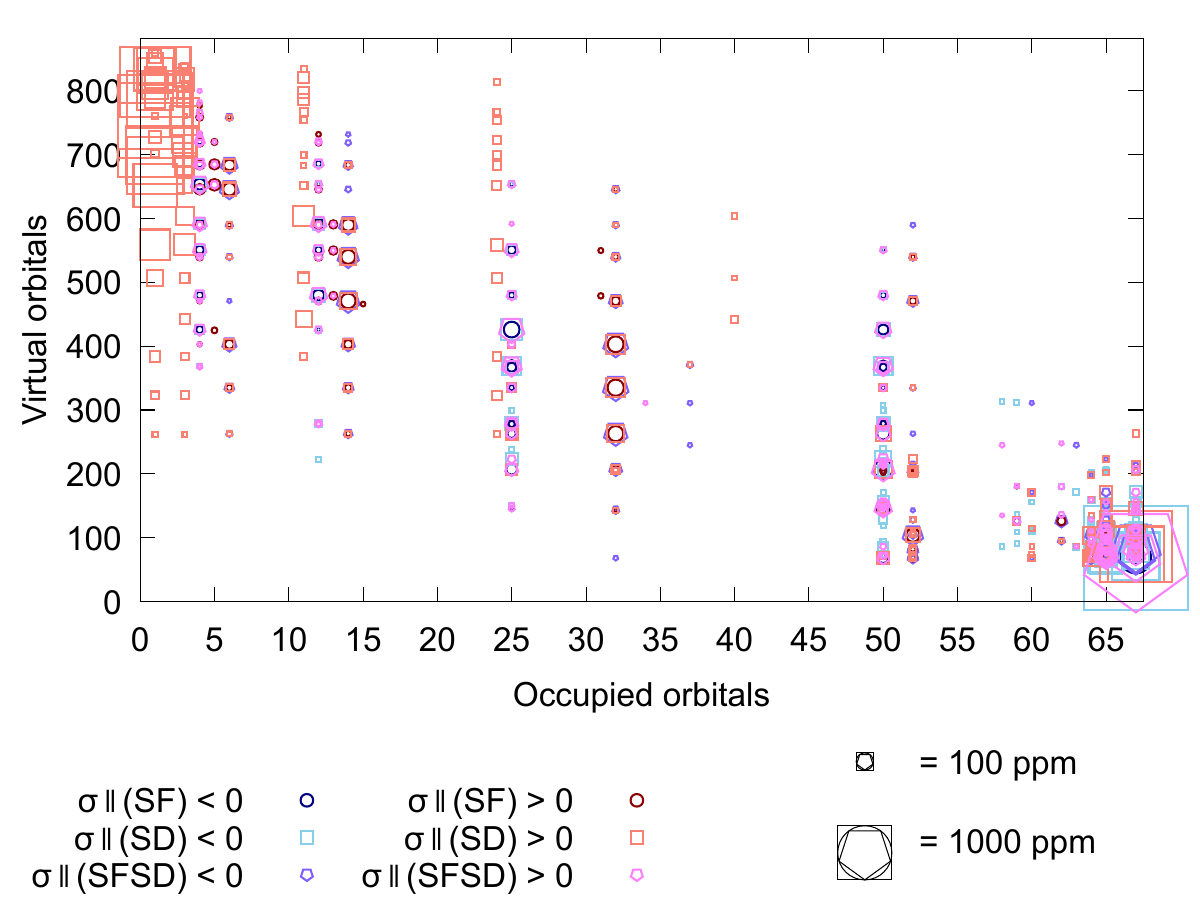}
      \end{minipage}%
      \caption{Pattern of excitations for $\sigma_\parallel$(Tl) (left)
                and $\sigma_\perp$(Tl) (right) for TlI shown separately for the SD and SF parts of the
                NMR shielding propagator at the level of rZORA+eX2C-HF. The magnitude of the
                amplitude of each type is proportional to marked area. The numbers that label the occupied and virtual orbitals correspond to the Kramers pairs, such that 1 refers to the lowest energy occupied pair and 67 corresponds to the highest energy occupied pair.}
      \label{fig:propagator_rZORA+eX2C-HF_TlI}
      \end{figure*}

\subsection{Pattern of virtual $ee$ and $pp$ excitations}
%
As mentioned in Sec.~\ref{sec:shield-rpp} one can shed some light on the complex behavior of $\sigma_\perp^{ee}$ through the analysis of the pattern of virtual excitations arising from the occupied orbitals and the unoccupied ones within the RelPPT theory. 
The pattern of excitations for $\sigma$(Tl$^+$) is presented in Fig.~\ref{fig:ex_plot_Tl}. The major part of the $ee$ term originates from virtual excitations that arise from s-type Kramer's pairs: 1s, 2s, 3s, 4s occupied orbitals (labeled 1, 2, 6, and 15 on Fig.~\ref{fig:ex_plot_Tl}, respectively).

Subsequently, we monitor the trends when moving from thallium atomic system to the thallium halide molecules. The pattern of excitations for parallel and perpendicular contributions to ${\bm \sigma}$(Tl) in TlI are presented in Fig.~\ref{fig:ex_plot_TlI}. One can see that the pattern of excitations for $\sigma_\parallel^{ee}$(Tl) in TlI is similar to the ones for Tl$^+$, with some small contribution of excitations arising from the highest occupied molecular orbital (HOMO), the highest occupied atomic orbital, or orbitals lying energetically near to HOMO.

The contributions of virtual $ee$ excitations arising from HOMO have opposite sign compared to that of the contributing excitations arising from s-type orbitals. For $\sigma_\perp^{ee}$, the negative contributions of excitations arising from HOMO or near-to-HOMO orbitals are much larger in absolute values, being comparable in magnitude to the contributions of the excitations arising from s-type orbitals. On the other hand, the pattern of $pp$ contributing excitations is similar for both perpendicular and parallel tensor elements of ${\bm \sigma}$(Tl) in TlI, but also in all other thallium halides.

One should expect that the pattern of excitations would depend on the framework of calculations, being it relativistic or non--relativistic. As shown in Fig.~\ref{fig:ex_plot_TlI_nr}, the positive $ee$ amplitudes linked to excitations that arise from s-type orbitals vanish within the NR framework, but the negative contributions to $ee$ that are linked to excitations that arise from HOMO or near-to-HOMO orbitals, survive for the perpendicular component of ${\bm \sigma}$(Tl). This is not the case for its parallel component.

As shown in other figures stored in the Supplementary Material, when considering molecules with higher-$Z$ ligands (meaning $X$ = Cl, Br, I, and At), the negative contributions to $\sigma_\perp^{ee}$ turn larger and larger in their absolute values, becoming greater than the positive contributions linked to excitations that arises from s-type orbitals for $X$=I. A similar situation, though still higher in the relation among positive and negative contributions, appears in the case of TlH molecule. So, this pronounced change in the perpendicular component is the main reason why $\sigma_\text{iso}$(Tl) in the Tl$X$ ($X$ = F, Cl, Br, I, At, H) series changes a lot, starting from a large positive value and then becoming negative.

\subsection{ZORA analysis of mechanisms}
%
As expected from earlier studies,\cite{wodynski:2012,jankowska:2016} ZORA underestimates NMR absolute shieldings by more than 10\,\% for perpendicular components of the thallium monohalides. A renormalization of orbitals and orbital energies within ZORA can not reduce this huge deviation considerably. It has to be noted that the deviation is particularly large for the perpendicular components of the shielding tensor which depends strongly on the response contributions to the shielding. Moreover a general increase of this deviation can be observed for increasing nuclear charge number of the halide $X$ in Tl$X$, with the exception of TlAt.

An analysis of the ZORA orbital energies shows that a renormalization can correct all occupied and low-lying virtual orbital energies to agree with X2C orbitals within a few percent. Energies of high-lying virtual orbitals, however, which have a pronounced influence on the NMR shielding tensor due to core-electron excitations, are dramatically overestimated by up to three orders of magnitude within the ZORA approach, which leads to a wrong weighting of polarization propagator. We therefore replaced all ZORA orbital energies with X2C orbital energies and renormalized the ZORA orbitals subsequently. This rather simple re-scaling leads to a much better agreement of ZORA with 4C results with relative deviations being about 10\,\% or smaller as shown in Table \ref{tab:sigma_Tl_ZORA},  except for $\sigma_\perp$(Tl) in TlBr and TlI, as in these two cases partial cancellations of paramagnetic-like and diamagnetic-like parts lead to very small absolute values.

We use this X2C-rescaled ZORA approach to analyze the contributions of the (electronic-)spin-dependent (SD) and (electronic-)spin-free (SF) parts of the NMR operators by splitting up the polarization propagator, i.e.~the paramagnetic-like contribution to the shielding tensor ${\bm \sigma}^\mathrm{p}$, into the following parts:
\begin{widetext}
\begin{align} 
\bm{\sigma}^{\mathrm{p}-\mathrm{SF}} (K) &=\frac{\mu_0\,e^2}{8\pi} \Braket{\Braket{\left\{\hat{\mathbf{p}} \, \overset{\times}{,} \, c^2\omega\frac{\mathbf{r}_K}{r_K^3}\right\};\left\{\hat{\mathbf{p}} \, \overset{\times}{,} \, c^2\omega\mathbf{r}_G\right\}}}\\
\bm{\sigma}^{\mathrm{p}-\mathrm{SD}} (K) &= \frac{\mu_0\,e^2}{8\pi} \left\langle\left\langle\left[-\hat{\mathbf{p}}\otimes\mathrm{i}\mathbf{\bm{\sigma}}+\mathrm{i}\mathbf{\bm{\sigma}}\otimes\hat{\mathbf{p}} \, \overset{\cdot}{,} \, c^2\omega\frac{\mathbf{r}_K}{r_K^3}\right];
\left[-\hat{\mathbf{p}}\otimes\mathrm{i}\mathbf{\bm{\sigma}}+\mathrm{i}\mathbf{\bm{\sigma}}\otimes\hat{\mathbf{p}} \, \overset{\cdot}{,} \, c^2\omega\mathbf{r}_G\right]\right\rangle\right\rangle\\
\bm{\sigma}^{\mathrm{p}-\mathrm{SFSD}} (K) &= \frac{\mu_0\,e^2}{8\pi}
\Braket{\Braket{\left\{\hat{\mathbf{p}} \, \overset{\times}{,} \, c^2\omega\frac{\mathbf{r}_K}{r_K^3}\right\};\left[-\hat{\mathbf{p}}\otimes\mathrm{i}\mathbf{\bm{\sigma}}+\mathrm{i}\mathbf{\bm{\sigma}}\otimes\hat{\mathbf{p}} \, \overset{\cdot}{,} \, c^2\omega\mathbf{r}_G\right]}}
\nonumber\\
&\;\;\;
+ \frac{\mu_0\,e^2}{8\pi} \Braket{\Braket{\left[-\hat{\mathbf{p}}\otimes\mathrm{i}\mathbf{\bm{\sigma}}+\mathrm{i}\mathbf{\bm{\sigma}}\otimes\hat{\mathbf{p}} \, \overset{\cdot}{,} \, c^2\omega\frac{\mathbf{r}_K}{r_K^3}\right];\left\{\hat{\mathbf{p}} \, \overset{\times}{,} \, c^2\omega\mathbf{r}_G\right\}}}
, \label{eq:zora-sfsd} 
\end{align}
\end{widetext}
where $\otimes$ denotes the outer product, $\mathrm{i}=\sqrt{-1}$ is the imaginary unit, $\omega=(2m_\mathrm{e}c^2-\tilde{V})^{-1}$ is the ZORA factor with van W\"ullen's model potential $\tilde{V}$ (see Sec.~\ref{sec:comp-det}), $m_\mathrm{e}$ is the electron mass, $\hat{\mathbf{p}}$ is the electronic linear momentum operator, $\bm{\sigma}$ is the Pauli vector, $\left[\mathbf{A}\,\overset{\cdot}{,}\,\mathbf{B}\right]=\mathbf{A}\cdot \mathbf{B}-\mathbf{B}\cdot \mathbf{A}$ and $\left\{\mathbf{A} \, \overset{\times}{,} \, \mathbf{B}\right\}=\mathbf{A}\times \mathbf{B}+\mathbf{B}\times \mathbf{A}$ are the commutator and anti-commutator of two operators, respectively. As opposed to the response terms, the diamagnetic-like expectation value contribution to the shielding tensor ${\bm \sigma}^\mathrm{d}$ is purely of SF-type:
\begin{align}
\bm{\sigma}^{\mathrm{d}} (K) &=-\frac{\mu_0\,e^2}{4\pi} \Braket{c^2\omega\frac{\mathbf{r}_K\otimes\mathbf{r}_G-\bm{1}\mathbf{r}_K\cdot\mathbf{r}_G}{r_K^3}}.
\end{align}

\begin{figure*}[htb!]
\includegraphics[width=.7\textwidth]{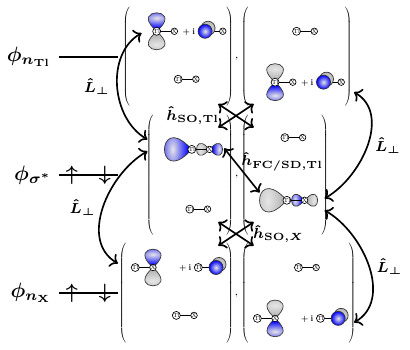}
\caption{Schematics of the SO enhancement mechanism of NMR shieldings of the Tl nuclei in Tl$X$ compounds. We follow closely the schematic explanations of Ref.~\citenum{vicha:2020}. The scheme includes three orbitals which are mainly involved in the HAVHA mechanism, shown as the 2C Kramers partners with the upper and lower components being the spin-up and spin-down parts of the spinor respectively. Spin-orbit coupling between the orbitals is indicated by one-electron SO coupling operators centered at Tl, $\hat{h}_\mathrm{SO,Tl}$, and at $X$, $\hat{h}_{\mathrm{SO},X}$; the electron spin-dependent hyperfine coupling at the Tl is indicated by the operator $\hat{h}_\mathrm{FC/SD,Tl}$ (in leading order spin-dipole and Fermi contact interactions) and the coupling by the external magnetic field is indicated by the perpendicular components of the orbital angular momentum operator $\hat{L}_\perp$. A detailed explanation is provided in the text.}
\label{fig:HAVHA_mechanism}
\end{figure*}

\begin{figure*}[htb!]
\includegraphics[width=\textwidth]{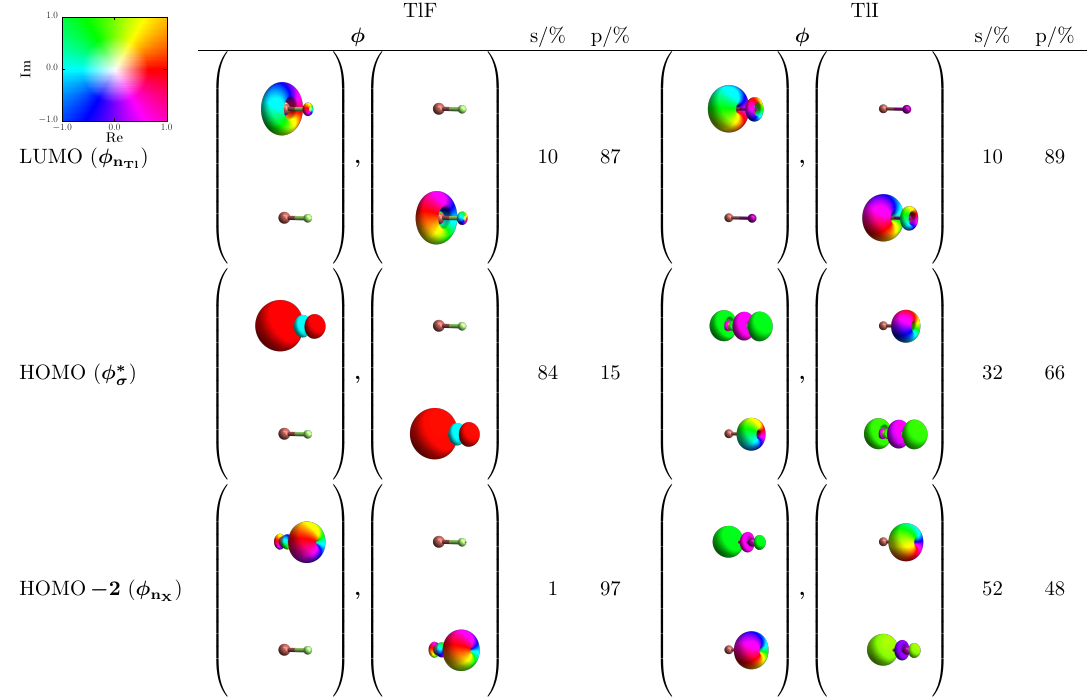}
\caption{Molecular 2C spinors at the level of ZORA which are relevant for the HAVHA mechanism described in Fig.~\ref{fig:HAVHA_mechanism}. The complex 2C orbitals are visualized as two Kramers paired spinors, where the upper and lower components correspond to spin-up and spin-down respectively. Orbital amplitudes were calculated on a three-dimensional grid and plotted with the help of \textsc{Mathematica} version 11 \cite{mathematica11} by mapping the phase in the complex plane via a color code on the contour surface with value 0.03 of the absolute value of each spinor. Note that the absolute phase of every single orbital is arbitrary. For each MO, the contribution from s-type and p-type basis functions is given.}
\label{fig:orbitals}
\end{figure*}

In the absence of spin-orbit (SO) coupling, the combination of SF and SD operators in Eq.~\eqref{eq:zora-sfsd} is zero by symmetry. Therefore, the SFSD contributions can be interpreted as a pure SO coupling contribution to the NMR shieldings. In Figs.~\ref{fig:propagator_rZORA+eX2C-HF_Tl+} and \ref{fig:propagator_rZORA+eX2C-HF_TlI}, and also in figures in the Supplementary Material, we show the values of the SD, SF, and mixed SFSD contributions to the propagator. We find that all important core-excitations are of SDSD type. This is expected as the leading order contribution to the hyperfine operator is mainly given by the Fermi contact interaction, which is largest for core orbitals. On the contrary, valence contributions are of SFSF and SFSD type. At the HF level of theory SFSD contributions are larger in magnitude than SFSF contributions, whereas at the DFT level of theory both contributions are of similar magnitudes. A closer inspection shows that these SFSD contributions stem from the second term in Eq.~\eqref{eq:zora-sfsd} (see also Table IX in the Supplementary Material). Other SO contributions to SDSD parts of the shielding tensor do not contribute as strongly and will therefore not be discussed in detail in the following. Even though contributions of SFSF type can in principle appear without SO coupling, the strongly deshielding SFSD contributions to the perpendicular component for increasing $Z$ of $X$ agrees with a strong increase in SO coupling from Tl$^+$ to TlAt and previous studies of the heavy-atom on the light atom (HALA) (see Reviews in Refs.\citenum{vicha:2020} and \citenum{kaupp:2004}) and heavy-atom on the vicinal heavy-atom (HAVHA) enhancement effects.\cite{Maldonado2009,Maldonado2015,zapata-escobar:2022,zapata-escobar:2023} The SFSF contributions are best analysed within the linear response within the elimination of small components with localized molecular orbitals (LRESC-Loc) approach.\cite{zapata-escobar:2022, zapata-escobar:2023}
From the patterns of excitations in Fig. \ref{fig:propagator_rZORA+eX2C-HF_TlI} and the Supplementary Material it can be seen that the strongly $X$-dependent contribution to the perpendicular component of the NMR shielding tensor observed in Tables \ref{tab:sigma_Tl} and \ref{tab:sigma_Tl_ZORA} is due to valence excitations. A closer inspection of contributions from individual excitations shows that the dominant negative contribution is related to excitations from the highest occupied molecular orbital to lowest unoccupied molecular orbital (HOMO-LUMO).
The underlying mechanism of the SO coupling dependent SFSD contributions can be understood in a simple NR three-orbital picture analogously to the analysis of the HALA effect in Ref.~\citenum{vicha:2020} and is sketched in Fig.~\ref{fig:HAVHA_mechanism}.

For this purpose we start from a doubly occupied orbital of p-type that is perpendicular to the bond axis and forms a lone pair on $X$, an unoccupied p-type orbital perpendicular to the bond axis located at Tl and a doubly occupied $\sigma^*$ orbital that forms the HOMO in Tl$X$. SO coupling is now introduced perturbatively at both centers. SO coupling on the Tl atom (vicinal heavy atom, VHA) mixes the unoccupied non-bonding $\mathrm{p}_x+\mathrm{i}\mathrm{p}_y$ orbital at Tl $\phi_{n_\mathrm{Tl}}$ with the anti-bonding doubly occupied $\sigma^*$ orbital which is of s and p$_z$ character where $z$ is chosen along the bond axes. In contrast to the atom, this coupling is allowed as the orbital at Tl is no longer spherically symmetric, but received a p$_z$-like contribution. Moreover, SO coupling on the halogen atom mixes the occupied non-bonding $\mathrm{p}_x+\mathrm{i}\mathrm{p}_y$ orbital (a lone pair) at $X$, $\phi_{n_X}$, with the occupied $\sigma^*$ orbital. The resulting SO coupled unoccupied ($\phi_a$) and occupied ($\phi_i$) orbitals are of the form $\phi^\alpha_a = N^\alpha_a ( \phi^\alpha_{n_\mathrm{Tl}} + c^\alpha_{\mathrm{SO,Tl},a} \phi^\beta_{\sigma^*} + c^\alpha_{\mathrm{SO,Tl}X,a} \phi^\alpha_{n_X})$, $\phi^\alpha_i=N^\alpha_i (\phi^\alpha_{\sigma^*} + c^\alpha_{\mathrm{SO,Tl},i} \phi^\beta_{n_\mathrm{Tl}} + c^\alpha_{\mathrm{SO,}X,i} \phi^\beta_{n_X})$, where $\alpha$ and $\beta$ denote the spin components and with all spins reverted for the Kramers partner. Here $N^\alpha_a$ and $N^\alpha_i$ are normalization constants, $c^\gamma_{\mathrm{SO},A,j}$ are coefficients for the admixture of other orbitals due to SO coupling on $A$, with $A$ being Tl, $X$ or simultaneous SO coupling on both centers ($A$ being Tl$X$). Here $\gamma$ denotes the spin-component $\alpha$ or $\beta$ of the SO coupled orbital and $j$ denotes the occupied orbital ($j=i$) or unoccupied orbital ($j=a$). Note that admixtures, which originate from simultaneous SO coupling on both centers, appear to be non-linear in the SO operator and therefore mix orbitals of same spin. The coefficients $c^\gamma_{\mathrm{SO},A,j}$ are obtained from diagonalisation of the SO Hamiltonian in the chosen basis of six spin orbitals. Here we assumed that interference terms between SO coupling at Tl with SO coupling on $X$ are less important than SO coupling from a single center for mixing the $\sigma^*$ orbital with non-bonding orbitals as these interference terms will be of higher order. In contrast the mixing of the two non-bonding orbitals appears solely due to this interference.

In leading order, the integrals of the electron spin-dependent hyperfine operator $\hat{h}_{\mathrm{FC/SD}}$ appearing in the dominant second term of eq. (\ref{eq:zora-sfsd}) reduce to Fermi contact type integrals, which require s-type basis function contribution to the orbital to give a non-zero matrix element. The spin-free magnetic field integrals are in good approximation integrals of the orbital angular momentum operator $\hat{\vec{L}}$. Thus, the perpendicular component of the NMR shielding tensor is proportional to $\sigma_\perp \sim \frac{\Braket{\phi^\alpha_{i}|\hat{h}_{\mathrm{FC/SD,Tl},\perp}|\phi^\alpha_{a}}\Braket{\phi^\alpha_{a}|\hat{L}_\perp|\phi^\alpha_{i}}}{\epsilon_i-\epsilon_a}+\,\mathrm{cc}$. Within the approximations introduced above we obtain the SO coupling contribution to the perpendicular components as function of the SO mixing coefficients:
\begin{multline}
    \sigma_\perp\sim \frac{4|N_a|^2|N_i|^2}{\epsilon_i-\epsilon_a} c_{\mathrm{SO,Tl},a}\Braket{\phi_{\sigma^*}|\hat{h}_{\mathrm{FC/SD,Tl},\perp}|\phi_{\sigma^*}}\\
    \times\left((1+c_{\mathrm{SO,Tl},a}c_{\mathrm{SO,Tl},i})\Braket{\phi_{\sigma^*}|\hat{L}_\perp|\phi_{n_\mathrm{Tl}}} \right.\\
    +\left. \left(c_{\mathrm{SO,Tl}X,a}+c_{\mathrm{SO,Tl},a}c_{\mathrm{SO,X},i}\right)\Braket{\phi_{\sigma^*}|\hat{L}_\perp|\phi_{n_\mathrm{X}}}\right)\,,
\label{eq:HAVHA}
\end{multline}
\noindent where we dropped the spin indices assuming contributions from $\alpha$ and $\beta$ spin to be equal, i.e. $c_{\mathrm{SO},A,j}=c^\alpha_{\mathrm{SO},A,j}=c^\beta_{\mathrm{SO},A,j}$ and $N_j=N^\alpha_j=N^\beta_j$. Here the first term in parenthesis is a heavy atom effect on the heavy atom (HAHA) \cite{Pyykko87} and the second term is a HAVHA effect which increases with increasing SO coupling at $X$. In this picture the contribution of the latter term to $\sigma_\perp$ is expected to be negative for an occupied $\sigma^*$ orbital (see Ref.~\citenum{vicha:2020}) and causes a strong decrease of the perpendicular component for increasing $Z$ of the $X$ atom as observed in the numerical calculations.

The ZORA 2C spinors which are actually involved in the simple model of Fig.~\ref{fig:HAVHA_mechanism} are shown exemplary for TlF and TlI in Fig.~\ref{fig:orbitals} at the HF level. Corresponding orbitals of all other molecules and also at the DFT level of theory are provided in the Supplementary Material. Note that at the DFT level for heavier homologous the $\sigma^*$ orbital becomes the HOMO$-1$. In any case the LUMO is dominated by a non-bonding $\mathrm{p}_x+\mathrm{i}\mathrm{p}_y$-type orbital of the Tl atom. Whereas the $\sigma^*$-type has only minor admixture of the $\mathrm{p}_x+\mathrm{i}\mathrm{p}_y$-type orbital located at the $X$ atom in case of light $X$ as e.g.~for TlF, large SO coupling in Tl$X$ with heavier $X$ such as TlI leads to a considerable mixture of these two orbitals, which is clearly visible in Fig.~\ref{fig:orbitals}. It has to be noted that the SO coupling on Tl is even increased by increasing SO coupling on $X$ (this is particularly visible for TlAt and TlTs, which are shown in the Supplementary Material). These observations are in agreement with a HAVHA enhancement mechanism for p-block elements described in Fig.~\ref{fig:HAVHA_mechanism}, which was reported previously e.g.~in Refs.~\citenum{Pyykko87,Maldonado2009,Maldonado2015,vicha:2020,zapata-escobar:2022,zapata-escobar:2023}.

\subsection{QED effects}

As the excitation pattern of the Tl$X$ series of molecules has a much more complex character than that of the Tl$^+$ ion, we have sought to improve our procedure for estimating the QED effects on NMR shieldings, with respect to that we have used and described in Ref.~\citenum{Koziol2019}. Because our procedure is based on the modification of the matrix elements that consider excitations from occupied s-type orbitals to virtual s-type ones, the crucial issue is to determine which ones within virtual MOs have appropriate ``s-type atomic-character''. 

Starting from Mulliken's projection analysis of the MOs associated to the Tl$X$ molecules, which are provided by the \textsc{Dirac} code, we separated the orbital expansion coefficients into four parts (see more details in the Appendix). In this way, the $m$-th MO of each molecule can be written as
\begin{equation}\label{eq:mulliken}
\begin{array}{>{\displaystyle}l>{\displaystyle}l}
%
%
\int |\Psi_m^\text{MO} ({\bm r})|^2 \, d^3{\bm r} 
&= n_{m\text{s}}^\text{Tl s, large}
+ \sum_i n_{mi}^\text{Tl, small} \\
& \quad + \sum_j n_{mj}^\text{Tl, other large}
+ \sum_k n_{mk}^X.
\end{array}
\end{equation}
\noindent Here, $n_{m\text{s}}^\text{Tl s, large}$ indicate the contributions arising from the large component parts of relativistic s-type orbitals centered on the Tl nucleus, while $n_{mi}^\text{Tl, small}$ refer to the contributions of the small component parts of the wave functions related to the Tl atom. Besides, $n_{mj}^\text{Tl, other large}$ collect the contributions to $\Psi_m^\text{MO} ({\bm r})$ arising from the large components of other, non s-type, orbitals related to the Tl atom, and finally $n_{mk}^\text{X}$ are the contributions related to the ligand atom (they include both large and small components). The fractional electronic populations are normalized, i.e. $n_{m\text{s}}^\text{Tl s, large} + \sum_i n_{mi}^\text{Tl, small} + \sum_j n_{mj}^\text{Tl, other large} + \sum_k n_{mk}^\text{X} =1$. In general, the values of $n_{m\text{s}}^\text{Tl s, large}$, $n_{mi}^\text{Tl, small}$, $n_{mj}^\text{Tl, other large}$, and $n_{mk}^\text{X}$ can be positive or negative. 

Then, we introduce two sums of selected coefficients. The first one is 
$\text{S}_1 = |n_{m\text{s}}^\text{Tl s, large}| + \sum_i n_{mi}^\text{Tl, small}$ and the second one is $\text{S}_2 = \sum_j |n_{mj}^\text{Tl, other large}| + \sum_k |n_{mk}^\text{X}|$. 
The first sum determines the s-type character of MO. Because the small components of relativistic MOs are important in the case of s-type orbitals, we also included them in our s-type orbital characterization. 
The second sum determines the non-s-type parts of MOs, and they are considered here as ``contaminations'' for the selection of ``pure'' s-type MOs. 
Furthermore, considering the absolute values of $n_{mj}^\text{Tl, other large}$ and $n_{mk}^\text{X}$ we realize that, if $\text{S}_2$ is high enough, then the contribution of non s-type orbitals in MOs is high even if some of them have negative values of $n_{mj}^\text{Tl, other large}$ and $n_{mk}^\text{X}$ in population expansion (what is common in bonding-like MOs). 

In order to find a right selection of the s-type MOs we propose two conditions to be fulfilled simultaneously: (1) $\text{S}_1 \ge \text{T}_1$, and 
(2) $\text{S}_1 + \text{S}_2 \le \text{T}_2$. 
The optimal, however arbitrary, chosen value for the threshold $\text{T}_1$ is 0.9. This value ensures that a selected MO has large enough s-type character. For the threshold $\text{T}_2$ we used 1.011. The values have been chosen based on results convergence rule, i.e. by taking the threshold values for which the QED corrections to the shieldings does not change drastically. The importance of condition 2 is that it ensures that selected orbitals have not too many contributions from expansion coefficients with negative sign -- that situation means that MO is far from the pure atomic character and it is more like bonding type MO, so it is not fitted to our method of estimating QED corrections to the shielding. 

\begin{figure}[!htb]
\begin{minipage}{.48\textwidth}
\includegraphics[width=\linewidth]{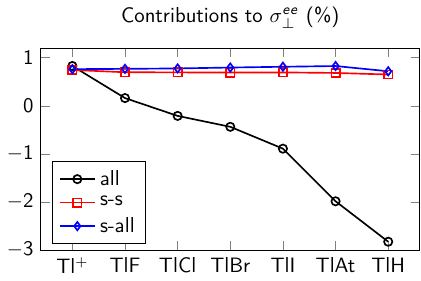}
\end{minipage}
\hfill
\begin{minipage}{.48\textwidth}
\includegraphics[width=\linewidth]{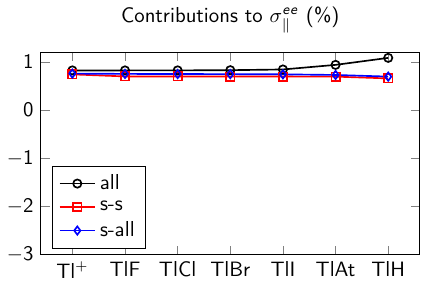}
\end{minipage}
\caption{Different type of contributions to ${\bm \sigma}$(Tl), related to: i) excitations from occupied s-type orbitals to positive-energy s-type virtual orbitals (s-s); ii) excitations from occupied s-type orbitals to all positive-energy virtual orbitals (s-all); and iii) all type of excitations from occupied orbitals to positive-energy virtual orbitals. Both perpendicular (top) and parallel (bottom) contributions to the NMR shielding tensor are presented.}
\label{fig:s-contrib}
\end{figure}

It is interesting to see how much the amplitudes related to excitations from occupied s-type orbitals to virtual s-type orbitals contribute to the total value of shielding. On Fig.~\ref{fig:s-contrib}, the sum of amplitudes related to excitations from occupied s-type orbitals to s-type virtual orbitals (labeled s-s), from occupied s-type orbitals to all virtual orbitals (s-all) and all excitations are presented for Tl$^+$ ion and Tl$X$ molecules (both perpendicular and parallel contributions to ${\bm \sigma}$). One can see that, in the case of parallel contributions the s-s amplitudes contribute about 70--90\,\% to the total value, what gives the major part. In the case of perpendicular contribution a different behavior is observed. In this case the s-s and s-all amplitudes are on similar levels than in the case of parallel contributions, but the sum of amplitudes related to excitations from non-s-type orbitals changes descending from positive value for Tl$^+$ to large negative value for TlAt. 

\begin{table*}
\caption{QED effects on ${\bm \sigma}$ for the Tl nucleus in Tl$^+$ ion and in Tl$X$ ($X$ = H, F, Cl, Br, I, At) molecules, calculated using the DHF approach. Numbers in parentheses were obtained using the ``s-all'' approximation (see text for details). Values in ppm.}
\label{tab:sigma_QED}
\tabcolsep=0.5\tabcolsep
\begin{tabular*}{\linewidth}{@{} @{\extracolsep{\fill}}l @{\hspace{1em}} *{10}{S[parse-numbers=false]} @{}}
\toprule
 & {Tl$^+$} & \multicolumn{3}{c}{TlF} & \multicolumn{3}{c}{TlCl} & \multicolumn{3}{c}{TlBr} \\
\cmidrule{3-5}\cmidrule{6-8}\cmidrule{9-11}
 & {$\sigma_\text{iso}$} & {$\sigma_\perp$} &  {$\sigma_\parallel$} & {$\sigma_\text{iso}$} & {$\sigma_\perp$} &  {$\sigma_\parallel$} & {$\sigma_\text{iso}$} & {$\sigma_\perp$} &  {$\sigma_\parallel$} & {$\sigma_\text{iso}$} \\
\midrule
1s & -41.4 & -40.6 & -40.8 & -40.7 & -40.6 & -40.8 & -40.6 & -40.4 & -40.7 & -40.5 \\
 & (-42.8) & (-42.5) & (-42.7) & (-42.6) & (-42.4) & (-42.7) & (-42.5) & (-42.3) & (-42.7) & (-42.4) \\
2s & -16.1 & -15.2 & -15.4 & -15.3 & -15.1 & -15.4 & -15.2 & -15.1 & -15.3 & -15.1 \\
 & (-16.3) & (-16.2) & (-16.2) & (-16.2) & (-16.2) & (-16.2) & (-16.2) & (-16.2) & (-16.2) & (-16.2) \\
3s & -7.1 & -6.0 & -6.2 & -6.1 & -5.9 & -6.2 & -6.0 & -5.9 & -6.1 & -6.0 \\
 & (-7.2) & (-7.1) & (-7.0) & (-7.1) & (-7.1) & (-7.0) & (-7.1) & (-7.2) & (-7.0) & (-7.2) \\
4s & -3.1 & -2.1 & -2.1 & -2.1 & -2.1 & -2.1 & -2.1 & -2.0 & -2.1 & -2.1 \\
 & (-3.2) & (-3.1) & (-3.0) & (-3.1) & (-3.3) & (-3.0) & (-3.2) & (-3.7) & (-2.9) & (-3.5) \\
5s & -1.1 & -0.4 & -0.5 & -0.4 & -0.4 & -0.5 & -0.4 & -0.4 & -0.5 & -0.4 \\
 & (-1.1) & (-2.0) & (-0.9) & (-1.6) & (-2.8) & (-0.7) & (-2.1) & (-4.0) & (-0.6) & (-2.9) \\
Total & -68.9 & -64.4 & -64.9 & -64.6 & -64.1 & -64.9 & -64.4 & -63.8 & -64.7 & -64.1 \\
 & (-70.5) & (-70.9) & (-69.8) & (-70.6) & (-71.7) & (-69.5) & (-71.0) & (-73.5) & (-69.3) & (-72.1) \\
\midrule
 & & \multicolumn{3}{c}{TlI} & \multicolumn{3}{c}{TlAt} & \multicolumn{3}{c}{TlH} \\
\cmidrule{3-5}\cmidrule{6-8}\cmidrule{9-11}
 & & {$\sigma_\perp$} &  {$\sigma_\parallel$} & {$\sigma_\text{iso}$} & {$\sigma_\perp$} &  {$\sigma_\parallel$} & {$\sigma_\text{iso}$} & {$\sigma_\perp$} &  {$\sigma_\parallel$} & {$\sigma_\text{iso}$} \\
\midrule
1s &  & -40.0 & -40.7 & -40.2 & -39.5 & -40.8 & -39.9 & -39.0 & -39.9 & -39.3 \\
 &  & (-42.1) & (-42.7) & (-42.3) & (-41.4) & (-42.6) & (-41.8) & (-41.3) & (-42.5) & (-41.7) \\
2s &  & -15.2 & -15.3 & -15.3 & -15.1 & -15.2 & -15.1 & -14.1 & -14.2 & -14.1 \\
 &  & (-16.2) & (-16.1) & (-16.2) & (-16.1) & (-16.1) & (-16.1) & (-15.4) & (-15.8) & (-15.5) \\
3s &  & -6.2 & -6.1 & -6.2 & -6.0 & -6.0 & -6.0 & -5.1 & -5.1 & -5.1 \\
 &  & (-7.4) & (-6.9) & (-7.2) & (-7.4) & (-6.9) & (-7.2) & (-6.1) & (-6.5) & (-6.2) \\
4s &  & -2.3 & -2.1 & -2.2 & -2.2 & -2.0 & -2.1 & -1.6 & -1.6 & -1.6 \\
 &  & (-4.1) & (-2.8) & (-3.7) & (-4.4) & (-2.7) & (-3.8) & (-1.7) & (-2.0) & (-1.8) \\
5s &  & -0.4 & -0.5 & -0.4 & -0.3 & -0.5 & -0.4 & -0.3 & -0.3 & -0.3 \\
 &  & (-5.2) & (-0.3) & (-3.6) & (-7.0) & 0.5 & (-4.5) & (-1.9) & 2.3 & (-0.5) \\
Total &  & -64.1 & -64.7 & -64.3 & -63.1 & -64.5 & -63.5 & -60.1 & -61.1 & -60.4 \\
 &  & (-75.0) & (-68.9) & (-72.9) & (-76.3) & (-67.7) & (-73.4) & (-66.3) & (-64.6) & (-65.7) \\
\bottomrule
\end{tabular*}
\end{table*}

Table \ref{tab:sigma_QED} shows the contributions of QED effects (following the model discussed in this work) to the NMR shielding tensor elements for the Tl nuclei in the molecules studied in this work, distinguishing their total values from the contributions by main mechanisms. 
Small (about 4--8 ppm) but clear ligand effect is observed. This means that the atoms bonded to the thallium atom modifies the electron density related to $n$s-type MO and, as a results, also modifies QED effects on nuclear magnetic shielding. There is no big difference between QED effect on parallel and perpendicular contributions to shielding -- that is because the s-type MOs are both, nucleus centered and almost spherical orbitals. Besides, being such difference small it proves that our procedure to extract s-type MOs to estimating QED effect is a reliable one.

The numbers inside parentheses in Table~\ref{tab:sigma_QED} are obtained for ``s-all'' approximation, it means all excitations from occupied s-type orbital to any virtual orbital are treated as s-to-s excitations. This approximation has been used in our previous paper (Ref.~\citenum{Koziol2019}) with the note that the expected difference between calculations using ``s-all'' and ``s-s'' models (so-called ``s-all'' approximation error) is about 5\,\%. This claim is confirmed in the present study for Tl$^+$ ion, where the ``s-all'' approximation error is about 2\,\%. However, for the Tl$X$ ($X$ = H, F, Cl, Br, I, At) molecules such error is 9--15\,\% for the total value of QED correction to $\sigma_\text{iso}$. 

\subsection{\changed{Other small contributions to NMR shielding constant}\label{small_contributions}}

\changed{
In addition to including QED effects in NMR shielding calculations, the influence of Breit interactions must be considered. However, so far in the \textsc{Dirac} code it is not possible to take into account Breit or Gaunt interactions in the four-component calculation of linear response functions involving time-reversal antisymmetric operators. Then, these effects cannot be included in the NMR shielding calculations presented in this work.
From a previous work on Breit effects on NMR shieldings of noble-gas atoms \cite{Ishikawa1998} we expect the Breit correction to be less than 1~\%\ of the total relativistic effect for atoms heavier than Xe. 
%
}

\changed{On the other hand, we checked whether a self-consistent treatment of the Breit interaction alters the Lamb shift for atoms significantly. We did it by performing calculations with codes in which it is possible to switch between self-consistent (SC) and perturbation theory (PT) treatment of the Breit interaction. The Lamb shift contribution to the 1s orbital energy of Tl atom calculated by means of the \textsc{Mcdfgme} code \cite{mcdfgme} is 166.1843~eV (SC treatment of the Breit interaction) and 166.8630~eV (PT treatment); so, the difference between SC and PT treatments is on the level of just -0.41\%. In the case of calculations performed with the \textsc{Dbsr\_hf} \cite{Zatsarinny2016} plus \textsc{Qedmod} \cite{Shabaev2015} codes the numbers are (different than above QED model used) 167.3385~eV (SC treatment) and 167.6045~eV (PT treatment), so that for this last model the effect is even smaller in magnitude: -0.16\%. Based on these results one would expect that the Breit interaction enhances other ``nuclear-centered'' atomic/molecular properties no more than one percent.}

\changed{
Besides QED and Breit effects, the impact of the finite size of the nucleus, i.e.~the BR and BW effects, usually play a similarly important role. As mentioned in the computational details, the BR effect was included in all calculations reported in this manuscript by employing a Gaussian nuclear charge density distribution. To estimate the size of BR effects on NMR shieldings, we performed calculations of NMR shieldings in which nuclei are modeled as point-charges at the 2C level for Tl$^+$ and 4C level for Tl$^+$ and Tl$X$. Moreover, we mentioned that we assumed in 4C calculations a Gaussian magnetization density distribution to model the hyperfine operator, whereas point-like magnetic dipoles were assumed in 2C calculations for all numbers reported in the previous section. In order to estimate the size of this so-called BW effect, we compared 4C and 2C calculations using the point-like and Gaussian magnetization densities of Tl$^+$. Moreover, the BW effects were studied for all Tl$X$ at the 4C level.

We find the BR and BW nuclear-size effects on $\sigma_\text{iso}$(Tl) of Tl$^+$ to be \SI{-192}{ppm} (\SI{-193}{ppm} on the 2C level, rZORA) and \SI{-178}{ppm} (\SI{-128}{ppm} on the 2C level, rZORA), respectively. As seen, these effects are of a similar size as QED contributions and must be properly taken into account if highly-accurate predictions of NMR shieldings are targeted. Whereas a Gaussian model is well approved for the nuclear charge density distribution, the nuclear magnetization density of most nuclei is not known and an accurate treatment of the BW effect may require a very different model to obtain quantitative results.\cite{andrae:2000,andrae:2017} In Table \ref{tab:nuc-eff} we report the 4C estimates of these effects for all the systems studied in this work. Both the BR and BW effects are in all cases below 2~\% of the total isotropic values of this property, except for the case of TlAt, where they slightly overpass the level of 5~\%. Moreover, it has been observed that these two effects are essentially additive in all the studied systems. As the BW effect is below 3~\% of the total shielding for all studied systems, the comparison of 4C and 2C calculations is not hampered by the fact the the BW effect was included at the 4C level but not at the 2C level in the previous sections.
}

\begin{table}[h]
\centering
\caption{Bohr--Weisskopf and Breit--Rosenthal effects on $\sigma_\text{iso}$(Tl) in Tl$^+$ and Tl$X$ ($X$ = H, F, Cl, Br, I, At), calculated using the DHF approach. Values in ppm.}
\label{tab:nuc-eff}
\renewcommand*{\arraystretch}{1.3}
\begin{tabular*}{\linewidth}{@{} l @{\extracolsep{\fill}} SS @{}} 
 \toprule
 System & {Bohr--Weisskopf} & {Breit--Rosenthal} \\
 \midrule
Tl$^+$& -177.91  &  -192.02 \\
TlH   &  104.13  &   126.09 \\
TlF   & -123.22  &  -133.19 \\
TlCl  &  -97.40  &  -104.46 \\
TlBr  &  -80.49  &   -83.95 \\
TlI   &  -42.74  &   -40.25 \\
TlAt  &   52.87  &    96.65 \\
 \bottomrule
\end{tabular*}
\end{table}

\section{Conclusions}
When looking for highly accurate theoretical calculations of atomic and molecular response properties, physical effects that were considered extremely small a few years ago must be considered today. Among them, one must include QED effects and Breit interactions. In this work, we apply our own model to estimate QED corrections to the NMR shielding constant \cite{Gimenez2016,Koziol2019} to the Tl$^+$ ion and Tl$X$ ($X$ = H, F, Cl, Br, I, At) molecules, besides the calculation and the analysis of the relativistic effects on the NMR magnetic shieldings of the nucleus of thallium atom and the halogen atom bonded to it. These relativistic effects were studied by applying the RelPPT at RPA level of approach. This approach gave us the opportunity to learn in some details about how large those effects are and which are the patterns of virtual excitations that most contribute to the shielding tensor. 

We found a strong ligand effect on the shielding tensor ${\bm \sigma}$(Tl), resulting in changing its character from being diamagnetic (which diminish from TlF) to paramagnetic (TlAt). This ligand effect can be explained by the patterns of virtual excitations between occupied MOs and the unoccupied ones within the polarization propagator theory. While in the case of atomic systems the $ee$ amplitudes are mainly given by $n$s$\to n'$s excitation types, in the case of Tl$X$ molecules the $ee$ amplitudes behave more complex. The parallel tensor element of ${\bm \sigma}$(Tl) is constructed in a similar manner as in the atomic case, from positive $n$s$\to n'$s excitation types, but the perpendicular element of ${\bm \sigma}$(Tl) is strongly affected by the negative amplitudes arising from the Tl--$X$ bonds. In addition to it, the analysis with a 2C ZORA approach of the SFSD contributions involving MOs shows that a HAVHA mechanism driven by SO is responsible for the observed relativistic effects on the perpendicular component of the tensor ${\bm \sigma}$(Tl).

The order of magnitude of QED effects on the shielding of thallium nucleus, as described within the atomic-based model of this work, is in the family of compounds analyzed herein almost independent of the halogen atom bonded to Tl, though their values show a tiny dependence with the atom bonded to Tl when the virtual ``s-all'' excitations are considered. The absolute values of QED effects for the whole set of molecules, including only ``s-s'' virtual excitations, is negative (paramagnetic-like) and close to \SI{-64}{ppm}, comparing to \SI{-69}{ppm} of sole Tl$^+$ ion. The dependence of QED effects with the bonded atom to Tl deserves to be investigated with a non-atomic model. There is work in progress in our groups along these lines.

\section*{Supplementary Material}
In the Supplementary Material we display the 4C values of the different contributions to ${\bm \sigma}(\text{Tl})$ and ${\bm \sigma}($X$)$ in Tl$X$ ($X$ = H, F, Cl, Br, I, At, Ts), including those arising from the two-electron small component integrals of ($SS$$\mid$$SS$) type at the DHF, DFT-LDA, and DFT-PBE0 levels of theory. Results using the CGO and GIAO schemes are displayed as well, and in some cases we also show results obtained in the NR limit (scaling the speed of light in vacuum to 30 times its real value). Figures displaying patterns of excitations are also shown. Additional ZORA results are shown at the level of HF for TlTs and at the level of DFT-LDA and DFT-PBE0 for all Tl$X$ and Tl$^+$. Separate SD, SF and SFSD contributions to the parallel and perpendicular components of the shielding tensors are provided for HF and DFT-PBE0 results. Plots of propagators at the level of rZORA+X2Ce-HF for Tl$X$ ($X$ = H, F, Cl, Br, At, and Ts) are provided alongside propagator plots at the levels of rZORA-HF, rZORA-PBE0 and rZORA-LDA for all Tl$X$ molecules. Two-component ZORA-orbitals relevant for the HAVHA mechanism  are provided for all Tl$X$ at the HF, DFT-LDA and DFT-PBE0 levels of theory. The coefficients $C_{QR}^\text{QED/DC}$ for atoms with atomic number $Z$ = 10--92 are also presented.

\section*{Acknowledgments}

This research was funded in part by National Science Centre, Poland (NCN) grant MINIATURA 6, No. 2022/06/X/ST4/00845. 
For the purpose of Open Access, the authors have applied a CC-BY public copyright licence to any Author Accepted Manuscript (AAM) version arising from this submission. 
IAA acknowledges partial support from FONCYT through grant PICT-2020-SerieA-0052 and from CONICET through grant PIBAA-2022-0125CO. GAA and IAA acknowledge support from FONCYT by grant PICT-2021-I-A-0933.
We would also like to thank the Institute for Modeling and Innovation in Technologies (IMIT) and the University of Groningen's Center for Information Technology for their support and for providing access to the IMIT and Hábrók high-performance computing clusters. 
KG and RB gratefully acknowledge the computing time provided at the NHR Center NHR@SW at Goethe-University Frankfurt. This is funded by the Federal Ministry of Education and Research, and the state governments participating on the basis of the resolutions of the GWK for national high performance computing at universities (\url{http://www.nhr-verein.de/unsere-partner}).

\appendix

\section{Mulliken population analysis}

The four-component one-electron molecular orbitals $\Psi^\text{MO}_m(\bm{r})$ obtained within the Born-Oppenheimer approximation as a solution of the time-independent Dirac-Coulomb equation by applying the DHF procedure can be written as the bispinors
\begin{equation}\label{eq:psi-MO}
\Psi^\text{MO}_m (\bm{r}) =
\begin{pmatrix}
\phi^{\alpha,L}_m (\bm{r})\\
\phi^{\beta,L}_m (\bm{r})\\
\phi^{\alpha,S}_m (\bm{r})\\
\phi^{\beta,S}_m (\bm{r})
\end{pmatrix},
\end{equation}
\noindent where
\begin{equation}\label{eq:WFbasisexp}
\phi^{\omega,Y}_m(\bm{r}) = \sum_A^{N_\text{nuc}} \sum_j^{N_\text{basis}} \left( c^{\omega,Y}_{mj} + i d^{\omega,Y}_{mj} \right) \chi_j(\bm{r}-\bm{R}_A).
\end{equation}

\noindent Here, $\alpha$ and $\beta$ stand for spin up and down, respectively, whereas $L$ and $S$ refer to the large and small components of the wave function, respectively. The primitive real functions $\chi_j(\bm{r}-\bm{R}_A)$ are elements of the basis sets, and in our particular case they are a set of cartesian Gaussian functions centered at the position of either of the two nuclei of the diatomic molecules studied in this work ($\bm{R}_A$ can be the position of either of the nuclei, i.e., Tl or $X$), $i=\sqrt{-1}$ is the imaginary unit, and $c^{\omega,Y}_{mj}$ and $d^{\omega,Y}_{mj}$ are real constants obtained as solutions of the DHF equations.

The electron probability density at a given position $\bm{r}$ from the $m$-th single one-electron MO ($\Psi^\text{MO}_m$) is given by $\rho_m(\bm{r})=|\Psi^{MO}_m(\bm{r})|^2$. Therefore, integrating and summing over all occupied MOs gives the total number of electrons ($N_\text{elec}$) in the system,
\begin{widetext}
\begin{eqnarray}\label{eq:densMO}
\sum_m^{N_\text{occ}} \int |\Psi_m^\text{MO} ({\bm r})|^2 \, d^3{\bm r} &=& \sum_m^{N_\text{occ}}  \sum_{jk}^{N_\text{basis}} \sum_{\omega Y} (c^{\omega,Y}_{mj} c^{\omega,Y}_{mk} + d^{\omega,Y}_{mj} d^{\omega,Y}_{mk}) \int \chi_j(\bm{r}-\bm{R}_A) \chi_k(\bm{r}-\bm{R}_B) \, d^3{\bm r}  \nonumber \\
&=& \sum_m^{N_\text{occ}} \sum_{jk}^{N_\text{basis}} \sum_{\omega Y} P^{\omega,Y}_{mjk} \; S_{jk}(\bm{R}_A,\bm{R}_B) = N_\text{elec},
\end{eqnarray}
\end{widetext}
\noindent with $S_{jk}(\bm{R}_A,\bm{R}_B)$ being the overlap matrix element related to the primitive functions $\chi_j(\bm{r}-\bm{R}_A)$ and $\chi_k(\bm{r}-\bm{R}_B)$ (it should be kept in mind that for some particular $jk$ matrix elements $\bm{R}_A \neq \bm{R}_B$, whereas for others these position vectors are the same). In Eq.~\eqref{eq:densMO}, the sums over $\omega$ run over $\alpha$ and $\beta$, whereas the sums over $Y$ run over the large (L) and small (S) components of the one-electron wave function (see Eqs.~\eqref{eq:psi-MO} and \eqref{eq:WFbasisexp}).

Equation~\eqref{eq:densMO} can be generalized by introducing the occupation number (number of electrons) of each $m$-th MO. For a single-determinant wave function corresponding to a closed-shell system, as it is the case for all of our DHF and DFT calculations, this number ($\eta_m$) can be either 0 or 1 (it should be kept in mind that as we employ the restricted DHF method, each orbital $\Psi^{MO}_m$ has a Kramers pair $\bar{\Psi}^{MO}_m$ such that both have the same one-electron energy).

\begin{widetext}
\begin{eqnarray}\label{eq:densMO-allorb}
\sum_m^{N_\text{orb}} \eta_m \int |\Psi_m^\text{MO} ({\bm r})|^2 \, d^3{\bm r}
&=& \sum_{jk}^{N_\text{basis}} \sum_m^{N_\text{orb}} \eta_m \sum_{\omega Y} P^{\omega,Y}_{mjk} \; S_{jk}(\bm{R}_A,\bm{R}_B) = N_\text{elec},
\end{eqnarray}
\end{widetext}
where the sum
\begin{equation}
D_{jk}=\sum_m^{N_\text{orb}} \eta_m \sum_{\omega Y} P^{\omega,Y}_{mjk}
\end{equation}
corresponds to the $jk$ element of the four-component density matrix $D$.

We can also analyze the contributions arising from the integrals in each of the individual terms at the left hand side of Eq.~\eqref{eq:densMO-allorb}, writing them as
\begin{widetext}
\begin{eqnarray}\label{eq:mulliken2}
\int |\Psi_m^\text{MO} ({\bm r})|^2 \, d^3{\bm r} &=& \int \rho_m^\text{MO} ({\bm r}) \, d^3{\bm r}
= \sum_j^{N_\text{basis}} \left( \sum_k^{N_\text{basis}} \sum_{\omega Y} P^{\omega,Y}_{mjk} \; S_{jk}(\bm{R}_A,\bm{R}_B) \right)
\nonumber \\
&=& \sum_{j \, \in \, A} \left( \sum_k^{N_\text{basis}} \sum_{\omega Y} P^{\omega,Y}_{mjk} \; S_{jk}(\bm{R}_A,\bm{R}_B) \right)
+
\sum_{j \, \in \, B} \left( \sum_k^{N_\text{basis}} \sum_{\omega Y} P^{\omega,Y}_{mjk} \; S_{jk}(\bm{R}_A,\bm{R}_B) \right).
\end{eqnarray}
\end{widetext}

In Eq.~\eqref{eq:mulliken2}, the sums over $\left\{j \, \in \, A\right\}$ and $\left\{j \, \in \, B\right\}$ run over the primitives that are centered at $\bm{R}_A$ and $\bm{R}_B$, respectively, while the sums over $k$ run over all the $N_\text{basis}$ primitive functions of the basis set, centered at both the positions $\bm{R}_A$ and $\bm{R}_B$. In our study of the Tl$X$ molecules, we can analyze these contributions by splitting up some of these terms and rewriting this equation as
\begin{widetext}
\begin{eqnarray}\label{eq:densMO2}
\int |\Psi_m^\text{MO} ({\bm r})|^2 \, d^3{\bm r} &=& 
\sum_{j \, \in \, \text{Tl}} \left( \sum_{k \, \in \, \text{Tl}} \sum_{\omega Y} P^{\omega,Y}_{mjk} \; S_{jk}(\bm{R}_\text{Tl},\bm{R}_\text{Tl}) + \sum_{k \, \in \, X} \sum_{\omega Y} P^{\omega,Y}_{mjk} \; S_{jk}(\bm{R}_\text{Tl},\bm{R}_X) \right)
\nonumber \\ &&
+
\sum_{j \, \in \, X} \left( \sum_{k \, \in \, X} \sum_{\omega Y} P^{\omega,Y}_{mjk} \; S_{jk}(\bm{R}_X,\bm{R}_X) + \sum_{k \, \in \, \text{Tl}} \sum_{\omega Y} P^{\omega,Y}_{mjk} \; S_{jk}(\bm{R}_\text{Tl},\bm{R}_X) \right).
\end{eqnarray}
\end{widetext}

Furthermore, we also split up the particular contributions arising from the large components of the MO that depend on $\bm{R}_\text{Tl}$ (i.e., $Y=L$ in Eq.~\eqref{eq:WFbasisexp}) writing them as the sum of two different linear combinations of primitives. In one of them, we keep only those Gaussian functions of $s$ type that are centered at the Tl nucleus (labeled as $\chi_\zeta(\bm{r}-\bm{R}_\text{Tl})$). On the second term, we collect all the rest of primitives centered at Tl that have symmetries other than that of $s$ type (labeled $\chi_\mu(\bm{r}-\bm{R}_\text{Tl})$). Then, we can rewrite Eq.~\eqref{eq:densMO2} as
\begin{widetext}
\begin{eqnarray}\label{eq:densMO3}
\int |\Psi_m^\text{MO} ({\bm r})|^2 \, d^3{\bm r} &=& 
\sum_{\zeta \, \in \, \text{Tl}} \left( \sum_{k \, \in \, \text{Tl}} \sum_{\omega} P^{\omega,L}_{m\zeta k} \; S_{\zeta k}(\bm{R}_\text{Tl},\bm{R}_\text{Tl}) + \sum_{k \, \in \, X} \sum_{\omega} P^{\omega,L}_{m\zeta k} \; S_{\zeta k}(\bm{R}_\text{Tl},\bm{R}_X) \right)
\nonumber \\ &&
+ \sum_{\mu \, \in \, \text{Tl}} \left( \sum_{k \, \in \, \text{Tl}} \sum_{\omega} P^{\omega,L}_{m\mu k} \; S_{\mu k}(\bm{R}_\text{Tl},\bm{R}_\text{Tl}) + \sum_{k \, \in \, X} \sum_{\omega} P^{\omega,L}_{m\mu k} \; S_{\mu k}(\bm{R}_\text{Tl},\bm{R}_X) \right)
\nonumber \\ &&
+ \sum_{j \, \in \, \text{Tl}} \left( \sum_{k \, \in \, \text{Tl}} \sum_{\omega} P^{\omega,S}_{mjk} \; S_{jk}(\bm{R}_\text{Tl},\bm{R}_\text{Tl}) + \sum_{k \, \in \, X} \sum_{\omega} P^{\omega,S}_{mjk} \; S_{jk}(\bm{R}_\text{Tl},\bm{R}_X) \right)
\nonumber \\ &&
+ \sum_{j \, \in \, X} \left( \sum_{k \, \in \, X} \sum_{\omega Y} P^{\omega,Y}_{mjk} \; S_{jk}(\bm{R}_X,\bm{R}_X) + \sum_{k \, \in \, \text{Tl}} \sum_{\omega Y} P^{\omega,Y}_{mjk} \; S_{jk}(\bm{R}_\text{Tl},\bm{R}_X) \right).
\end{eqnarray}
\end{widetext}

We can connect each of the four terms in Eq.~\eqref{eq:densMO3} to each of the terms at the right hand side of Eq.~\eqref{eq:mulliken}. The first term in Eq.~\eqref{eq:densMO3} is equal to $n_{m\text{s}}^\text{Tl s, large}$; besides, the second one is equal to the sum $\sum_j n_{mj}^\text{Tl, other large}$, where each individual value $n_{mj}^\text{Tl, other large}$ is obtained as the sum over the primitives $\chi_\mu(\bm{r}-\bm{R}_\text{Tl})$ with particular symmetries (e.g., $\mathrm{p}_x$, $\mathrm{p}_y$, $\mathrm{p}_z$, $\mathrm{d}_{xx}$, $\mathrm{d}_{xy}$, etc). 

The third term in Eq.~\eqref{eq:densMO3} is equivalent to $\sum_i n_{mi}^\text{Tl, small}$. Finally, the last term in Eq.~\eqref{eq:densMO3} is equal to $\sum_k n_{mk}^X$, where each value $n_{mk}^X$ correspond in the fourth term of Eq.~\eqref{eq:densMO3} to the partial sum over the primitive functions $\chi_j(\bm{r}-\bm{R}_X)$ that have a particular symmetry (e.g., $\mathrm{p}_x$, $\mathrm{p}_y$, $\mathrm{p}_z$, $\mathrm{d}_{xx}$, $\mathrm{d}_{xy}$, etc).

\bibliographystyle{aipnum4-2}
\bibliography{refs}

\end{document}


\title{
Relativistic and quantum electrodynamics effects on NMR shielding tensors of Tl$X$ ($X$ = H, F, Cl, Br, I, At) molecules
}

\author{Karol Kozio{\l}}
\affiliation{\NCBJ}

\author{I. Agust\'{\i}n Aucar}
\affiliation{\RUG}
\affiliation{\IMIT}

\author{Konstantin Gaul}
\author{Robert Berger}
\affiliation{\Philipps}

\author{Gustavo A. Aucar}
\affiliation{\IMIT}

\newcommand{\NCBJ}{
Narodowe Centrum Bada\'{n} J\k{a}drowych (NCBJ), Andrzeja So{\l}tana 7, 05-400 Otwock-\'{S}wierk, Poland\\
\vspace*{0.3cm}
}

\newcommand{\IMIT}{
Instituto de Modelado e Innovación Tecnológica (UNNE-CONICET),
Facultad de Ciencias Exactas y Naturales y Agrimensura,
Universidad Nacional del Nordeste, Avda. Libertad 5460, Corrientes, Argentina\\
\vspace*{0.3cm}
}

\newcommand{\Philipps}{
Fachbereich Chemie, Philipps–Universit\"at Marburg, Hans-Meerwein-Stra{\ss}e 4, 35032 Marburg, Germany\\
\vspace*{0.3cm}
}

\newcommand{\RUG}{
Van Swinderen Institute for Particle Physics and Gravity,
University of Groningen, Nijenborgh 4, 9747 AG Groningen, The Netherlands\\
\vspace*{0.3cm}
}

\date{\today}

\maketitle

\tableofcontents

\section{Various computational approaches}

All the values in the Tables given in this Supplementary Material were obtained following the computational details of the main text of this manuscript. In particular, for TlTs we used an internuclear distance that was obtained by performing a structure optimization using the \textsc{Dirac} code and employing Dirac-Kohn-Sham DFT with the PBE0 functional. This distance is 3.043407~\AA{}. To get NR values, we scaled the speed of light in vacuum to 30 times its actual value. Two-component ZORA results were obtained as detailed in the computational details of the main text of this manuscript.

\begin{table}[!htb]
\small
\caption{Contributions to ${\bm \sigma}$(Tl) in Tl$^+$ and Tl$X$ ($X$ = H, F, Cl, Br, I, At, Ts), calculated at the DKS-PBE0/RPA level of approach, including (or not) 
two-electron integrals of ($SS$$\mid$$SS$) type. If not otherwise specified, CGO scheme is used with the GO placed at the CM. The label DC-$30c_0$ indicates calculations tending to the NR limit. Values in ppm.}
\begin{tabular*}{\linewidth}{@{}l @{\extracolsep{\fill}} l *{7}{S[parse-numbers=false]} @{}}
\toprule
 & &  \multicolumn{2}{c}{$pp$} & \multicolumn{2}{c}{$ee$} & \multicolumn{3}{c}{Total} \\
\cmidrule{3-4}\cmidrule{5-6}\cmidrule{7-9}
System & Method & {$\sigma_\perp$} &  {$\sigma_\parallel$} &  {$\sigma_\perp$} &  {$\sigma_\parallel$} &  {$\sigma_\perp$} &  {$\sigma_\parallel$} &  {$\sigma_\text{iso}$} \\
\midrule
Tl$^+$ & DC & \multicolumn{2}{S[parse-numbers=false]}{8161.90} & \multicolumn{2}{S[parse-numbers=false]}{8447.07} & 16608.98 & 16608.98 & 16608.98 \\
 & DC-GIAO & \multicolumn{2}{S[parse-numbers=false]}{8161.90} & \multicolumn{2}{S[parse-numbers=false]}{8447.07} & 16608.98 & 16608.98 & 16608.98 \\
 & DC-SSSS & \multicolumn{2}{S[parse-numbers=false]}{8161.09} & \multicolumn{2}{S[parse-numbers=false]}{8437.32} & 16598.42 & 16598.42 & 16598.42 \\
 & DC-$30c_0$ & \multicolumn{2}{S[parse-numbers=false]}{9884.35} & \multicolumn{2}{S[parse-numbers=false]}{8.81} & 9893.16 & 9893.16 & 9893.16 \\
\\
TlH & DC & 8171.52 & 8166.08 & -19224.17 & 11248.99 & -11052.65 & 19415.07 & -896.75 \\
 & DC-GIAO & - & - & - & - & -11053.06 & 19415.07 & -897.02 \\
 & DC-SSSS & 8170.72 & 8165.27 & -19160.49 & 11226.98 & -10989.77 & 19392.25 & -862.43 \\
 & DC-$30c_0$ & 9893.27 & 9882.91 & -8048.75 & 9.74 & 1844.52 & 9892.65 & 4527.23 \\
\\
TlF & DC & 8221.72 & 8166.99 & 1188.90 & 8678.09 & 9410.62 & 16845.08 & 11888.78 \\
 & DC-GIAO & - & - & - & - & 9410.81 & 16845.08 & 11888.90 \\
 & DC-SSSS & 8220.92 & 8166.18 & 1196.61 & 8667.46 & 9417.53 & 16833.64 & 11889.57 \\
 & DC-$30c_0$ & 9940.12 & 9884.79 & -2506.56 & 9.13 & 7433.56 & 9893.93 & 8253.69 \\
\\
TlCl & DC & 8246.95 & 8168.95 & -1998.49 & 8790.19 & 6248.46 & 16959.14 & 9818.69 \\
 & DC-GIAO & - & - & - & - & 6248.70 & 16959.14 & 9818.85 \\
 & DC-SSSS & 8246.15 & 8168.14 & -1983.19 & 8778.95 & 6262.96 & 16947.09 & 9824.34 \\
 & DC-$30c_0$ & 9965.50 & 9886.76 & -3641.36 & 9.21 & 6324.14 & 9895.97 & 7514.75 \\
\\
TlBr & DC & 8299.59 & 8170.26 & -4081.39 & 8948.15 & 4218.20 & 17118.41 & 8518.27 \\
 & DC-GIAO & - & - & - & - & 4218.36 & 17118.41 & 8518.38 \\
 & DC-SSSS & 8298.79 & 8169.45 & -4059.50 & 8936.09 & 4239.29 & 17105.54 & 8528.04 \\
 & DC-$30c_0$ & 10018.45 & 9888.04 & -4158.88 & 9.26 & 5859.57 & 9897.30 & 7205.48 \\
\\
TlI & DC & 8325.73 & 8171.90 & -7734.75 & 9286.26 & 590.98 & 17458.16 & 6213.37 \\
 & DC-GIAO & - & - & - & - & 591.28 & 17458.16 & 6213.57 \\
 & DC-SSSS & 8324.93 & 8171.10 & -7700.46 & 9272.35 & 624.47 & 17443.44 & 6230.80 \\
 & DC-$30c_0$ & 10045.52 & 9889.66 & -4935.71 & 9.50 & 5109.82 & 9899.16 & 6706.26 \\
\\
TlAt & DC & 8361.64 & 8173.24 & -14001.56 & 10355.15 & -5639.92 & 18528.39 & 2416.19 \\
 & DC-GIAO & - & - & - & - & -5639.58 & 18528.39 & 2416.41 \\
 & DC-SSSS & 8360.86 & 8172.43 & -13934.75 & 10334.49 & -5573.89 & 18506.93 & 2453.05 \\
 & DC-$30c_0$ & 10084.40 & 9890.93 & -5408.20 & 9.20 & 4676.21 & 9900.13 & 6417.51 \\
\\
TlTs & DC & 8369.08 & 8175.16 & -29801.46 & 13841.40 & -21432.37 & 22016.56 & -6949.40 \\
 & DC-GIAO & - & - & - & - & -21431.59 & 22016.56 & -6948.87 \\
 & DC-SSSS & 8368.36 & 8174.36 & -29626.04 & 13795.22 & -21257.68 & 21969.58 & -6848.60 \\
& DC-$30c_0$ & 10097.52 & 9892.41 & -6066.43 & 9.40 & 4031.09 & 9901.81 & 5987.99 \\
\bottomrule
\end{tabular*}
\end{table}

\newpage

\begin{table}[!htb]
\caption{Contributions to ${\bm \sigma}$($X$) in Tl$X$ ($X$ = H, F, Cl, Br, I, At, Ts), calculated at the DKS-PBE0/RPA level of approach, including (or not) 
two-electron integrals of ($SS$$\mid$$SS$) type. If not otherwise specified, CGO scheme is used with the GO placed at the CM. The label DC-$30c_0$ indicates calculations tending to the NR limit. Values in ppm.}
\begin{tabular*}{\linewidth}{@{}l @{\extracolsep{\fill}} l *{7}{S[parse-numbers=false]} @{}}
\toprule
 & &  \multicolumn{2}{c}{$pp$} & \multicolumn{2}{c}{$ee$} & \multicolumn{3}{c}{Total} \\
\cmidrule{3-4}\cmidrule{5-6}\cmidrule{7-9}
System & Method & {$\sigma_\perp$} &  {$\sigma_\parallel$} &  {$\sigma_\perp$} &  {$\sigma_\parallel$} &  {$\sigma_\perp$} &  {$\sigma_\parallel$} &  {$\sigma_\text{iso}$} \\
\midrule
TlH & DC & 11.07 & 44.93 & -252.62 & -61.28 & -241.54 & -16.35 & -166.48 \\
 & DC-GIAO & - & - & - & - & -241.43 & -16.35 & -166.40 \\
 & DC-SSSS & 11.08 & 44.94 & -251.80 & -60.97 & -240.72 & -16.03 & -165.83 \\
 & DC-$30c_0$ & 11.68 & 45.97 & -2.01 & 0.00 & 9.68 & 45.97 & 21.77 \\
\\
TlF & DC & 505.45 & 489.23 & -618.61 & 19.60 & -113.16 & 508.83 & 94.17 \\
 & DC-GIAO & - & - & - & - & -113.18 & 508.83 & 94.15 \\
 & DC-SSSS & 505.45 & 489.23 & -618.12 & 19.60 & -112.67 & 508.83 & 94.49 \\
 & DC-$30c_0$ & 499.91 & 483.78 & -497.17 & 0.03 & 2.74 & 483.81 & 163.09 \\
\\
TlCl & DC & 1175.86 & 1128.42 & -966.76 & 76.04 & 209.11 & 1204.46 & 540.89 \\
 & DC-GIAO & - & - & - & - & 209.09 & 1204.46 & 540.88 \\
 & DC-SSSS & 1175.87 & 1128.42 & -966.16 & 76.02 & 209.71 & 1204.45 & 541.29 \\
 & DC-$30c_0$ & 1196.33 & 1148.68 & -866.80 & 0.10 & 329.53 & 1148.79 & 602.62 \\
\\
TlBr & DC & 3023.93 & 2920.23 & -1684.30 & 565.10 & 1339.62 & 3485.32 & 2054.86 \\
 & DC-GIAO & - & - & - & - & 1339.55 & 3485.32 & 2054.81 \\
 & DC-SSSS & 3023.91 & 2920.21 & -1683.28 & 564.94 & 1340.63 & 3485.14 & 2055.47 \\
& DC-$30c_0$ & 3225.46 & 3121.21 & -1915.35 & 0.75 & 1310.11 & 3121.96 & 1914.06 \\
\\
TlI & DC & 5032.79 & 4892.33 & -1588.84 & 1972.89 & 3443.95 & 6865.21 & 4584.37 \\
 & DC-GIAO & - & - & - & - & 3443.80 & 6865.21 & 4584.27 \\
 & DC-SSSS & 5032.67 & 4892.20 & -1588.01 & 1971.96 & 3444.66 & 6864.15 & 4584.49 \\
& DC-$30c_0$ & 5648.53 & 5507.24 & -3023.35 & 2.63 & 2625.18 & 5509.86 & 3586.74 \\
\\
TlAt & DC & 8828.07 & 8642.02 & 5534.45 & 10378.97 & 14362.52 & 19020.99 & 15915.34 \\
 & DC-GIAO & - & - & - & - & 14362.43 & 19020.99 & 15915.28 \\
 & DC-SSSS & 8827.08 & 8641.01 & 5517.50 & 10365.82 & 14344.57 & 19006.82 & 15898.65 \\
 & DC-$30c_0$ & 10742.81 & 10556.67 & -5063.84 & 10.31 & 5678.97 & 10566.98 & 7308.30 \\
\\
TlTs & DC & 13082.63 & 12862.74 & 70043.33 & 53015.79 & 83125.96 & 65878.53 & 77376.81 \\
 & DC-GIAO & - & - & - & - & 83124.31 & 65878.53 & 77375.72 \\
 & DC-SSSS & 13076.70 & 12855.93 & 69689.60 & 52856.37 & 82766.30 & 65712.30 & 77081.64 \\
 & DC-$30c_0$ & 16504.72 & 16291.02 & -7260.88 & 26.71 & 9243.84 & 16317.73 & 11601.80 \\
\bottomrule
\end{tabular*}
\end{table}

\newpage

\begin{table}[!htb]
\small
\caption{Contributions to ${\bm \sigma}$(Tl) in Tl$^+$ and Tl$X$ ($X$ = H, F, Cl, Br, I, At, Ts), calculated at the DKS-LDA/RPA level of approach, including (or not) two-electron integrals of ($SS$$\mid$$SS$) type. If not otherwise specified, CGO scheme is used with the GO placed at the CM. The label DC-$30c_0$ indicates calculations tending to the NR limit. Values in ppm.}
\begin{tabular*}{\linewidth}{@{}l @{\extracolsep{\fill}} l *{7}{S[parse-numbers=false]} @{}}
\toprule
 & &  \multicolumn{2}{c}{$pp$} & \multicolumn{2}{c}{$ee$} & \multicolumn{3}{c}{Total} \\
\cmidrule{3-4}\cmidrule{5-6}\cmidrule{7-9}
System & Method & {$\sigma_\perp$} &  {$\sigma_\parallel$} &  {$\sigma_\perp$} &  {$\sigma_\parallel$} &  {$\sigma_\perp$} &  {$\sigma_\parallel$} &  {$\sigma_\text{iso}$} \\
\midrule
Tl$^+$ & DC & \multicolumn{2}{S[parse-numbers=false]}{8162.01} & \multicolumn{2}{S[parse-numbers=false]}{8429.36} & 16591.37 & 16591.37 & 16591.37 \\
 & DC-GIAO & \multicolumn{2}{S[parse-numbers=false]}{8162.01} & \multicolumn{2}{S[parse-numbers=false]}{8429.36} & 16591.37 & 16591.37 & 16591.37 \\
 & DC-SSSS & \multicolumn{2}{S[parse-numbers=false]}{8161.15} & \multicolumn{2}{S[parse-numbers=false]}{8417.06} & 16578.21 & 16578.21 & 16578.21 \\
  & DC-$30c_0$ & \multicolumn{2}{S[parse-numbers=false]}{9881.62} & \multicolumn{2}{S[parse-numbers=false]}{8.80} & 9890.42 & 9890.42 & 9890.42 \\
\\
TlH & DC & 8171.42 & 8166.42 & -20145.33 & 11955.02 & -11973.91 & 20121.44 & -1275.46 \\
 & DC-GIAO & - & - & - & - & -11974.64 & 20121.45 & -1275.94 \\
 & DC-SSSS & 8170.57 & 8165.55 & -20093.86 & 11927.94 & -11923.29 & 20093.49 & -1251.03 \\
 & DC-$30c_0$ & 9890.45 & 9880.38 & -9480.38 & 9.81 & 410.07 & 9890.19 & 3570.11 \\
 \\
TlF & DC & 8221.64 & 8167.43 & 705.25 & 8787.01 & 8926.88 & 16954.44 & 11602.73 \\
 & DC-GIAO & - & - & - & - & 8927.13 & 16954.44 & 11602.90 \\
 & DC-SSSS & 8220.78 & 8166.57 & 710.83 & 8773.22 & 8931.60 & 16939.79 & 11601.00 \\
 & DC-$30c_0$ & 9937.29 & 9882.38 & -2885.78 & 8.95 & 7051.50 & 9891.33 & 7998.11 \\
 \\
TlCl & DC & 8246.81 & 8169.37 & -2654.79 & 8937.43 & 5592.02 & 17106.80 & 9430.28 \\
 & DC-GIAO & - & - & - & - & 5592.34 & 17106.80 & 9430.50 \\
 & DC-SSSS & 8245.95 & 8168.51 & -2641.54 & 8922.85 & 5604.41 & 17091.36 & 9433.39 \\
 & DC-$30c_0$ & 9962.70 & 9884.32 & -4142.87 & 9.00 & 5819.83 & 9893.31 & 7177.66 \\
 \\
TlBr & DC & 8299.35 & 8170.67 & -4671.97 & 9136.83 & 3627.38 & 17307.50 & 8187.42 \\
 & DC-GIAO & - & - & - & - & 3627.62 & 17307.50 & 8187.58 \\
 & DC-SSSS & 8298.50 & 8169.81 & -4653.21 & 9121.31 & 3645.29 & 17291.12 & 8193.90 \\
 & DC-$30c_0$ & 10015.67 & 9885.57 & -4691.46 & 9.03 & 5324.21 & 9894.61 & 6847.68 \\
 \\
TlI & DC & 8325.44 & 8172.31 & -8126.15 & 9548.04 & 199.29 & 17720.35 & 6039.64 \\
 & DC-GIAO & - & - & - & - & 199.81 & 17720.35 & 6039.99 \\
 & DC-SSSS & 8324.58 & 8171.45 & -8097.62 & 9530.48 & 226.96 & 17701.92 & 6051.95 \\
 & DC-$30c_0$ & 10042.76 & 9887.18 & -5525.13 & 9.11 & 4517.63 & 9896.29 & 6310.52 \\
 \\
TlAt & DC & 8361.51 & 8173.61 & -13010.29 & 10625.56 & -4648.78 & 18799.17 & 3167.21 \\
 & DC-GIAO & - & - & - & - & -4648.28 & 18799.17 & 3167.54 \\
 & DC-SSSS & 8360.68 & 8172.75 & -12963.03 & 10602.25 & -4602.35 & 18775.00 & 3190.10 \\
 & DC-$30c_0$ & 10081.66 & 9888.45 & -6024.77 & 9.24 & 4056.88 & 9897.70 & 6003.82 \\
 \\
TlTs & DC & 8370.21 & 8175.21 & -21900.36 & 13040.53 & -13530.15 & 21215.74 & -1948.18 \\
 & DC-GIAO & - & - & - & - & -13529.82 & 21215.74 & -1947.96 \\
 & DC-SSSS & 8369.43 & 8174.35 & -21813.18 & 13004.32 & -13443.75 & 21178.67 & -1902.95 \\
 & DC-$30c_0$ & 10094.77 & 9889.93 & -6728.00 & 9.47 & 3366.77 & 9899.40 & 5544.31 \\
\bottomrule
\end{tabular*}
\end{table}

\newpage

\begin{table}[!htb]
\caption{Contributions to ${\bm \sigma}$($X$) in Tl$X$ ($X$ = H, F, Cl, Br, I, At, Ts), calculated at the DKS-LDA/RPA level of approach, including (or not) two-electron integrals of ($SS$$\mid$$SS$) type. If not otherwise specified, CGO scheme is used with the GO placed at the CM. The label DC-$30c_0$ indicates calculations tending to the NR limit. Values in ppm.}
\begin{tabular*}{\linewidth}{@{}l @{\extracolsep{\fill}} l *{7}{S[parse-numbers=false]} @{}}
\toprule
 & &  \multicolumn{2}{c}{$pp$} & \multicolumn{2}{c}{$ee$} & \multicolumn{3}{c}{Total} \\
\cmidrule{3-4}\cmidrule{5-6}\cmidrule{7-9}
System & Method & {$\sigma_\perp$} &  {$\sigma_\parallel$} &  {$\sigma_\perp$} &  {$\sigma_\parallel$} &  {$\sigma_\perp$} &  {$\sigma_\parallel$} &  {$\sigma_\text{iso}$} \\
\midrule
TlH & DC & 10.42 & 44.32 & -250.64 & -61.90 & -240.22 & -17.57 & -166.00 \\
 & DC-GIAO & - & - & - & - & -240.12 & -17.57 & -165.94 \\
 & DC-SSSS & 10.43 & 44.33 & -249.92 & -61.60 & -239.49 & -17.27 & -165.42 \\
 & DC-$30c_0$ & 11.10 & 45.41 & -3.30 & 0.00 & 7.80 & 45.41 & 20.33 \\
 \\
TlF & DC & 503.55 & 487.24 & -731.75 & 27.58 & -228.20 & 514.82 & 19.47 \\
 & DC-GIAO & - & - & - & - & -228.24 & 514.82 & 19.44 \\
 & DC-SSSS & 503.55 & 487.24 & -731.27 & 27.53 & -227.72 & 514.78 & 19.78 \\
 & DC-$30c_0$ & 498.09 & 481.86 & -619.24 & 0.03 & -121.16 & 481.89 & 79.86 \\
 \\
TlCl & DC & 1173.94 & 1126.38 & -1166.40 & 86.72 & 7.53 & 1213.11 & 409.39 \\
 & DC-GIAO & - & - & - & - & 7.55 & 1213.11 & 409.40 \\
 & DC-SSSS & 1173.94 & 1126.38 & -1165.79 & 86.66 & 8.16 & 1213.04 & 409.78 \\
 & DC-$30c_0$ & 1194.37 & 1146.55 & -1071.89 & 0.10 & 122.47 & 1146.66 & 463.87 \\
 \\
TlBr & DC & 3022.10 & 2918.41 & -2132.33 & 584.35 & 889.76 & 3502.75 & 1760.76 \\
 & DC-GIAO & - & - & - & - & 889.72 & 3502.75 & 1760.73 \\
 & DC-SSSS & 3022.09 & 2918.39 & -2131.27 & 584.05 & 890.82 & 3502.45 & 1761.36 \\
 & DC-$30c_0$ & 3223.14 & 3118.71 & -2355.69 & 0.74 & 867.45 & 3119.44 & 1618.11 \\
 \\
TlI & DC & 5031.42 & 4891.18 & -2294.23 & 1994.54 & 2737.19 & 6885.72 & 4120.03 \\
 & DC-GIAO & - & - & - & - & 2737.02 & 6885.72 & 4119.92 \\
 & DC-SSSS & 5031.31 & 4891.07 & -2293.54 & 1993.22 & 2737.77 & 6884.29 & 4119.95 \\
 & DC-$30c_0$ & 5646.17 & 5504.68 & -3698.06 & 2.53 & 1948.11 & 5507.21 & 3134.48 \\
 \\
TlAt & DC & 8827.23 & 8642.24 & 3443.21 & 10244.99 & 12270.43 & 18887.23 & 14476.03 \\
 & DC-GIAO & - & - & - & - & 12270.33 & 18887.23 & 14475.97 \\
 & DC-SSSS & 8826.13 & 8641.13 & 3428.12 & 10229.21 & 12254.26 & 18870.34 & 14459.62 \\
 & DC-$30c_0$ & 10740.19 & 10553.86 & -6132.48 & 10.38 & 4607.71 & 10564.24 & 6593.22 \\
 \\
TlTs & DC & 13080.73 & 12864.62 & 49322.47 & 51665.34 & 62403.20 & 64529.95 & 63112.12 \\
 & DC-GIAO & - & - & - & - & 62402.72 & 64529.95 & 63111.79 \\
 & DC-SSSS & 13072.05 & 12855.94 & 49079.43 & 51474.77 & 62151.48 & 64330.71 & 62877.89 \\
 & DC-$30c_0$ & 16502.17 & 16288.27 & -8748.62 & 26.96 & 7753.55 & 16315.23 & 10607.44 \\
\bottomrule
\end{tabular*}
\end{table}

\newpage

\begin{table}[!htb]
\small
\caption{Contributions to ${\bm \sigma}$(Tl) in Tl$^+$ and Tl$X$ ($X$ = H, F, Cl, Br, I, At, Ts), calculated at the DHF/RPA level of approach, including (or not) selectively 
two-electron integrals of ($SS$$\mid$$SS$) type. If not otherwise specified, CGO scheme is used with the GO placed at the CM. Values in ppm.}
\begin{tabular*}{\linewidth}{@{}l @{\extracolsep{\fill}} l *{7}{S[parse-numbers=false]} @{}}
\toprule
 & &  \multicolumn{2}{c}{$pp$} & \multicolumn{2}{c}{$ee$} & \multicolumn{3}{c}{Total} \\
\cmidrule{3-4}\cmidrule{5-6}\cmidrule{7-9}
System & Method & {$\sigma_\perp$} &  {$\sigma_\parallel$} &  {$\sigma_\perp$} &  {$\sigma_\parallel$} &  {$\sigma_\perp$} &  {$\sigma_\parallel$} &  {$\sigma_\text{iso}$} \\
\midrule
Tl$^+$ & DC & \multicolumn{2}{S[parse-numbers=false]}{8159.10} & \multicolumn{2}{S[parse-numbers=false]}{8433.73} & 16592.83 & 16592.83 & 16592.83 \\
 & DC-GIAO & \multicolumn{2}{S[parse-numbers=false]}{8159.10} & \multicolumn{2}{S[parse-numbers=false]}{8433.73} & 16592.83 & 16592.83 & 16592.83 \\
 & DC-SSSS & \multicolumn{2}{S[parse-numbers=false]}{8158.45} & \multicolumn{2}{S[parse-numbers=false]}{8431.69} & 16590.14 & 16590.14 & 16590.14 \\
 & DC-$30c_0$ & \multicolumn{2}{S[parse-numbers=false]}{9884.56} & \multicolumn{2}{S[parse-numbers=false]}{8.64} & 9893.20 & 9893.20 & 9893.20 \\[1.5ex]
\\
%
TlH & DC & 8173.47 & 8162.10 & -28076.00 & 10988.56 & -19902.53 & 19150.66 & -6884.80 \\
 & DC-GIAO & - & - & - & - & -19902.88 & 19150.66 & -6885.03 \\
 & DC-SSSS & 8172.91 & 8161.55 & -27966.46 & 10974.16 & -19793.55 & 19135.71 & -6817.13 \\
 & DC-$30c_0$ & 9893.50 & 9882.78 & -6932.15 & 0.34 & 2961.36 & 9883.12 & 5268.61 \\
\\
%
TlF & DC & 8220.21 & 8163.55 & 1589.06 & 8349.86 & 9809.27 & 16513.41 & 12043.98 \\
 & DC-GIAO & - & - & - & - & 9809.23 & 16513.41 & 12043.96 \\
 & DC-SSSS & 8219.58 & 8162.91 & 1601.53 & 8348.26 & 9821.11 & 16511.17 & 12051.13 \\
 & DC-$30c_0$ & 9940.42 & 9884.45 & -1964.67 & 9.13 & 7975.75 & 9893.57 & 8615.03 \\
\\
%
TlCl & DC & 8246.05 & 8165.62 & -2140.00 & 8364.76 & 6106.05 & 16530.38 & 9580.83 \\
 & DC-GIAO & - & - & - & - & 6105.92 & 16530.38 & 9580.74 \\
 & DC-SSSS & 8245.43 & 8164.98 & -2118.37 & 8362.98 & 6127.06 & 16527.96 & 9594.03 \\
 & DC-$30c_0$ & 9965.69 & 9886.56 & -3040.67 & 9.10 & 6925.02 & 9895.66 & 7915.23 \\
\\
%
TlBr & DC & 8299.33 & 8166.89 & -4428.62 & 8414.16 & 3870.72 & 16581.05 & 8107.49 \\
 & DC-GIAO & - & - & - & - & 3870.48 & 16581.05 & 8107.33 \\
 & DC-SSSS & 8298.72 & 8166.25 & -4399.59 & 8412.06 & 3899.13 & 16578.31 & 8125.52 \\
 & DC-$30c_0$ & 10018.61 & 9887.85 & -3500.92 & 9.34 & 6517.69 & 9897.19 & 7644.19 \\
\\
%
TlI & DC & 8326.37 & 8168.46 & -9078.85 & 8568.99 & -752.48 & 16737.45 & 5077.50 \\
 & DC-GIAO & - & - & - & - & -752.66 & 16737.45 & 5077.38 \\
 & DC-SSSS & 8325.77 & 8167.84 & -9033.20 & 8565.86 & -707.43 & 16733.69 & 5106.28 \\
 & DC-$30c_0$ & 10045.65 & 9889.50 & -4246.27 & 9.43 & 5799.38 & 9898.93 & 7165.90 \\
\\
%
TlAt & DC & 8363.70 & 8169.55 & -20124.32 & 9519.46 & -11760.62 & 17689.01 & -1944.08 \\
 & DC-GIAO & - & - & - & - & -11760.47 & 17689.01 & -1943.98 \\
 & DC-SSSS & 8363.14 & 8168.94 & -20008.30 & 9509.61 & -11645.15 & 17678.55 & -1870.59 \\
 & DC-$30c_0$ & 10084.51 & 9890.79 & -4702.16 & 10.26 & 5382.36 & 9901.05 & 6888.59 \\
\\
%
TlTs & DC & 8374.19 & 8170.99 & -303862.19 & 157016.39 & -295488.00 & 165187.38 & -141929.54 \\
 & DC-GIAO & - & - & - & - & -295478.80 & 165187.40 & -141923.40 \\
 & DC-SSSS & 8374.27 & 8172.23 & -291136.97 & 116162.05 & -282762.70 & 124334.28 & -147063.71 \\
 & DC-$30c_0$ & 10097.61 & 9892.29 & -5376.46 & 10.81 & 4721.15 & 9903.11 & 6448.47 \\
\bottomrule
\end{tabular*}
\end{table}

\newpage

\begin{table}[!htb]
\caption{Contributions to ${\bm \sigma}$($X$) in Tl$X$ ($X$ = H, F, Cl, Br, I, At, Ts), calculated at the DHF/RPA level of approach, including (or not) selectively 
two-electron integrals of ($SS$$\mid$$SS$) type. If not otherwise specified, CGO scheme is used with the GO placed at the CM. Values in ppm.}
\begin{tabular*}{\linewidth}{@{}l @{\extracolsep{\fill}} l *{7}{S[parse-numbers=false]} @{}}
\toprule
 & &  \multicolumn{2}{c}{$pp$} & \multicolumn{2}{c}{$ee$} & \multicolumn{3}{c}{Total} \\
\cmidrule{3-4}\cmidrule{5-6}\cmidrule{7-9}
System & Method & {$\sigma_\perp$} &  {$\sigma_\parallel$} &  {$\sigma_\perp$} &  {$\sigma_\parallel$} &  {$\sigma_\perp$} &  {$\sigma_\parallel$} &  {$\sigma_\text{iso}$} \\
\midrule
TlH & DC & 10.91 & 45.08 & -383.52 & -130.31 & -372.61 & -85.24 & -276.82 \\
 & DC-GIAO & - & - & - & - & -372.50 & -85.24 & -276.74 \\
 & DC-SSSS & 10.91 & 45.08 & -382.19 & -129.67 & -371.28 & -84.59 & -275.72 \\
 & DC-$30c_0$ & 11.70 & 46.14 & -0.12 & -0.01 & 11.58 & 46.13 & 23.10 \\
 \\
TlF & DC & 505.80 & 489.78 & -463.47 & 7.25 & 42.32 & 497.03 & 193.89 \\
 & DC-GIAO & - & - & - & - & 42.26 & 497.03 & 193.85 \\
 & DC-SSSS & 505.80 & 489.78 & -463.06 & 7.29 & 42.74 & 497.07 & 194.18 \\
 & DC-$30c_0$ & 500.13 & 484.22 & -327.48 & 0.04 & 172.65 & 484.26 & 276.52 \\
\\
TlCl & DC & 1175.88 & 1128.63 & -813.77 & 59.15 & 362.11 & 1187.77 & 637.33 \\
 & DC-GIAO & - & - & - & - & 362.09 & 1187.77 & 637.32 \\
 & DC-SSSS & 1175.88 & 1128.62 & -813.12 & 59.21 & 362.76 & 1187.84 & 637.78 \\
 & DC-$30c_0$ & 1196.21 & 1148.83 & -652.02 & 0.13 & 544.20 & 1148.96 & 745.79 \\
\\
TlBr & DC & 3024.04 & 2920.41 & -1362.23 & 528.24 & 1661.81 & 3448.64 & 2257.42 \\
 & DC-GIAO & - & - & - & - & 1661.68 & 3448.64 & 2257.34 \\
 & DC-SSSS & 3024.02 & 2920.37 & -1360.88 & 528.40 & 1663.14 & 3448.77 & 2258.35 \\
 & DC-$30c_0$ & 3225.47 & 3121.52 & -1437.09 & 0.80 & 1788.38 & 3122.31 & 2233.02 \\
\\
TlI & DC & 5032.24 & 4891.55 & -1218.88 & 1914.46 & 3813.36 & 6806.01 & 4810.91 \\
 & DC-GIAO & - & - & - & - & 3813.16 & 6806.01 & 4810.78 \\
 & DC-SSSS & 5032.08 & 4891.38 & -1216.56 & 1914.76 & 3815.52 & 6806.13 & 4812.39 \\
 & DC-$30c_0$ & 5648.10 & 5507.13 & -2329.44 & 2.63 & 3318.66 & 5509.76 & 4049.03 \\
\\
TlAt & DC & 8825.75 & 8638.56 & 8146.39 & 10682.45 & 16972.14 & 19321.01 & 17755.09 \\
 & DC-GIAO & - & - & - & - & 16971.92 & 19321.01 & 17754.95 \\
 & DC-SSSS & 8825.07 & 8637.83 & 8132.71 & 10678.13 & 16957.78 & 19315.96 & 17743.84 \\
 & DC-$30c_0$ & 10742.62 & 10556.81 & -3927.91 & 10.98 & 6814.71 & 10567.79 & 8065.73 \\
\\
TlTs & DC & 12994.13 & 12762.94 & 570242.14 & 656632.27 & 583236.27 & 669395.21 & 611955.92 \\
 & DC-GIAO & - & - & - & - & 583214.24 & 669395.29 & 611941.26 \\
 & DC-SSSS & 13055.34 & 12832.73 & 549097.72 & 485780.22 & 562153.07 & 498612.95 & 540973.03 \\
 & DC-$30c_0$ & 16504.02 & 16290.64 & -5733.20 & 27.63 & 10770.82 & 16318.27 & 12619.97 \\
%
\bottomrule
\end{tabular*}
\end{table}

\newpage

\begin{table*}[!htb]
\begin{threeparttable}
\caption{Deviation (dev.) of 2C ZORA (ZORA), ZORA with renormalized orbitals (rZORA) and ZORA with renormalized orbitals and re-scaled with X2C orbital energies (rZORA+X2Ce) results for the tensor elements of ${\bm \sigma}$(Tl) in TlTs molecules from 4C DC including integrals of $(SS|SS)$ type (devDC) calculations. Results for all other Tl$X$ can be found in the main text. All results were obtained at the Hartree-Fock level. Contributions of the diamagnetic-like expectation values (Expec) and the paramagnetic-like linear responses (LR) are shown separately. Values in ppm.}
\begin{tabular*}{\linewidth}{@{}l @{\extracolsep{\fill}} r @{\hspace{1.5em}} *{4}{S[parse-numbers=false, table-format=8.1]} r *{4}{S[parse-numbers=false, table-format=8.1]} @{}}
\toprule
Molecule & Method
 &   \multicolumn{4}{c}{$\sigma_\perp$}
 &&  \multicolumn{4}{c}{$\sigma_\parallel$}  \\
 \cmidrule{3-6}\cmidrule{7-11}
 &&  {Expec} & {LR} & {Total} & {devDC/\%} &&  {Expec} & {LR} & {Total} & {devDC/\%}  \\
\midrule
         \multirow{3}{*}{TlTs} &       ZORA &     10353.1 &  -303633.8 &  -293280.7 &  -16.5 &&    10146.9 &   147455.2 &   157602.1 & -106.1\\
                               &      rZORA &     10002.9 &  -304578.6 &  -294575.7 &  -17.0 &&     9798.9 &   152846.4 &   162645.3 & -112.6\\
                               & rZORA+X2Ce &     10002.9 &  -312327.6 &  -302324.7 &  -20.1 &&     9798.9 &   209414.4 &   219213.3 & -186.6\\
\\
\bottomrule
\end{tabular*}
\begin{tablenotes}
\item[*] Huge relative deviations are originated in partial cancellations of Exp and LR contributions that lead to much smaller total values. Relative deviations are therefore not meaningful for the contributions but only for the total isotropic value. However, still the great improvement of X2C orbital corrected results can be seen.
\end{tablenotes}
\end{threeparttable}
\end{table*}

\begin{table*}[!htb]
\begin{threeparttable}
\caption{Deviation (dev.) of two-component ZORA (ZORA), ZORA with renormalized orbitals (rZORA) and ZORA with renormalized orbitals and X2C orbital energies (rZORA+X2Ce) results for the tensor elements of ${\bm \sigma}$(Tl) in Tl$^+$ ion and Tl$X$ ($X$ = H, F, Cl, Br, I, At) molecules from four-component DC including integrals of $(SS|SS)$ type (devDC) calculations. All results were obtained at the ZORA-PBE0 level. Contributions of the diamagnetic-like expectation values (Expec) and the paramagnetic-like linear responses (LR) are shown separately. Values in ppm.}
\begin{tabular*}{\linewidth}{@{}l @{\extracolsep{\fill}} r @{\hspace{1.5em}} *{4}{S[parse-numbers=false, table-format=7.1]} r *{4}{S[parse-numbers=false, table-format=7.1]} @{}}
\toprule
Molecule & Method
 &  \multicolumn{4}{c}{$\sigma_\perp$} 
 &&  \multicolumn{4}{c}{$\sigma_\parallel$}  \\
 \cmidrule{3-6}\cmidrule{8-11}
 &&  {Expec} & {LR} & {Total} & {devDC/\%} &&  {Expec} & {LR} & {Total} & {devDC/\%} \\
\midrule
          \multirow{3}{*}{Tl+} &       ZORA &     10134.6 &     4886.9 &    15021.5 &    9.5 &&    10134.6 &     4886.9 &    15021.5 &    9.5\\
                               &      rZORA &      9786.6 &     5579.6 &    15366.2 &    7.4 &&     9786.6 &     5579.6 &    15366.2 &    7.4\\
                               & rZORA+X2Ce &      9786.6 &     7723.3 &    17509.9 &   -5.5 &&     9786.6 &     7723.3 &    17509.9 &   -5.5\\
\\
          \multirow{3}{*}{TlH} &       ZORA &     10147.1 &   -24116.8 &   -13969.7 &  -27.1 &&    10137.9 &     7524.2 &    17662.1 &    8.9\\
                               &      rZORA &      9799.2 &   -23473.9 &   -13674.7 &  -24.4 &&     9790.0 &     8212.1 &    18002.1 &    7.2\\
                               & rZORA+X2Ce &      9799.2 &   -21394.9 &   -11595.7 &   -5.5 &&     9790.0 &    10340.0 &    20130.0 &   -3.8\\                               
\\
          \multirow{3}{*}{TlF} &       ZORA &     10194.8 &    -2201.1 &     7993.7 &   15.1 &&    10139.6 &     5091.5 &    15231.1 &    9.5\\
                               &      rZORA &      9846.8 &    -1519.9 &     8326.9 &   11.6 &&     9791.6 &     5783.3 &    15574.9 &    7.5\\
                               & rZORA+X2Ce &      9846.8 &      608.5 &    10455.3 &  -11.0 &&     9791.6 &     7925.9 &    17717.6 &   -5.3\\
\\
         \multirow{3}{*}{TlCl} &       ZORA &     10219.8 &    -5254.5 &     4965.4 &   20.7 &&    10141.4 &     5188.2 &    15329.6 &    9.5\\
                               &      rZORA &      9871.9 &    -4577.3 &     5294.6 &   15.5 &&     9793.5 &     5879.7 &    15673.2 &    7.5\\
                               & rZORA+X2Ce &      9871.9 &    -2439.6 &     7432.2 &  -18.7 &&     9793.5 &     8039.4 &    17832.9 &   -5.2\\
\\
         \multirow{3}{*}{TlBr} &       ZORA &     10272.1 &    -7199.5 &     3072.6 &   27.5 &&    10142.8 &     5318.2 &    15461.0 &    9.6\\
                               &      rZORA &      9923.9 &    -6524.3 &     3399.5 &   19.8 &&     9794.8 &     6009.5 &    15804.3 &    7.6\\
                               & rZORA+X2Ce &      9923.9 &    -4364.5 &     5559.4 &  -31.1 &&     9794.8 &     8193.1 &    17987.9 &   -5.2\\
\\
          \multirow{3}{*}{TlI} &       ZORA &     10299.0 &   -10606.7 &     -307.7 &  149.3 &&    10144.3 &     5585.6 &    15729.9 &    9.8\\
                               &      rZORA &      9950.5 &    -9936.4 &       14.1 &   97.7 &&     9796.4 &     6276.7 &    16073.1 &    7.9\\
                               & rZORA+X2Ce &      9950.5 &    -7830.0 &     2120.5 & -239.6 &&     9796.4 &     8417.5 &    18213.9 &   -4.4\\
\\
         \multirow{3}{*}{TlAt} &       ZORA &     10334.5 &   -16443.8 &    -6109.3 &   -9.6 &&    10145.3 &     6368.4 &    16513.7 &   10.8\\
                               &      rZORA &      9985.2 &   -15782.6 &    -5797.3 &   -4.0 &&     9797.3 &     7060.2 &    16857.5 &    8.9\\
                               & rZORA+X2Ce &      9985.2 &   -13697.2 &    -3712.0 &   33.4 &&     9797.3 &     9200.2 &    18997.6 &   -2.7\\
\\
         \multirow{3}{*}{TlTs} &       ZORA &     10353.1 &   -30331.5 &   -19978.4 &    6.0 &&    10147.0 &     8868.6 &    19015.6 &   13.4\\
                               &      rZORA &     10002.9 &   -29692.2 &   -19689.3 &    7.4 &&     9799.1 &     9565.8 &    19364.9 &   11.9\\
                               & rZORA+X2Ce &     10002.9 &   -27670.0 &   -17667.1 &   16.9 &&     9799.1 &    11694.8 &    21493.9 &    2.2\\
\bottomrule
\end{tabular*}
\end{threeparttable}
\end{table*}

\begin{table*}[!htb]
\begin{threeparttable}
\caption{Deviation (dev.) of two-component ZORA with renormalized orbitals (rZORA) results for the tensor elements of ${\bm \sigma}$(Tl) in Tl$^+$ ion and Tl$X$ ($X$ = H, F, Cl, Br, I, At) molecules from four-component DC including integrals of $(SS|SS)$ type (devDC) calculations. All results were obtained at the ZORA-LDA level. Contributions of the diamagnetic-like expectation values (Expec) and the paramagnetic-like linear responses (LR) are shown separately. Values in ppm.}
\begin{tabular*}{\linewidth}{@{}l @{\extracolsep{\fill}} r @{\hspace{1.5em}} *{4}{S[parse-numbers=false, table-format=7.1]} r *{4}{S[parse-numbers=false, table-format=7.1]} @{}}
\toprule
Molecule & Method
 &  \multicolumn{4}{c}{$\sigma_\perp$} 
 &&  \multicolumn{4}{c}{$\sigma_\parallel$}  \\
 \cmidrule{3-6}\cmidrule{8-11}
 &&  {Expec} & {LR} & {Total} & {devDC/\%} &&  {Expec} & {LR} & {Total} & {devDC/\%}\\
\midrule
          \multirow{1}{*}{Tl+} &      rZORA &      9784.3 &     5545.1 &    15329.4 &    7.5 &&     9784.3 &     5545.1 &    15329.4 &    7.5\\
\\
          \multirow{1}{*}{TlH} &      rZORA &      9796.8 &   -22905.7 &   -13108.9 &   -9.9 &&     9787.9 &     9042.8 &    18830.7 &    6.3\\          
\\
          \multirow{1}{*}{TlF} &      rZORA &      9844.4 &    -2145.3 &     7699.1 &   13.8 &&     9789.6 &     5899.5 &    15689.1 &    7.4\\
\\
         \multirow{1}{*}{TlCl} &      rZORA &      9869.5 &    -5489.4 &     4380.1 &   21.8 &&     9791.5 &     6048.4 &    15839.8 &    7.3\\
\\
         \multirow{1}{*}{TlBr} &      rZORA &      9921.5 &    -7494.9 &     2426.6 &   33.4 &&     9792.7 &     6246.0 &    16038.8 &    7.2\\
\\
          \multirow{1}{*}{TlI} &      rZORA &      9948.1 &   -10931.0 &     -982.9 &  533.1 &&     9794.3 &     6653.7 &    16448.1 &    7.1\\
\\
         \multirow{1}{*}{TlAt} &      rZORA &      9982.9 &   -15961.6 &    -5978.8 &  -29.9 &&     9795.3 &     7756.9 &    17552.1 &    6.5\\
\\
         \multirow{1}{*}{TlTs} &      rZORA &     10000.5 &   -24605.7 &   -14605.1 &   -8.6 &&     9797.0 &    10116.6 &    19913.6 &    6.0\\
\bottomrule
\end{tabular*}
\end{threeparttable}
\end{table*}

\begin{sidewaystable}
\begin{threeparttable}
\footnotesize
\caption{Contributions from spin-free (SF) and spin-dependent (SD) parts to the response function of NMR shielding tensors at the two-component ZORA level (ZORA), with renormalized orbitals (rZORA) as well as with renormalized orbitals and X2C orbital energies (rZORA+X2Ce). Results for the tensor elements of ${\bm \sigma}$(Tl) in Tl$^+$ ion and Tl$X$ ($X$ = H, F, Cl, Br, I, At) are shown at the HF and DFT PBE0 level. Values in ppm.}
\begin{tabular*}{\linewidth}{@{}l @{\extracolsep{\fill}} r @{\hspace{1.5em}} *{3}{S[parse-numbers=false, table-format=7.1]} r *{3}{S[parse-numbers=false, table-format=7.1]} @{}}
\toprule
Molecule & Method
 &  \multicolumn{3}{c}{$\sigma_\perp$} 
 &&  \multicolumn{3}{c}{$\sigma_\parallel$}  \\
 \cmidrule{3-5}\cmidrule{7-9}
 &&  {SFSF} & {SDSD} & {SDSF} &&  {SFSF} & {SDSD} & {SDSF} \\
\midrule
          \multirow{7}{*}{Tl+} &              HF-ZORA &       436.0 &     4435.6 &        6.4 &&      436.0 &     4435.6 &        6.4 \\
                               &             HF-rZORA &       444.2 &     5101.2 &       11.7 &&      444.2 &     5101.2 &       11.7 \\
                               &        HF-rZORA+X2Ce &       466.4 &     7138.2 &       26.2 &&      466.4 &     7138.2 &       26.2 \\
\\
                               &            PBE0-ZORA &       438.1 &     4434.7 &       14.0 &&      438.1 &     4434.7 &       14.0 \\
                               &           PBE0-rZORA &       446.1 &     5114.0 &       19.5 &&      446.1 &     5114.0 &       19.5 \\
                               &      PBE0-rZORA+X2Ce &       468.2 &     7220.8 &       34.3 &&      468.2 &     7220.8 &       34.3 \\
                               &            LDA-rZORA &       447.2 &     5078.8 &       19.0 &&      447.2 &     5078.8 &       19.0 \\
\\
          \multirow{7}{*}{TlF} &              HF-ZORA &     -2211.2 &     3528.1 &    -3315.1 &&      388.6 &     3775.7 &      628.6 \\
                               &             HF-rZORA &     -2203.8 &     4191.2 &    -3319.3 &&      396.8 &     4438.4 &      635.7 \\
                               &        HF-rZORA+X2Ce &     -2179.9 &     6228.2 &    -3306.6 &&      418.9 &     6475.7 &      650.0 \\
\\
                               &            PBE0-ZORA &     -3325.6 &     3835.8 &    -2711.4 &&      313.2 &     4408.2 &      370.1 \\
                               &           PBE0-rZORA &     -3317.9 &     4513.3 &    -2715.3 &&      321.1 &     5085.4 &      376.8 \\
                               &      PBE0-rZORA+X2Ce &     -3294.2 &     6618.3 &    -2715.7 &&      343.2 &     7188.9 &      393.8 \\
                               &            LDA-rZORA &     -4060.8 &     4501.7 &    -2586.1 &&      235.5 &     5316.8 &      347.3 \\
\\
         \multirow{7}{*}{TlCl} &              HF-ZORA &     -3735.8 &     3176.7 &    -5182.8 &&      366.9 &     3557.2 &      884.0 \\
                               &             HF-rZORA &     -3728.8 &     3838.9 &    -5192.1 &&      375.0 &     4218.8 &      891.8 \\
                               &        HF-rZORA+X2Ce &     -3705.1 &     5888.6 &    -5186.6 &&      397.1 &     6268.9 &      906.8 \\
\\
                               &            PBE0-ZORA &     -5008.3 &     3634.7 &    -3880.9 &&      262.2 &     4414.3 &      511.6 \\
                               &           PBE0-rZORA &     -5000.4 &     4311.7 &    -3888.5 &&      270.2 &     5090.9 &      518.6 \\
                               &      PBE0-rZORA+X2Ce &     -4976.9 &     6433.5 &    -3896.2 &&      292.1 &     7211.0 &      536.3 \\
                               &            LDA-rZORA &     -6007.7 &     4294.6 &    -3776.2 &&      142.9 &     5406.4 &      499.1 \\
\\
         \multirow{7}{*}{TlBr} &              HF-ZORA &     -4353.7 &     2895.8 &    -6585.0 &&      346.6 &     3419.7 &     1092.2 \\
                               &             HF-rZORA &     -4346.2 &     3557.1 &    -6597.6 &&      354.7 &     4080.7 &     1100.5 \\
                               &        HF-rZORA+X2Ce &     -4321.7 &     5632.5 &    -6595.2 &&      377.0 &     6156.9 &     1116.4 \\
\\
                               &            PBE0-ZORA &     -5832.9 &     3427.3 &    -4793.8 &&      199.8 &     4436.8 &      681.6 \\
                               &           PBE0-rZORA &     -5824.6 &     4103.6 &    -4803.4 &&      207.7 &     5113.0 &      688.8 \\
                               &      PBE0-rZORA+X2Ce &     -5800.1 &     6248.1 &    -4812.5 &&      229.9 &     7256.0 &      707.2 \\
                               &            LDA-rZORA &     -6963.5 &     4091.4 &    -4622.9 &&       24.4 &     5533.7 &      687.9 \\
\\
          \multirow{7}{*}{TlI} &              HF-ZORA &     -5246.4 &     2335.1 &    -9808.7 &&      305.2 &     3183.8 &     1527.1 \\
                               &             HF-rZORA &     -5238.6 &     2994.9 &    -9829.3 &&      313.3 &     3843.6 &     1536.3 \\
                               &        HF-rZORA+X2Ce &     -5215.0 &     5026.6 &    -9836.3 &&      335.0 &     5877.0 &     1553.5 \\
\\
                               &            PBE0-ZORA &     -7011.6 &     3043.3 &    -6638.4 &&       76.3 &     4473.3 &     1036.0 \\
                               &           PBE0-rZORA &     -7002.9 &     3718.7 &    -6652.2 &&       84.2 &     5148.8 &     1043.7 \\
                               &      PBE0-rZORA+X2Ce &     -6979.6 &     5819.1 &    -6669.5 &&      105.7 &     7247.8 &     1064.0 \\
                               &            LDA-rZORA &     -8401.8 &     3724.4 &    -6253.6 &&     -218.6 &     5788.3 &     1084.0 \\
\\
         \multirow{7}{*}{TlAt} &              HF-ZORA &     -5457.3 &     1569.3 &   -20375.6 &&      191.2 &     3449.8 &     2293.8 \\
                               &             HF-rZORA &     -5449.1 &     2227.8 &   -20423.6 &&      199.2 &     4110.5 &     2303.3 \\
                               &        HF-rZORA+X2Ce &     -5722.6 &     4266.1 &   -20425.1 &&      221.2 &     6146.3 &     2320.0 \\
\\
                               &            PBE0-ZORA &     -7969.5 &     2377.7 &   -10852.0 &&     -263.9 &     4639.9 &     1992.4 \\
                               &           PBE0-rZORA &     -7960.6 &     3052.5 &   -10874.5 &&     -256.0 &     5316.5 &     1999.7 \\
                               &      PBE0-rZORA+X2Ce &     -7937.8 &     5150.1 &   -10909.6 &&     -234.3 &     7415.2 &     2019.3 \\
                               &            LDA-rZORA &     -9866.3 &     3185.0 &    -9280.3 &&     -847.2 &     6510.8 &     2093.3 \\
\\
         \multirow{7}{*}{TlTs} &              HF-ZORA &     -6228.1 &    -4130.5 &  -293275.0 &&     -214.6 &     8161.8 &   139508.0 \\
                               &             HF-rZORA &     -6220.6 &    -3487.1 &  -294871.0 &&     -206.8 &     8836.2 &   144217.0 \\
                               &        HF-rZORA+X2Ce &     -6204.6 &    -1545.1 &  -304578.0 &&     -185.2 &    10873.5 &   198726.0 \\
\\
                               &            PBE0-ZORA &     -9935.9 &     1161.4 &   -21557.1 &&    -1060.0 &     5614.0 &     4314.7 \\
                               &           PBE0-rZORA &     -9927.5 &     1838.1 &   -21602.7 &&    -1052.2 &     6299.7 &     4318.3 \\
                               &      PBE0-rZORA+X2Ce &     -9909.7 &     3927.0 &   -21687.3 &&    -1030.8 &     8391.5 &     4334.0 \\
                               &            LDA-rZORA &    -12395.2 &     2477.6 &   -14688.1 &&    -2061.8 &     8047.8 &     4130.7 \\
\\
          \multirow{7}{*}{TlH} &              HF-ZORA &     -7723.6 &    -2662.2 &   -21483.5 &&      159.9 &     2032.5 &     5294.7 \\
                               &             HF-rZORA &     -7719.5 &    -2013.7 &   -21531.5 &&      167.9 &     2683.4 &     5312.3 \\
                               &        HF-rZORA+X2Ce &     -7696.9 &       17.3 &   -21543.1 &&      190.0 &     4713.0 &     5326.3 \\
\\
                               &            PBE0-ZORA &     -9743.8 &      217.0 &   -14590.0 &&     -370.1 &     5071.9 &     2822.4 \\
                               &           PBE0-rZORA &     -9737.0 &      885.5 &   -14622.4 &&     -362.2 &     5740.0 &     2834.2 \\
                               &      PBE0-rZORA+X2Ce &     -9710.1 &     2975.8 &   -14660.6 &&     -340.0 &     7818.6 &     2861.3 \\
                               &            LDA-rZORA &    -11828.9 &      811.5 &   -11888.2 &&    -1364.2 &     7236.2 &     3170.7 \\

\bottomrule
\end{tabular*}
\end{threeparttable}
\end{sidewaystable}

\clearpage

\begin{figure*}[!htb]
\centering
\includegraphics[width=0.48\linewidth]{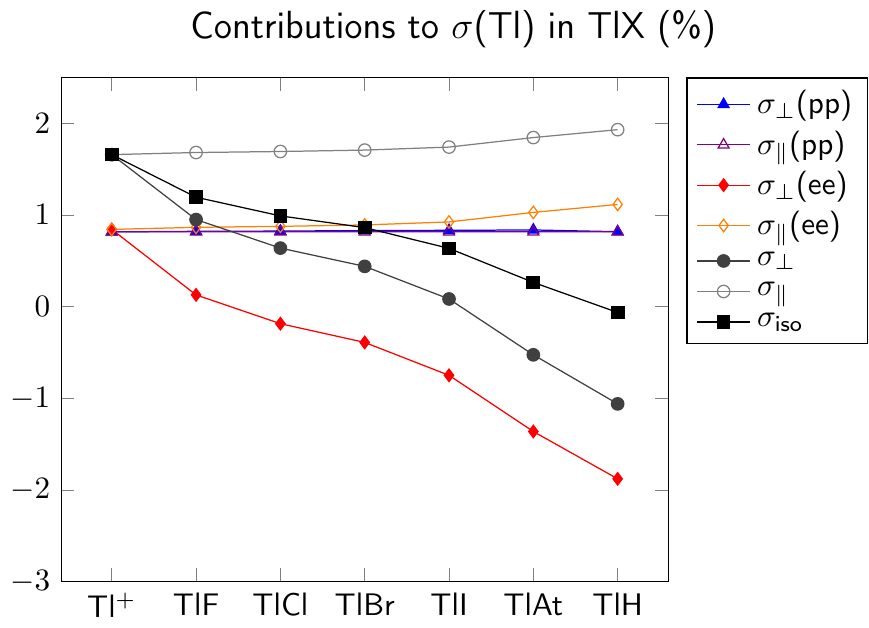}
\hfill
\includegraphics[width=0.48\linewidth]{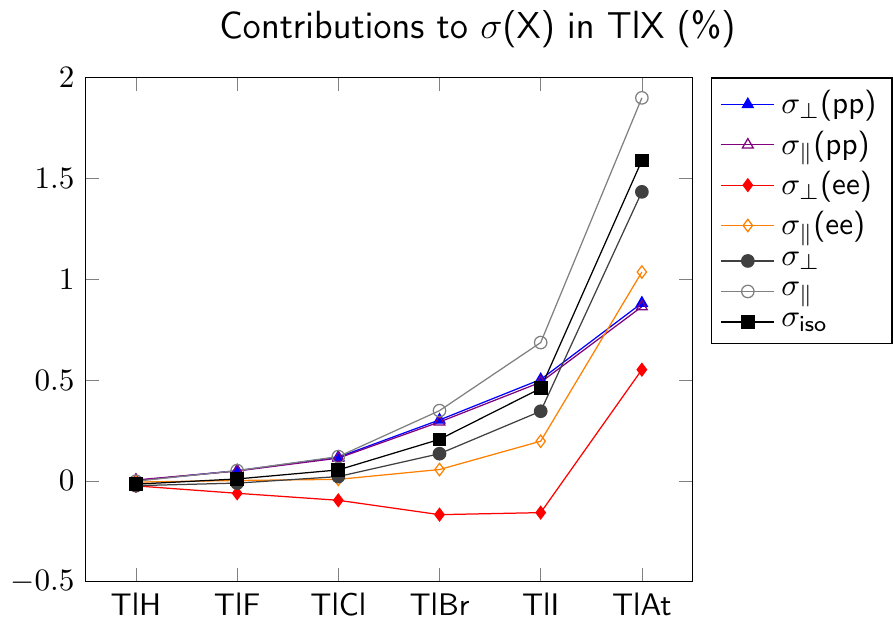}
\caption{Contributions to ${\bm \sigma}$(Tl) (left) and ${\bm \sigma}$($X$) (right) in Tl$^+$ ion and Tl$X$ ($X$ = H, F, Cl, Br, I, At) molecules, calculated using the DFT/PBE0 approach. }
\end{figure*}






\newpage

\section{Pattern of excitations - DHF approach}

\begin{figure*}[!htb]
\centering
\includegraphics[width=0.48\linewidth]{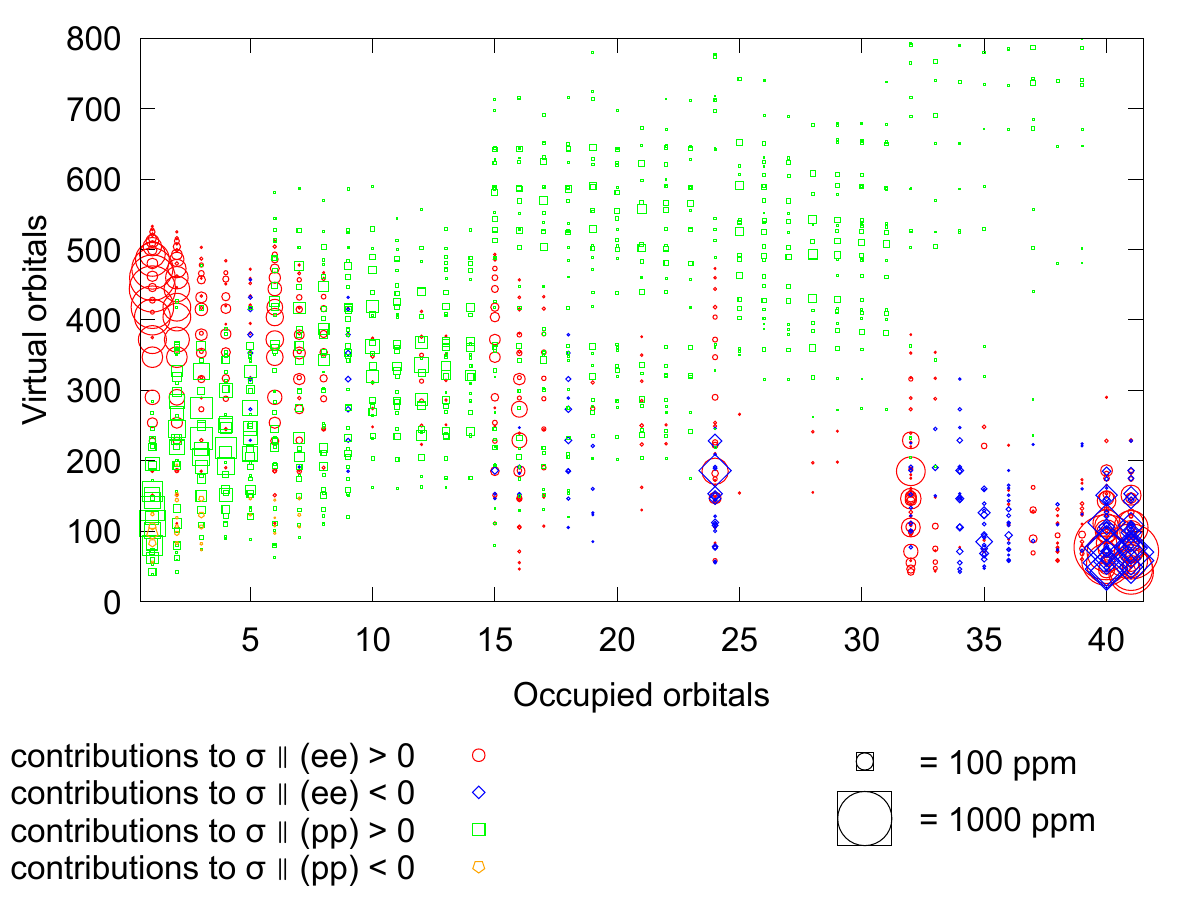}
\hfill
\includegraphics[width=0.48\linewidth]{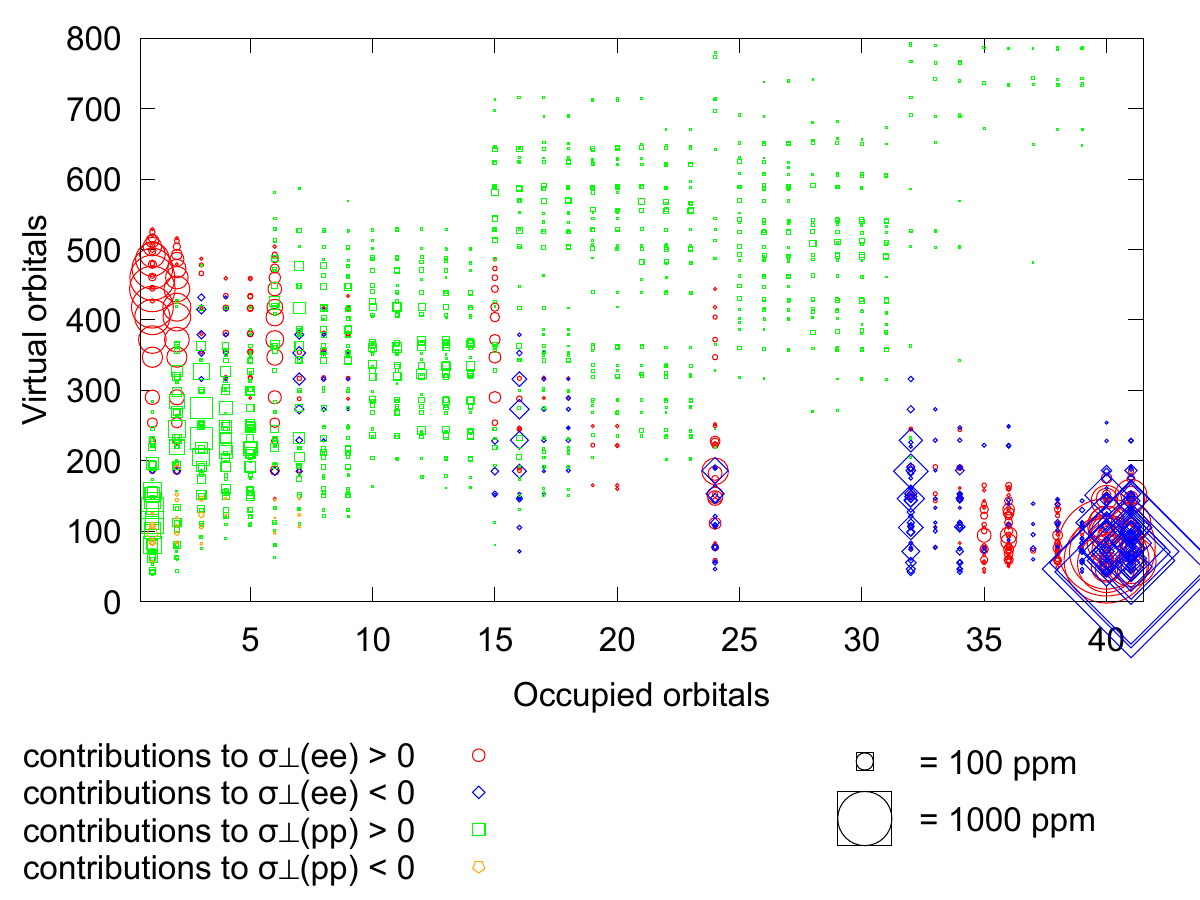}
\caption{Pattern of excitations for $\sigma_\parallel$(Tl) (left) and $\sigma_\perp$(Tl) (right) for TlH molecule at the DHF level of approach. The magnitude of the amplitude of each type is proportional to marked area. The numbers that label the occupied and virtual orbitals correspond to the Kramers pairs, such that 1 refers to the lowest energy occupied pair and 41 corresponds to the highest energy occupied pair.}
\label{fig:ex_plot_TlH}
\end{figure*}

\newpage

\begin{figure*}[!htb]
\centering
\includegraphics[width=0.48\linewidth]{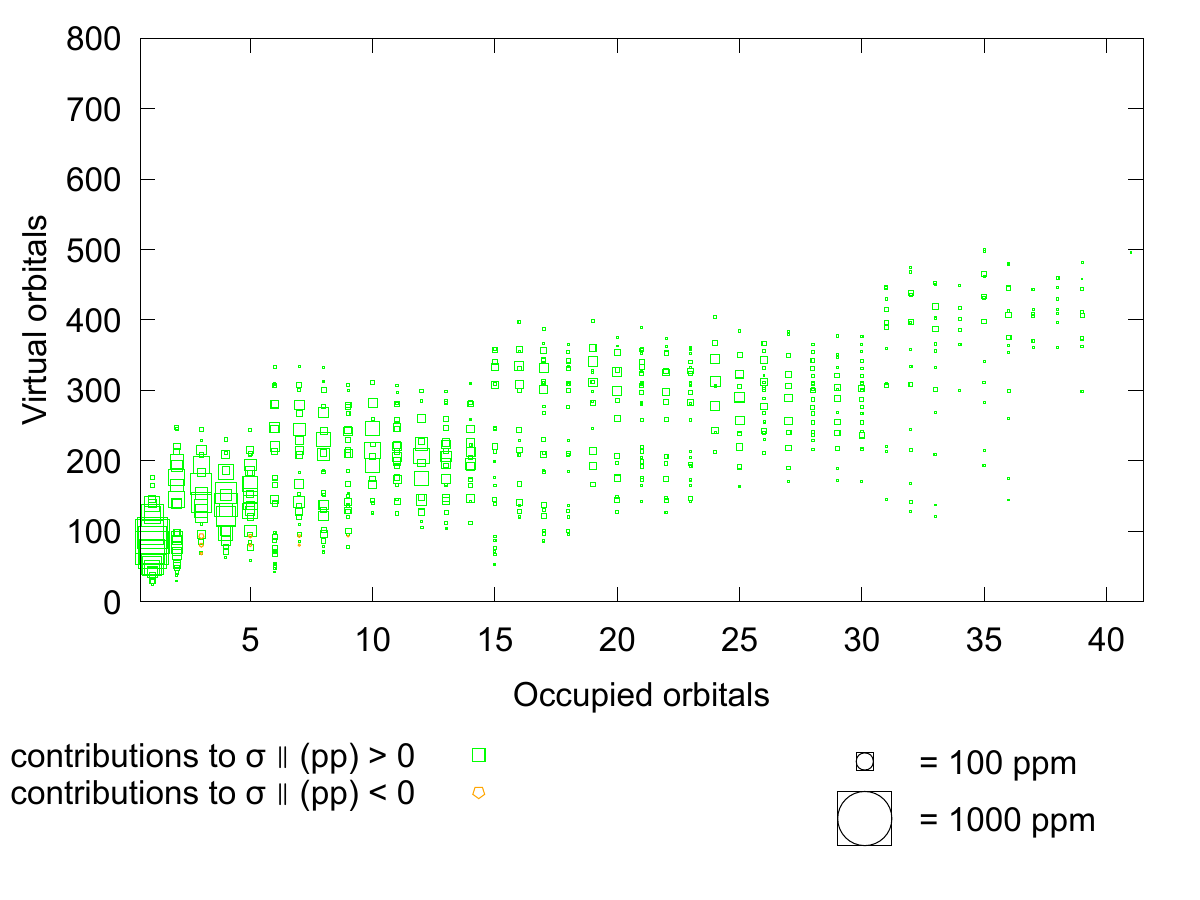}
\hfill
\includegraphics[width=0.48\linewidth]{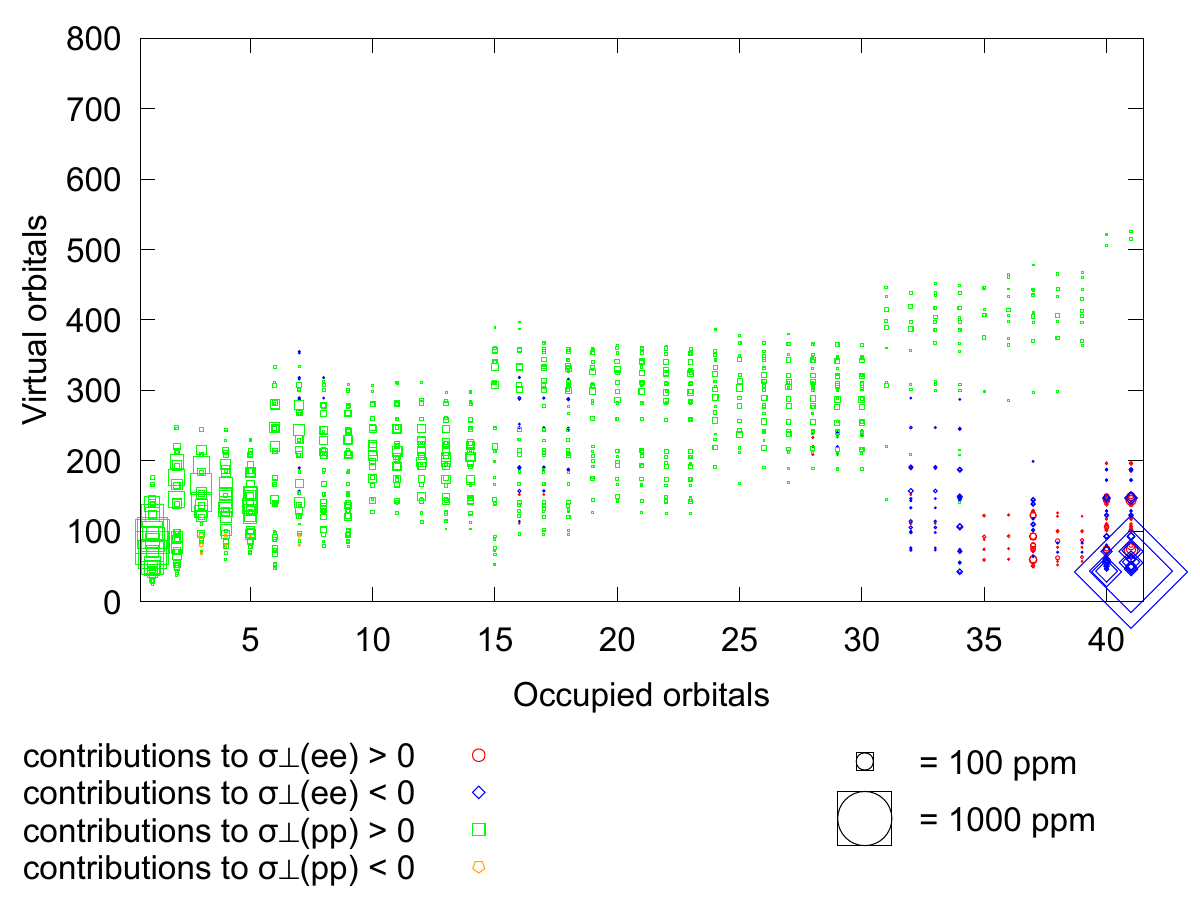}
\caption{Pattern of excitations for $\sigma_\parallel$(Tl) (left) and $\sigma_\perp$(Tl) (right) for TlH molecule (NR-limit approximation) at the DHF level of approach. The magnitude of the amplitude of each type is proportional to marked area. The numbers that label the occupied and virtual orbitals correspond to the Kramers pairs, such that 1 refers to the lowest energy occupied pair and 41 corresponds to the highest energy occupied pair.}
\label{fig:ex_plot_TlH_nr}
\end{figure*}

\newpage

\begin{figure*}[!htb]
\centering
\includegraphics[width=0.48\linewidth]{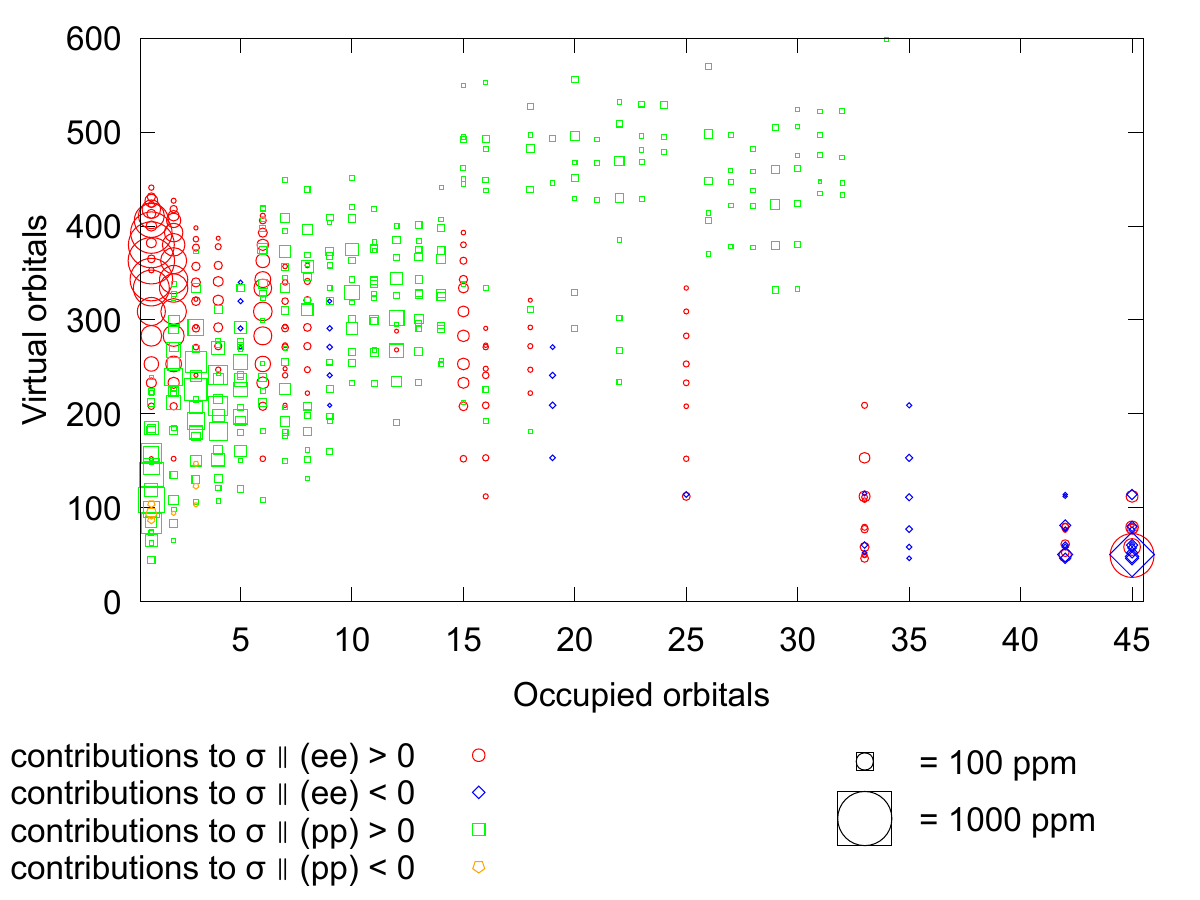}
\hfill
\includegraphics[width=0.48\linewidth]{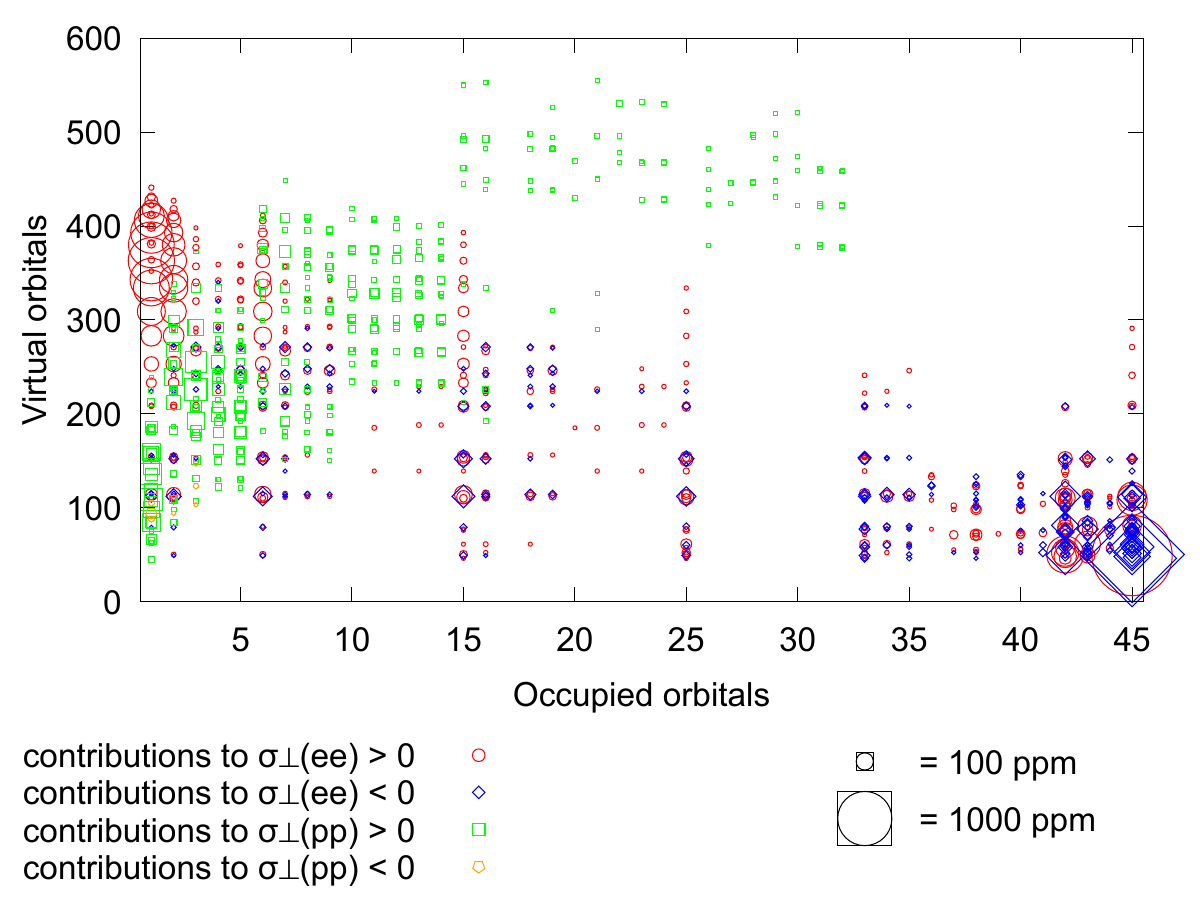}
\caption{Pattern of excitations for $\sigma_\parallel$(Tl) (left) and $\sigma_\perp$(Tl) (right) for TlF molecule at the DHF level of approach. The magnitude of the amplitude of each type is proportional to marked area. The numbers that label the occupied and virtual orbitals correspond to the Kramers pairs, such that 1 refers to the lowest energy occupied pair and 45 corresponds to the highest energy occupied pair.}
\label{fig:ex_plot_TlF}
\end{figure*}

\newpage

\begin{figure*}[!htb]
\centering
\includegraphics[width=0.48\linewidth]{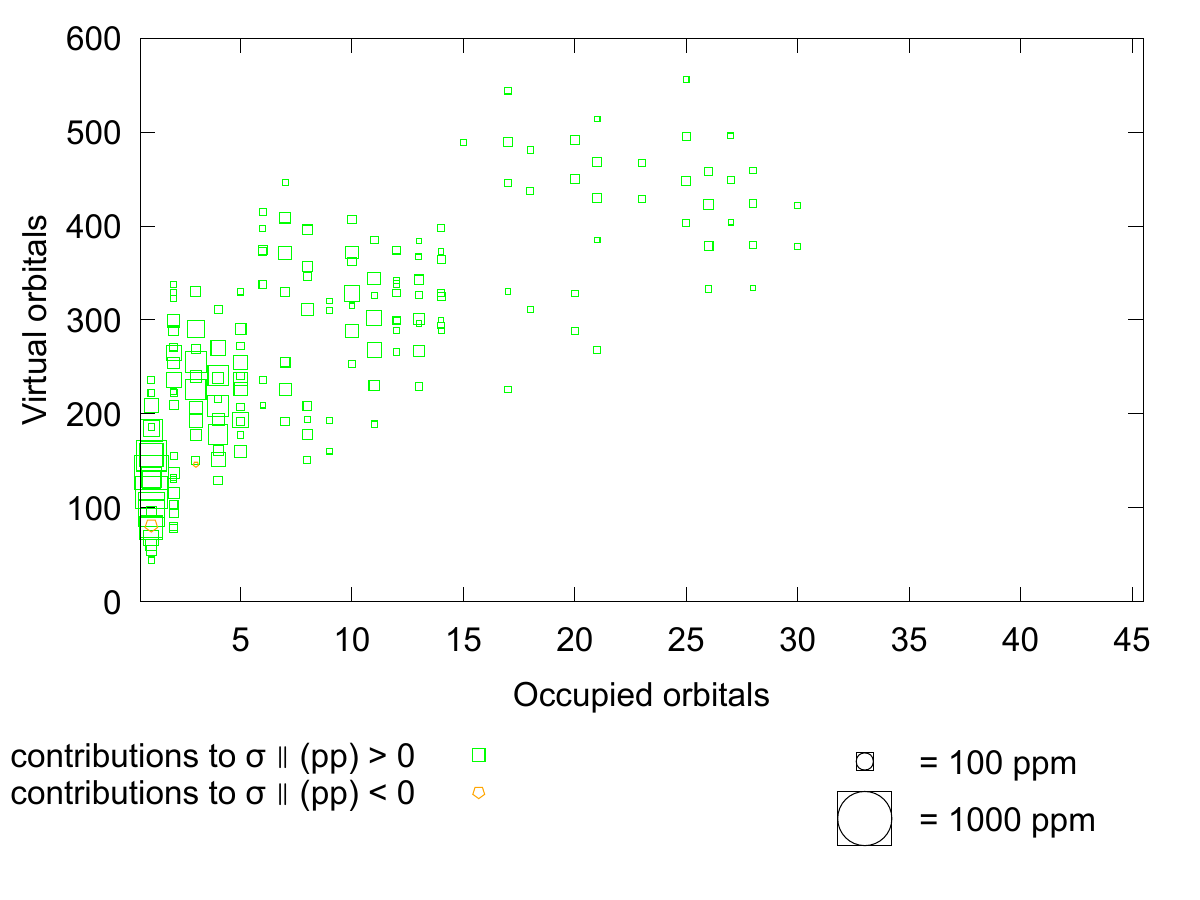}
\hfill
\includegraphics[width=0.48\linewidth]{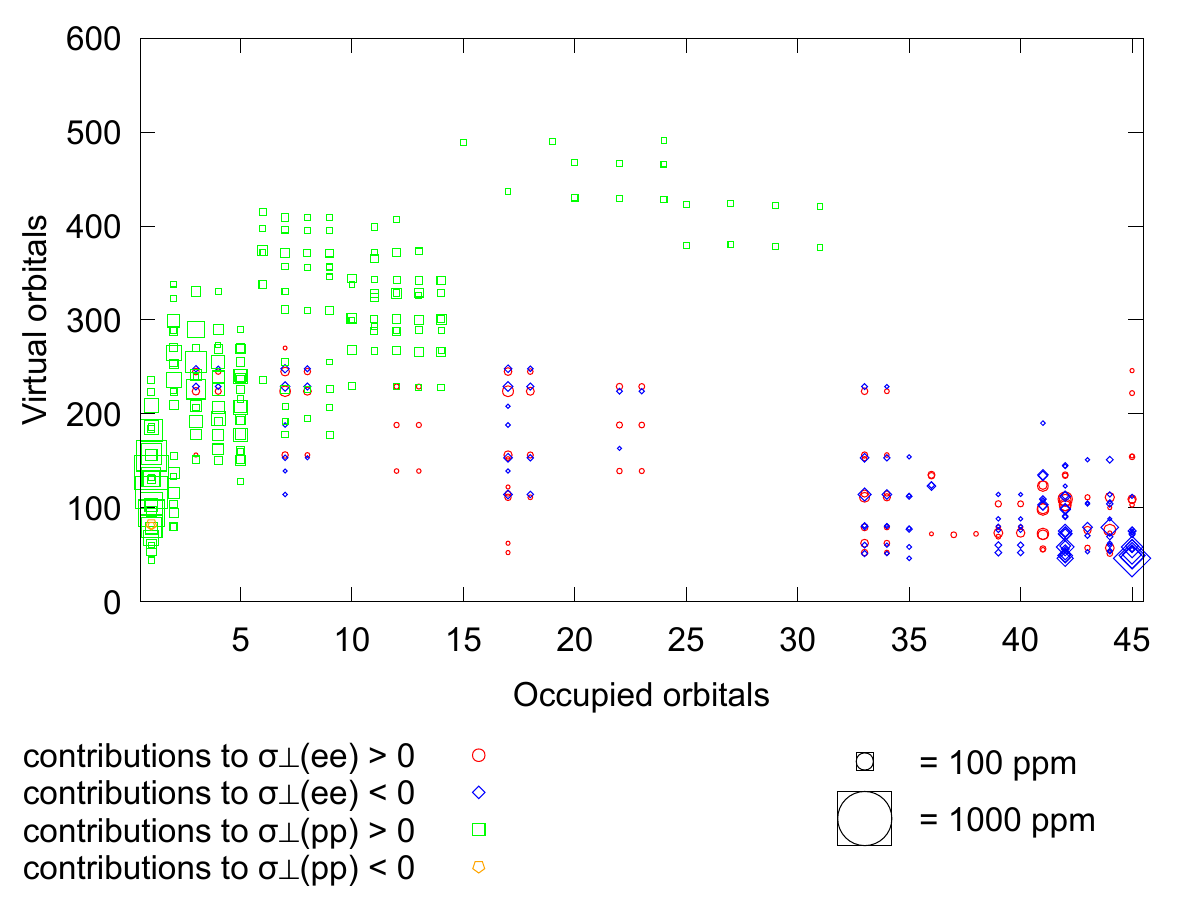}
\caption{Pattern of excitations for $\sigma_\parallel$(Tl) (left) and $\sigma_\perp$(Tl) (right) for TlF molecule (NR-limit approximation) at the DHF level of approach. The magnitude of the amplitude of each type is proportional to marked area. The numbers that label the occupied and virtual orbitals correspond to the Kramers pairs, such that 1 refers to the lowest energy occupied pair and 45 corresponds to the highest energy occupied pair.}
\label{fig:ex_plot_TlF_nr}
\end{figure*}

\newpage

\begin{figure*}[!htb]
\centering
\includegraphics[width=0.48\linewidth]{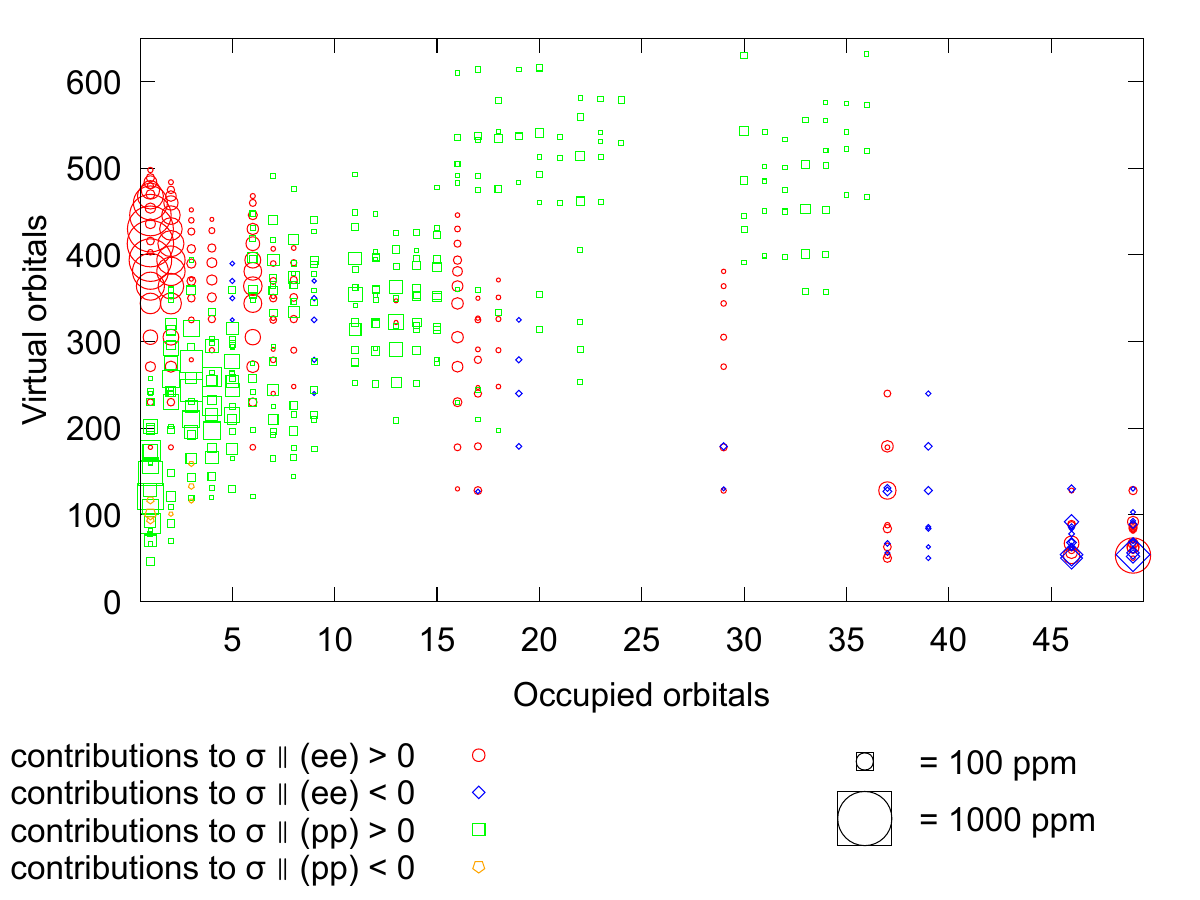}
\hfill
\includegraphics[width=0.48\linewidth]{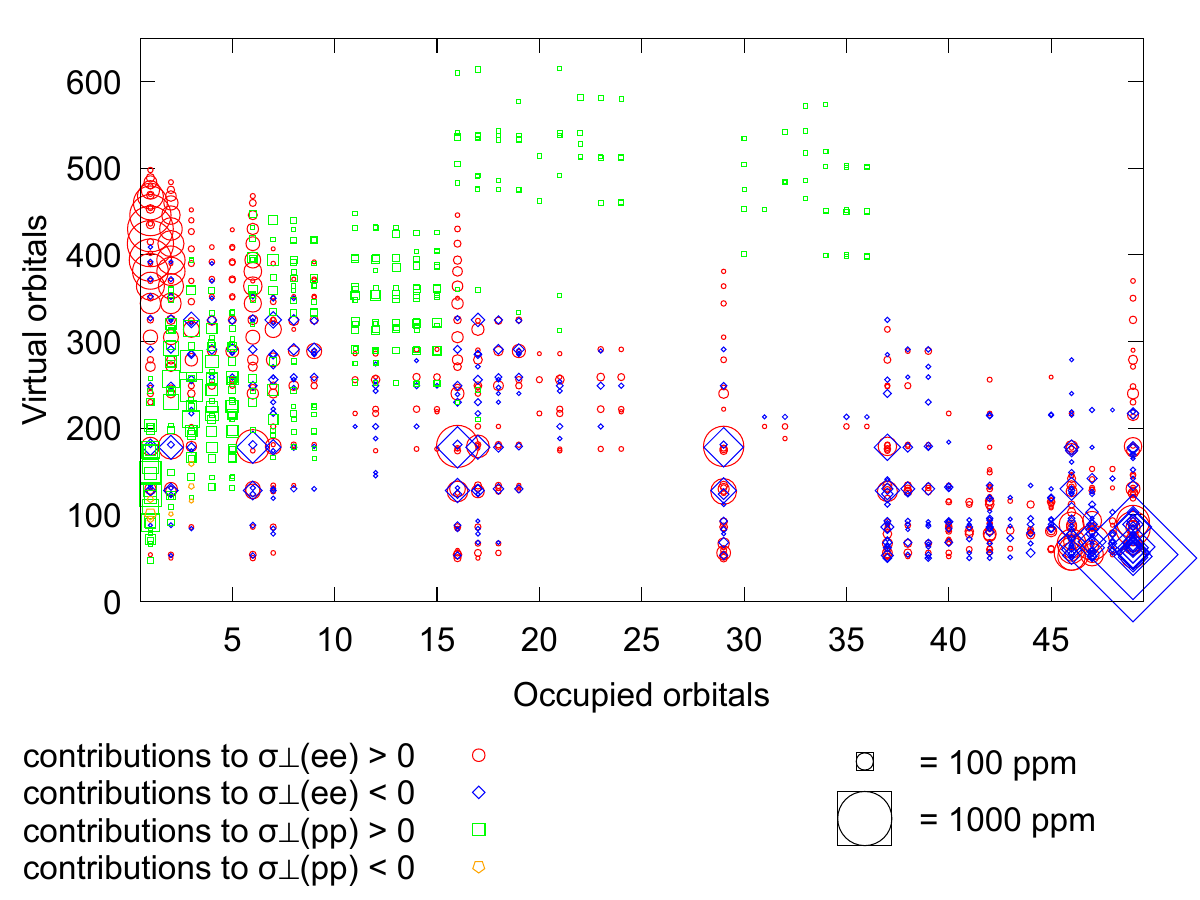}
\caption{Pattern of excitations for $\sigma_\parallel$(Tl) (left) and $\sigma_\perp$(Tl) (right) for TlCl molecule at the DHF level of approach. The magnitude of the amplitude of each type is proportional to marked area. The numbers that label the occupied and virtual orbitals correspond to the Kramers pairs, such that 1 refers to the lowest energy occupied pair and 49 corresponds to the highest energy occupied pair.}
\label{fig:ex_plot_TlCl}
\end{figure*}

\newpage

\begin{figure*}[!htb]
\centering
\includegraphics[width=0.48\linewidth]{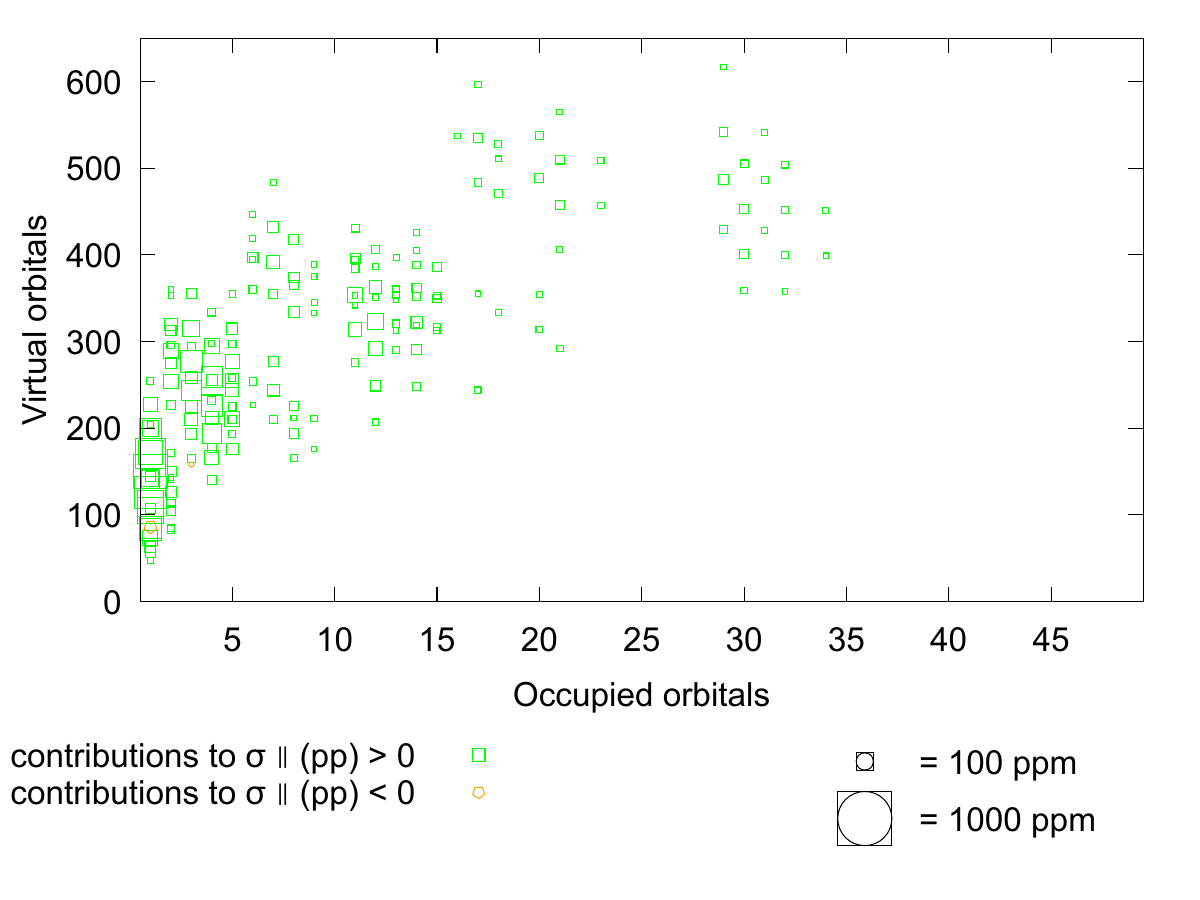}
\hfill
\includegraphics[width=0.48\linewidth]{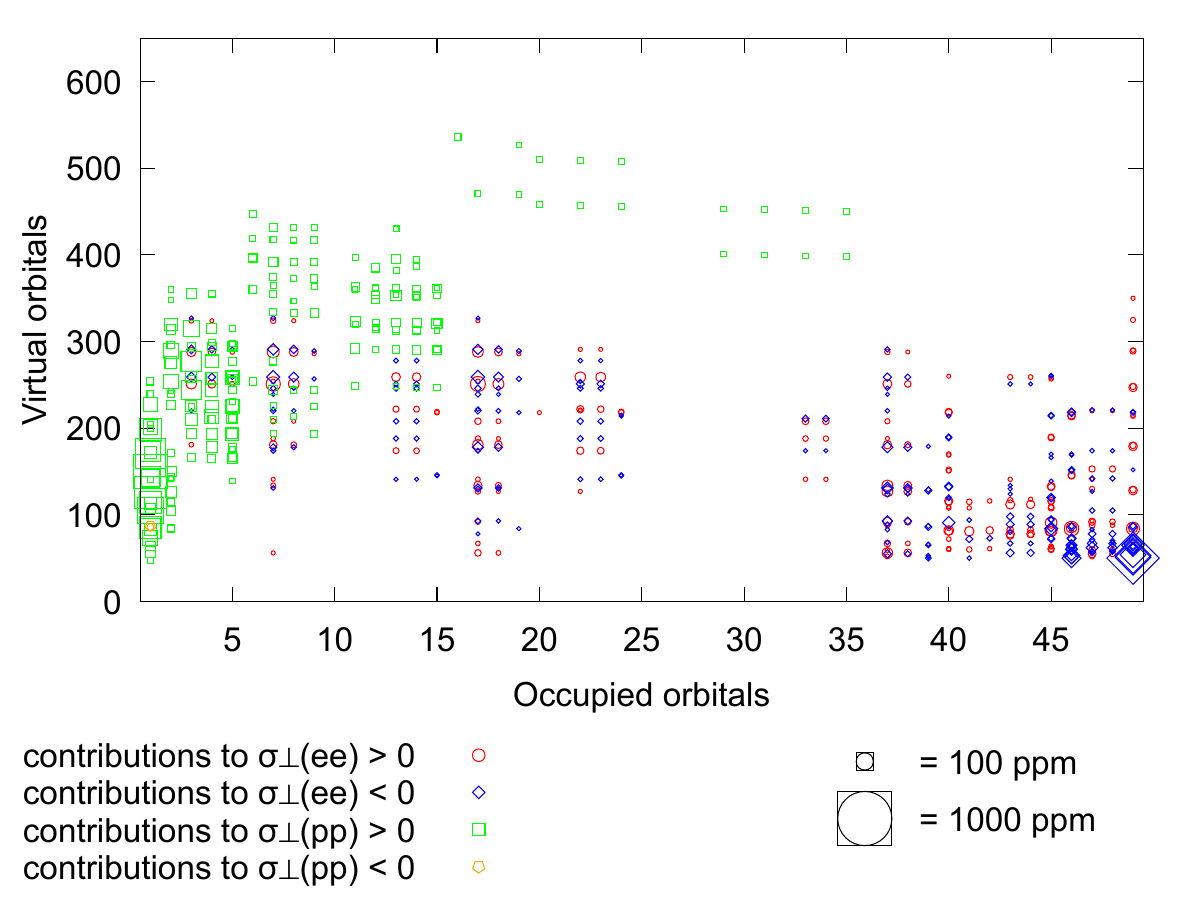}
\caption{Pattern of excitations for $\sigma_\parallel$(Tl) (left) and $\sigma_\perp$(Tl) (right) for TlCl molecule (NR-limit approximation) at the DHF level of approach. The magnitude of the amplitude of each type is proportional to marked area. The numbers that label the occupied and virtual orbitals correspond to the Kramers pairs, such that 1 refers to the lowest energy occupied pair and 49 corresponds to the highest energy occupied pair.}
\label{fig:ex_plot_TlCl_nr}
\end{figure*}

\newpage

\begin{figure*}[!htb]
\centering
\includegraphics[width=0.48\linewidth]{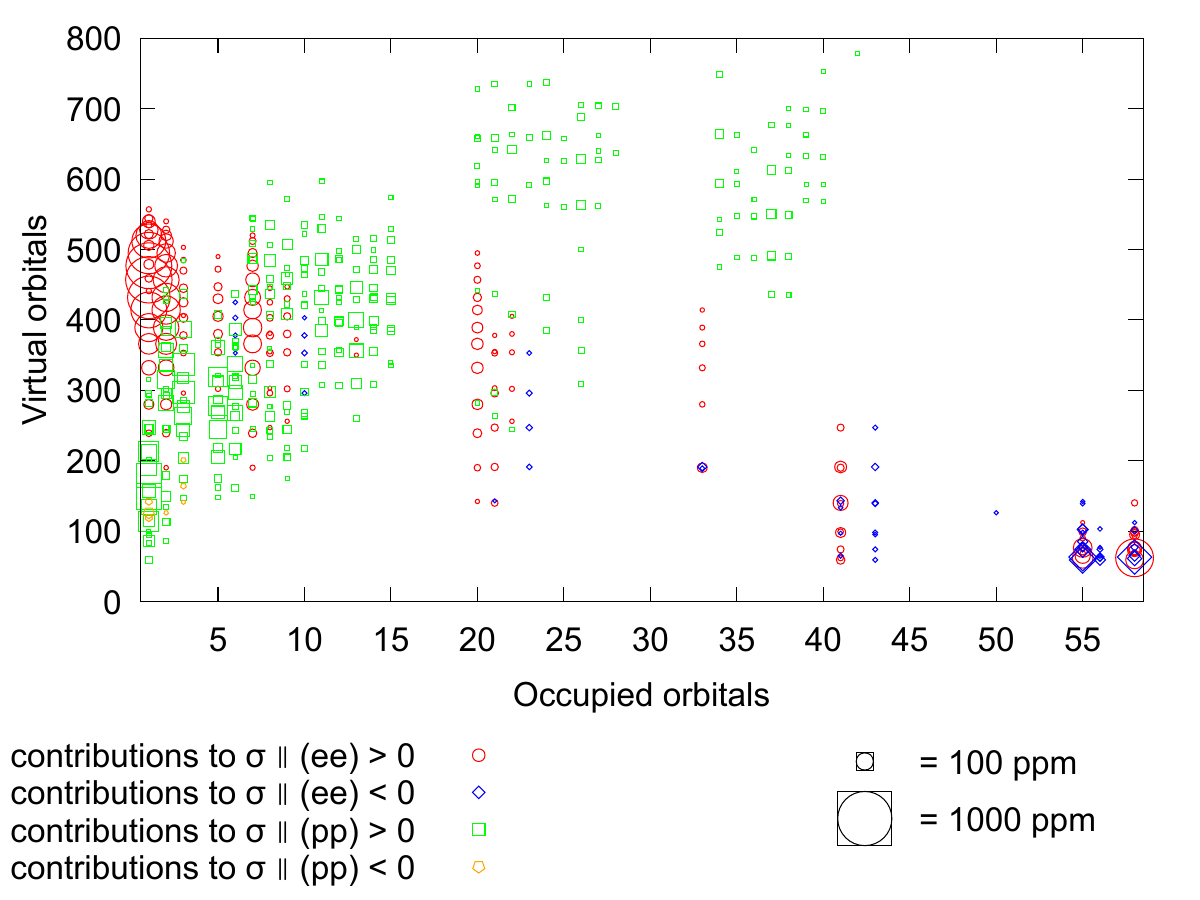}
\hfill
\includegraphics[width=0.48\linewidth]{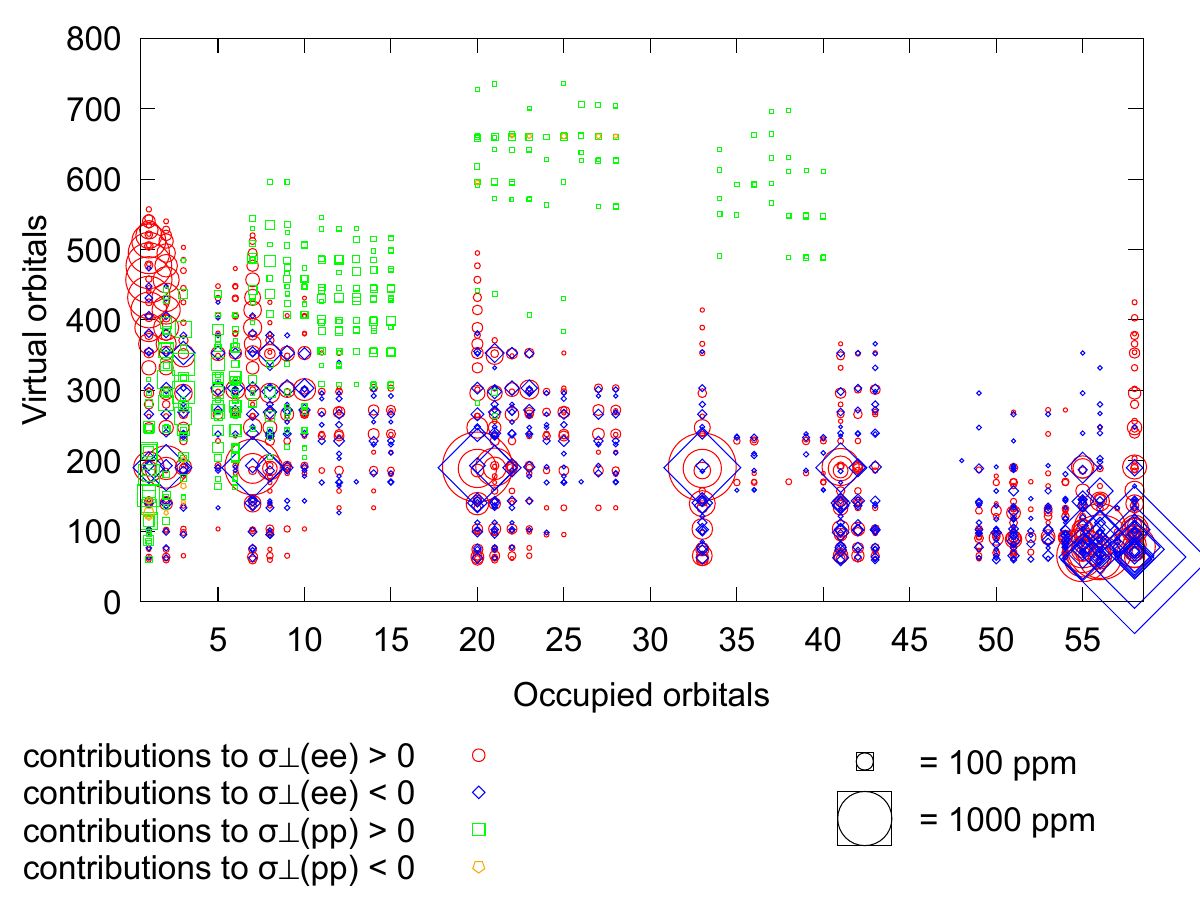}
\caption{Pattern of excitations for $\sigma_\parallel$(Tl) (left) and $\sigma_\perp$(Tl) (right) for TlBr molecule at the DHF level of approach. The magnitude of the amplitude of each type is proportional to marked area. The numbers that label the occupied and virtual orbitals correspond to the Kramers pairs, such that 1 refers to the lowest energy occupied pair and 58 corresponds to the highest energy occupied pair.}
\label{fig:ex_plot_TlBr}
\end{figure*}

\newpage

\begin{figure*}[!htb]
\centering
\includegraphics[width=0.48\linewidth]{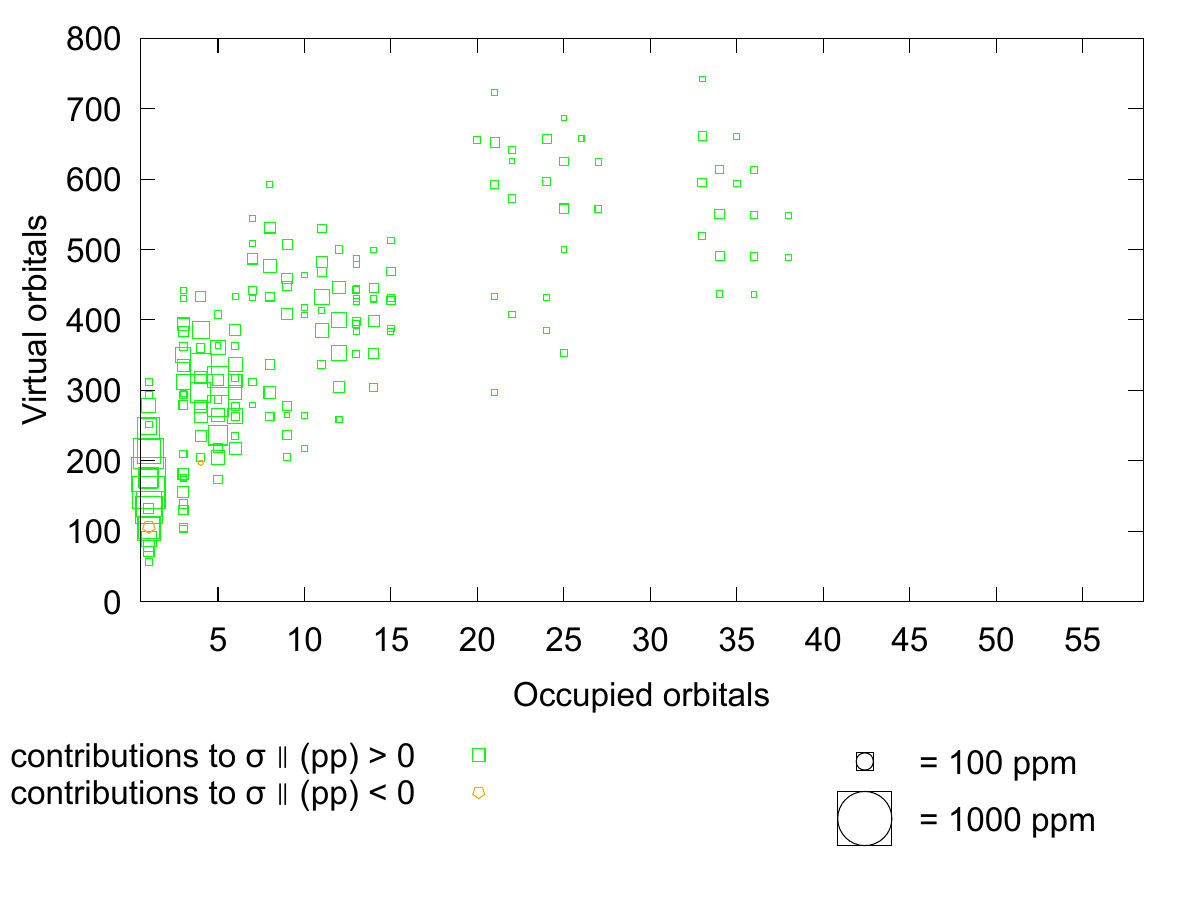}
\hfill
\includegraphics[width=0.48\linewidth]{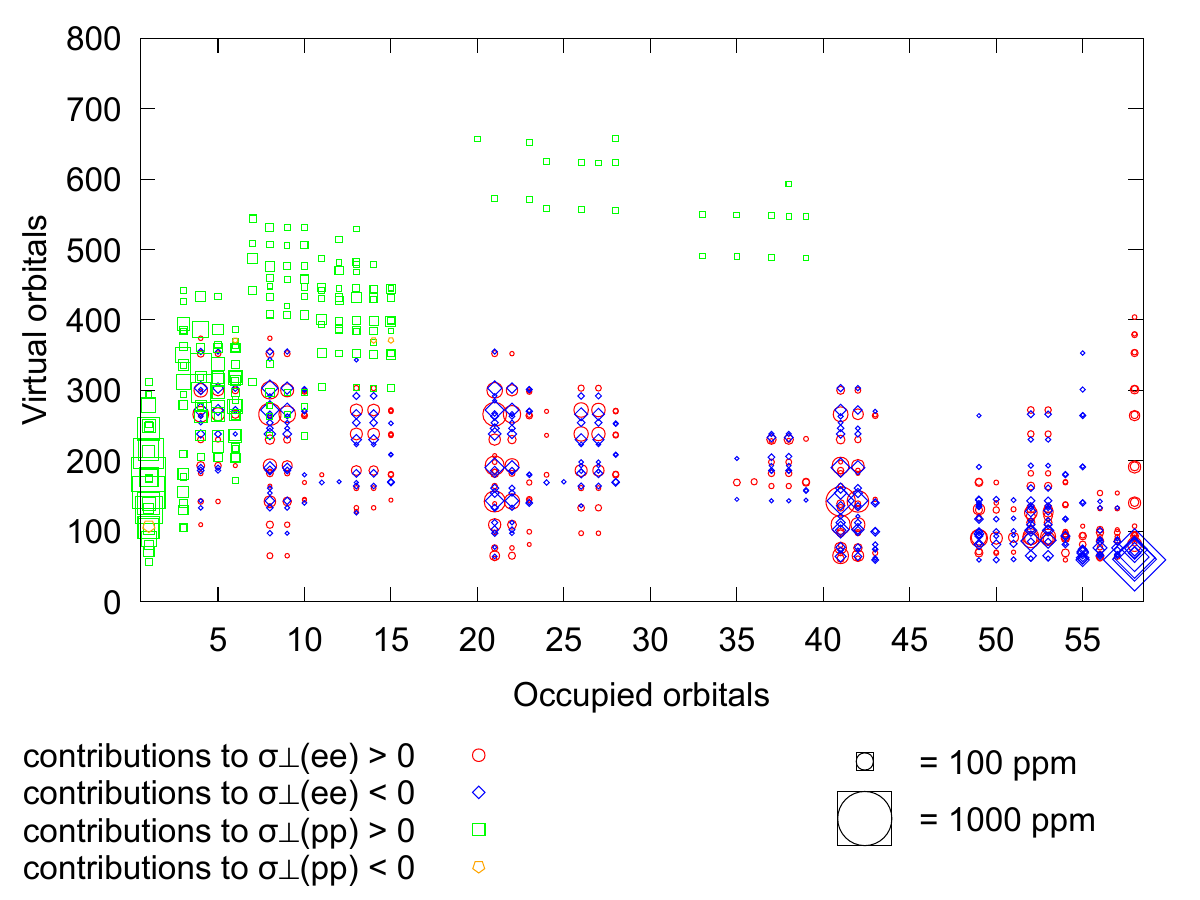}
\caption{Pattern of excitations for $\sigma_\parallel$(Tl) (left) and $\sigma_\perp$(Tl) (right) for TlBr molecule (NR-limit approximation) at the DHF level of approach. The magnitude of the amplitude of each type is proportional to marked area. The numbers that label the occupied and virtual orbitals correspond to the Kramers pairs, such that 1 refers to the lowest energy occupied pair and 58 corresponds to the highest energy occupied pair.}
\label{fig:ex_plot_TlBr_nr}
\end{figure*}

\newpage

%




\begin{figure*}[!htb]
\centering
\includegraphics[width=0.48\linewidth]{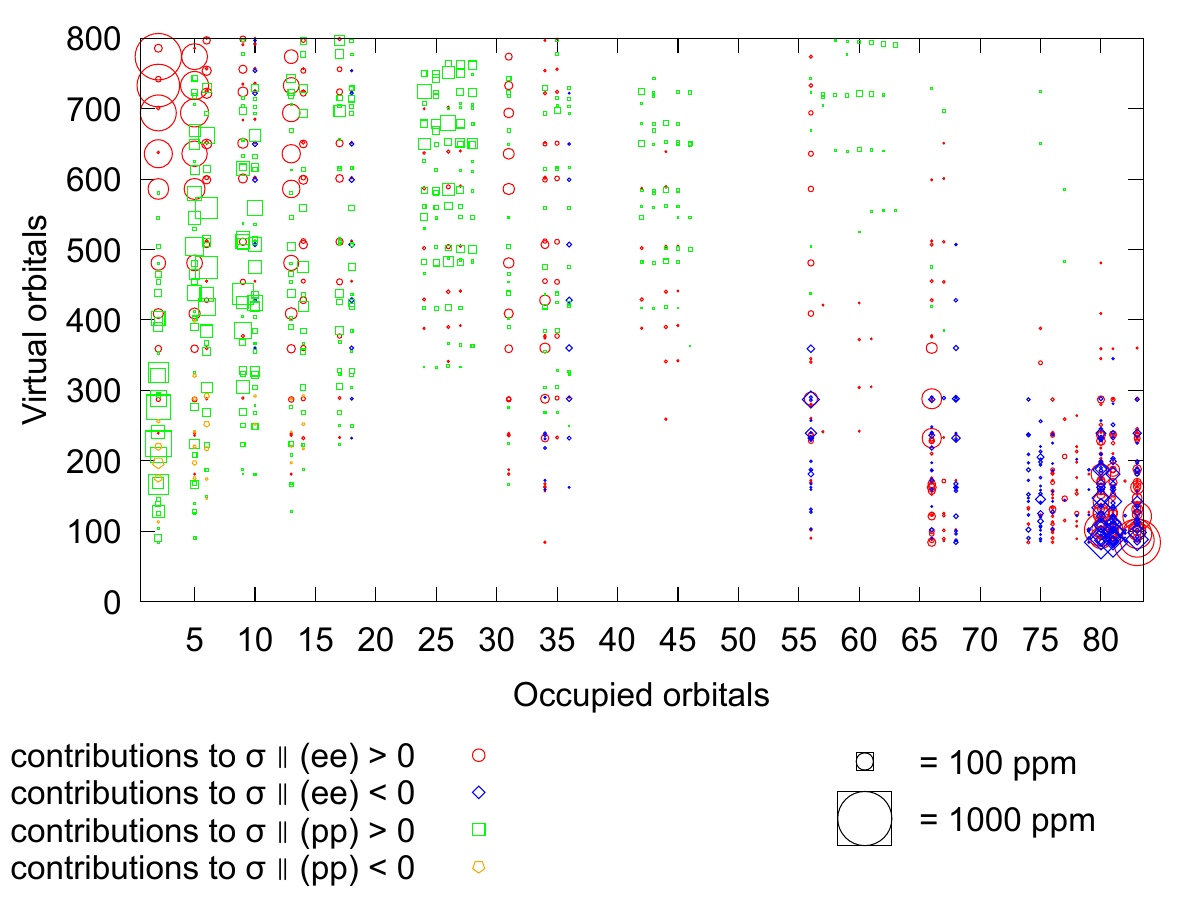}
\hfill
\includegraphics[width=0.48\linewidth]{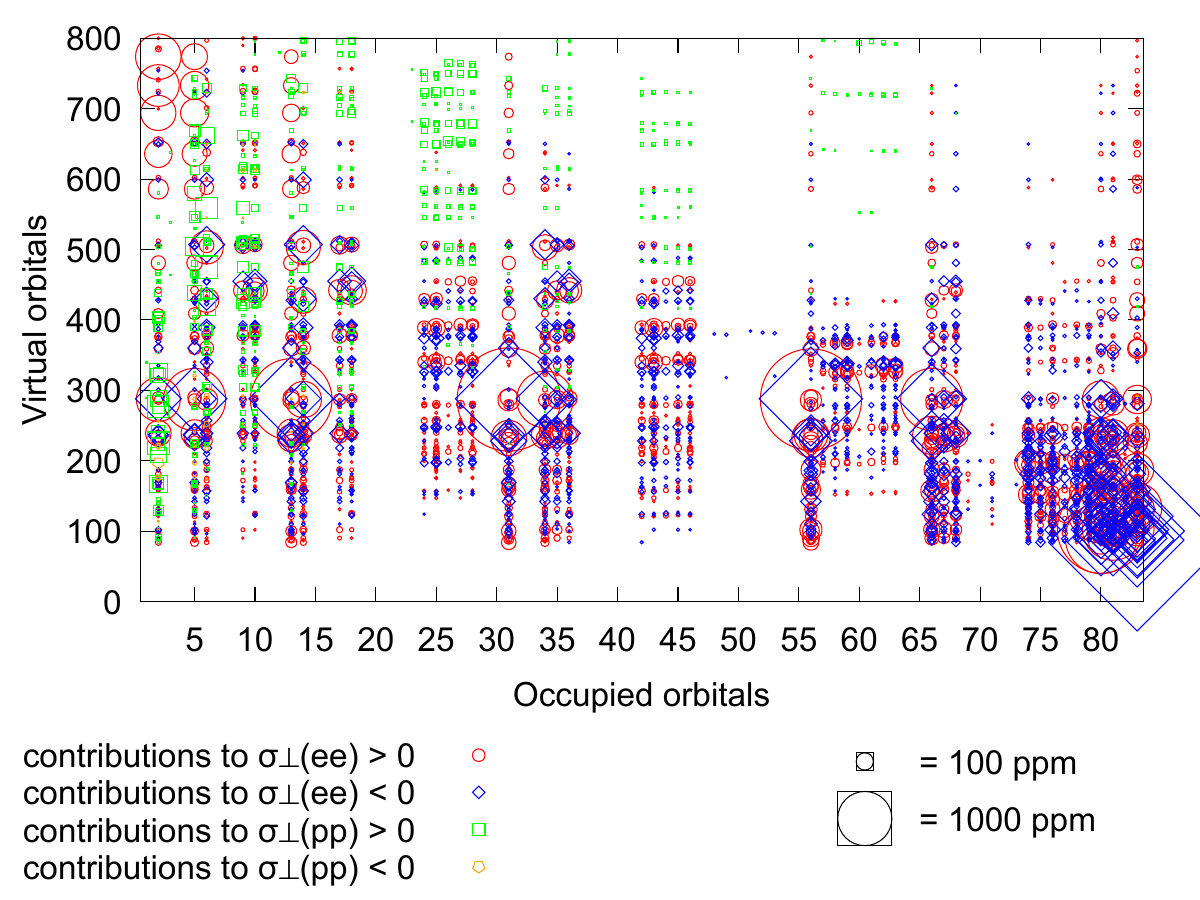}
\caption{Pattern of excitations for $\sigma_\parallel$(Tl) (left) and $\sigma_\perp$(Tl) (right) for TlAt molecule at the DHF level of approach. The magnitude of the amplitude of each type is proportional to marked area. The numbers that label the occupied and virtual orbitals correspond to the Kramers pairs, such that 1 refers to the lowest energy occupied pair and 83 corresponds to the highest energy occupied pair.}
\label{fig:ex_plot_TlAt}
\end{figure*}

\newpage

\begin{figure*}[!htb]
\centering
\includegraphics[width=0.48\linewidth]{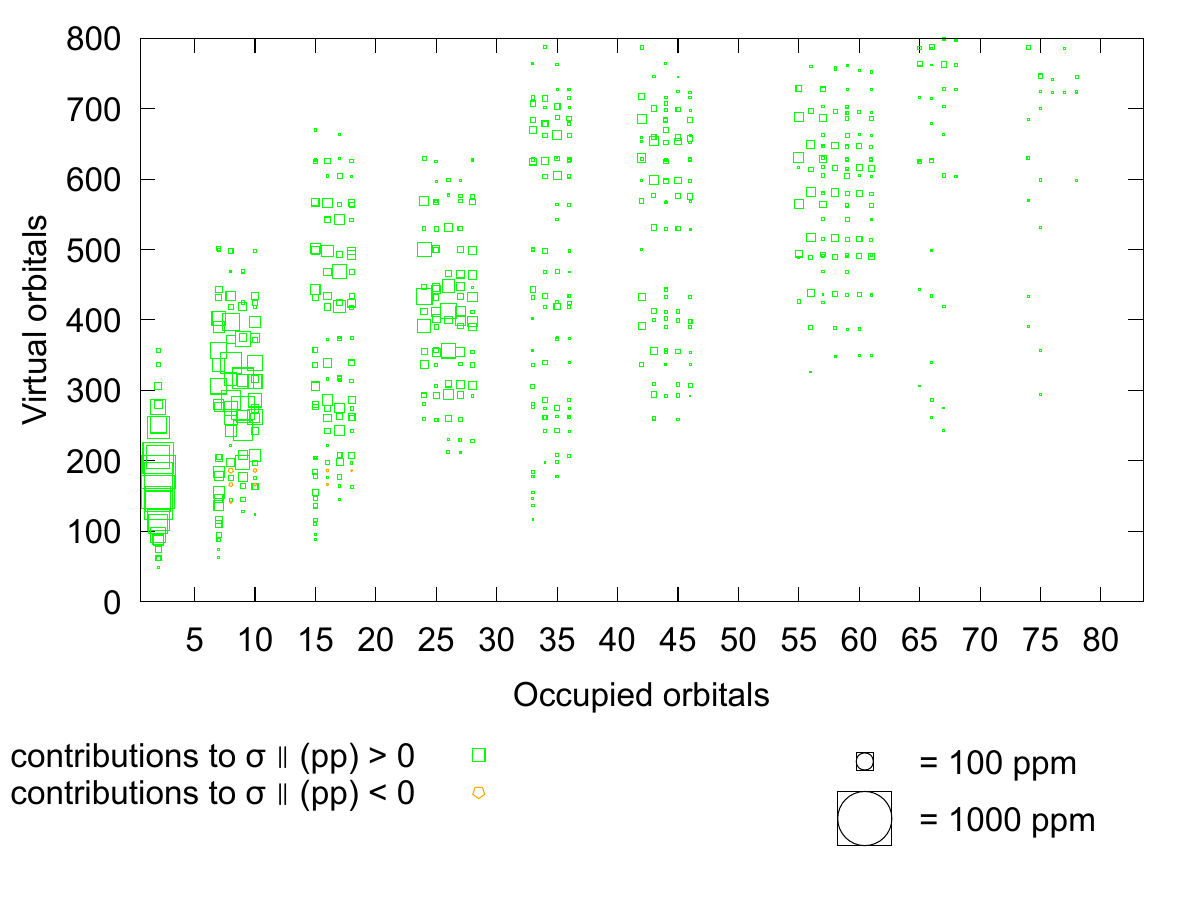}
\hfill
\includegraphics[width=0.48\linewidth]{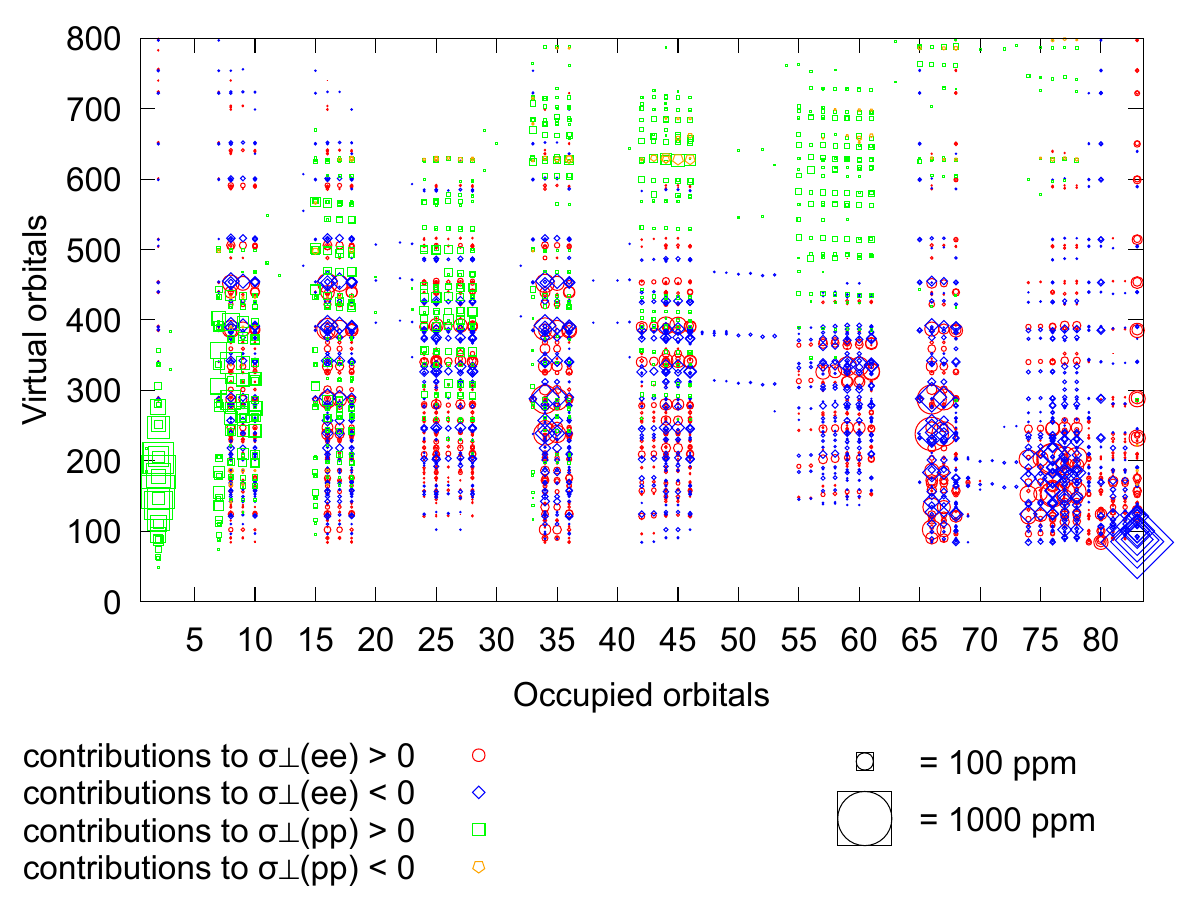}
\caption{Pattern of excitations for $\sigma_\parallel$(Tl) (left) and $\sigma_\perp$(Tl) (right) for TlAt molecule (NR-limit approximation) at the DHF level of approach. The magnitude of the amplitude of each type is proportional to marked area. The numbers that label the occupied and virtual orbitals correspond to the Kramers pairs, such that 1 refers to the lowest energy occupied pair and 83 corresponds to the highest energy occupied pair.}
\label{fig:ex_plot_TlAt_nr}
\end{figure*}

\clearpage

\section{Pattern of excitations - ZORA approach}
  \subsection{rZORA+eX2C-HF scheme}
 
    

      \begin{figure*}[h!]
      \begin{minipage}{.48\textwidth}
      \includegraphics[width=\textwidth]{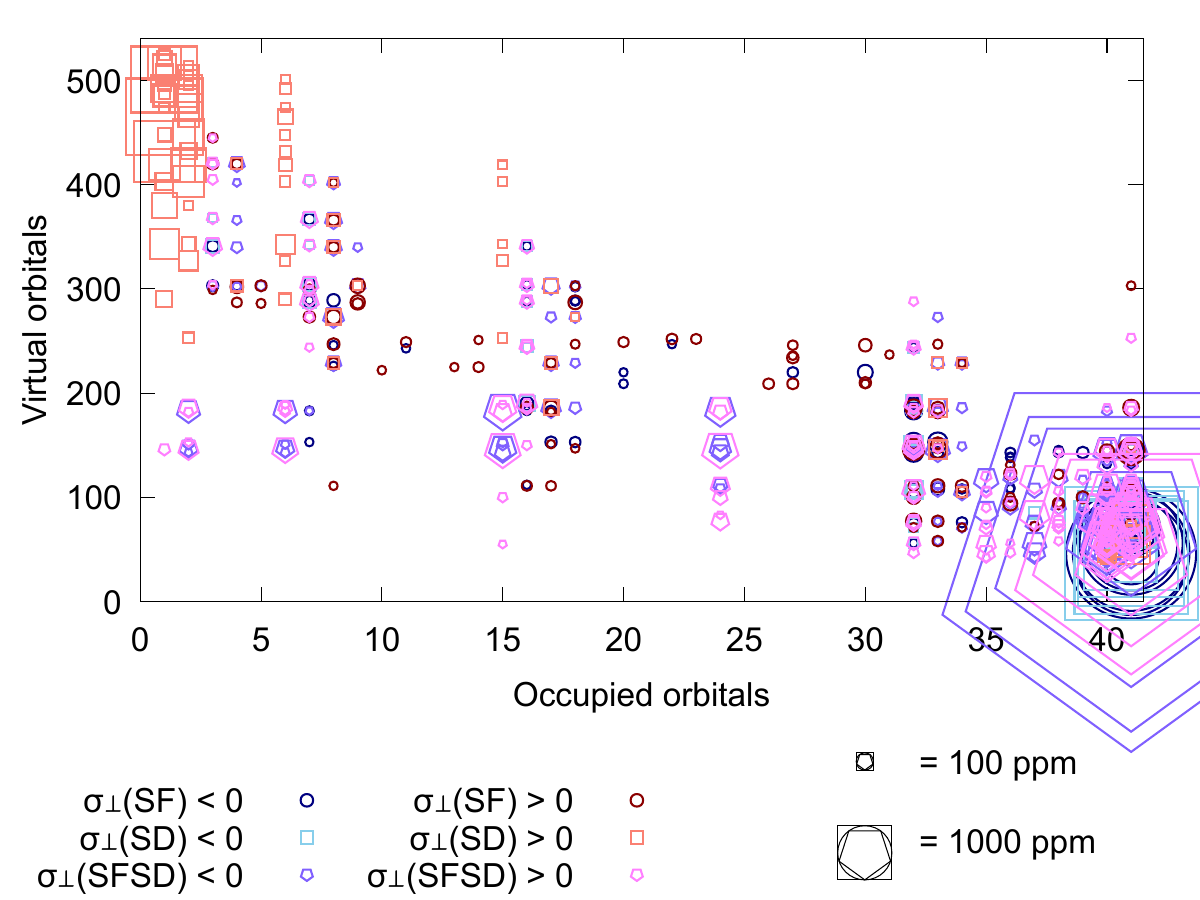}
      \end{minipage}
      \hfill
      \begin{minipage}{.48\textwidth}
      \includegraphics[width=\textwidth]{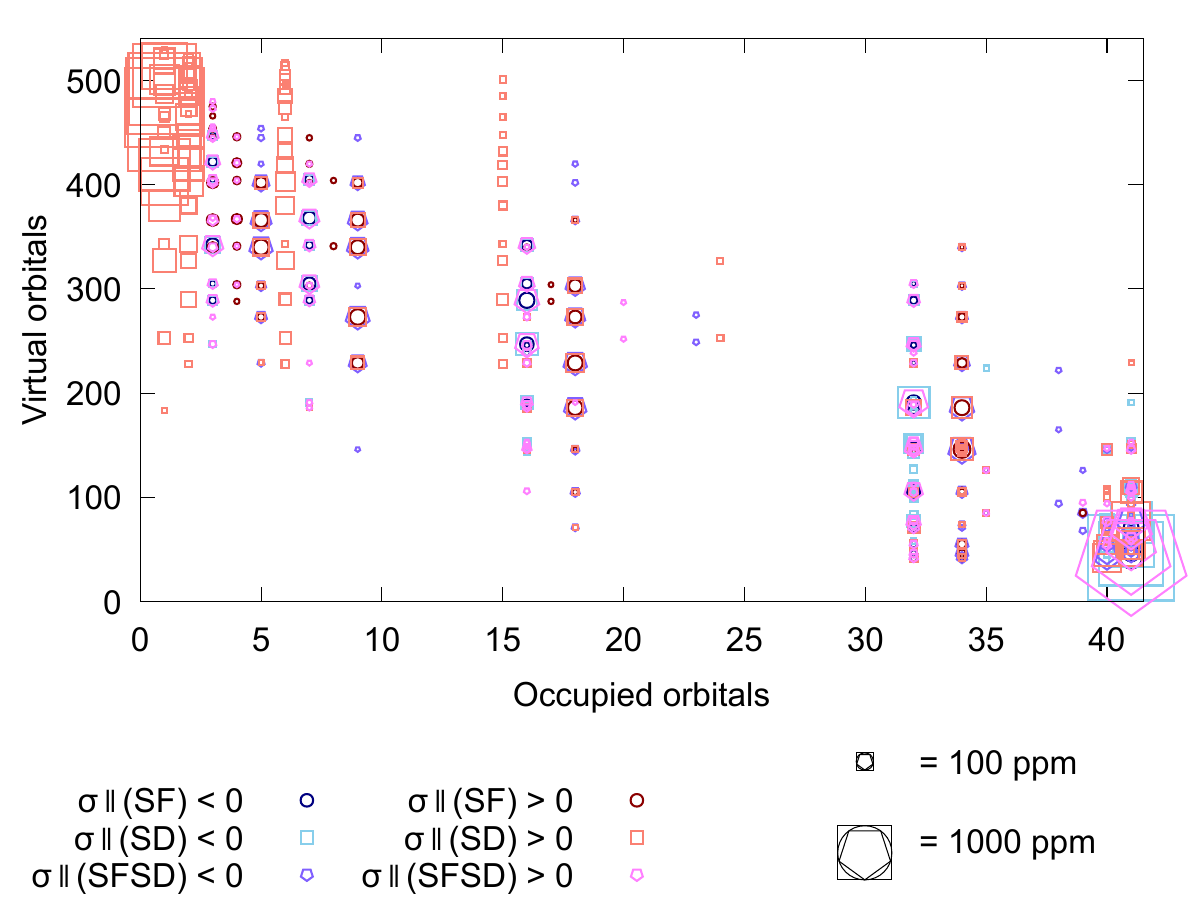}
      \end{minipage}%
      \caption{Pattern of excitations for $\sigma_\parallel$(Tl) (left)
                and $\sigma_\perp$(Tl) (right) for TlH shown separately for the SD and SF parts of the NMR shielding propagator at the level of rZORA+eX2C-HF. The magnitude of the amplitude of each type is proportional to marked area. The numbers that label the occupied and virtual orbitals correspond to the Kramers pairs, such that 1 refers to the lowest energy occupied pair and 41 corresponds to the highest energy occupied pair.}
      \label{fig:propagator_rZORA+eX2C-HF_TlH}
      \end{figure*}

      \newpage

      \begin{figure*}[h!]
      \begin{minipage}{.48\textwidth}
      \includegraphics[width=\textwidth]{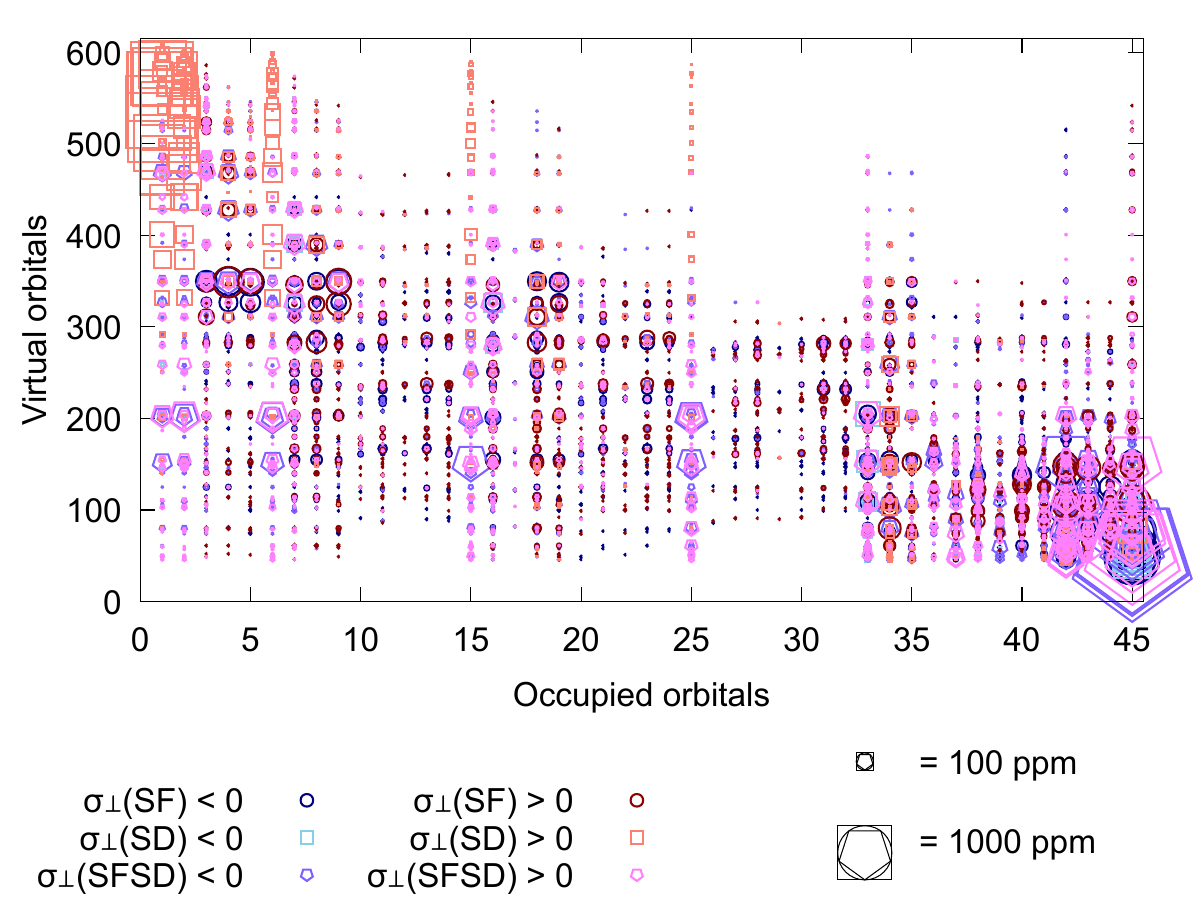}
      \end{minipage}
      \hfill
      \begin{minipage}{.48\textwidth}
      \includegraphics[width=\textwidth]{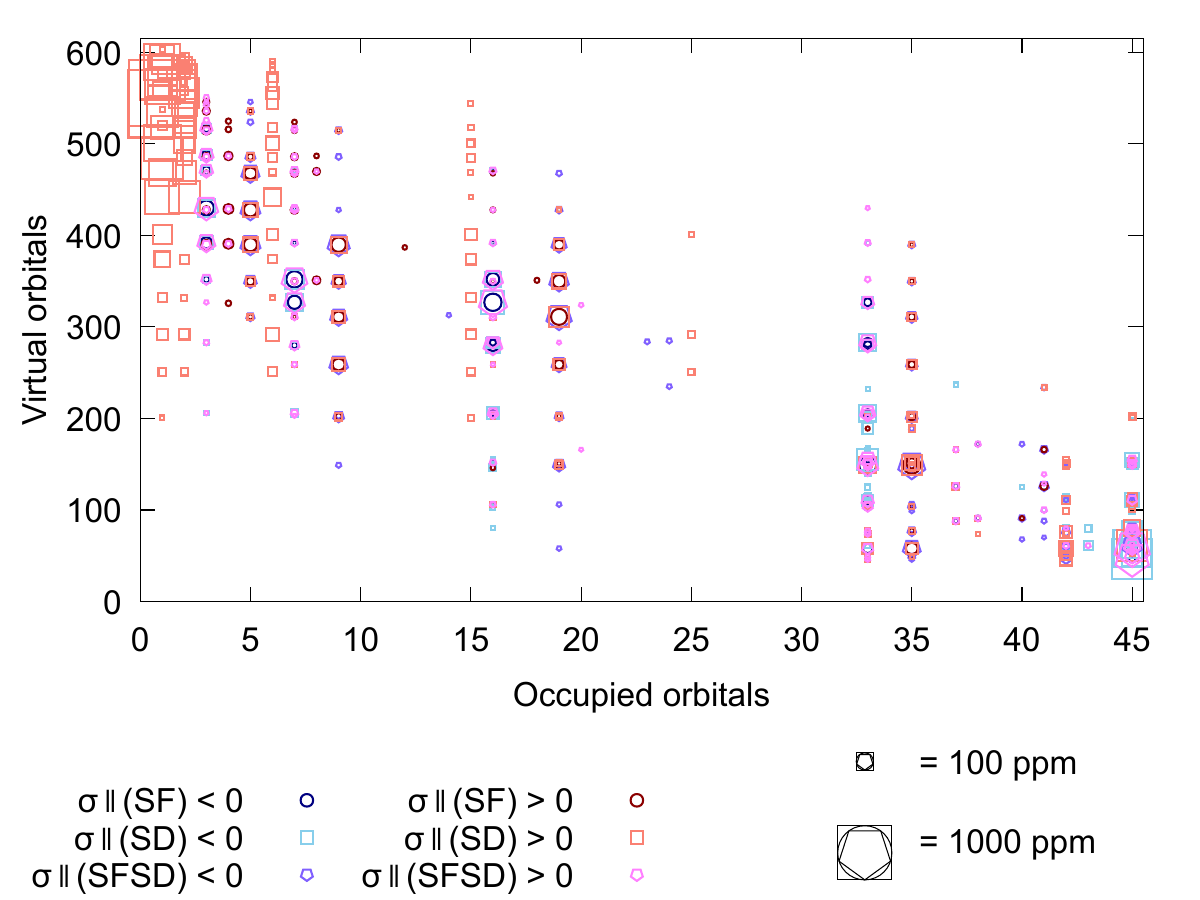}
      \end{minipage}%
      \caption{Pattern of excitations for $\sigma_\parallel$(Tl) (left)
                and $\sigma_\perp$(Tl) (right) for TlF shown separately for the SD and SF parts of the NMR shielding propagator at the level of rZORA+eX2C-HF. The magnitude of the amplitude of each type is proportional to marked area. The numbers that label the occupied and virtual orbitals correspond to the Kramers pairs, such that 1 refers to the lowest energy occupied pair and 45 corresponds to the highest energy occupied pair.}
      \label{fig:propagator_rZORA+eX2C-HF_TlF}
      \end{figure*}

      \newpage

      \begin{figure*}[h!]
      \begin{minipage}{.48\textwidth}
      \includegraphics[width=\textwidth]{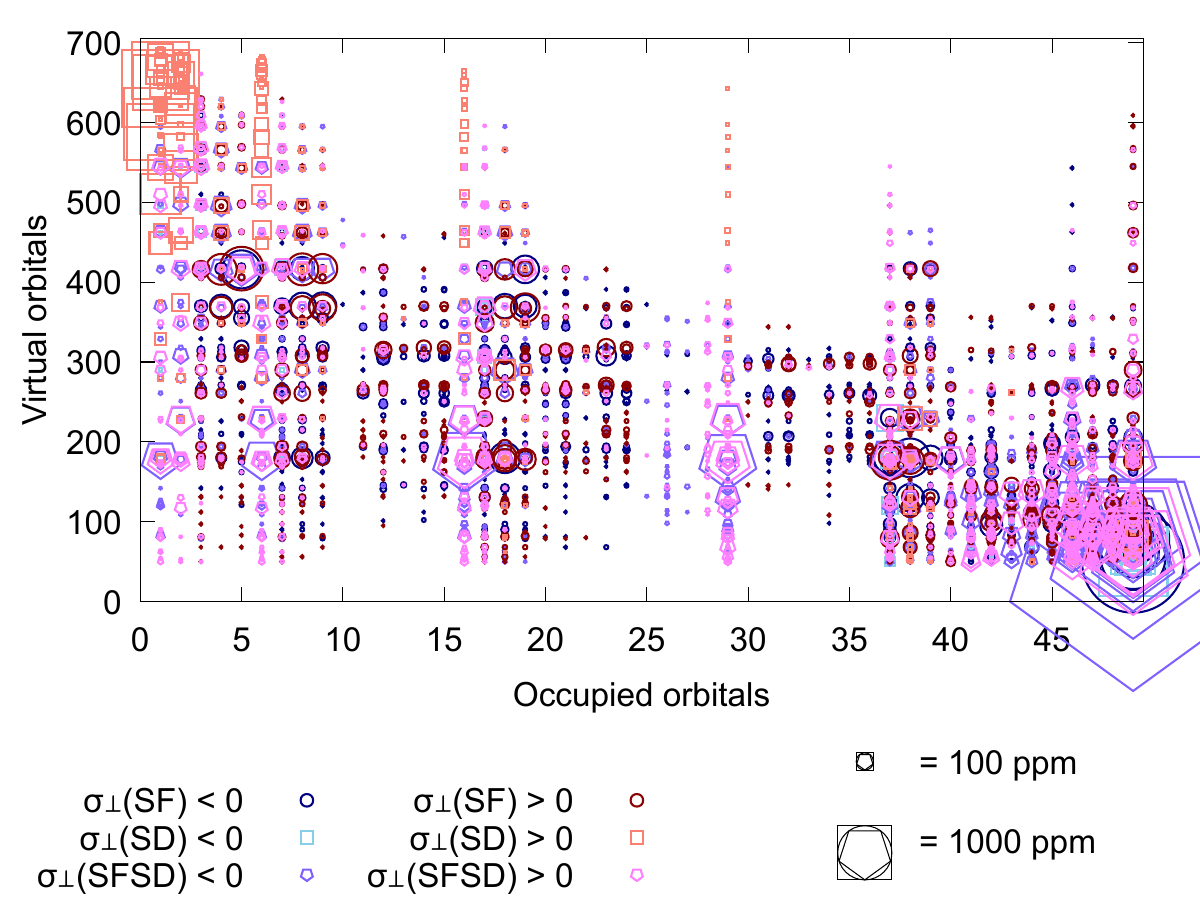}
      \end{minipage}
      \hfill
      \begin{minipage}{.48\textwidth}
      \includegraphics[width=\textwidth]{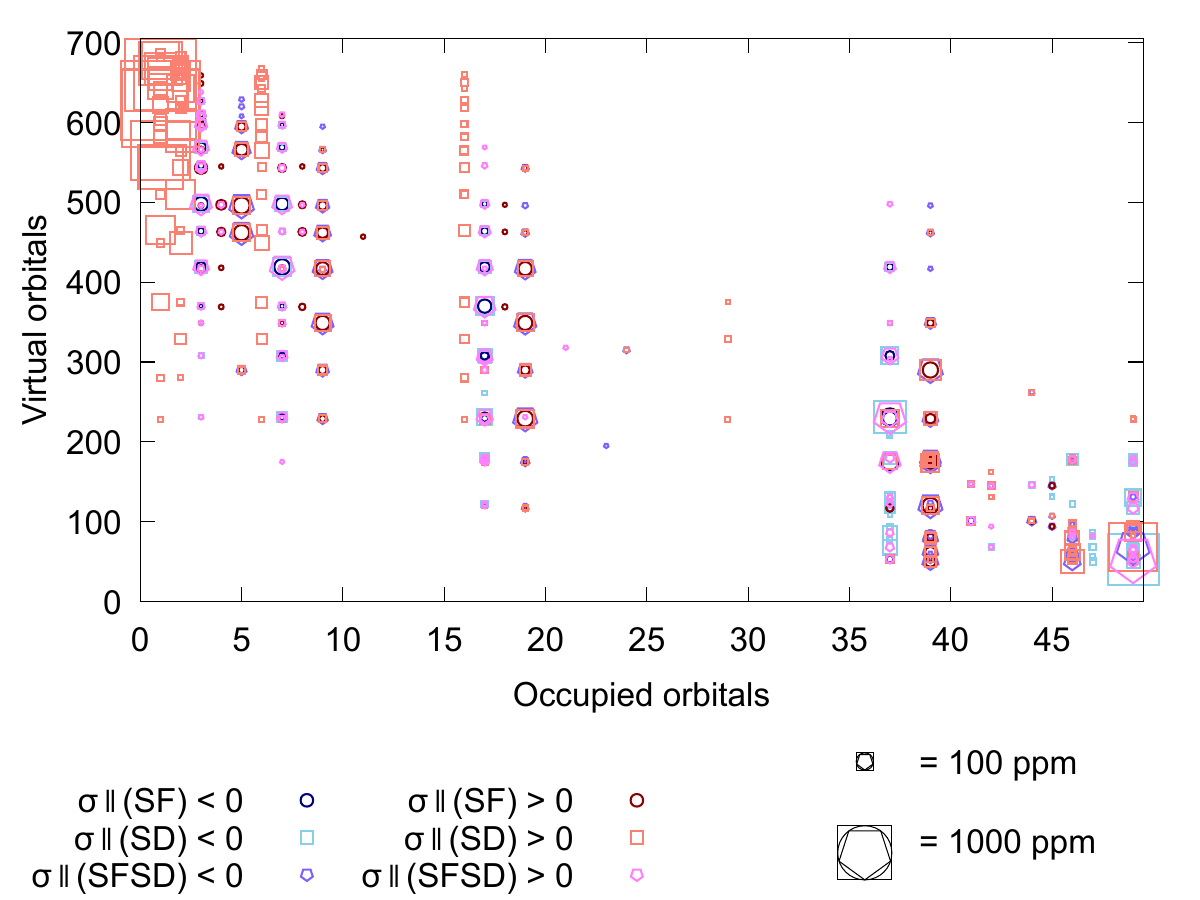}
      \end{minipage}%
      \caption{Pattern of excitations for $\sigma_\parallel$(Tl) (left)
                and $\sigma_\perp$(Tl) (right) for TlCl shown separately for the SD and SF parts of the NMR shielding propagator at the level of rZORA+eX2C-HF. The magnitude of the amplitude of each type is proportional to marked area. The numbers that label the occupied and virtual orbitals correspond to the Kramers pairs, such that 1 refers to the lowest energy occupied pair and 49 corresponds to the highest energy occupied pair.}
      \label{fig:propagator_rZORA+eX2C-HF_TlCl}
      \end{figure*}

      \newpage

      \begin{figure*}[h!]
      \begin{minipage}{.48\textwidth}
      \includegraphics[width=\textwidth]{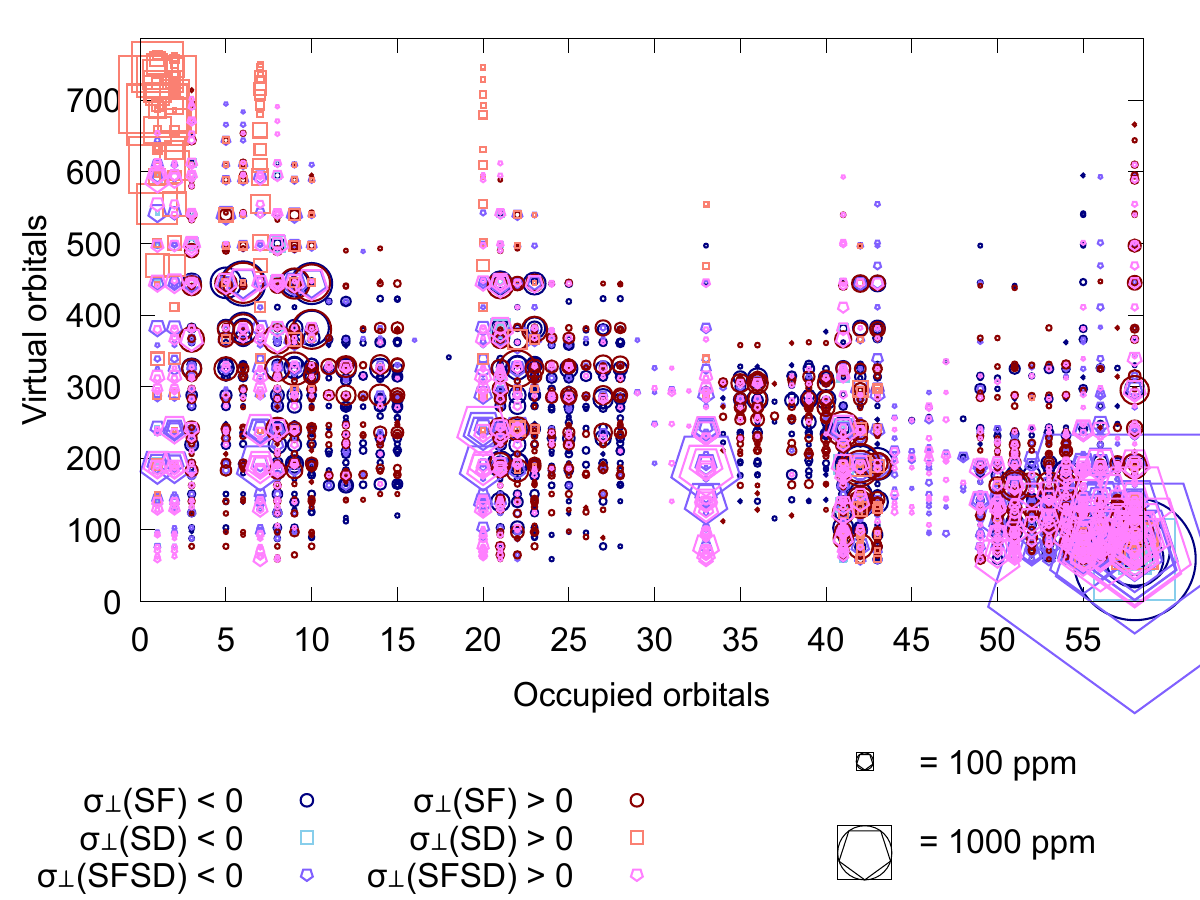}
      \end{minipage}
      \hfill
      \begin{minipage}{.48\textwidth}
      \includegraphics[width=\textwidth]{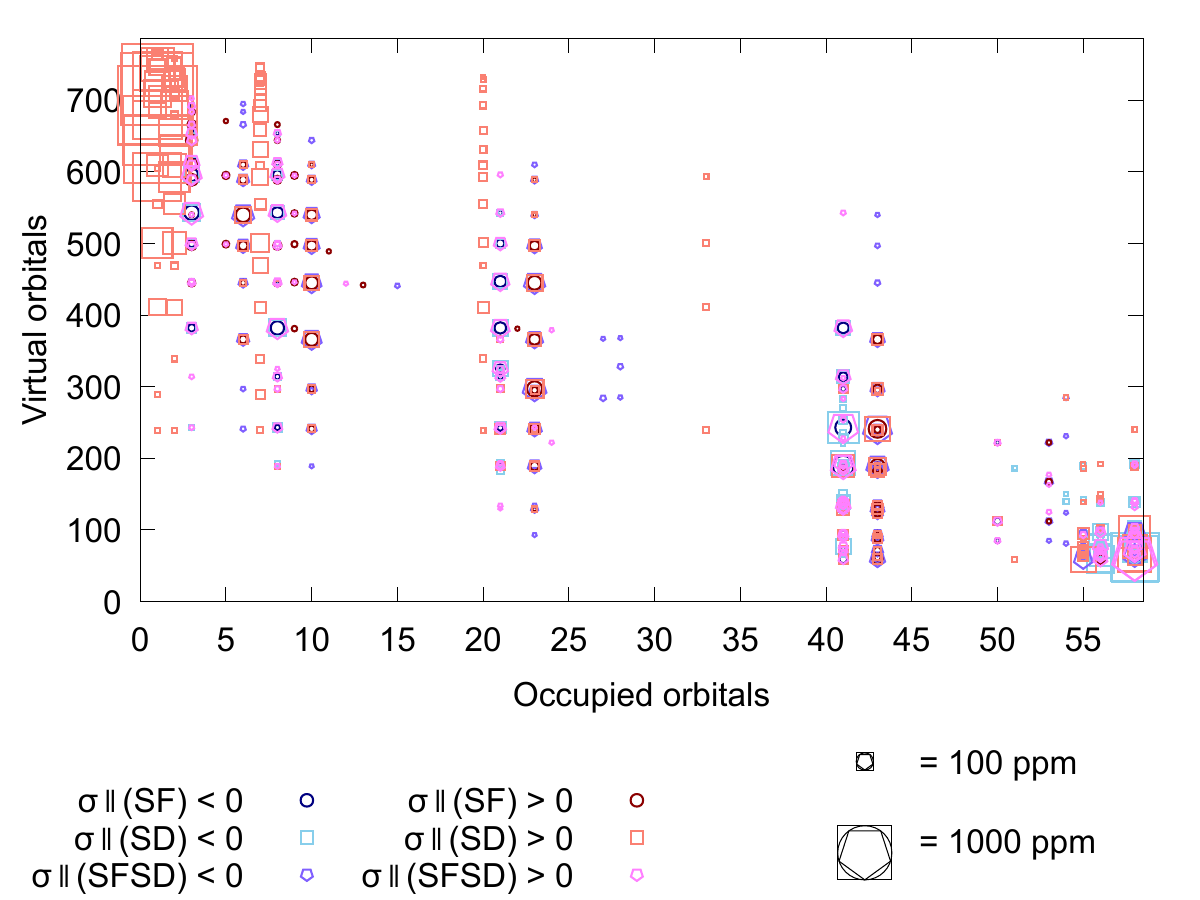}
      \end{minipage}%
      \caption{Pattern of excitations for $\sigma_\parallel$(Tl) (left)
                and $\sigma_\perp$(Tl) (right) for TlBr shown separately for the SD and SF parts of the NMR shielding propagator at the level of rZORA+eX2C-HF. The magnitude of the amplitude of each type is proportional to marked area. The numbers that label the occupied and virtual orbitals correspond to the Kramers pairs, such that 1 refers to the lowest energy occupied pair and 58 corresponds to the highest energy occupied pair.}
      \label{fig:propagator_rZORA+eX2C-HF_TlBr}
      \end{figure*}

      \newpage

%

      \begin{figure*}[h!]
      \begin{minipage}{.48\textwidth}
      \includegraphics[width=\textwidth]{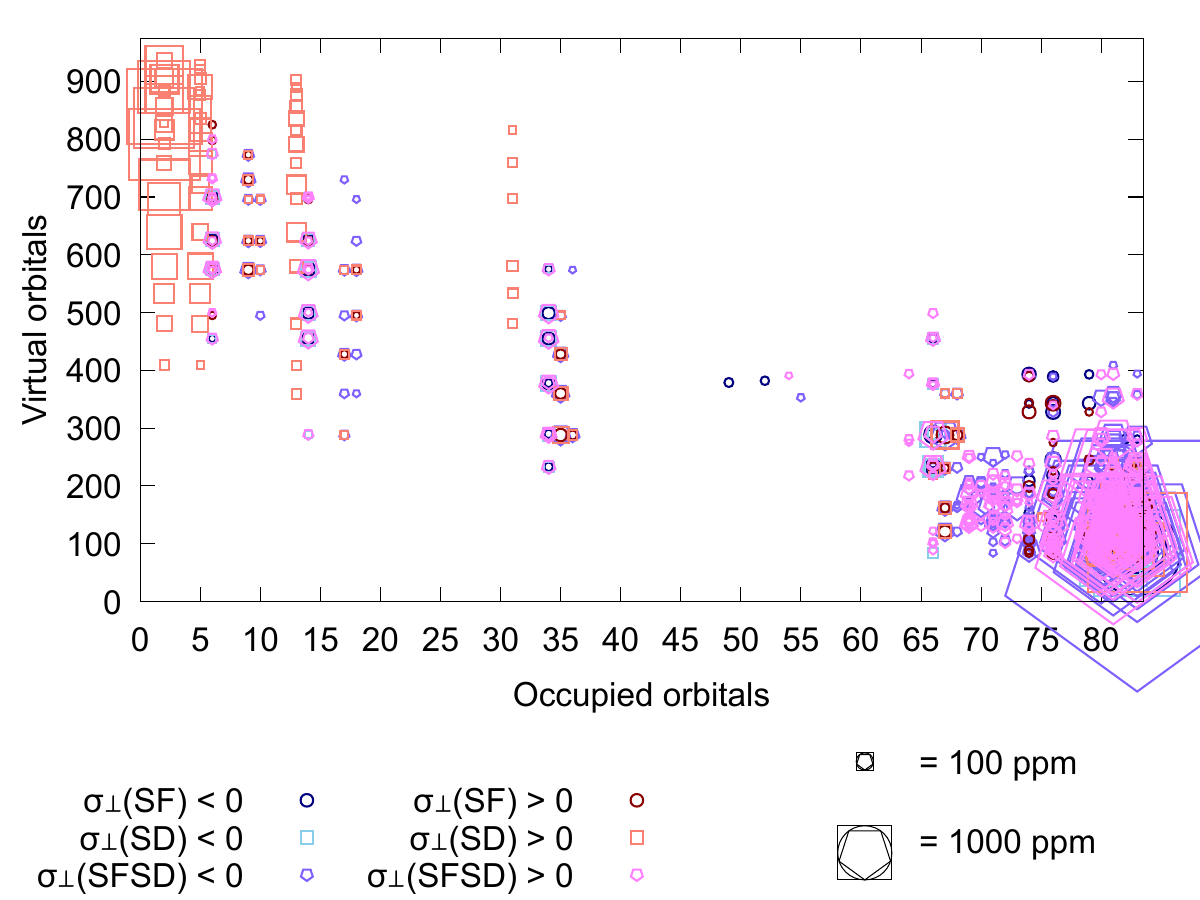}
      \end{minipage}
      \hfill
      \begin{minipage}{.48\textwidth}
      \includegraphics[width=\textwidth]{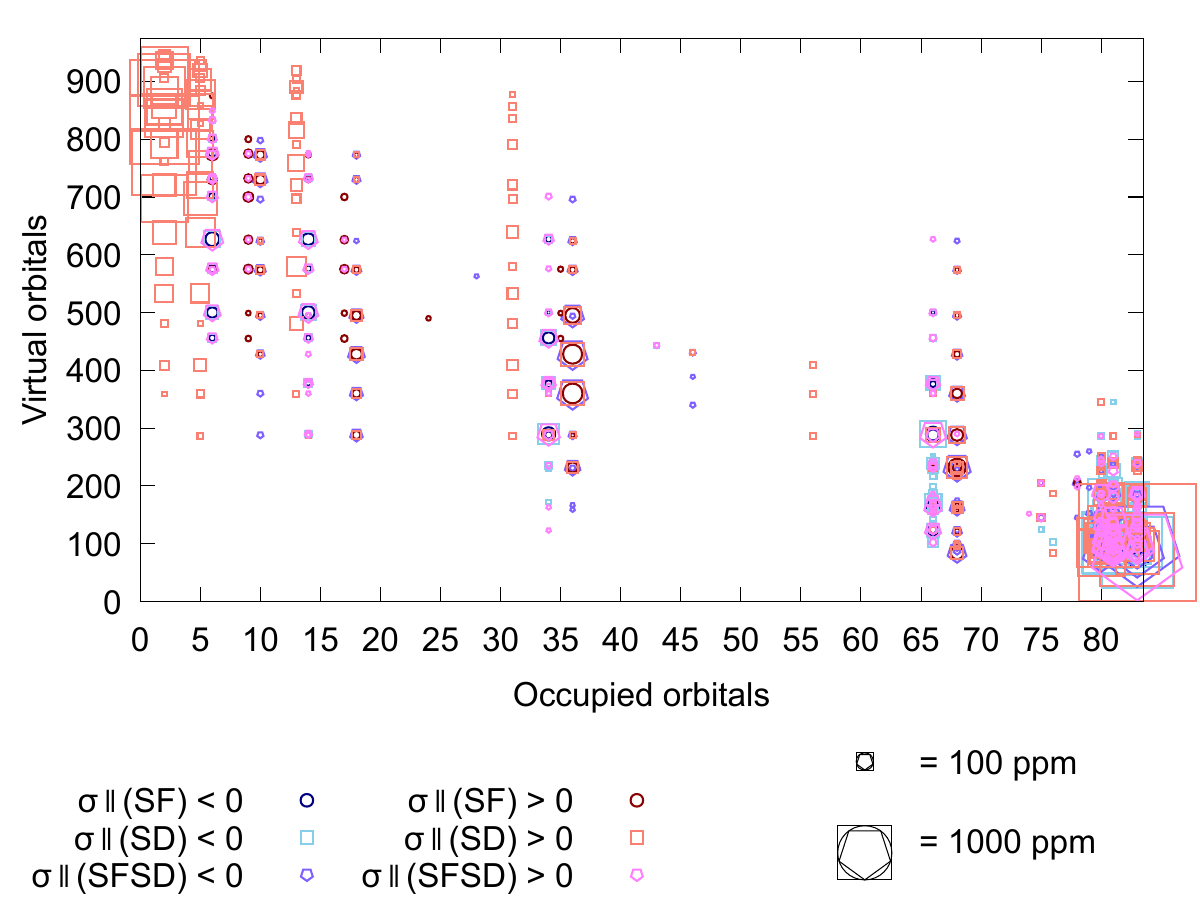}
      \end{minipage}%
      \caption{Pattern of excitations for $\sigma_\parallel$(Tl) (left)
                and $\sigma_\perp$(Tl) (right) for TlAt shown separately for the SD and SF parts of the NMR shielding propagator at the level of rZORA+eX2C-HF. The magnitude of the amplitude of each type is proportional to marked area. The numbers that label the occupied and virtual orbitals correspond to the Kramers pairs, such that 1 refers to the lowest energy occupied pair and 83 corresponds to the highest energy occupied pair.}
      \label{fig:propagator_rZORA+eX2C-HF_TlAt}
      \end{figure*}

      \newpage

      \begin{figure*}[h!]
      \begin{minipage}{.48\textwidth}
      \includegraphics[width=\textwidth]{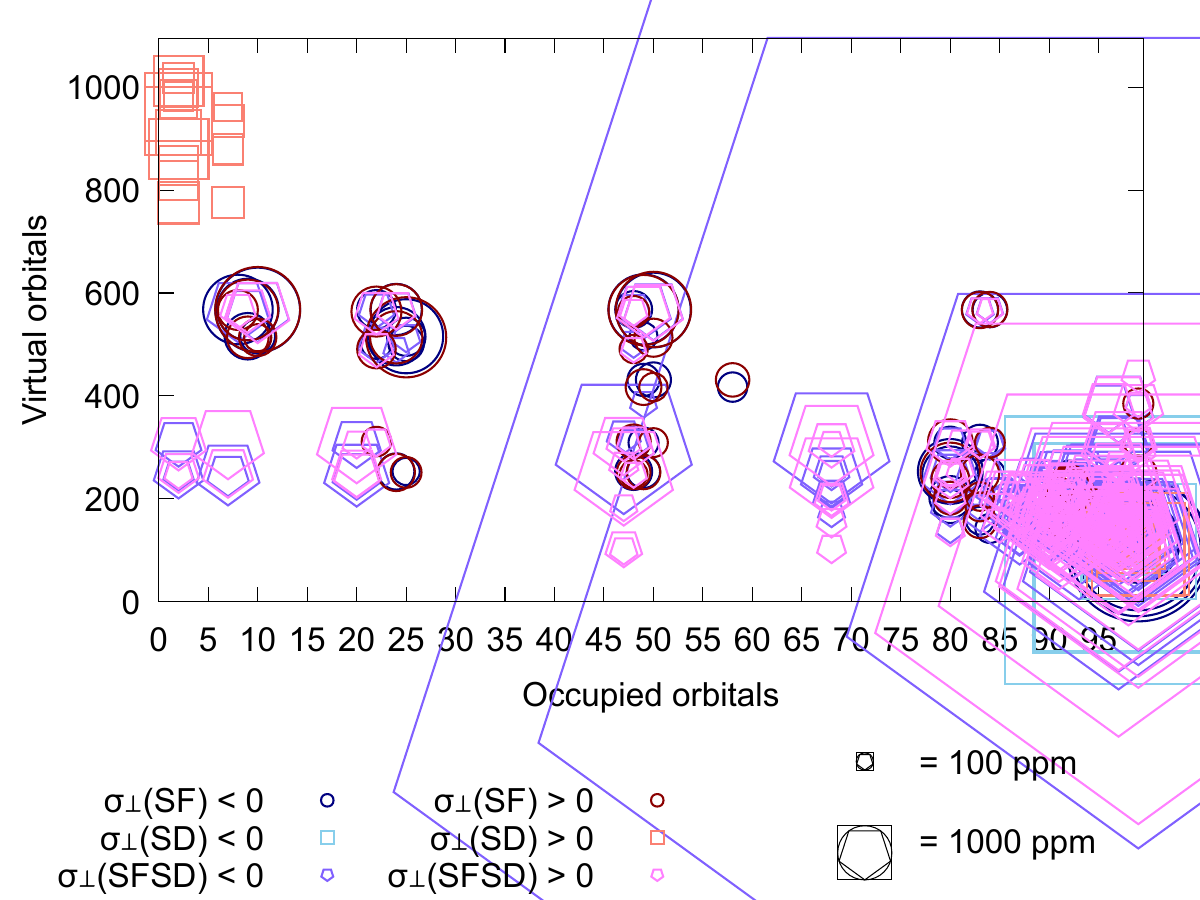}
      \end{minipage}
      \hfill
      \begin{minipage}{.48\textwidth}
      \includegraphics[width=\textwidth]{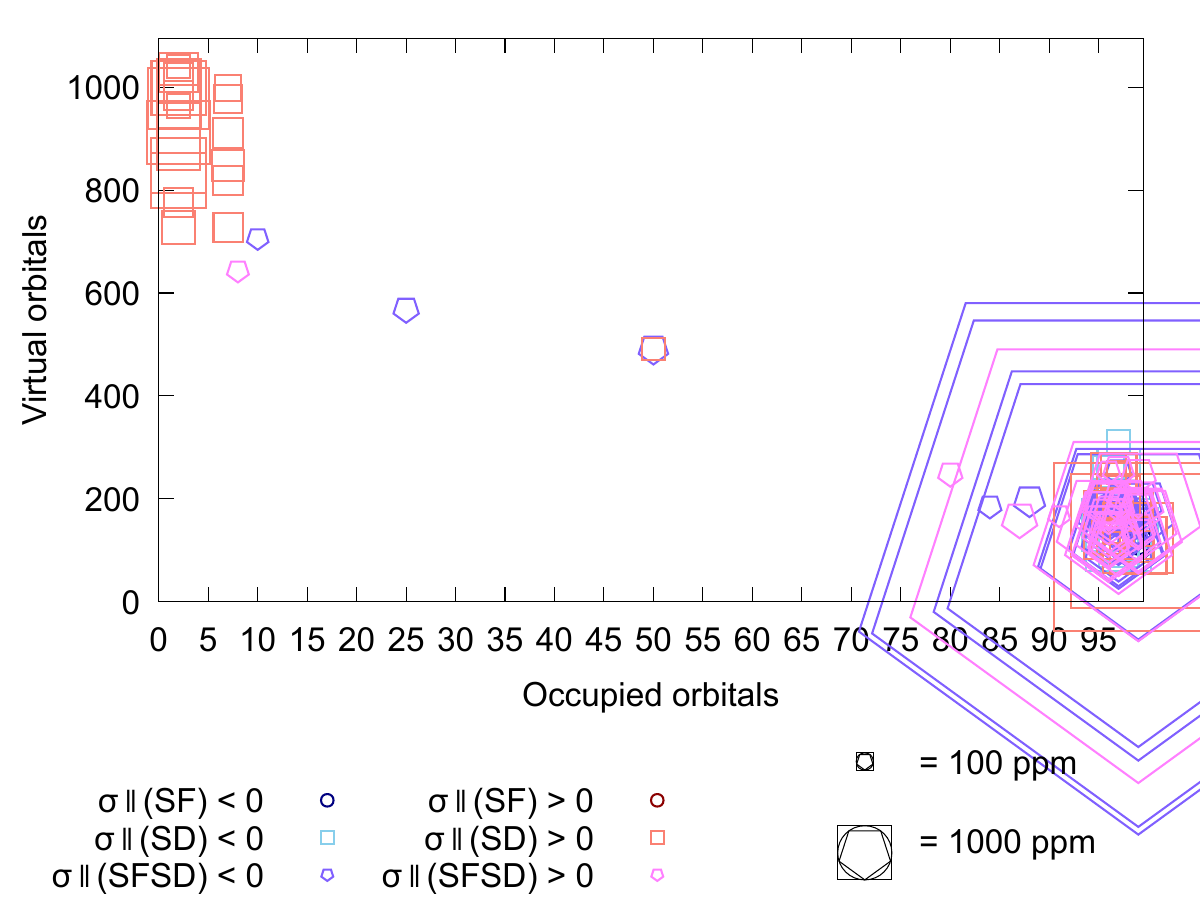}
      \end{minipage}%
      \caption{Pattern of excitations for $\sigma_\parallel$(Tl) (left)
                and $\sigma_\perp$(Tl) (right) for TlTs shown separately for the SD and SF parts of the NMR shielding propagator at the level of rZORA+eX2C-HF. The magnitude of the amplitude of each type is proportional to marked area. The numbers that label the occupied and virtual orbitals correspond to the Kramers pairs, such that 1 refers to the lowest energy occupied pair and 99 corresponds to the highest energy occupied pair.}
      \label{fig:propagator_rZORA+eX2C-HF_TlTs}
      \end{figure*}

      \newpage

  \subsection{rZORA-HF scheme}
 
    \begin{figure}[h!]
    \includegraphics[width=.5\textwidth]{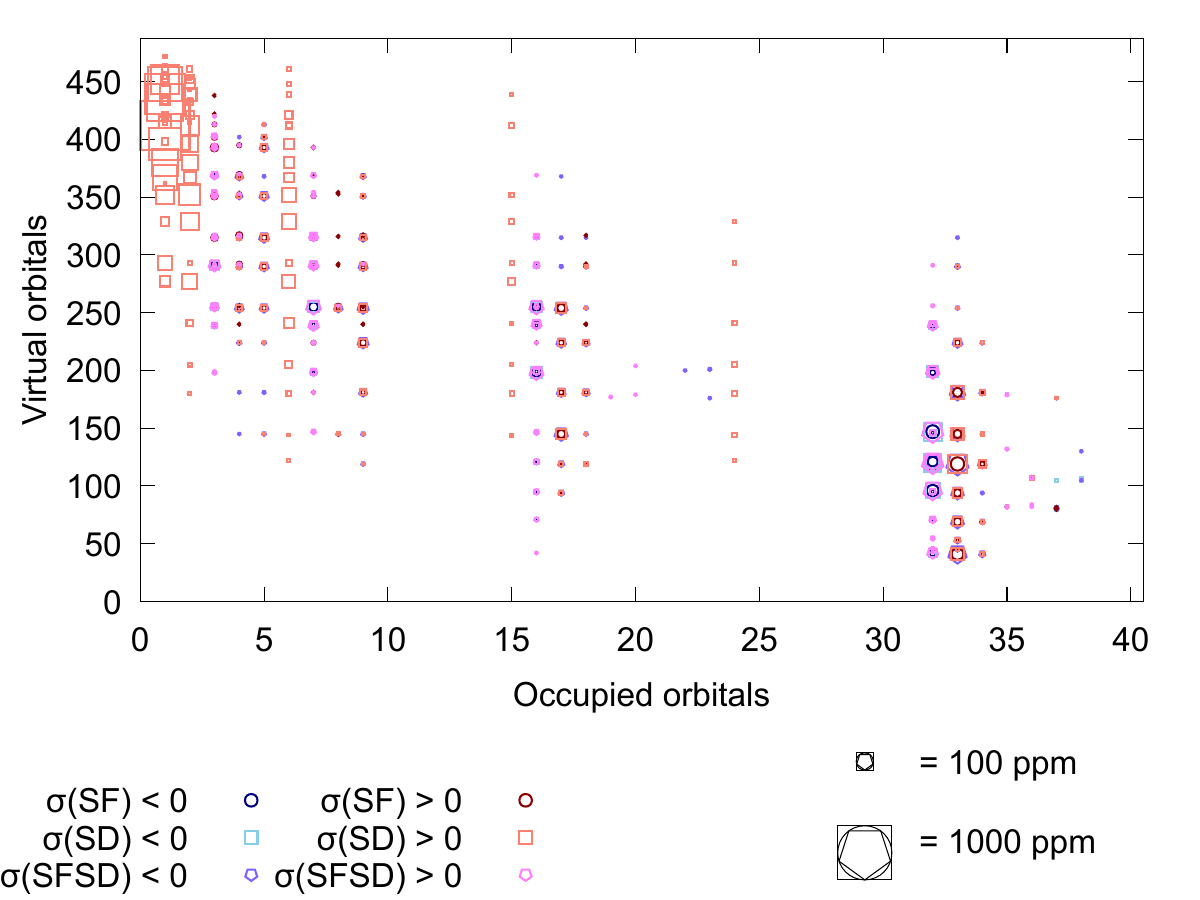}
    \caption{Pattern of excitations for $\sigma_\parallel$(Tl) (left)
              and $\sigma_\perp$(Tl) (right) for Tl$^+$ shown separately for the SD and SF parts of the NMR shielding propagator at the level of rZORA-HF. The magnitude of the amplitude of each type is proportional to marked area. The numbers that label the occupied and virtual orbitals correspond to the Kramers pairs, such that 1 refers to the lowest energy occupied pair and 40 corresponds to the highest energy occupied pair.}
    \label{fig:propagator_rZORA-HF_Tl+}
    \end{figure}
    
    \newpage

      \begin{figure*}[h!]
      \begin{minipage}{.48\textwidth}
      \includegraphics[width=\textwidth]{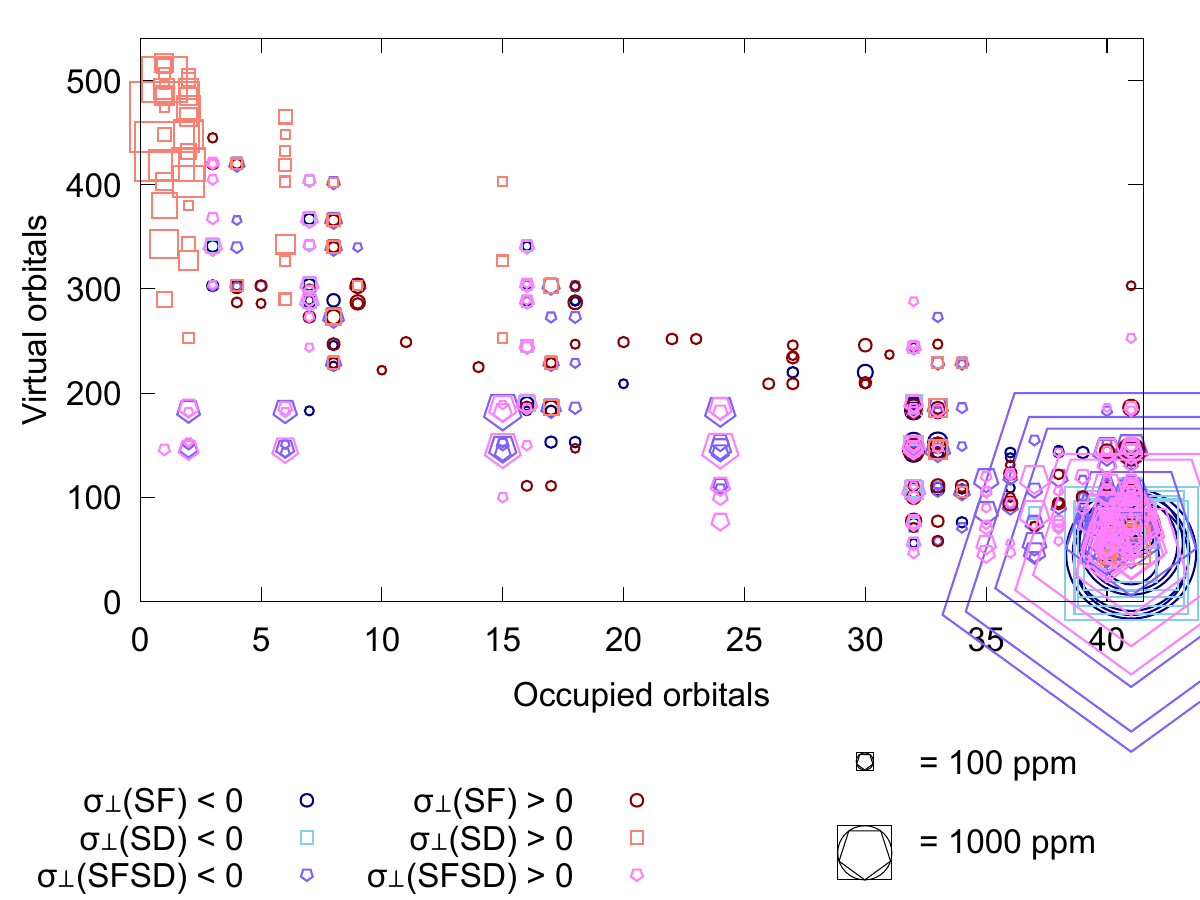}
      \end{minipage}
      \hfill
      \begin{minipage}{.48\textwidth}
      \includegraphics[width=\textwidth]{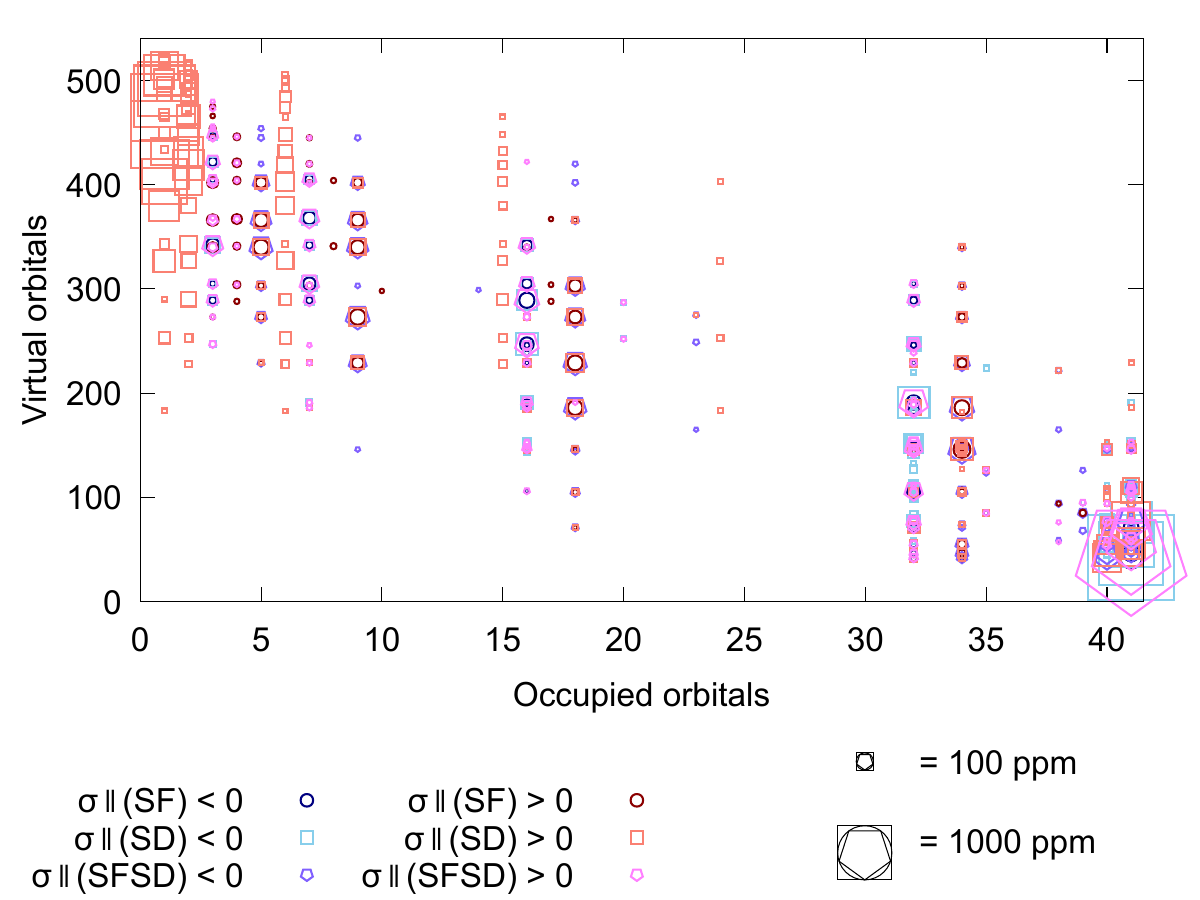}
      \end{minipage}%
      \caption{Pattern of excitations for $\sigma_\parallel$(Tl) (left)
                and $\sigma_\perp$(Tl) (right) for TlH shown separately for the SD and SF parts of the NMR shielding propagator at the level of rZORA-HF. The magnitude of the amplitude of each type is proportional to marked area. The numbers that label the occupied and virtual orbitals correspond to the Kramers pairs, such that 1 refers to the lowest energy occupied pair and 41 corresponds to the highest energy occupied pair.}
      \label{fig:propagator_rZORA-HF_TlH}
      \end{figure*}

      \newpage

      \begin{figure*}[h!]
      \begin{minipage}{.48\textwidth}
      \includegraphics[width=\textwidth]{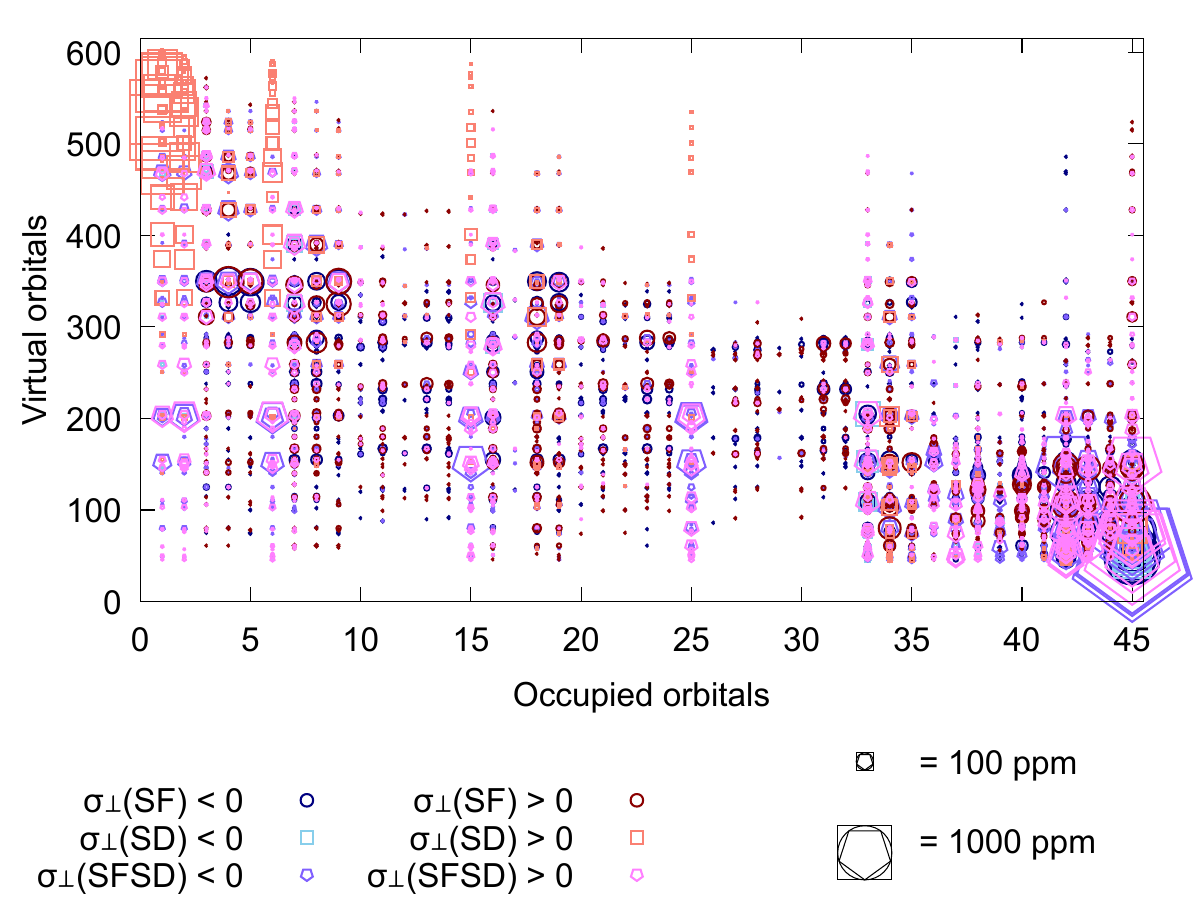}
      \end{minipage}
      \hfill
      \begin{minipage}{.48\textwidth}
      \includegraphics[width=\textwidth]{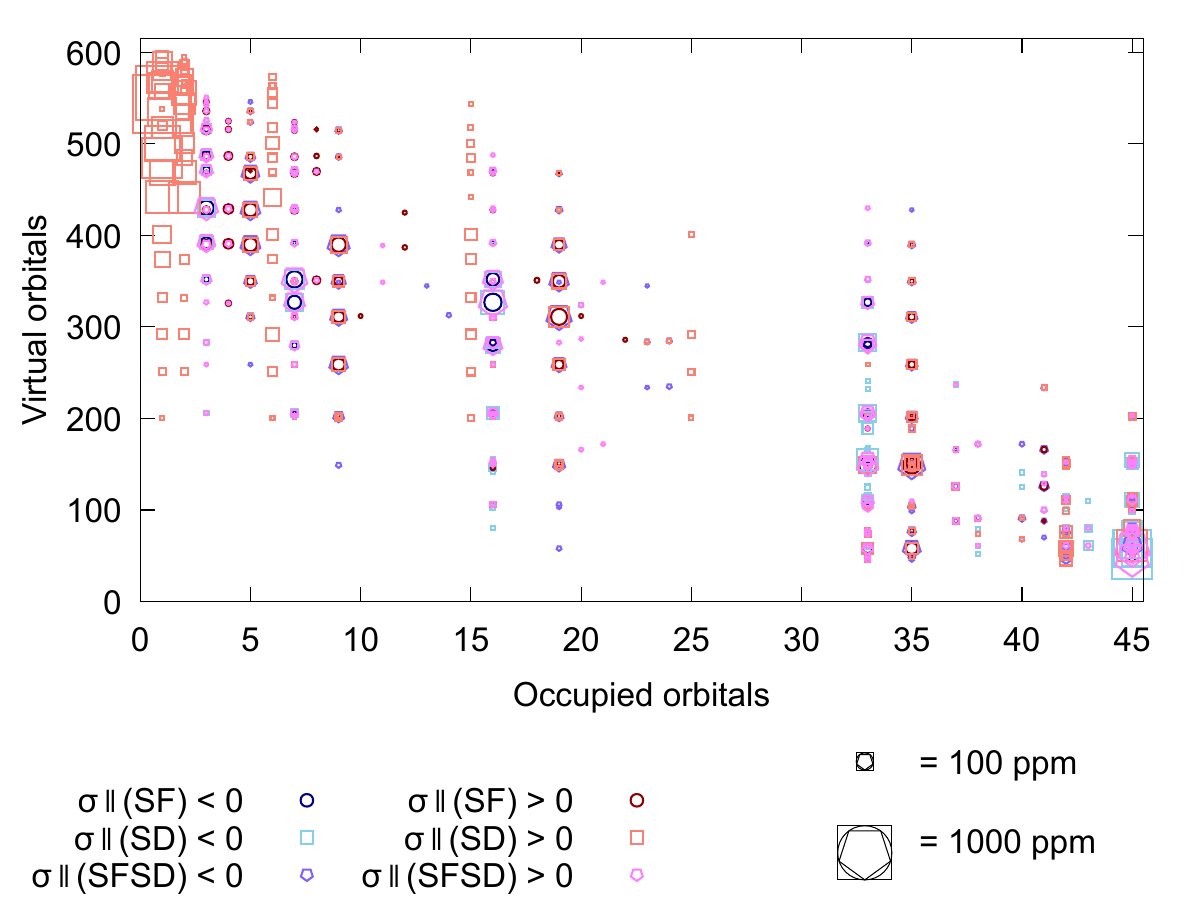}
      \end{minipage}%
      \caption{Pattern of excitations for $\sigma_\parallel$(Tl) (left)
                and $\sigma_\perp$(Tl) (right) for TlF shown separately for the SD and SF parts of the NMR shielding propagator at the level of rZORA-HF. The magnitude of the amplitude of each type is proportional to marked area. The numbers that label the occupied and virtual orbitals correspond to the Kramers pairs, such that 1 refers to the lowest energy occupied pair and 45 corresponds to the highest energy occupied pair.}
      \label{fig:propagator_rZORA-HF_TlF}
      \end{figure*}

      \newpage

      \begin{figure*}[h!]
      \begin{minipage}{.48\textwidth}
      \includegraphics[width=\textwidth]{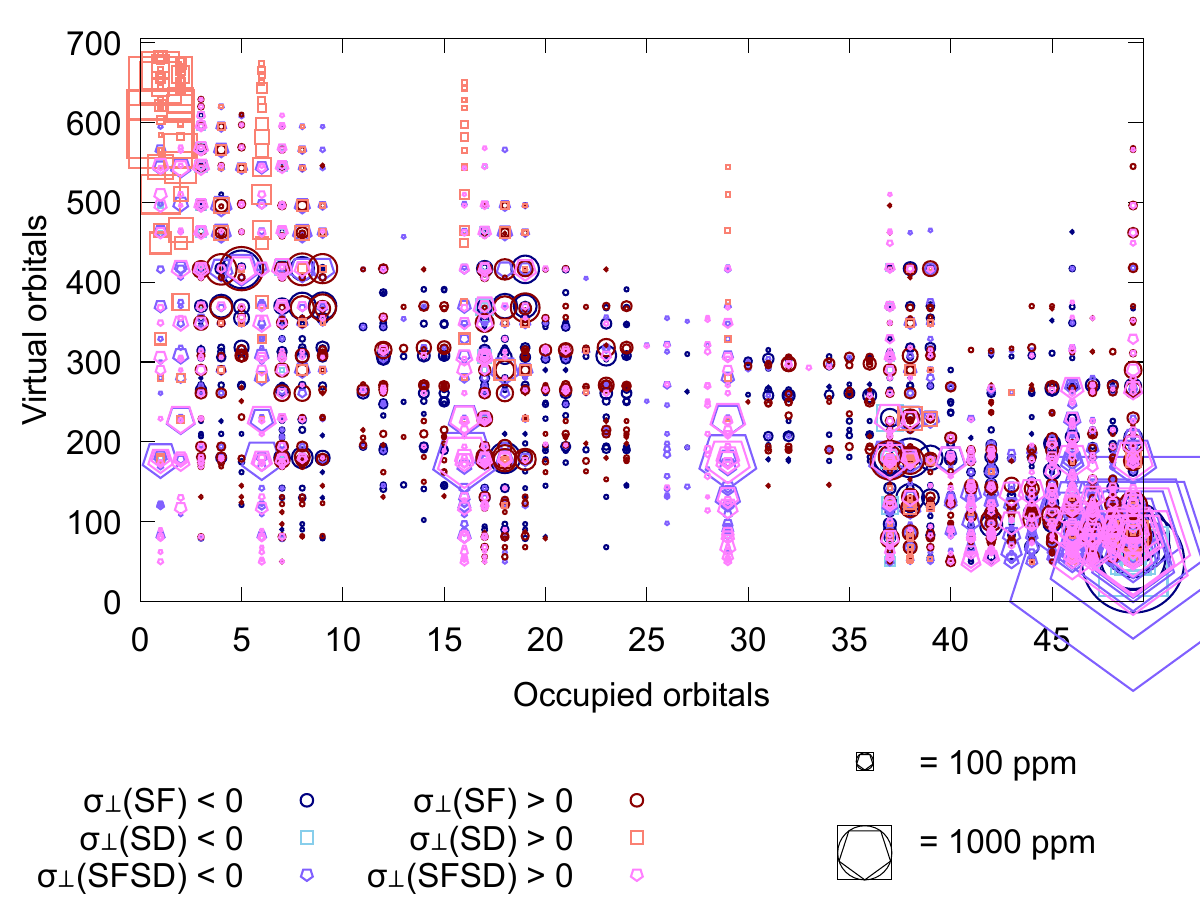}
      \end{minipage}
      \hfill
      \begin{minipage}{.48\textwidth}
      \includegraphics[width=\textwidth]{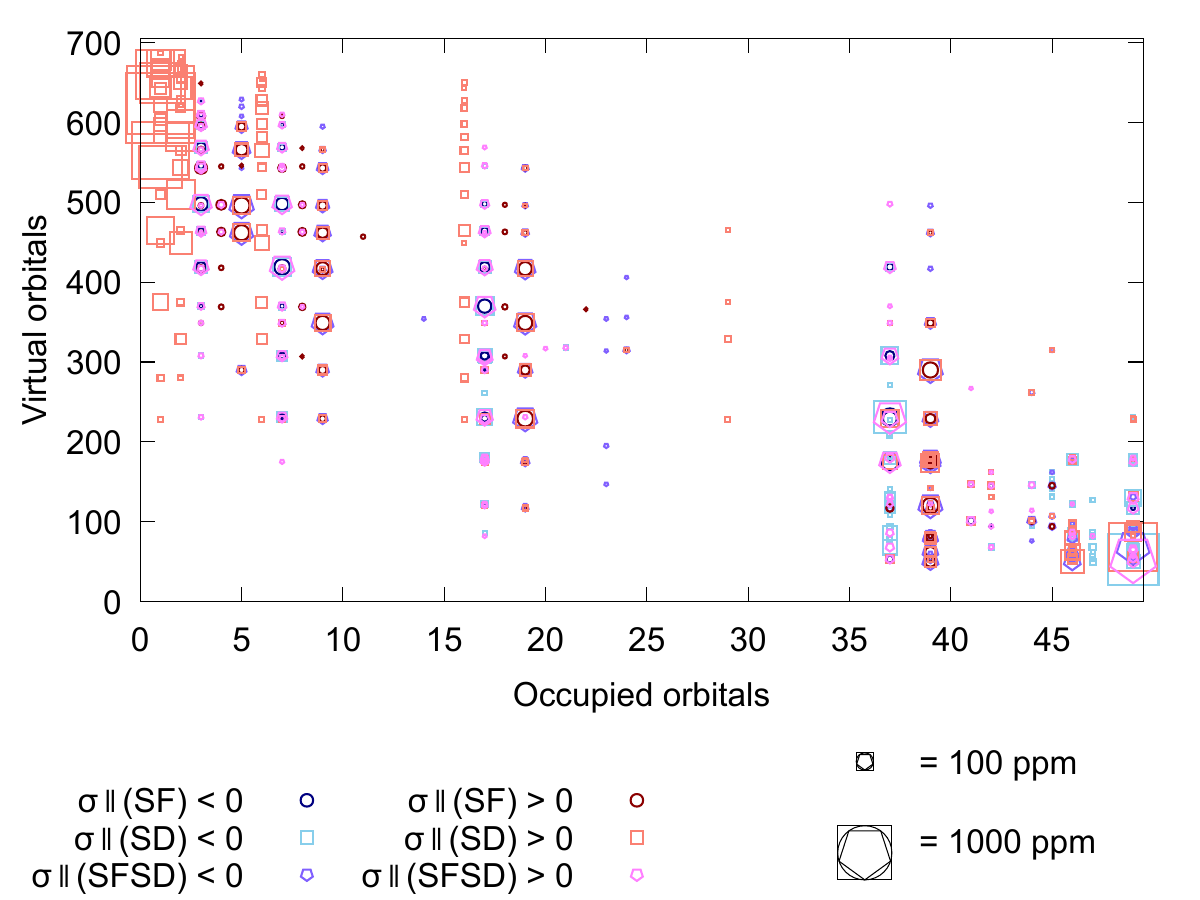}
      \end{minipage}%
      \caption{Pattern of excitations for $\sigma_\parallel$(Tl) (left)
                and $\sigma_\perp$(Tl) (right) for TlCl shown separately for the SD and SF parts of the NMR shielding propagator at the level of rZORA-HF. The magnitude of the amplitude of each type is proportional to marked area. The numbers that label the occupied and virtual orbitals correspond to the Kramers pairs, such that 1 refers to the lowest energy occupied pair and 49 corresponds to the highest energy occupied pair.}
      \label{fig:propagator_rZORA-HF_TlCl}
      \end{figure*}

      \newpage

      \begin{figure*}[h!]
      \begin{minipage}{.48\textwidth}
      \includegraphics[width=\textwidth]{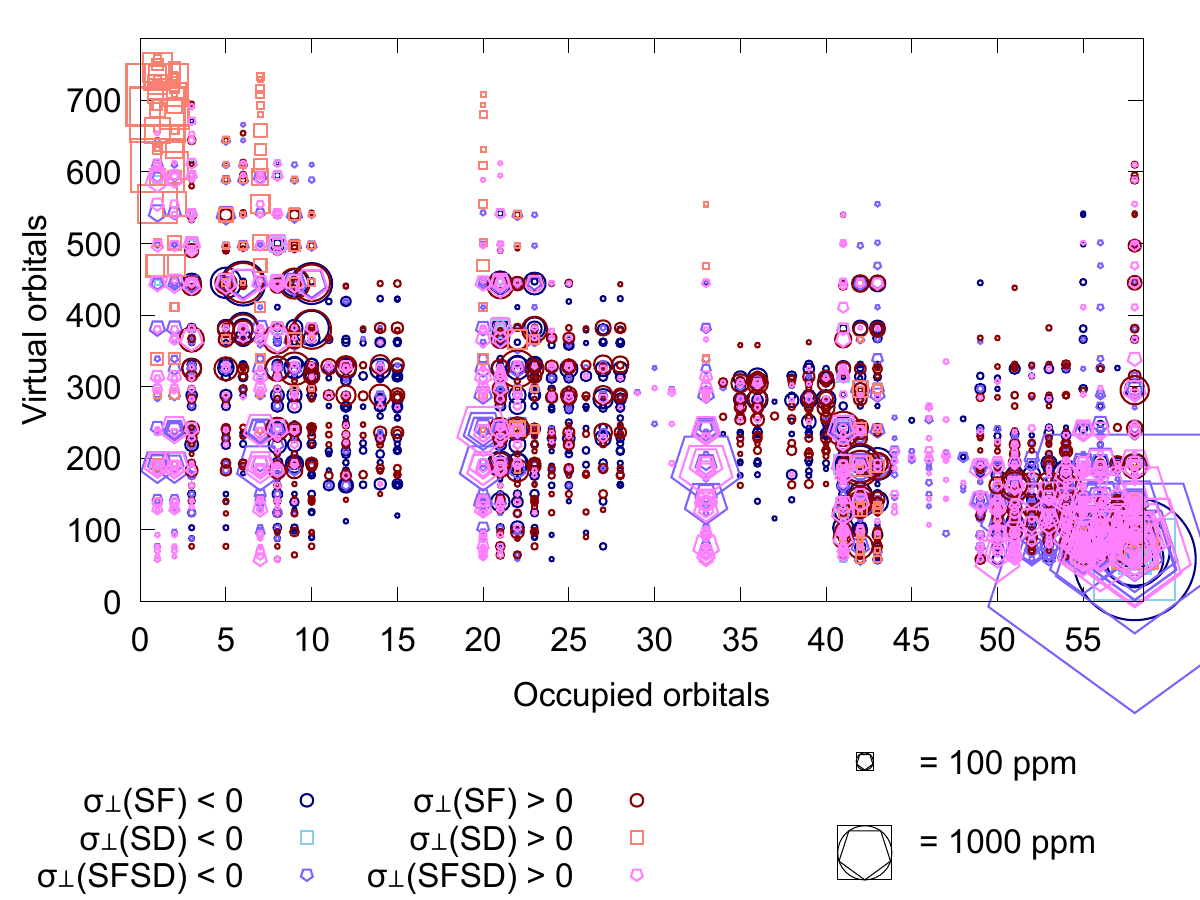}
      \end{minipage}
      \hfill
      \begin{minipage}{.48\textwidth}
      \includegraphics[width=\textwidth]{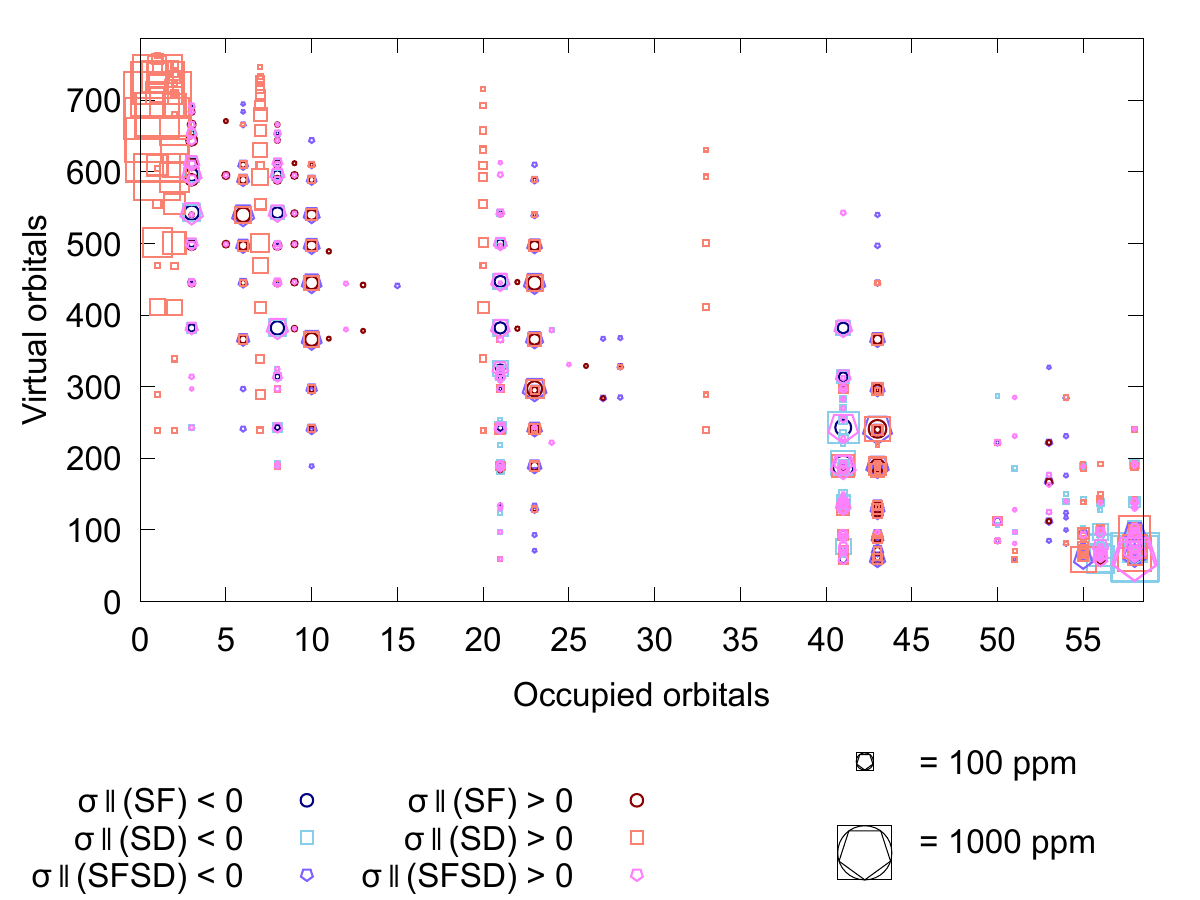}
      \end{minipage}%
      \caption{Pattern of excitations for $\sigma_\parallel$(Tl) (left)
                and $\sigma_\perp$(Tl) (right) for TlBr shown separately for the SD and SF parts of the NMR shielding propagator at the level of rZORA-HF. The magnitude of the amplitude of each type is proportional to marked area. The numbers that label the occupied and virtual orbitals correspond to the Kramers pairs, such that 1 refers to the lowest energy occupied pair and 58 corresponds to the highest energy occupied pair.}
      \label{fig:propagator_rZORA-HF_TlBr}
      \end{figure*}

      \newpage

      \begin{figure*}[h!]
      \begin{minipage}{.48\textwidth}
      \includegraphics[width=\textwidth]{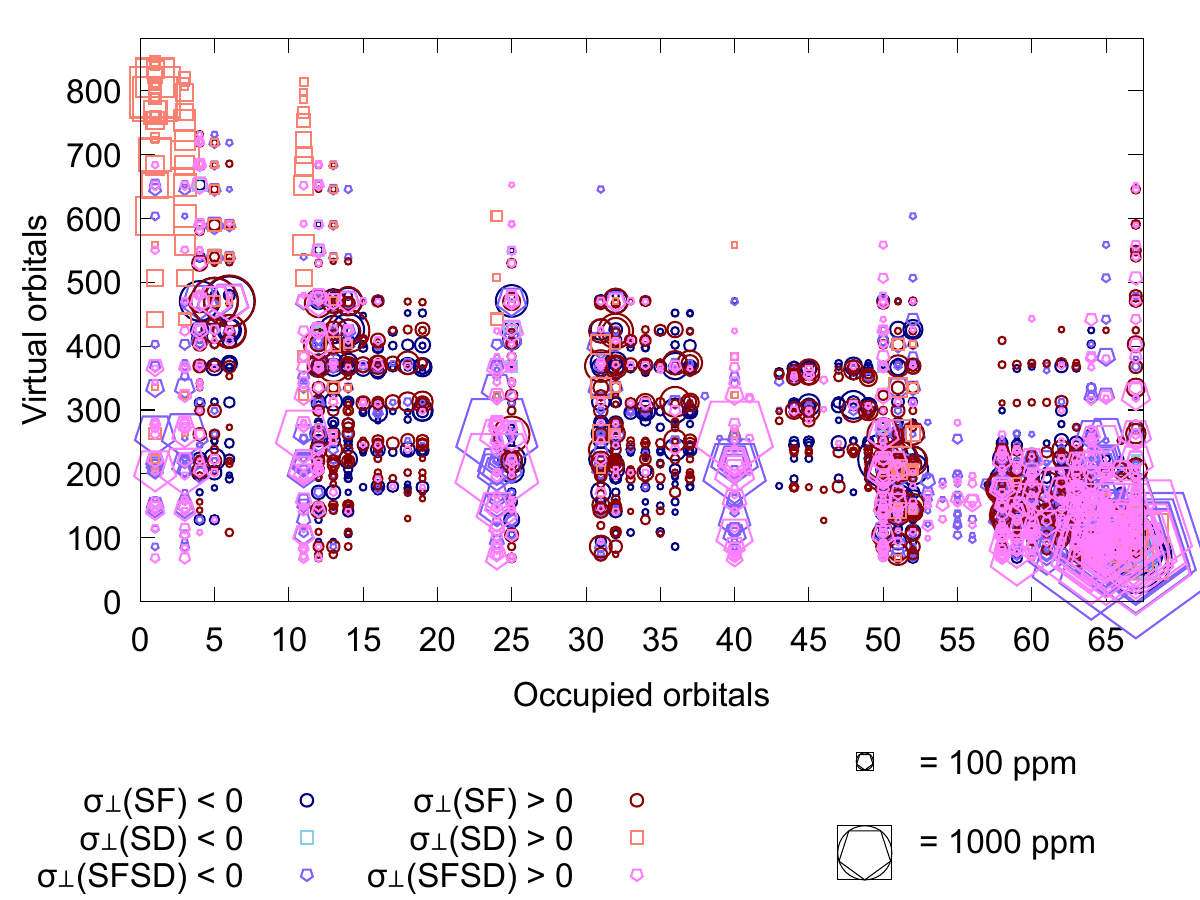}
      \end{minipage}
      \hfill
      \begin{minipage}{.48\textwidth}
      \includegraphics[width=\textwidth]{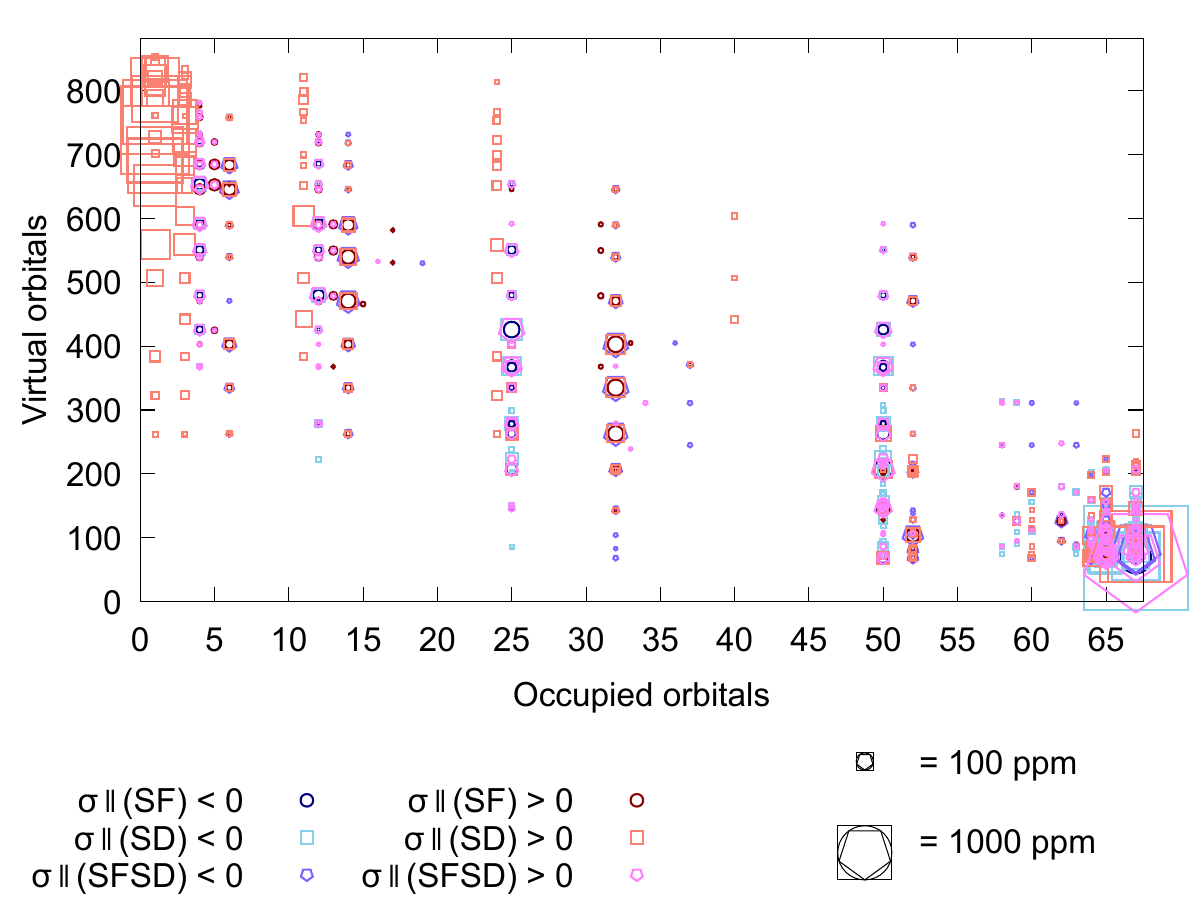}
      \end{minipage}%
      \caption{Pattern of excitations for $\sigma_\parallel$(Tl) (left)
                and $\sigma_\perp$(Tl) (right) for TlI shown separately for the SD and SF parts of the NMR shielding propagator at the level of rZORA-HF. The magnitude of the amplitude of each type is proportional to marked area. The numbers that label the occupied and virtual orbitals correspond to the Kramers pairs, such that 1 refers to the lowest energy occupied pair and 67 corresponds to the highest energy occupied pair.}
      \label{fig:propagator_rZORA-HF_TlI}
      \end{figure*}

      \newpage

      \begin{figure*}[h!]
      \begin{minipage}{.48\textwidth}
      \includegraphics[width=\textwidth]{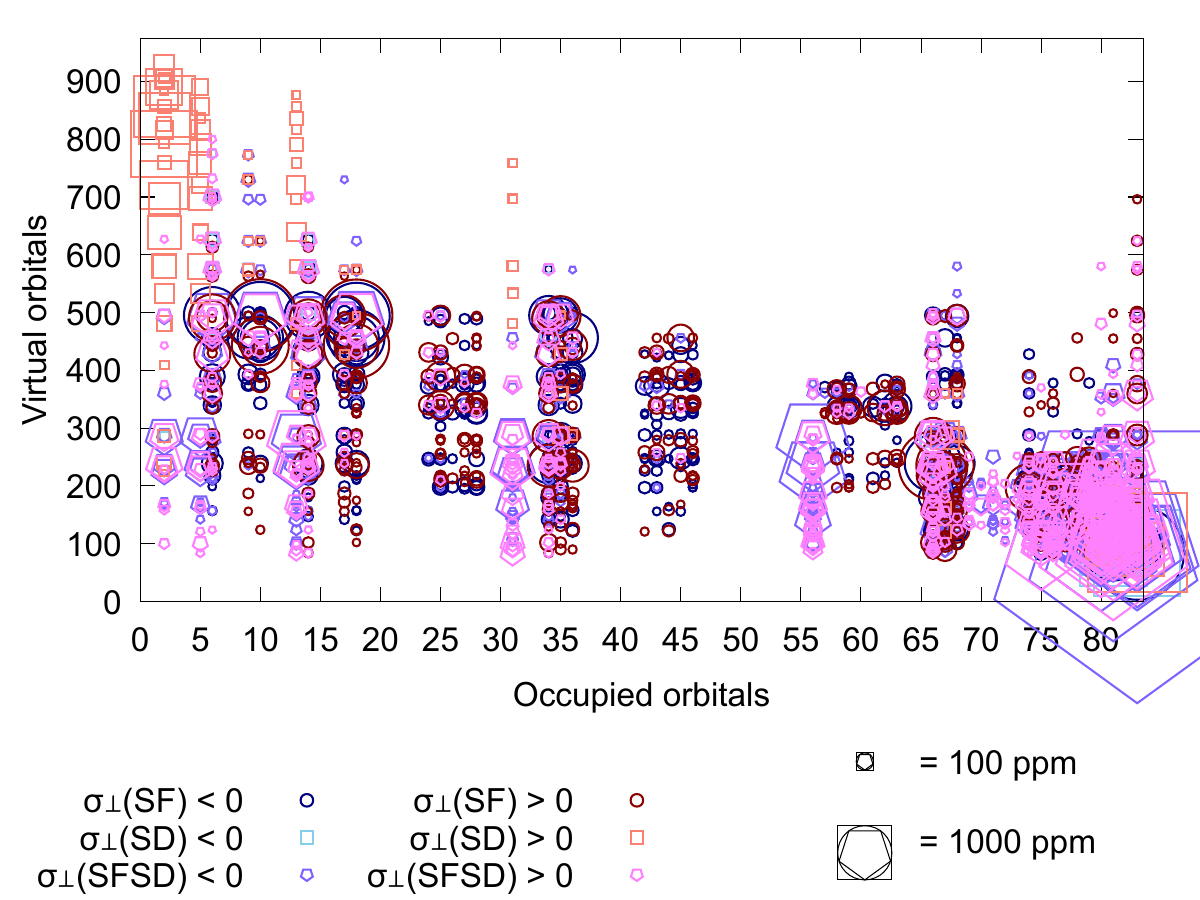}
      \end{minipage}
      \hfill
      \begin{minipage}{.48\textwidth}
      \includegraphics[width=\textwidth]{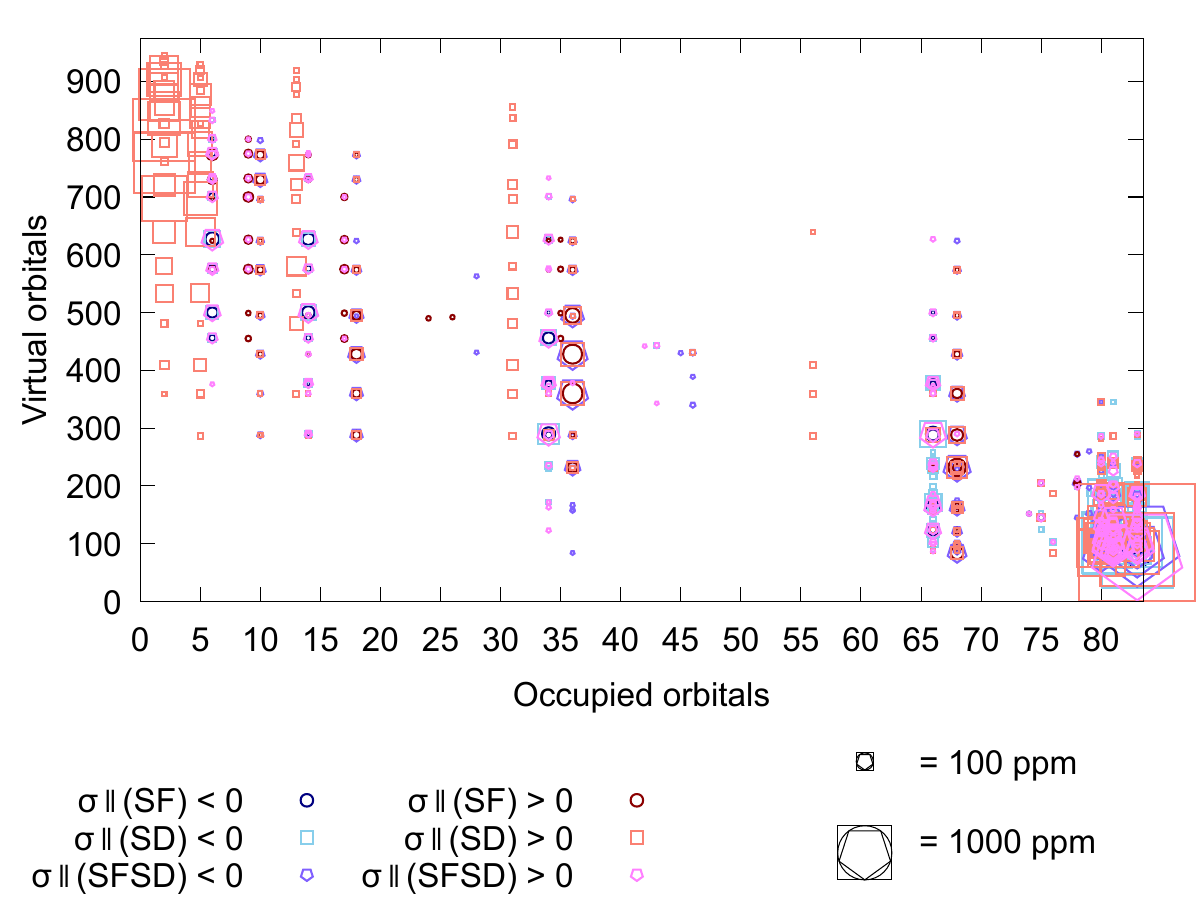}
      \end{minipage}%
      \caption{Pattern of excitations for $\sigma_\parallel$(Tl) (left)
                and $\sigma_\perp$(Tl) (right) for TlAt shown separately for the SD and SF parts of the NMR shielding propagator at the level of rZORA-HF. The magnitude of the amplitude of each type is proportional to marked area. The numbers that label the occupied and virtual orbitals correspond to the Kramers pairs, such that 1 refers to the lowest energy occupied pair and 83 corresponds to the highest energy occupied pair.}
      \label{fig:propagator_rZORA-HF_TlAt}
      \end{figure*}

      \newpage

      \begin{figure*}[h!]
      \begin{minipage}{.48\textwidth}
      \includegraphics[width=\textwidth]{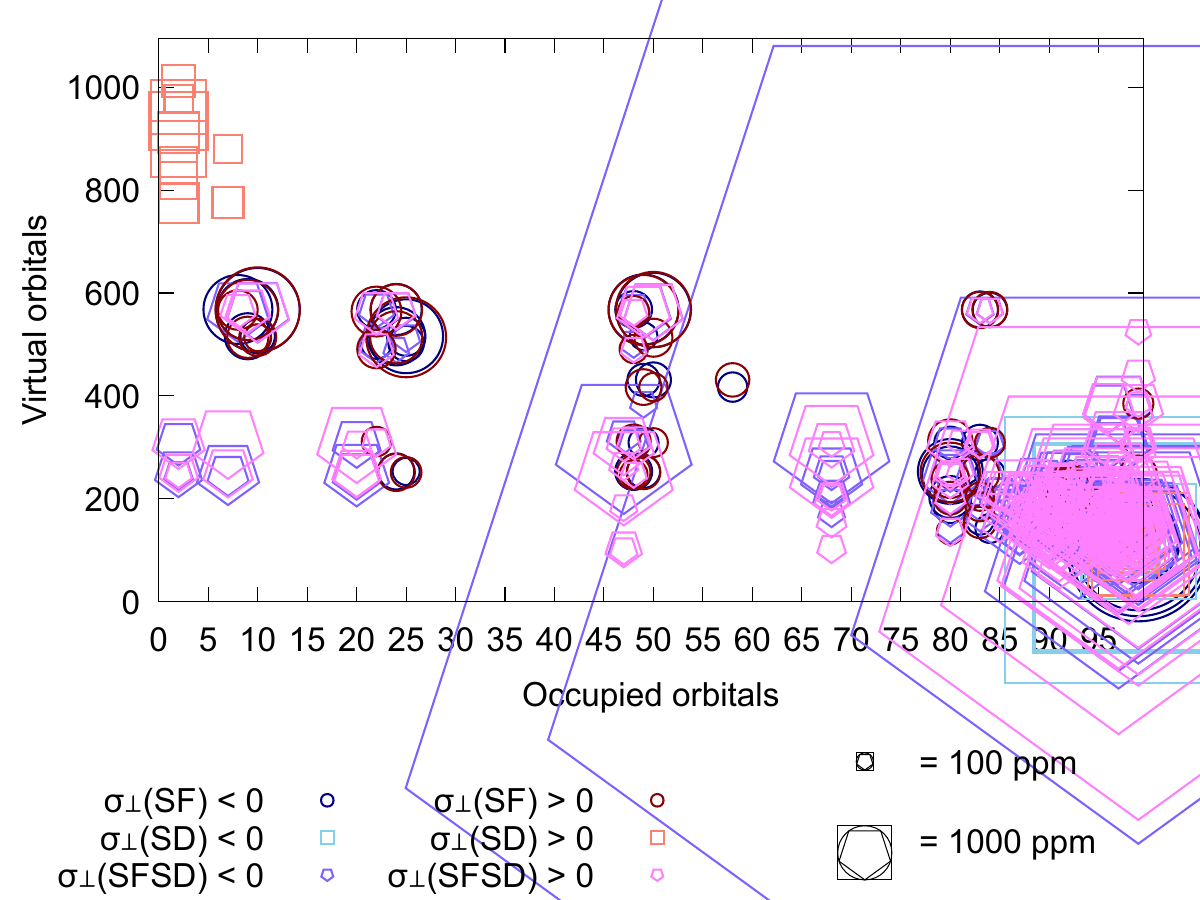}
      \end{minipage}
      \hfill
      \begin{minipage}{.48\textwidth}
      \includegraphics[width=\textwidth]{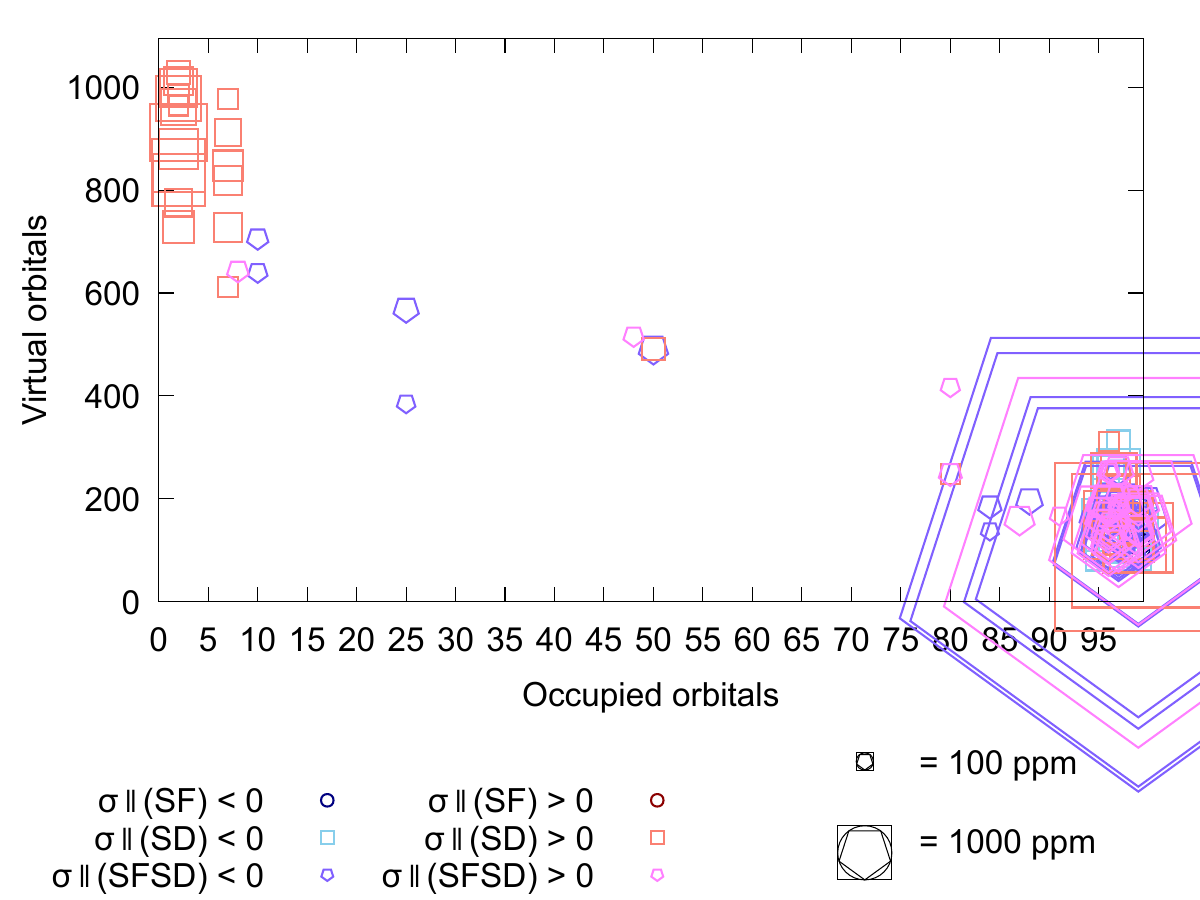}
      \end{minipage}%
      \caption{Pattern of excitations for $\sigma_\parallel$(Tl) (left)
                and $\sigma_\perp$(Tl) (right) for TlTs shown separately for the SD and SF parts of the NMR shielding propagator at the level of rZORA-HF. The magnitude of the amplitude of each type is proportional to marked area. The numbers that label the occupied and virtual orbitals correspond to the Kramers pairs, such that 1 refers to the lowest energy occupied pair and 99 corresponds to the highest energy occupied pair.}
      \label{fig:propagator_rZORA-HF_TlTs}
      \end{figure*}

      \newpage

  \subsection{rZORA-PBE0 scheme}
 
    \begin{figure}[h!]
    \includegraphics[width=.5\textwidth]{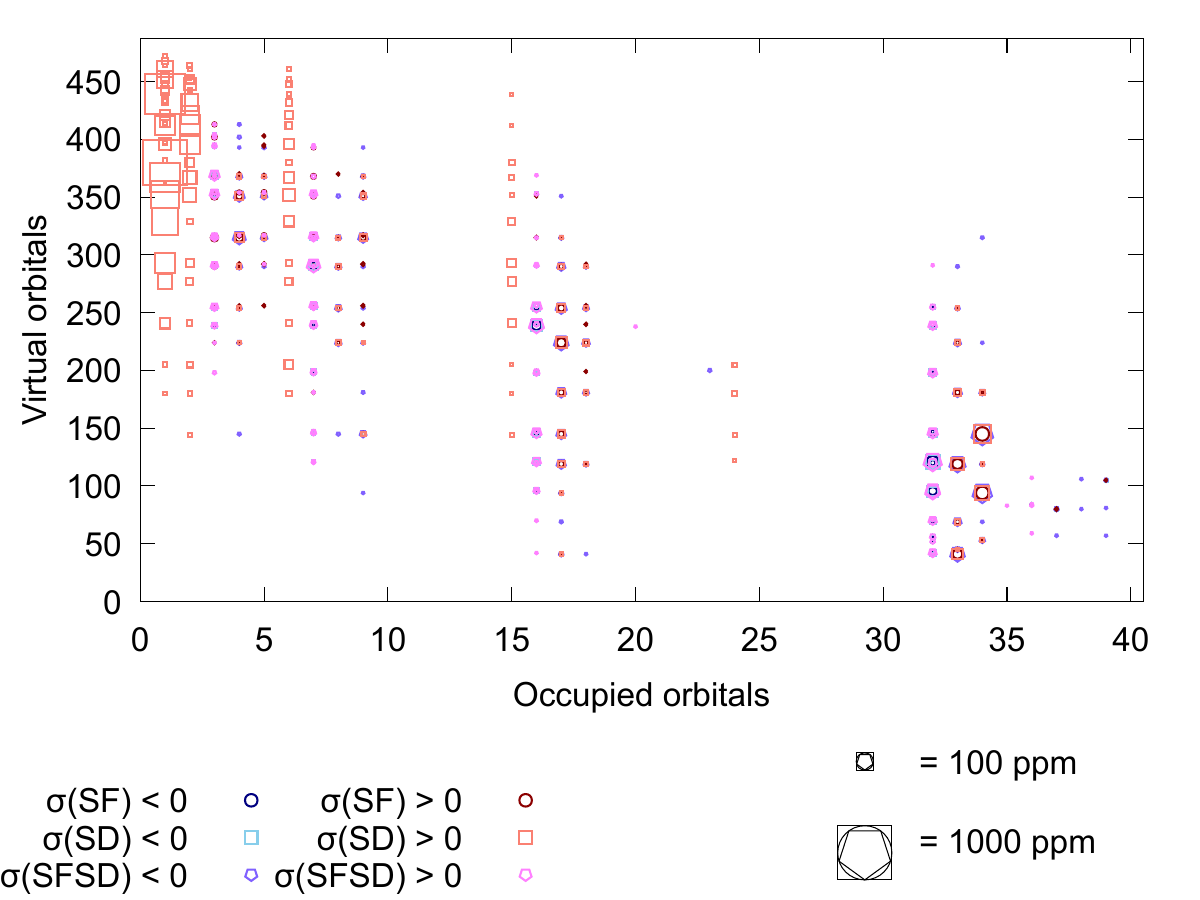}
    \caption{Pattern of excitations for $\sigma_\parallel$(Tl) (left)
              and $\sigma_\perp$(Tl) (right) for Tl$^+$ shown separately for the SD and SF parts of the NMR shielding propagator at the level of rZORA-PBE0. The magnitude of the amplitude of each type is proportional to marked area. The numbers that label the occupied and virtual orbitals correspond to the Kramers pairs, such that 1 refers to the lowest energy occupied pair and 40 corresponds to the highest energy occupied pair.}
    \label{fig:propagator_rZORA-PBE0_Tl+}
    \end{figure}
    
    \newpage

      \begin{figure*}[h!]
      \begin{minipage}{.48\textwidth}
      \includegraphics[width=\textwidth]{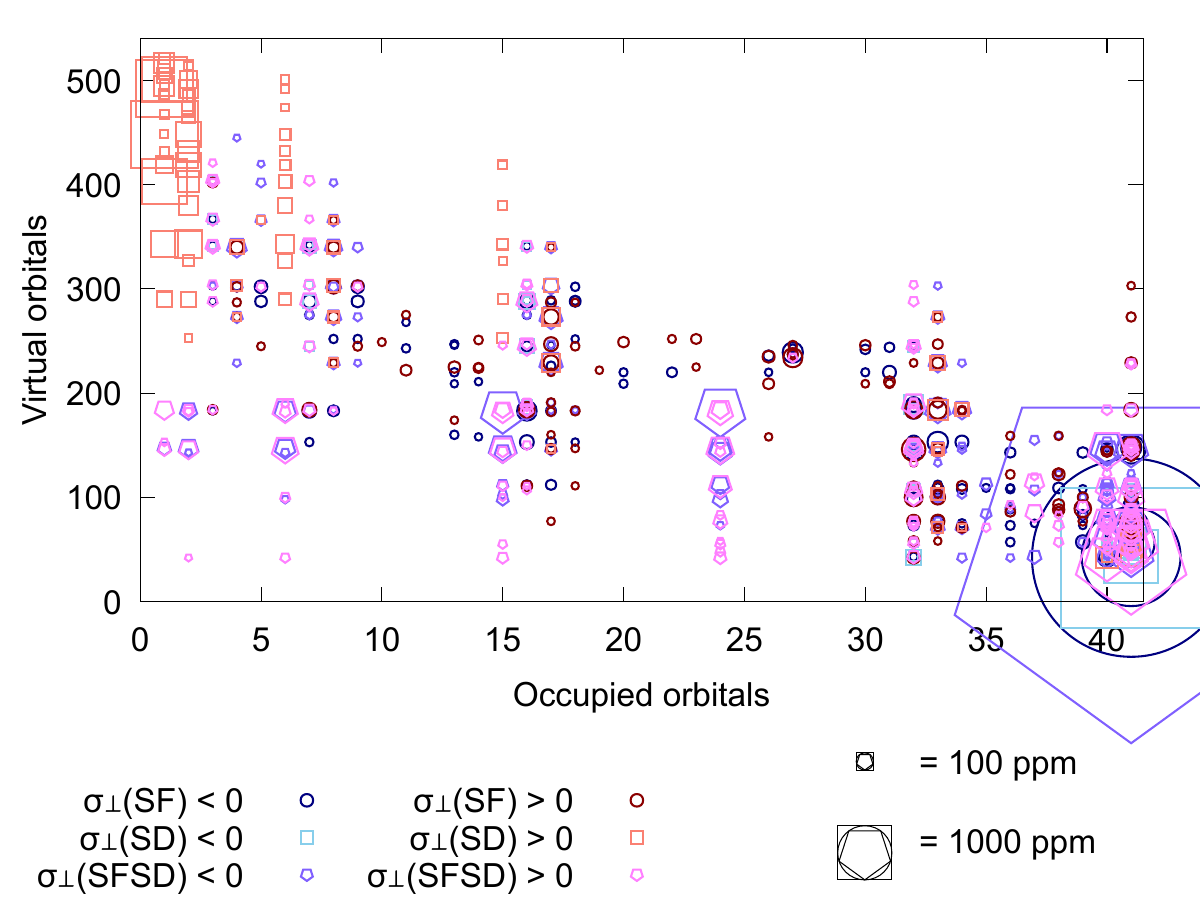}
      \end{minipage}
      \hfill
      \begin{minipage}{.48\textwidth}
      \includegraphics[width=\textwidth]{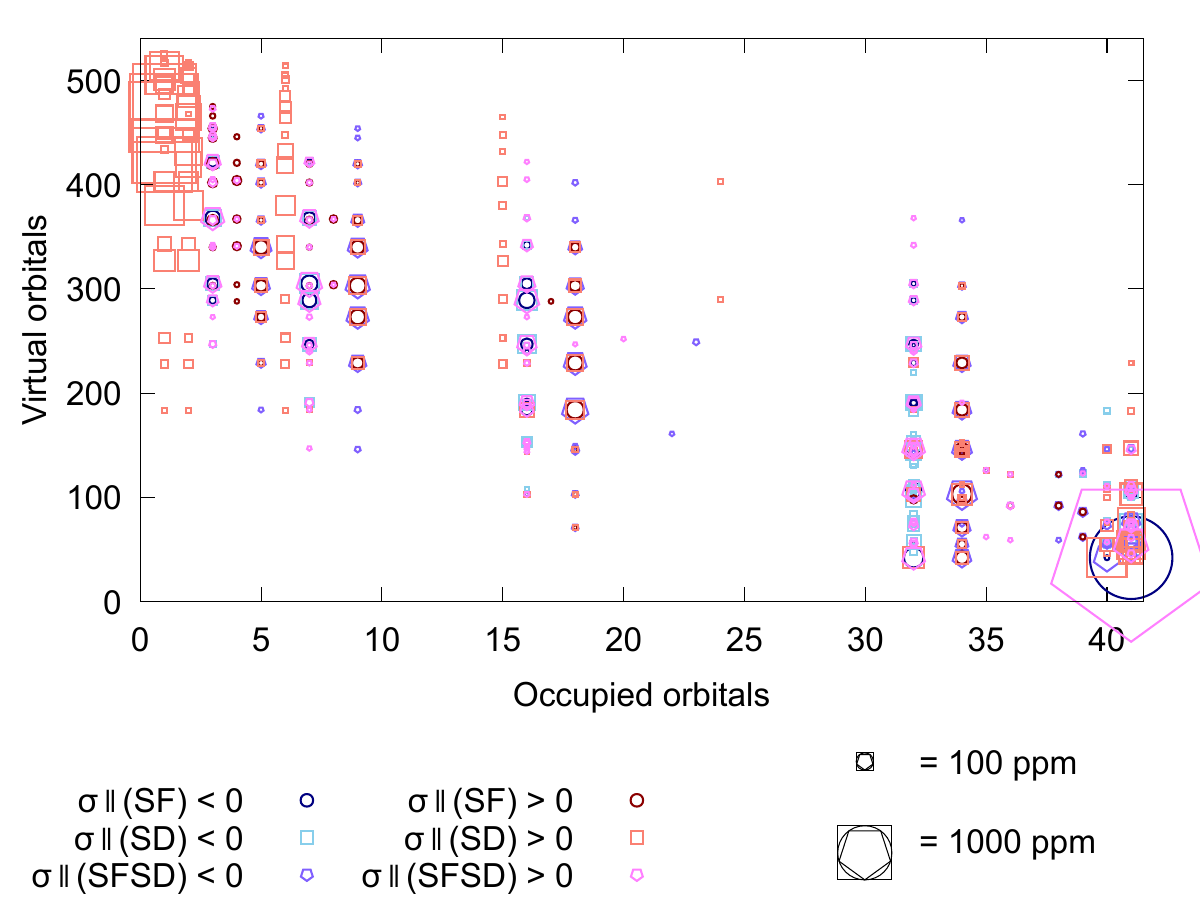}
      \end{minipage}%
      \caption{Pattern of excitations for $\sigma_\parallel$(Tl) (left)
                and $\sigma_\perp$(Tl) (right) for TlH shown separately for the SD and SF parts of the NMR shielding propagator at the level of rZORA-PBE0. The magnitude of the amplitude of each type is proportional to marked area. The numbers that label the occupied and virtual orbitals correspond to the Kramers pairs, such that 1 refers to the lowest energy occupied pair and 41 corresponds to the highest energy occupied pair.}
      \label{fig:propagator_rZORA-PBE0_TlH}
      \end{figure*}

      \newpage

      \begin{figure*}[h!]
      \begin{minipage}{.48\textwidth}
      \includegraphics[width=\textwidth]{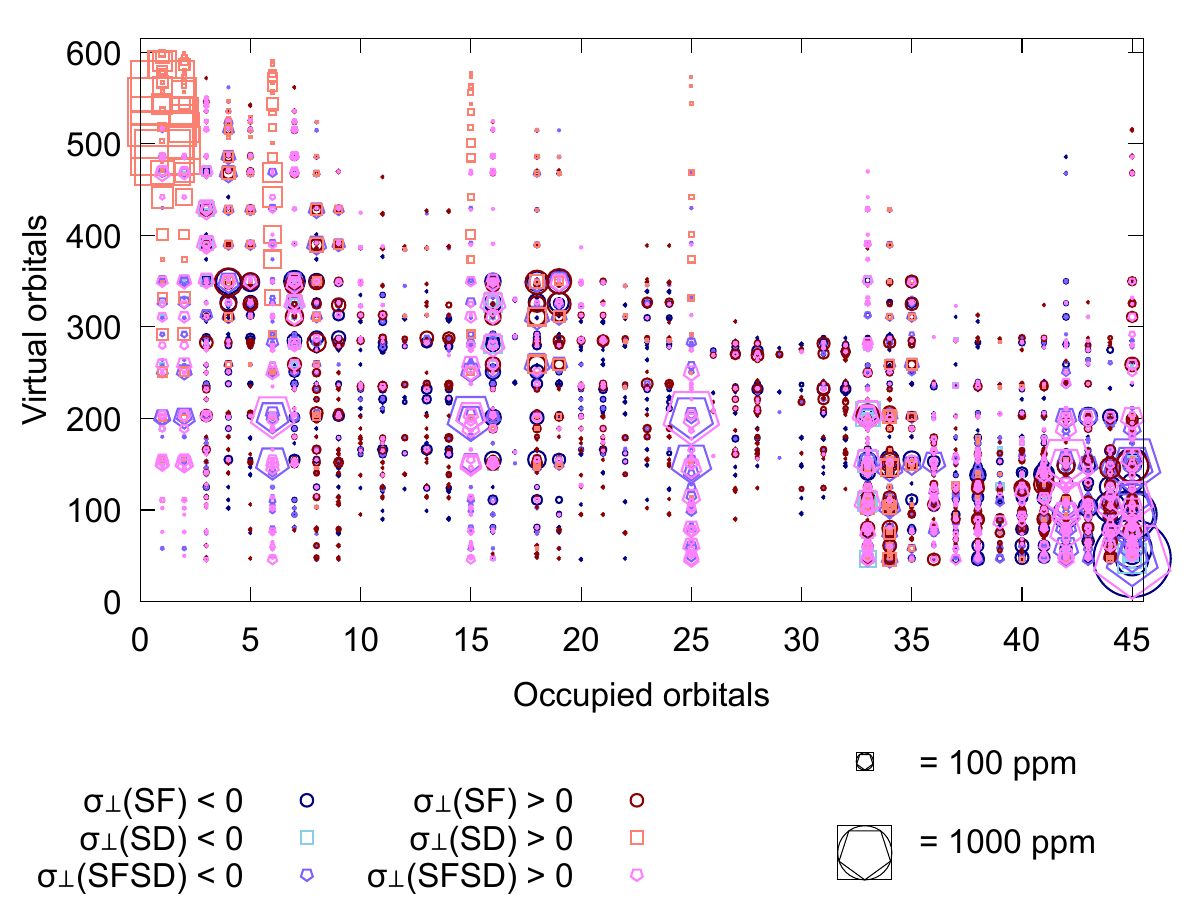}
      \end{minipage}
      \hfill
      \begin{minipage}{.48\textwidth}
      \includegraphics[width=\textwidth]{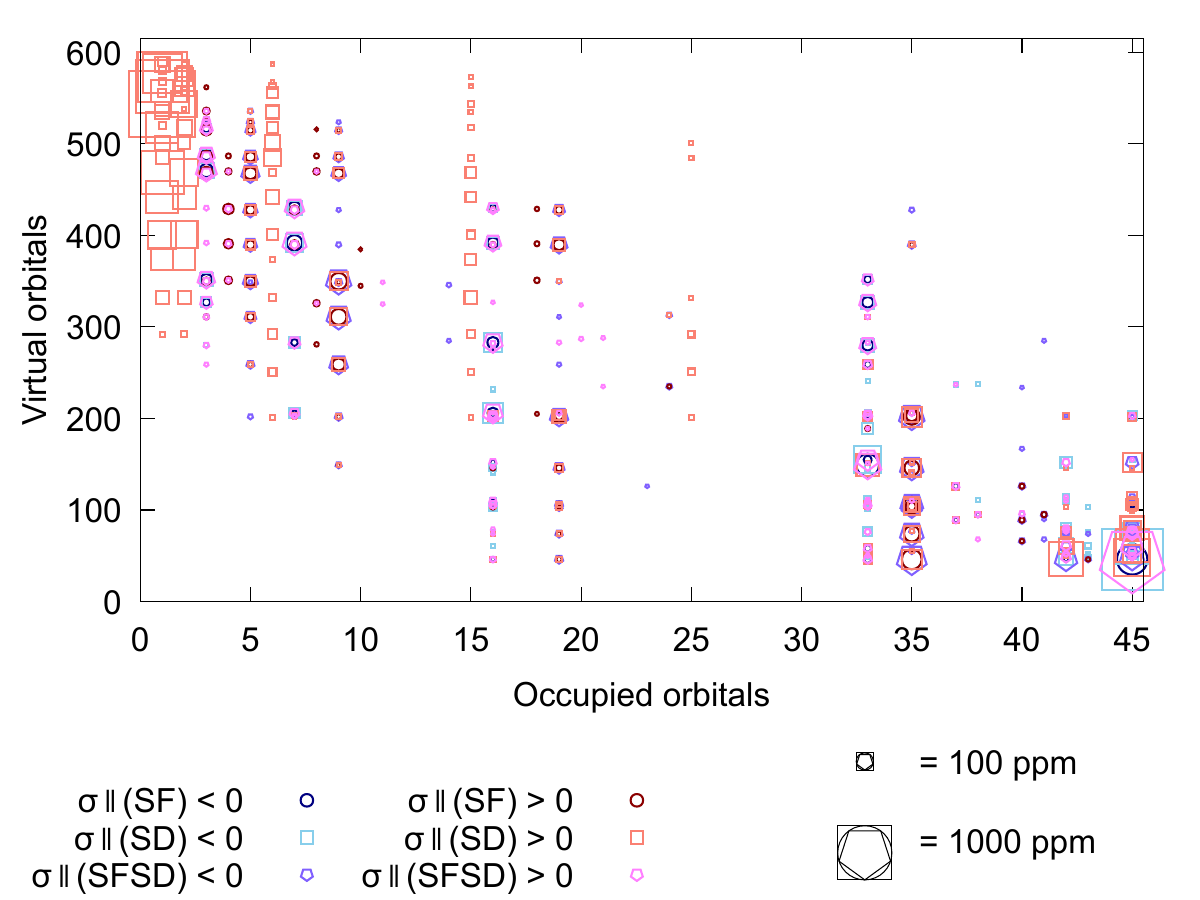}
      \end{minipage}%
      \caption{Pattern of excitations for $\sigma_\parallel$(Tl) (left)
                and $\sigma_\perp$(Tl) (right) for TlF shown separately for the SD and SF parts of the NMR shielding propagator at the level of rZORA-PBE0. The magnitude of the amplitude of each type is proportional to marked area. The numbers that label the occupied and virtual orbitals correspond to the Kramers pairs, such that 1 refers to the lowest energy occupied pair and 45 corresponds to the highest energy occupied pair.}
      \label{fig:propagator_rZORA-PBE0_TlF}
      \end{figure*}

      \newpage

      \begin{figure*}[h!]
      \begin{minipage}{.48\textwidth}
      \includegraphics[width=\textwidth]{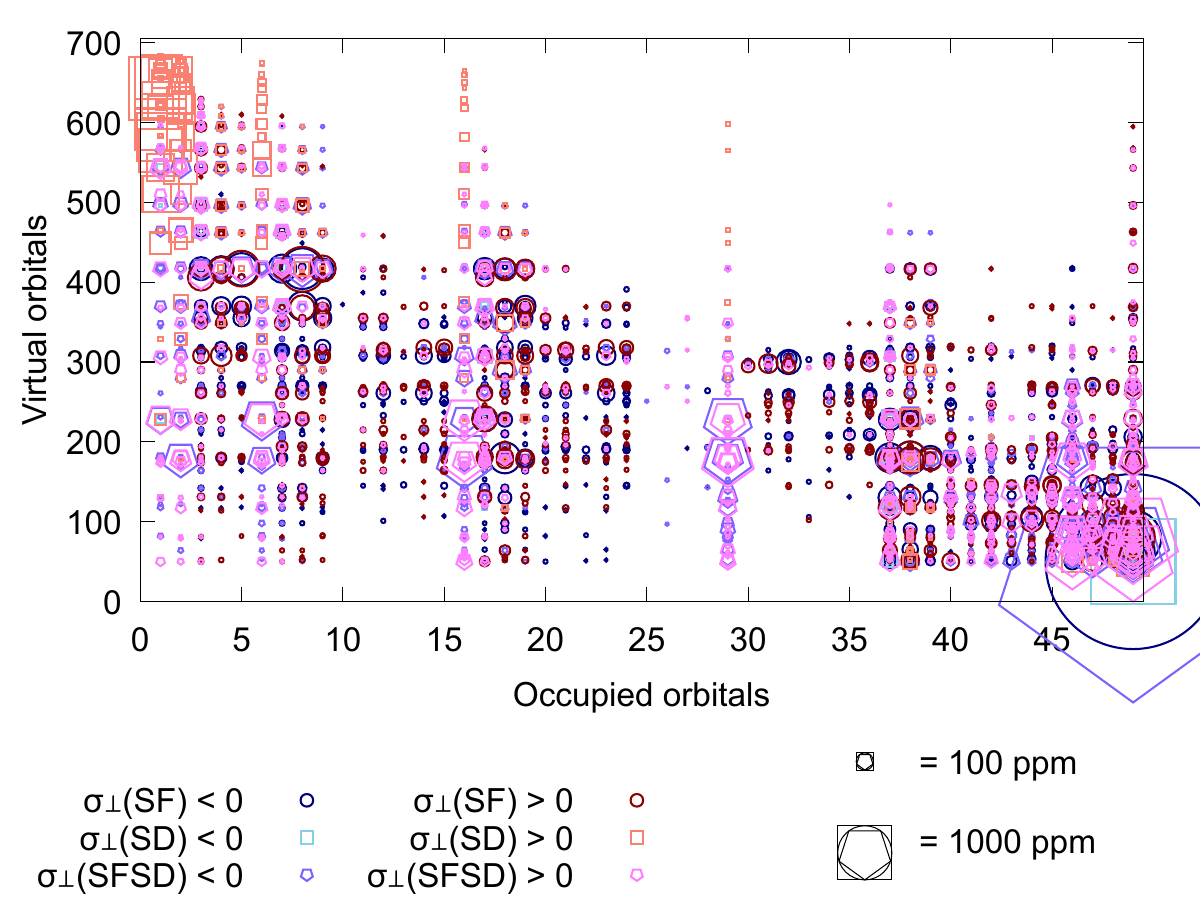}
      \end{minipage}
      \hfill
      \begin{minipage}{.48\textwidth}
      \includegraphics[width=\textwidth]{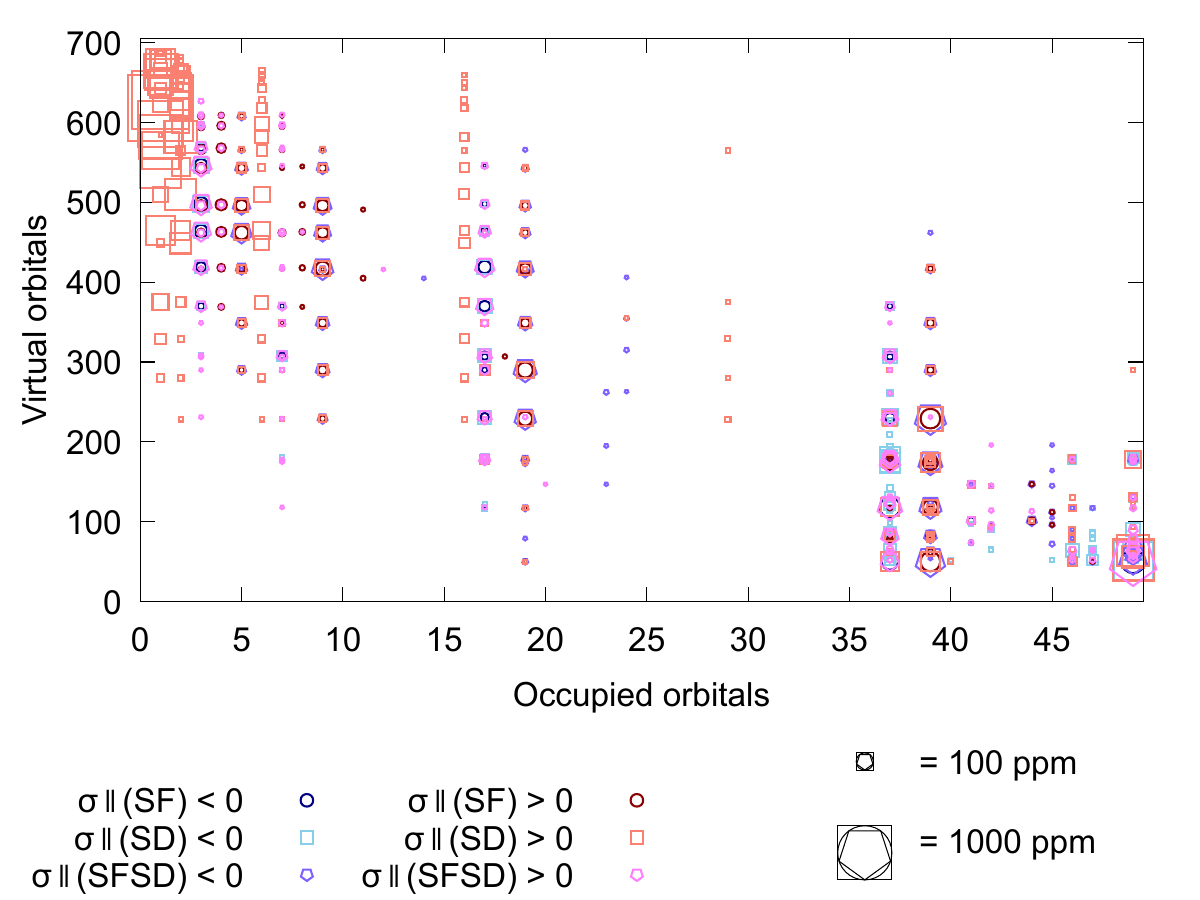}
      \end{minipage}%
      \caption{Pattern of excitations for $\sigma_\parallel$(Tl) (left)
                and $\sigma_\perp$(Tl) (right) for TlCl shown separately for the SD and SF parts of the NMR shielding propagator at the level of rZORA-PBE0. The magnitude of the amplitude of each type is proportional to marked area. The numbers that label the occupied and virtual orbitals correspond to the Kramers pairs, such that 1 refers to the lowest energy occupied pair and 49 corresponds to the highest energy occupied pair.}
      \label{fig:propagator_rZORA-PBE0_TlCl}
      \end{figure*}

      \newpage

      \begin{figure*}[h!]
      \begin{minipage}{.48\textwidth}
      \includegraphics[width=\textwidth]{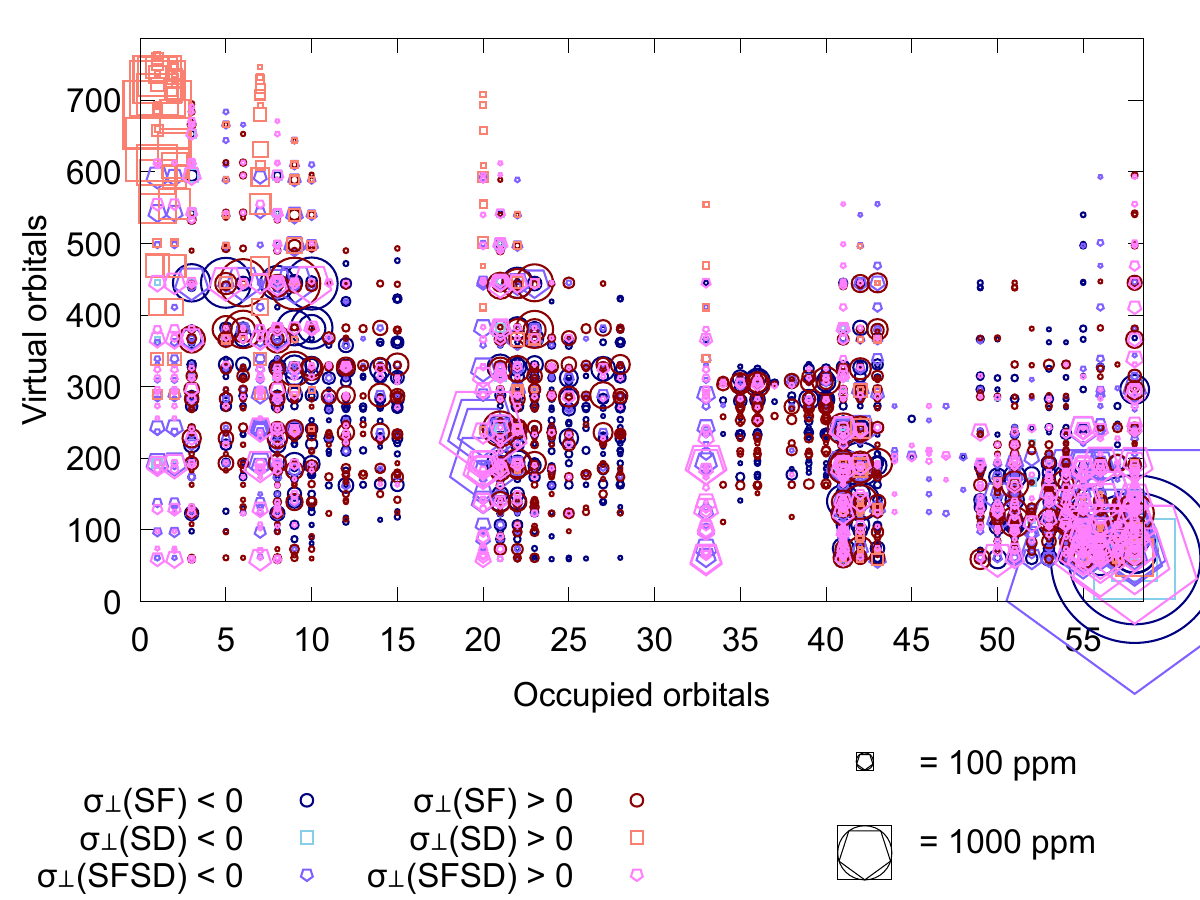}
      \end{minipage}
      \hfill
      \begin{minipage}{.48\textwidth}
      \includegraphics[width=\textwidth]{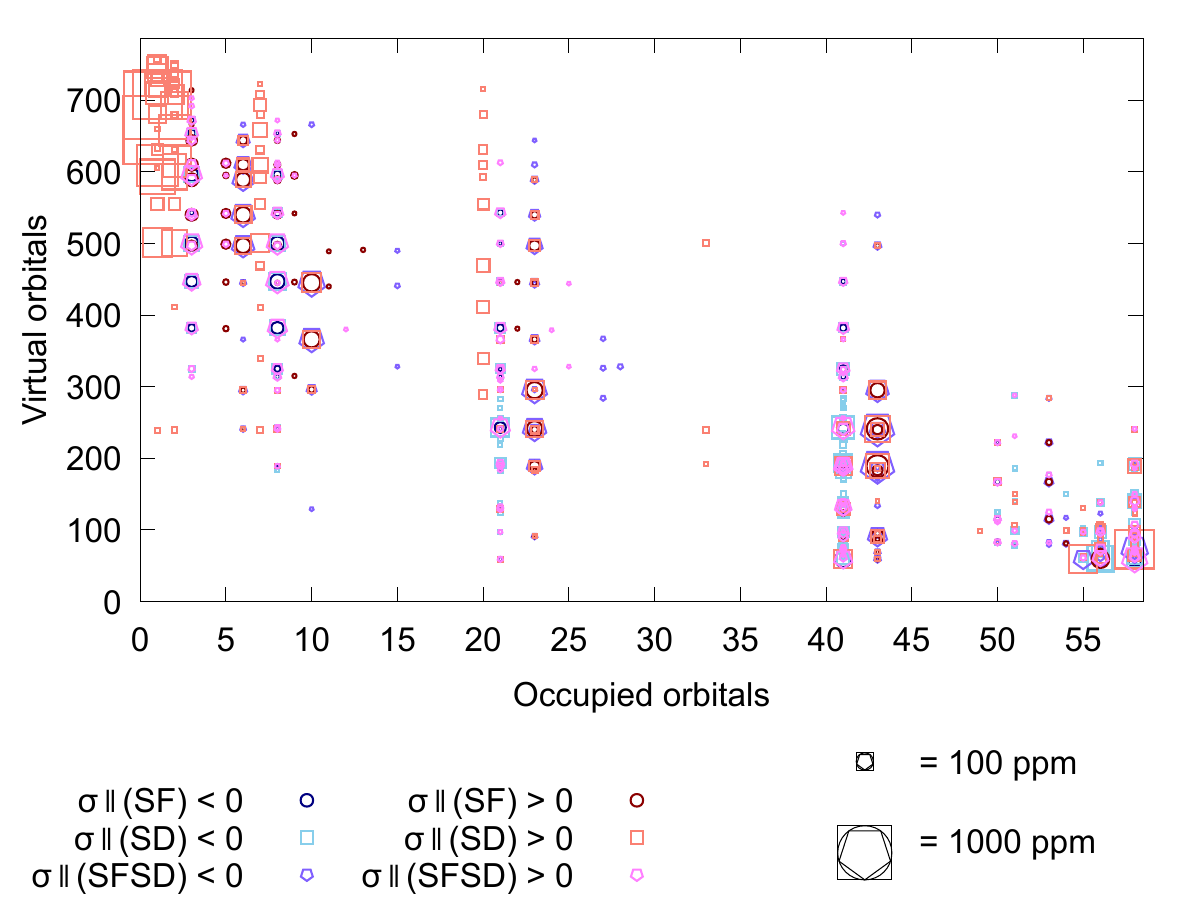}
      \end{minipage}%
      \caption{Pattern of excitations for $\sigma_\parallel$(Tl) (left)
                and $\sigma_\perp$(Tl) (right) for TlBr shown separately for the SD and SF parts of the NMR shielding propagator at the level of rZORA-PBE0. The magnitude of the amplitude of each type is proportional to marked area. The numbers that label the occupied and virtual orbitals correspond to the Kramers pairs, such that 1 refers to the lowest energy occupied pair and 58 corresponds to the highest energy occupied pair.}
      \label{fig:propagator_rZORA-PBE0_TlBr}
      \end{figure*}

      \newpage

      \begin{figure*}[h!]
      \begin{minipage}{.48\textwidth}
      \includegraphics[width=\textwidth]{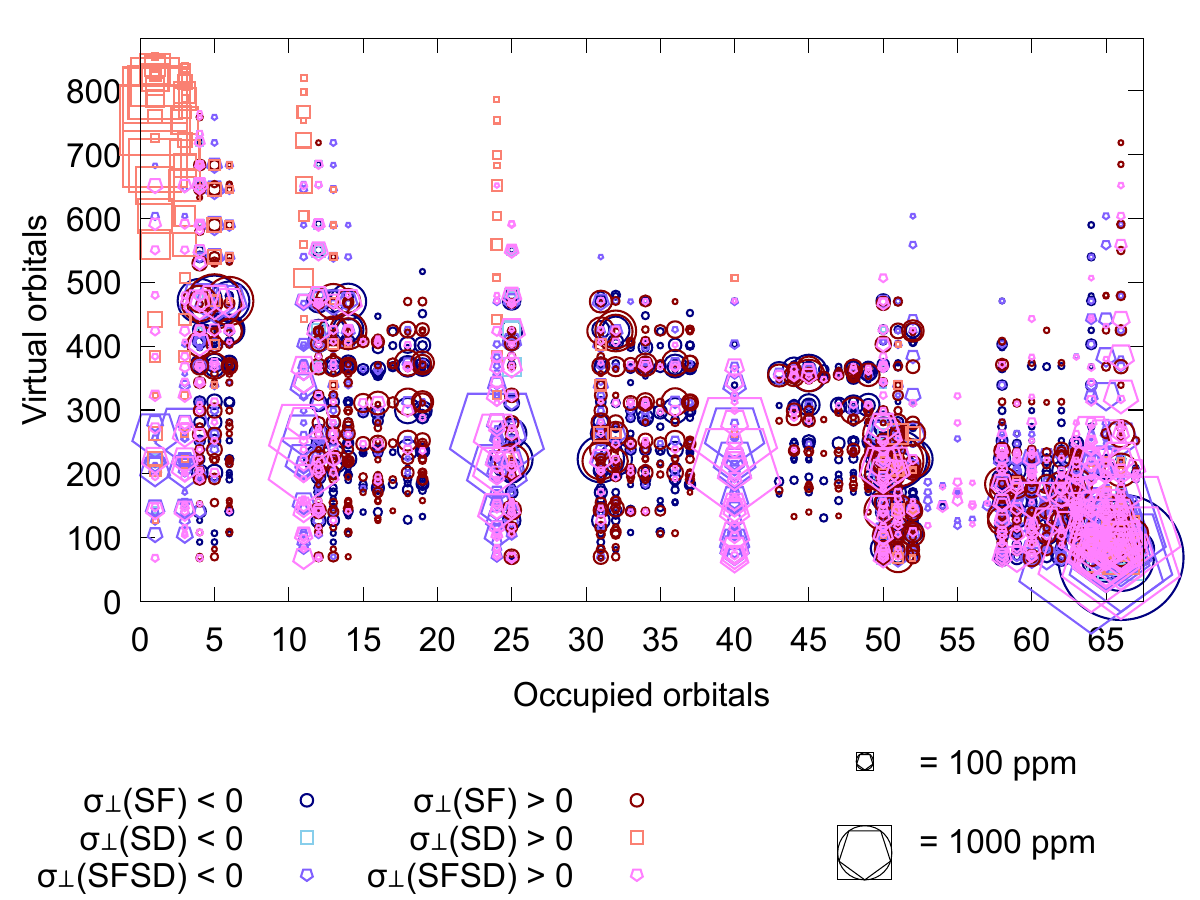}
      \end{minipage}
      \hfill
      \begin{minipage}{.48\textwidth}
      \includegraphics[width=\textwidth]{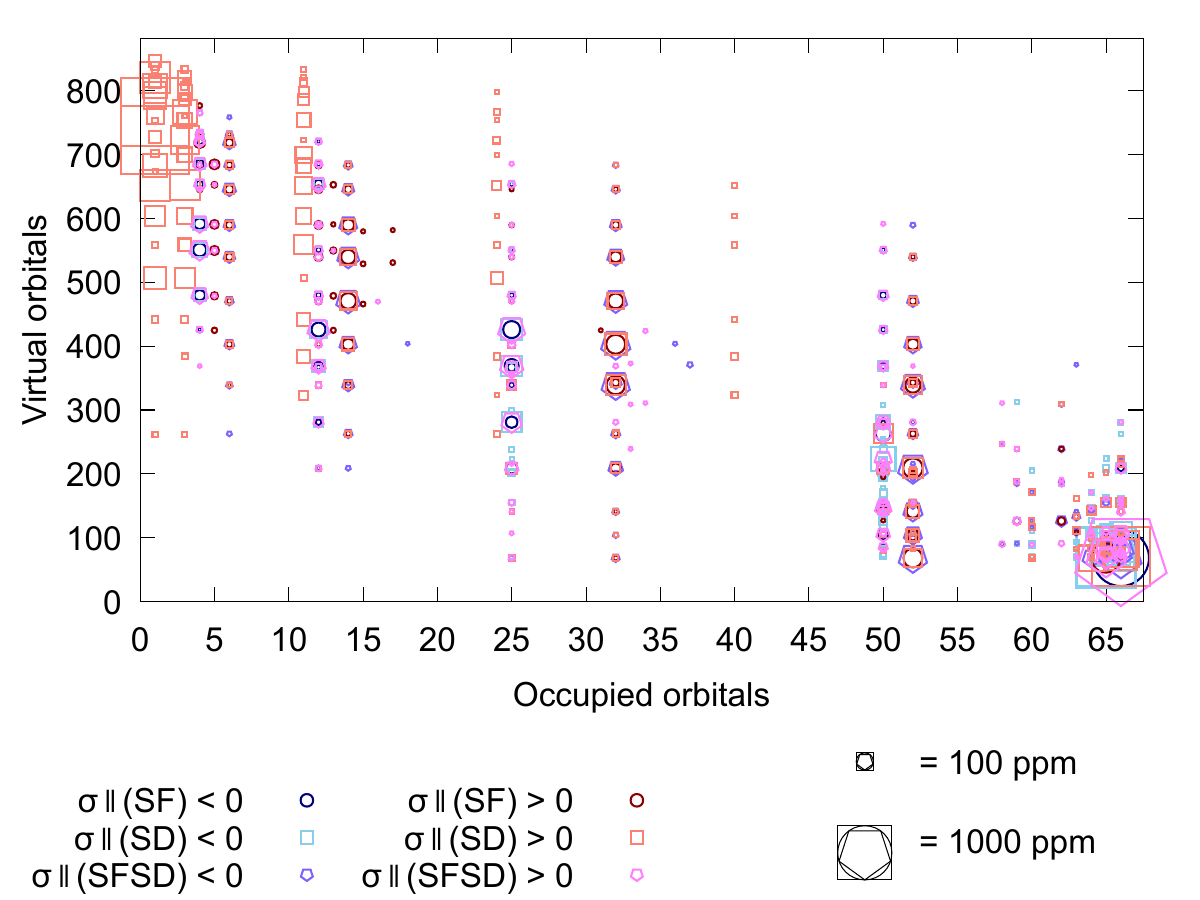}
      \end{minipage}%
      \caption{Pattern of excitations for $\sigma_\parallel$(Tl) (left)
                and $\sigma_\perp$(Tl) (right) for TlI shown separately for the SD and SF parts of the NMR shielding propagator at the level of rZORA-PBE0. The magnitude of the amplitude of each type is proportional to marked area. The numbers that label the occupied and virtual orbitals correspond to the Kramers pairs, such that 1 refers to the lowest energy occupied pair and 67 corresponds to the highest energy occupied pair.}
      \label{fig:propagator_rZORA-PBE0_TlI}
      \end{figure*}

      \newpage

      \begin{figure*}[h!]
      \begin{minipage}{.48\textwidth}
      \includegraphics[width=\textwidth]{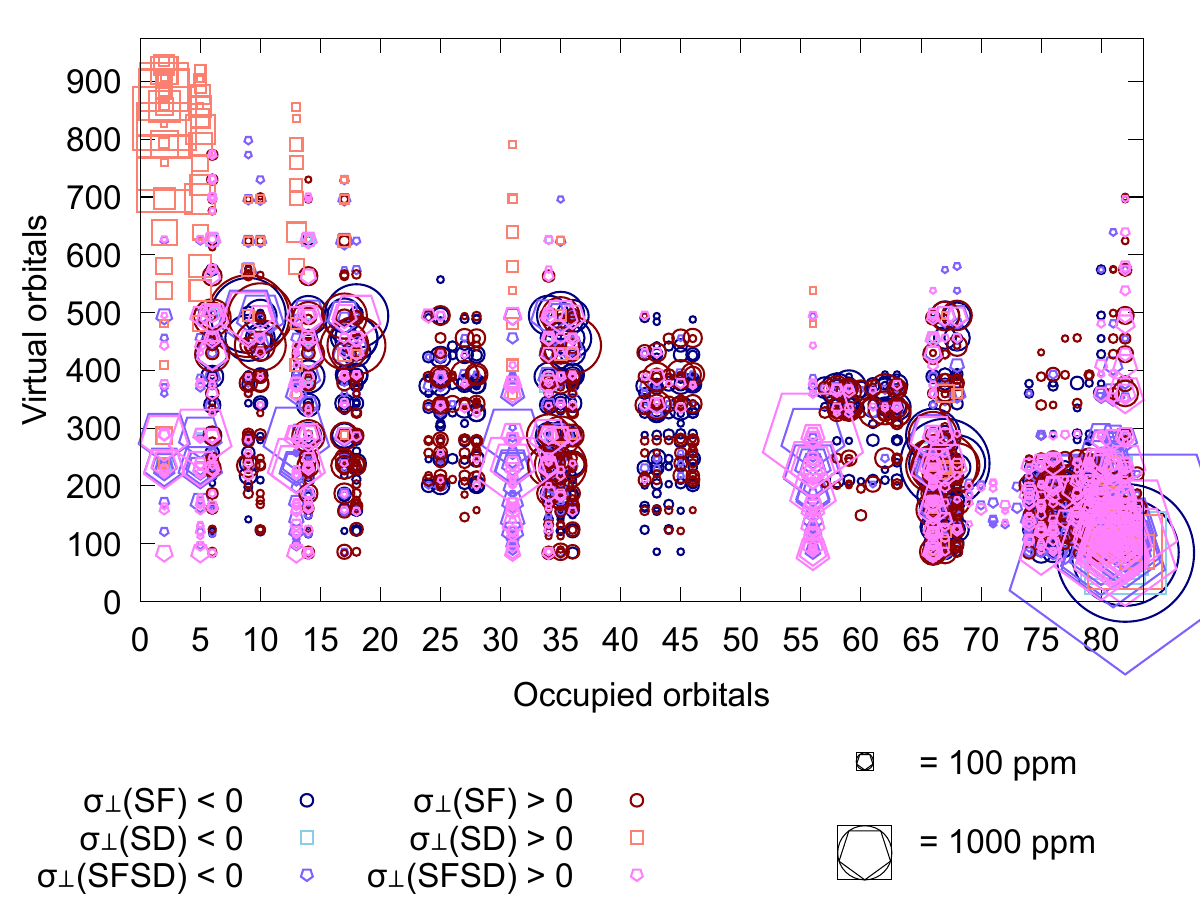}
      \end{minipage}
      \hfill
      \begin{minipage}{.48\textwidth}
      \includegraphics[width=\textwidth]{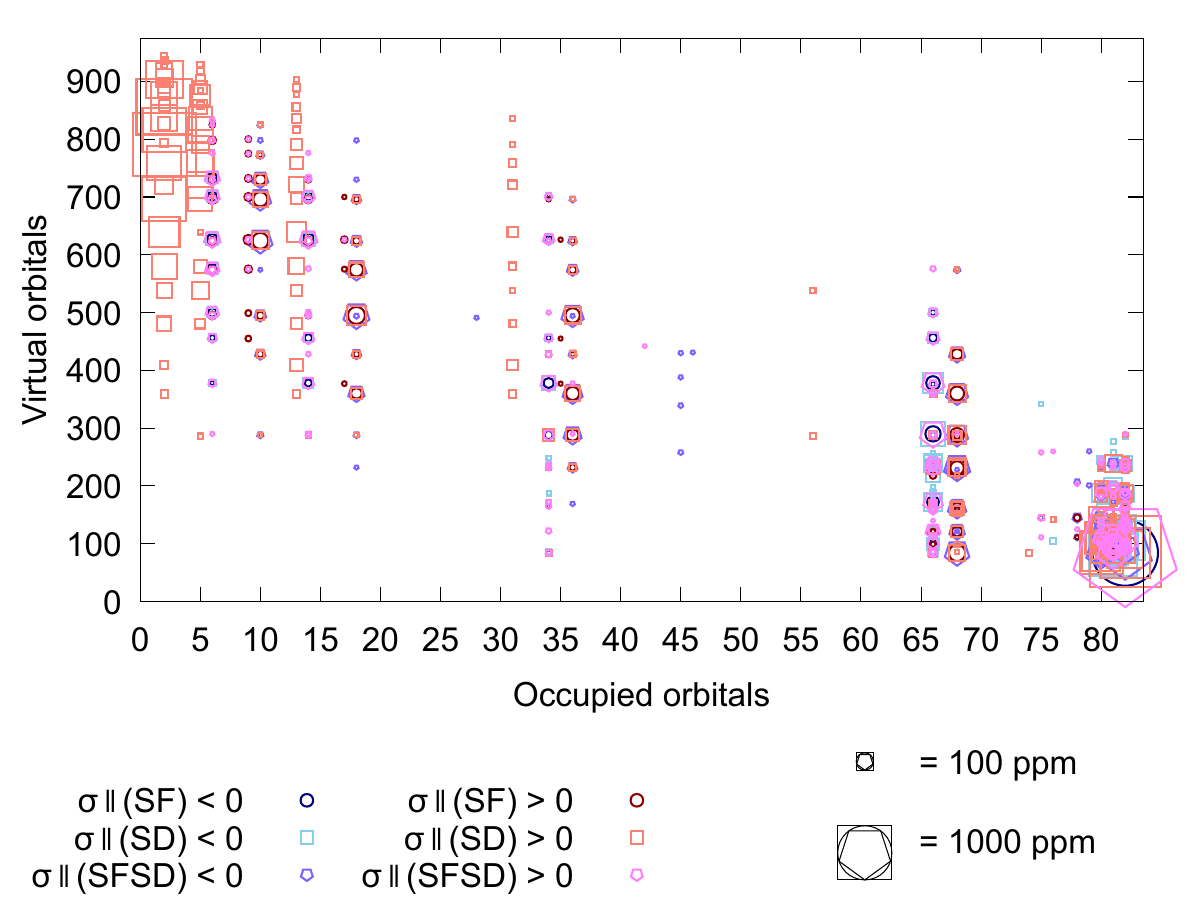}
      \end{minipage}%
      \caption{Pattern of excitations for $\sigma_\parallel$(Tl) (left)
                and $\sigma_\perp$(Tl) (right) for TlAt shown separately for the SD and SF parts of the NMR shielding propagator at the level of rZORA-PBE0. The magnitude of the amplitude of each type is proportional to marked area. The numbers that label the occupied and virtual orbitals correspond to the Kramers pairs, such that 1 refers to the lowest energy occupied pair and 83 corresponds to the highest energy occupied pair.}
      \label{fig:propagator_rZORA-PBE0_TlAt}
      \end{figure*}

      \newpage

      \begin{figure*}[h!]
      \begin{minipage}{.48\textwidth}
      \includegraphics[width=\textwidth]{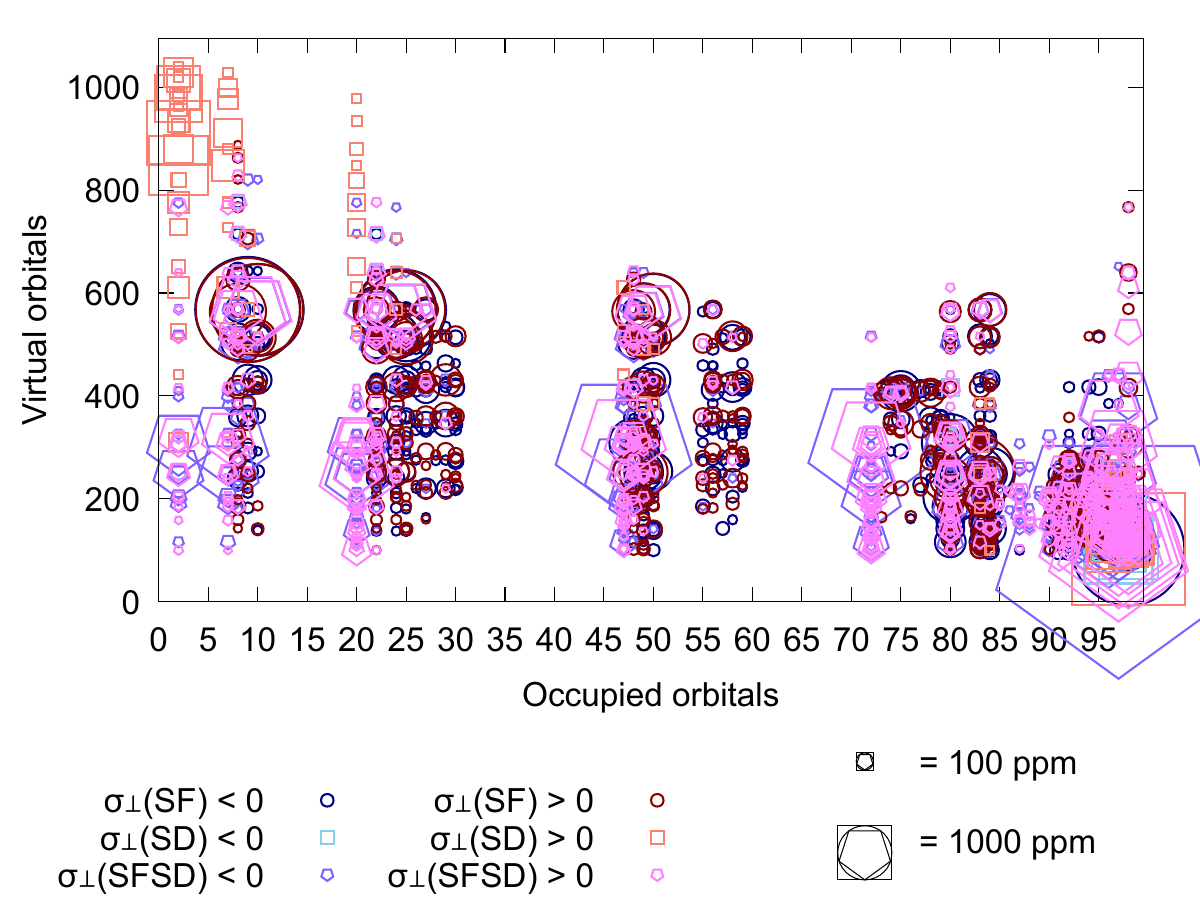}
      \end{minipage}
      \hfill
      \begin{minipage}{.48\textwidth}
      \includegraphics[width=\textwidth]{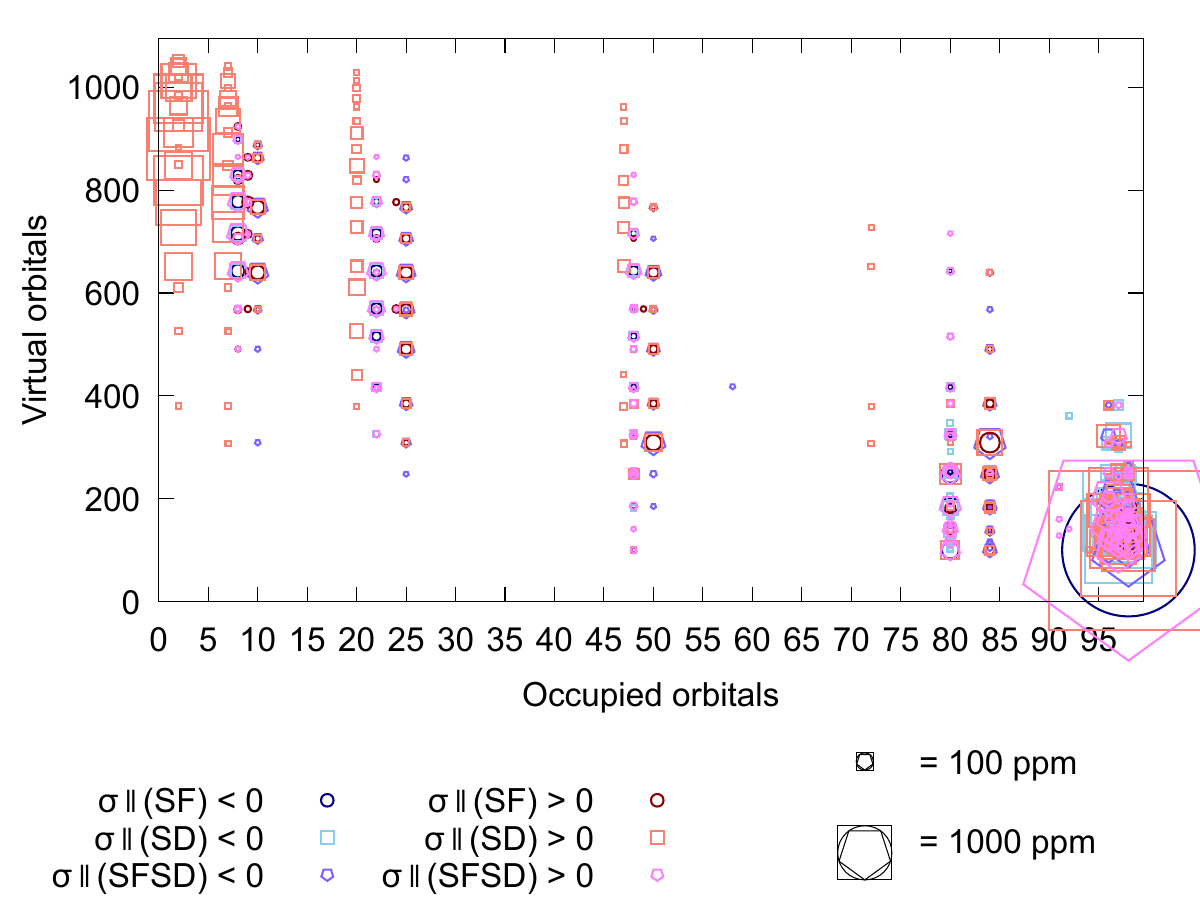}
      \end{minipage}%
      \caption{Pattern of excitations for $\sigma_\parallel$(Tl) (left)
                and $\sigma_\perp$(Tl) (right) for TlTs shown separately for the SD and SF parts of the NMR shielding propagator at the level of rZORA-PBE0. The magnitude of the amplitude of each type is proportional to marked area. The numbers that label the occupied and virtual orbitals correspond to the Kramers pairs, such that 1 refers to the lowest energy occupied pair and 99 corresponds to the highest energy occupied pair.}
      \label{fig:propagator_rZORA-PBE0_TlTs}
      \end{figure*}

      \newpage

  \subsection{rZORA-LDA scheme}
 
    \begin{figure}[h!]
    \includegraphics[width=.5\textwidth]{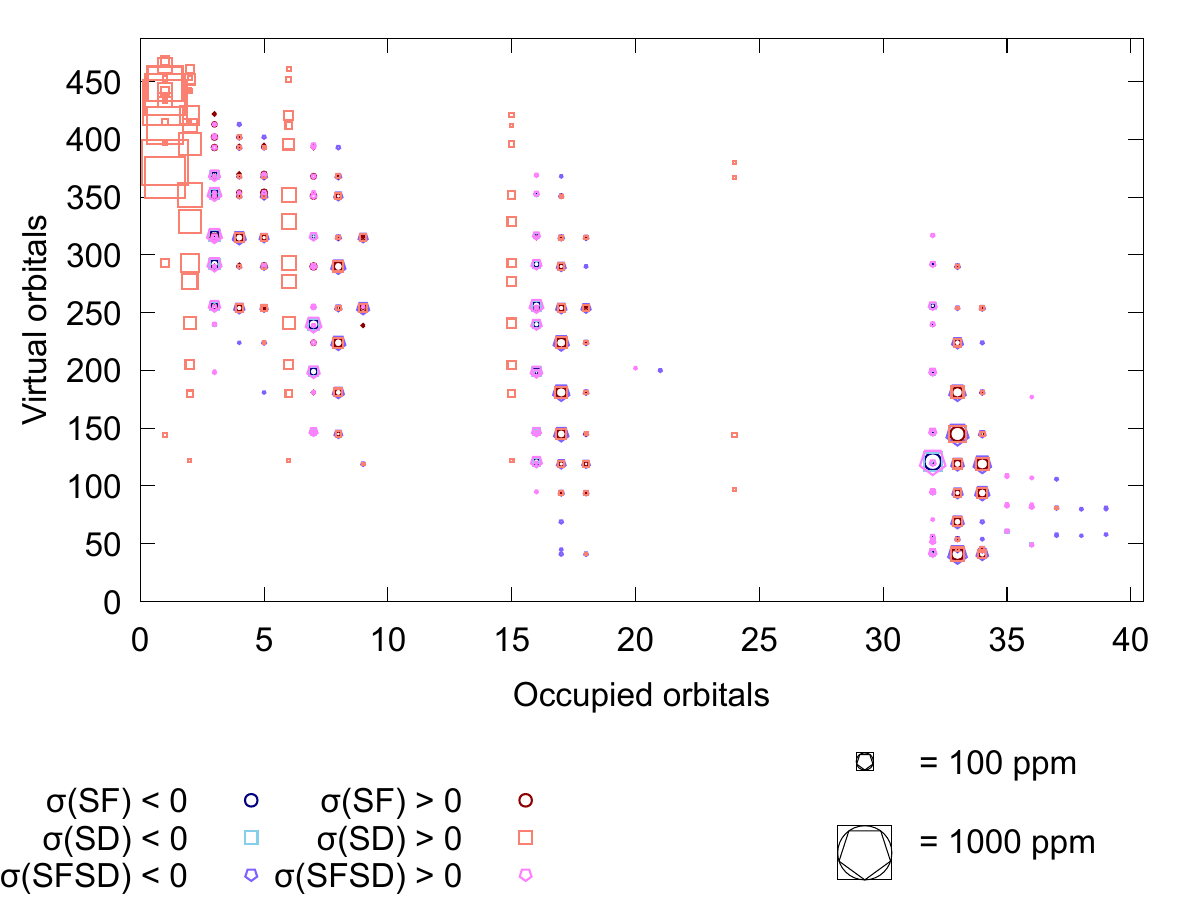}
    \caption{Pattern of excitations for $\sigma_\parallel$(Tl) (left)
              and $\sigma_\perp$(Tl) (right) for Tl$^+$ shown separately for the SD and SF parts of the NMR shielding propagator at the level of rZORA-LDA. The magnitude of the amplitude of each type is proportional to marked area. The numbers that label the occupied and virtual orbitals correspond to the Kramers pairs, such that 1 refers to the lowest energy occupied pair and 40 corresponds to the highest energy occupied pair.}
    \label{fig:propagator_rZORA-LDA_Tl+}
    \end{figure}
    
    \newpage

      \begin{figure*}[h!]
      \begin{minipage}{.48\textwidth}
      \includegraphics[width=\textwidth]{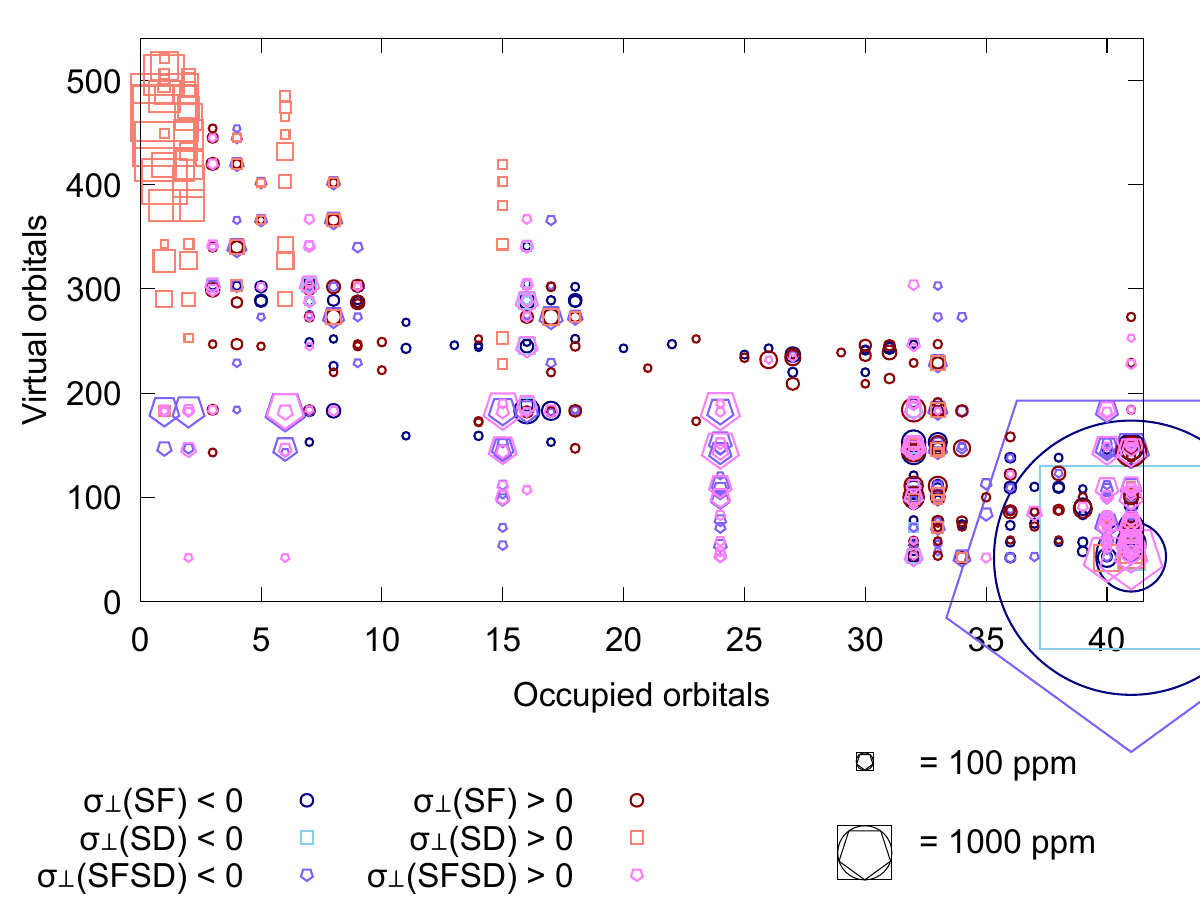}
      \end{minipage}
      \hfill
      \begin{minipage}{.48\textwidth}
      \includegraphics[width=\textwidth]{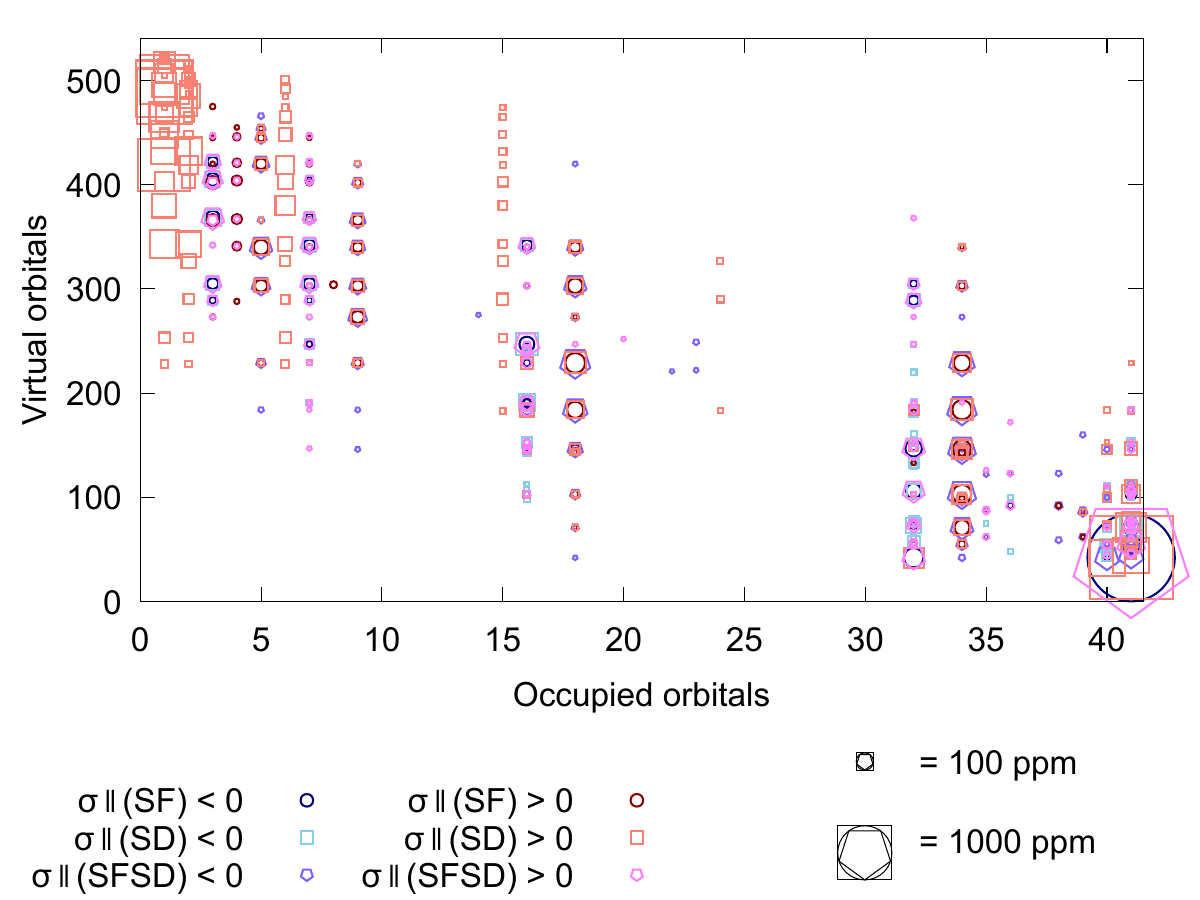}
      \end{minipage}%
      \caption{Pattern of excitations for $\sigma_\parallel$(Tl) (left)
                and $\sigma_\perp$(Tl) (right) for TlH shown separately for the SD and SF parts of the NMR shielding propagator at the level of rZORA-LDA. The magnitude of the amplitude of each type is proportional to marked area. The numbers that label the occupied and virtual orbitals correspond to the Kramers pairs, such that 1 refers to the lowest energy occupied pair and 41 corresponds to the highest energy occupied pair.}
      \label{fig:propagator_rZORA-LDA_TlH}
      \end{figure*}

      \newpage

      \begin{figure*}[h!]
      \begin{minipage}{.48\textwidth}
      \includegraphics[width=\textwidth]{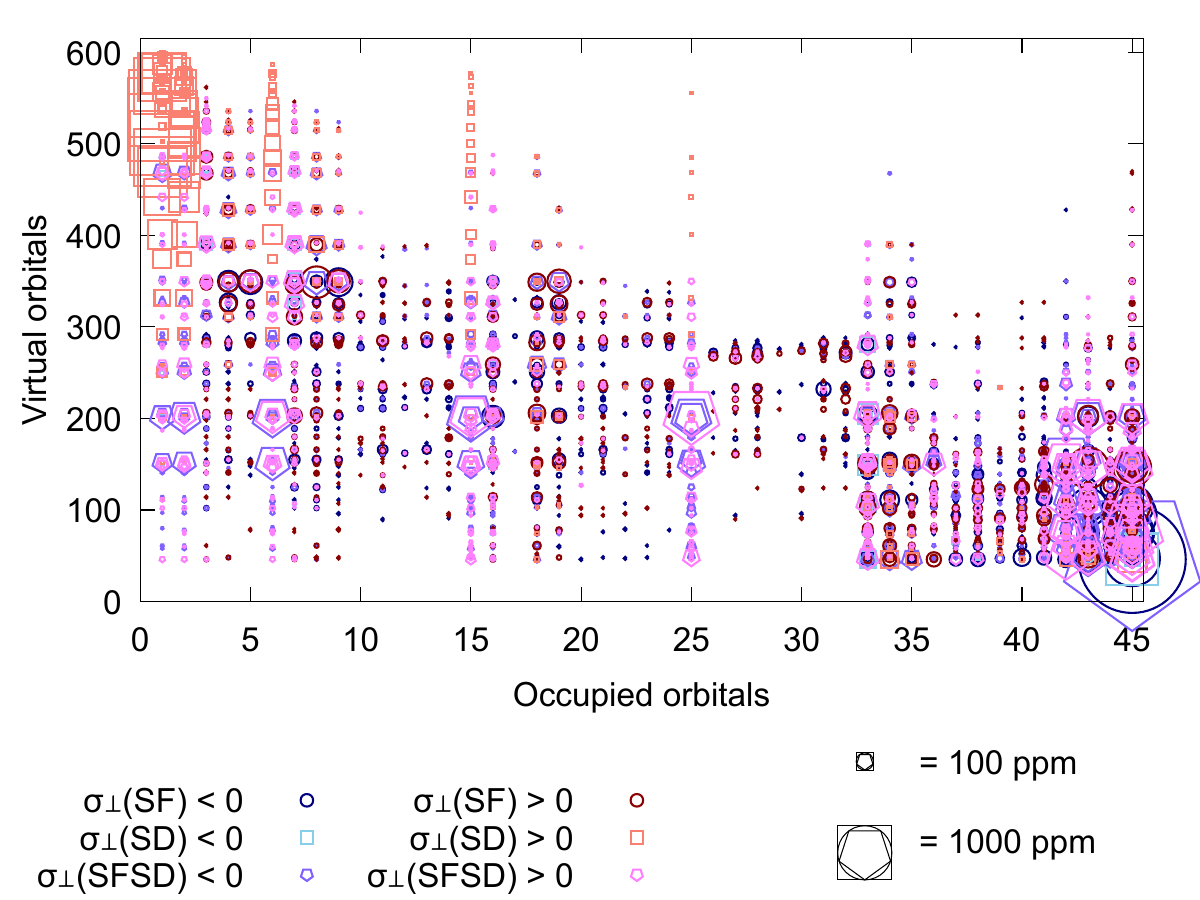}
      \end{minipage}
      \hfill
      \begin{minipage}{.48\textwidth}
      \includegraphics[width=\textwidth]{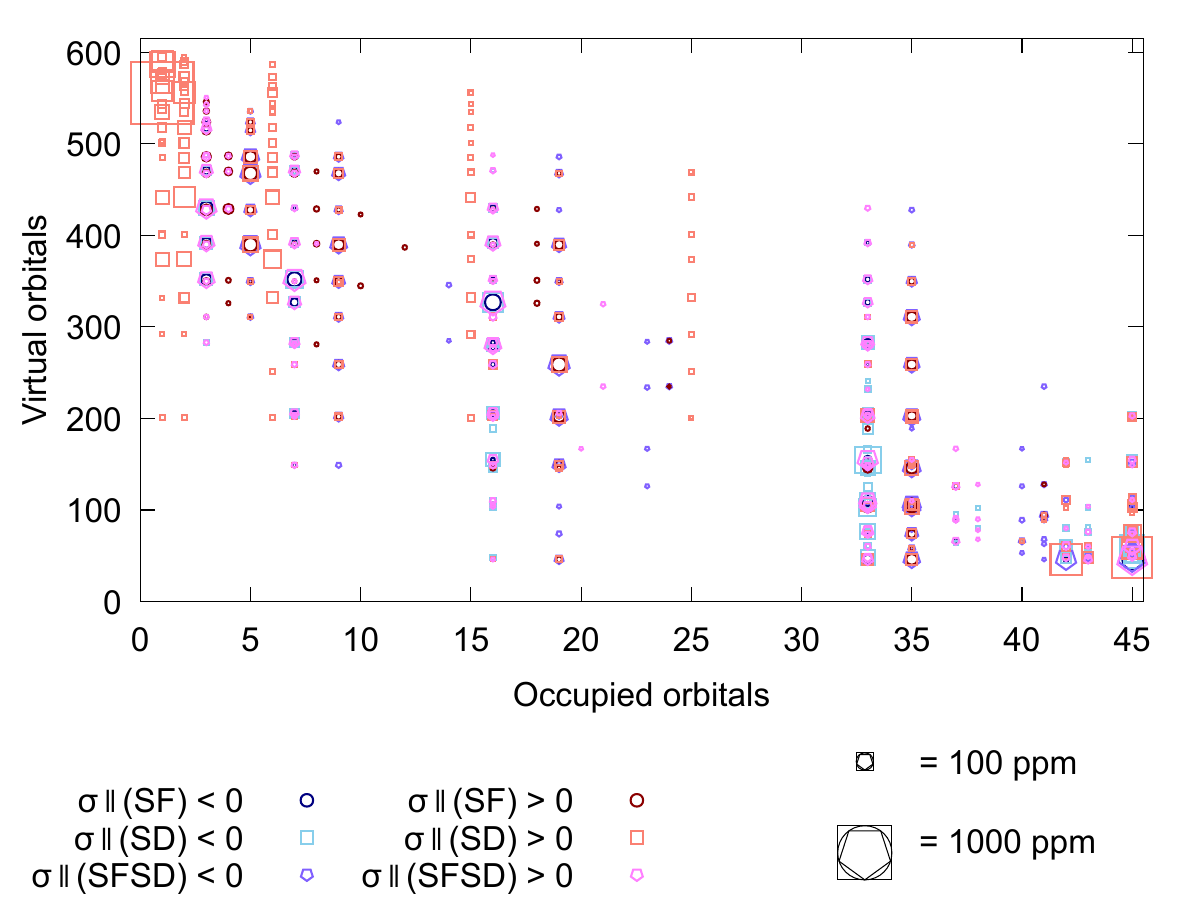}
      \end{minipage}%
      \caption{Pattern of excitations for $\sigma_\parallel$(Tl) (left)
                and $\sigma_\perp$(Tl) (right) for TlF shown separately for the SD and SF parts of the NMR shielding propagator at the level of rZORA-LDA. The magnitude of the amplitude of each type is proportional to marked area. The numbers that label the occupied and virtual orbitals correspond to the Kramers pairs, such that 1 refers to the lowest energy occupied pair and 45 corresponds to the highest energy occupied pair.}
      \label{fig:propagator_rZORA-LDA_TlF}
      \end{figure*}

      \newpage

      \begin{figure*}[h!]
      \begin{minipage}{.48\textwidth}
      \includegraphics[width=\textwidth]{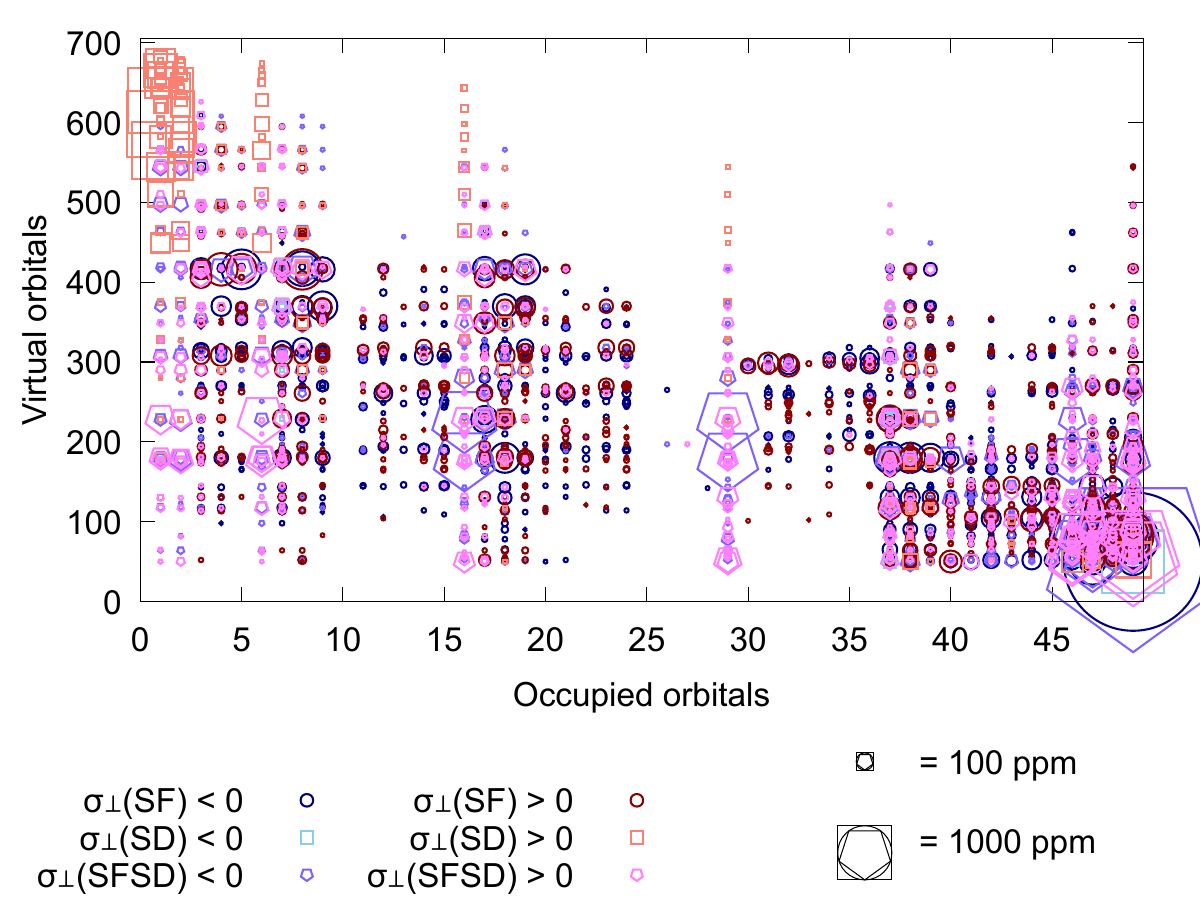}
      \end{minipage}
      \hfill
      \begin{minipage}{.48\textwidth}
      \includegraphics[width=\textwidth]{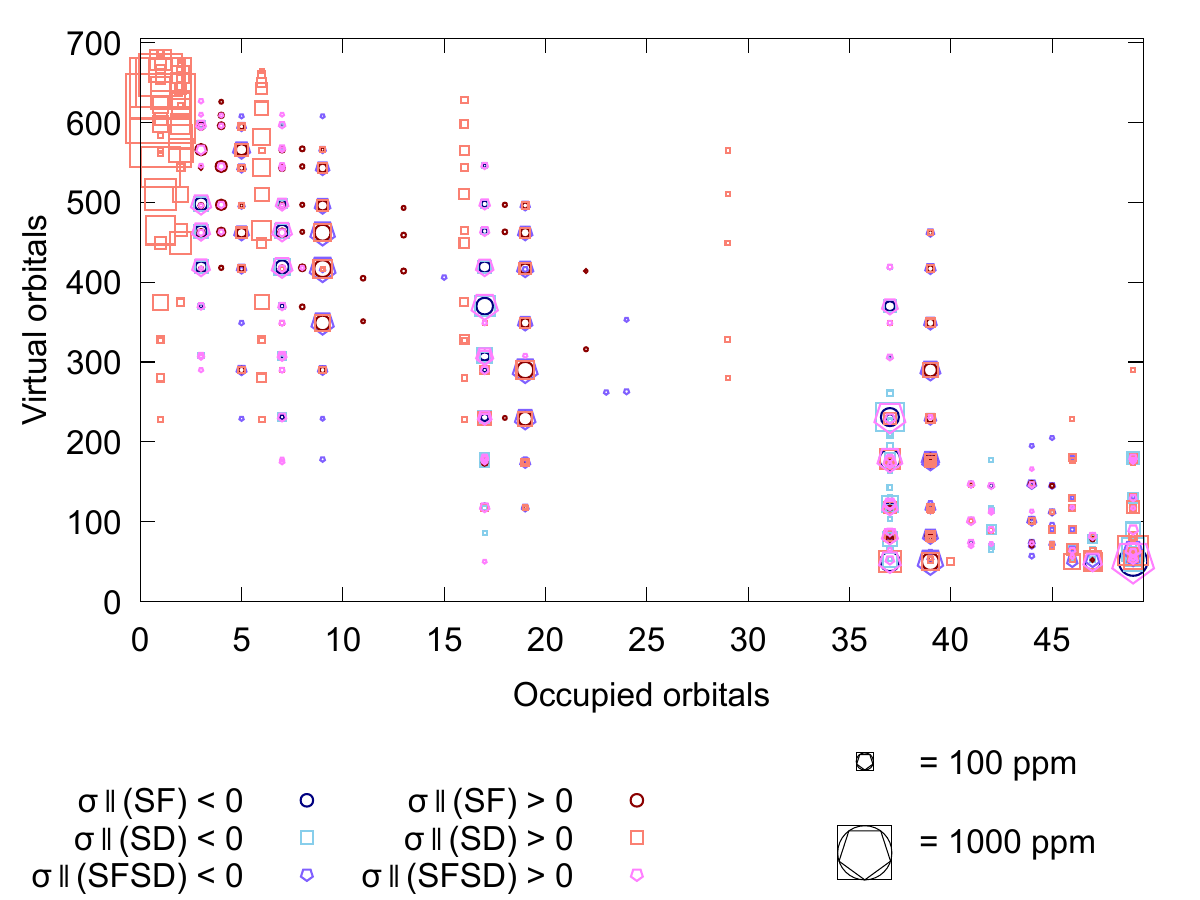}
      \end{minipage}%
      \caption{Pattern of excitations for $\sigma_\parallel$(Tl) (left)
                and $\sigma_\perp$(Tl) (right) for TlCl shown separately for the SD and SF parts of the NMR shielding propagator at the level of rZORA-LDA. The magnitude of the amplitude of each type is proportional to marked area. The numbers that label the occupied and virtual orbitals correspond to the Kramers pairs, such that 1 refers to the lowest energy occupied pair and 49 corresponds to the highest energy occupied pair.}
      \label{fig:propagator_rZORA-LDA_TlCl}
      \end{figure*}

      \newpage

      \begin{figure*}[h!]
      \begin{minipage}{.48\textwidth}
      \includegraphics[width=\textwidth]{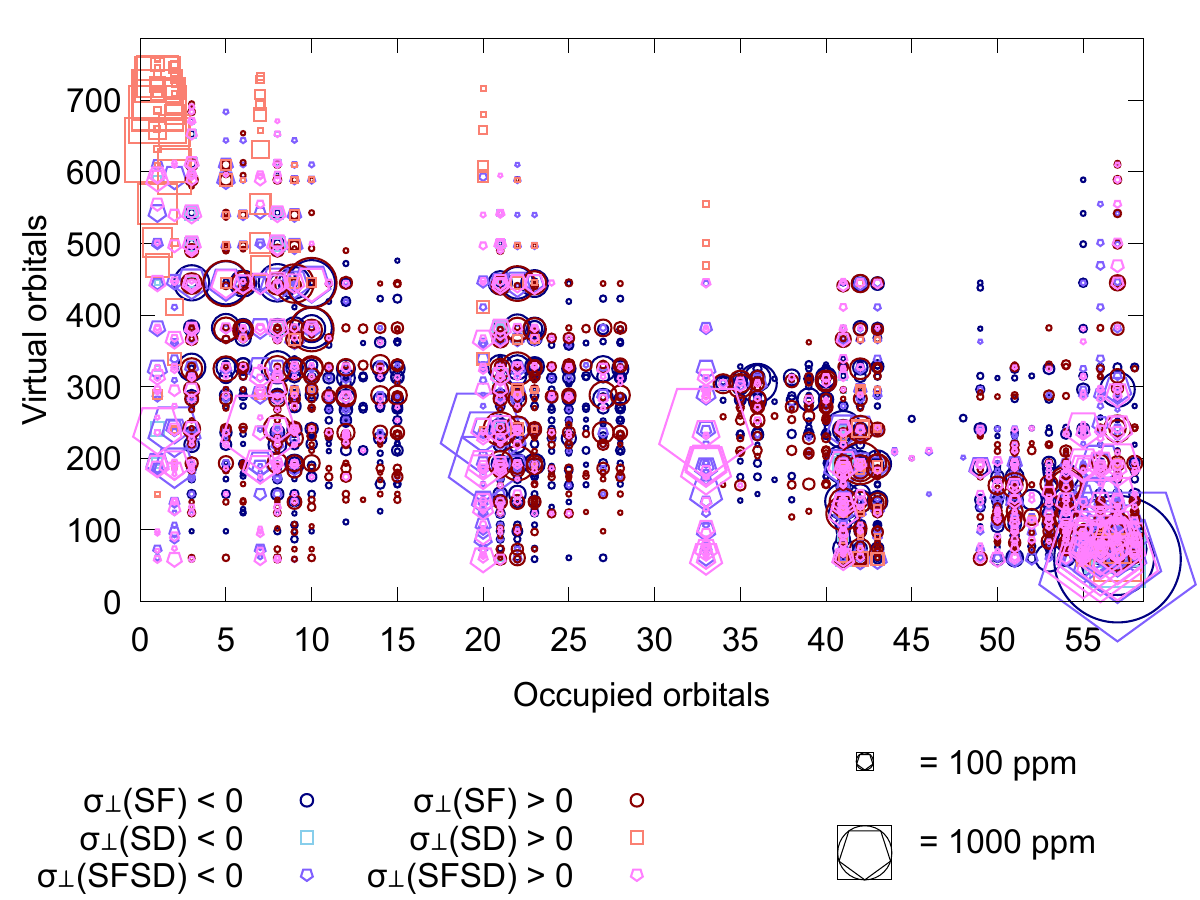}
      \end{minipage}
      \hfill
      \begin{minipage}{.48\textwidth}
      \includegraphics[width=\textwidth]{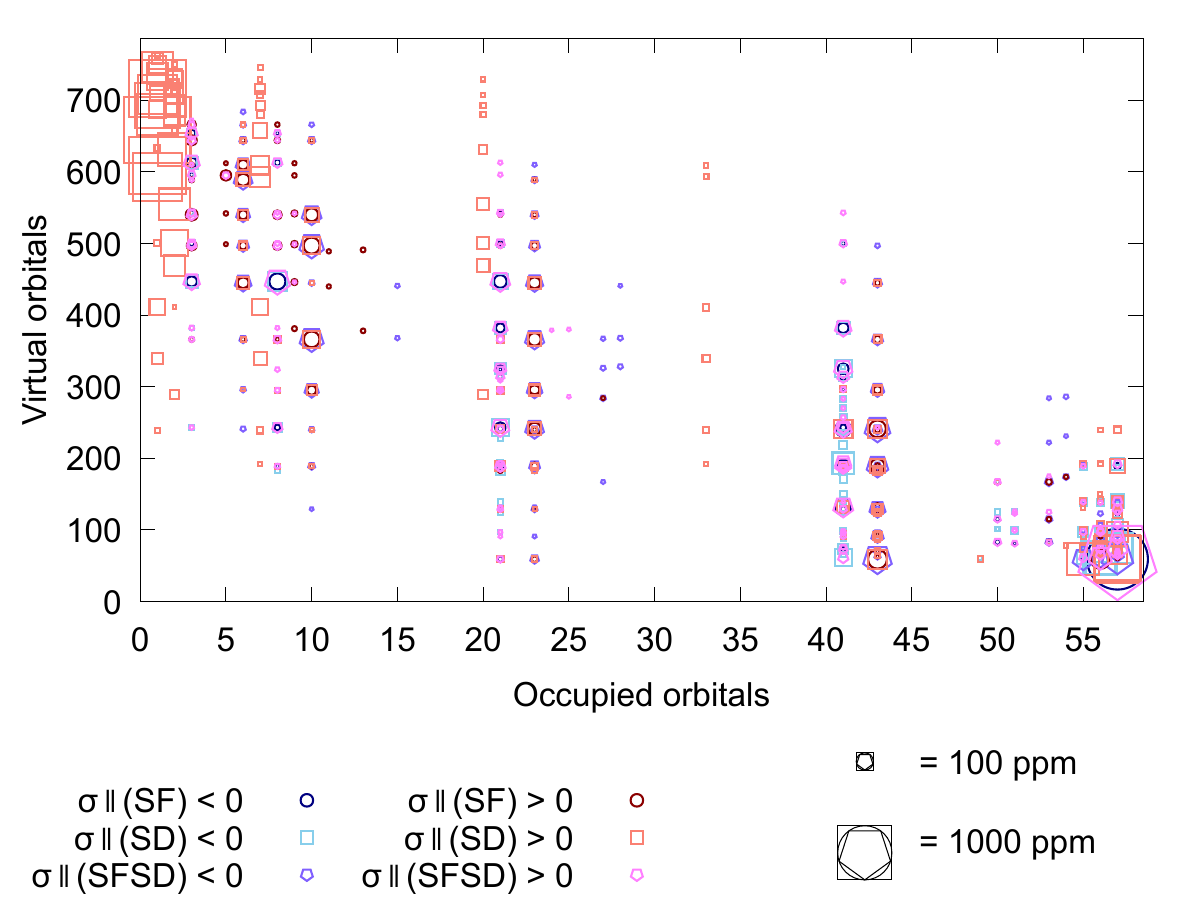}
      \end{minipage}%
      \caption{Pattern of excitations for $\sigma_\parallel$(Tl) (left)
                and $\sigma_\perp$(Tl) (right) for TlBr shown separately for the SD and SF parts of the NMR shielding propagator at the level of rZORA-LDA. The magnitude of the amplitude of each type is proportional to marked area. The numbers that label the occupied and virtual orbitals correspond to the Kramers pairs, such that 1 refers to the lowest energy occupied pair and 58 corresponds to the highest energy occupied pair.}
      \label{fig:propagator_rZORA-LDA_TlBr}
      \end{figure*}

      \newpage

      \begin{figure*}[h!]
      \begin{minipage}{.48\textwidth}
      \includegraphics[width=\textwidth]{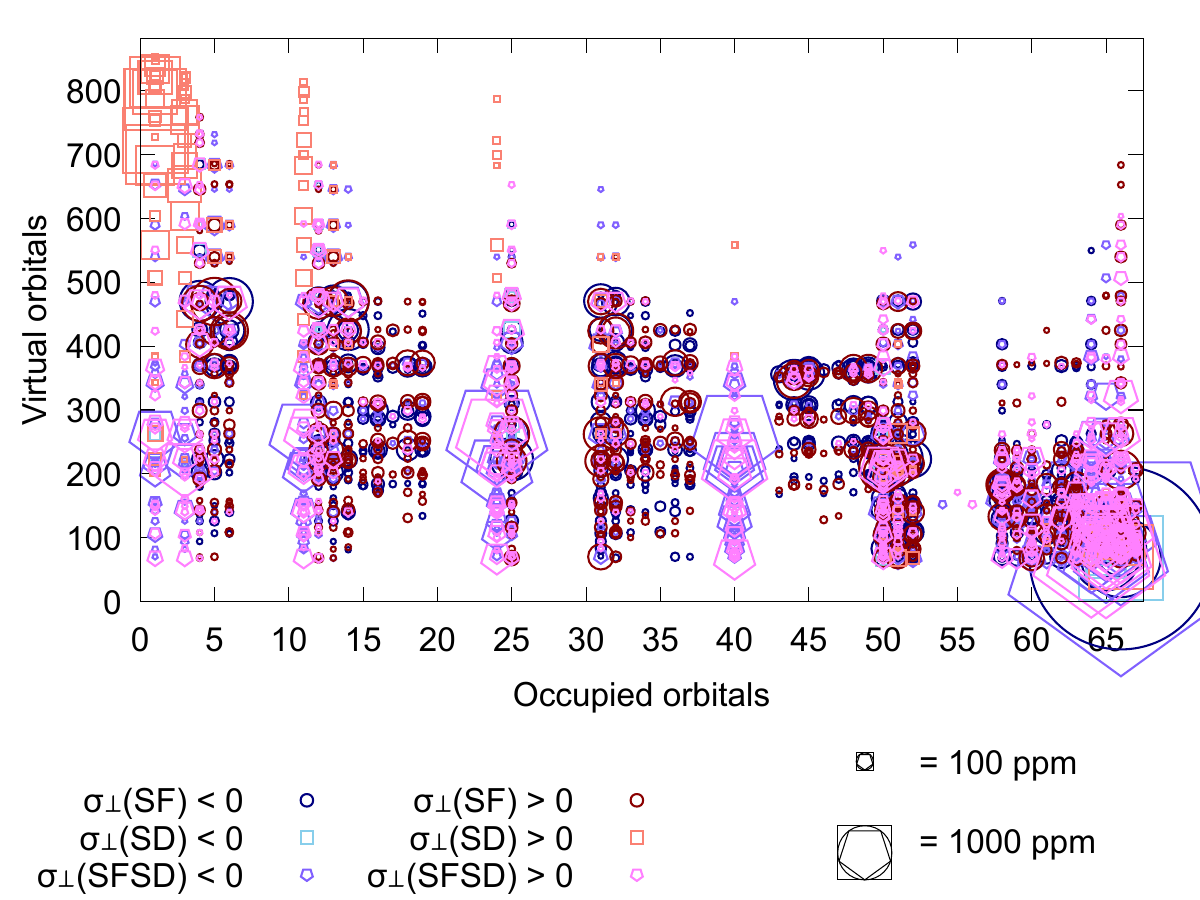}
      \end{minipage}
      \hfill
      \begin{minipage}{.48\textwidth}
      \includegraphics[width=\textwidth]{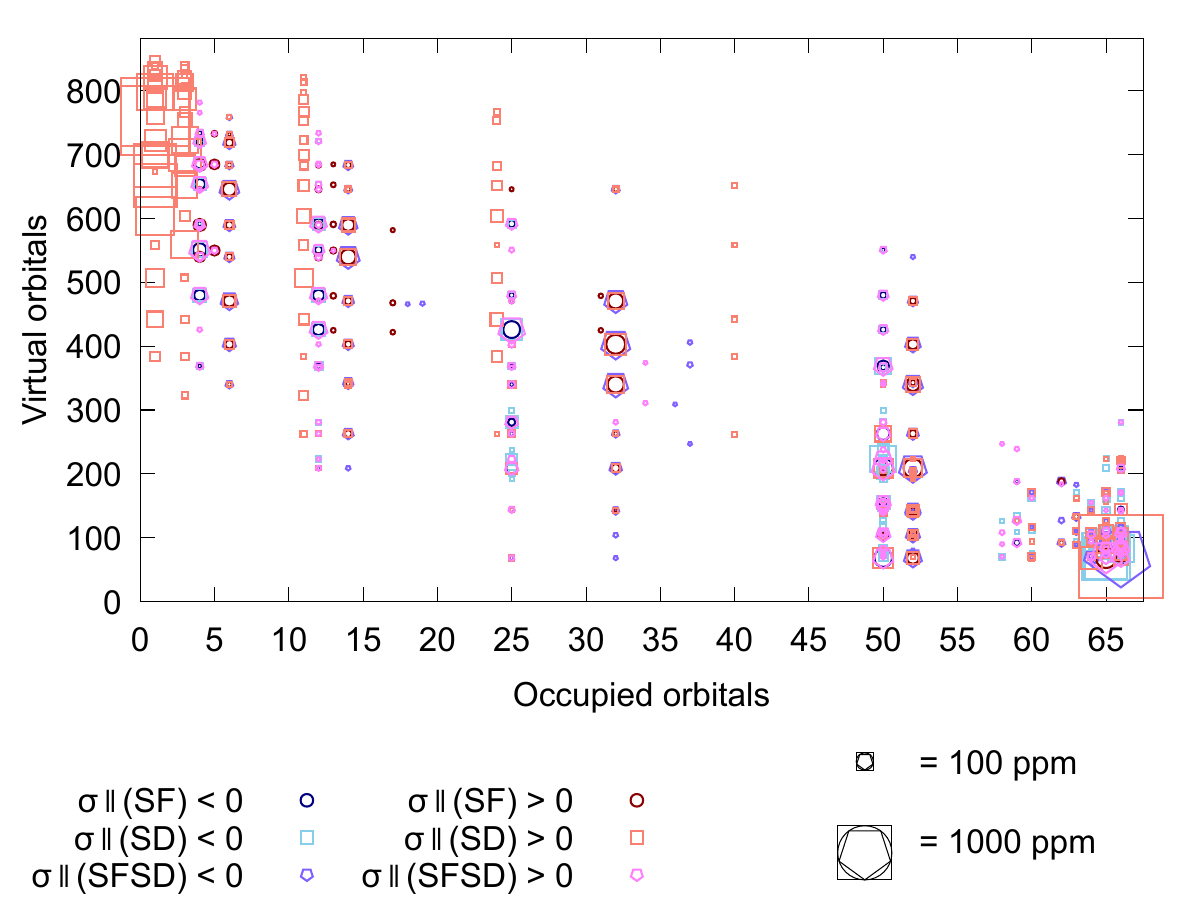}
      \end{minipage}%
      \caption{Pattern of excitations for $\sigma_\parallel$(Tl) (left)
                and $\sigma_\perp$(Tl) (right) for TlI shown separately for the SD and SF parts of the NMR shielding propagator at the level of rZORA-LDA. The magnitude of the amplitude of each type is proportional to marked area. The numbers that label the occupied and virtual orbitals correspond to the Kramers pairs, such that 1 refers to the lowest energy occupied pair and 67 corresponds to the highest energy occupied pair.}
      \label{fig:propagator_rZORA-LDA_TlI}
      \end{figure*}

      \newpage

      \begin{figure*}[h!]
      \begin{minipage}{.48\textwidth}
      \includegraphics[width=\textwidth]{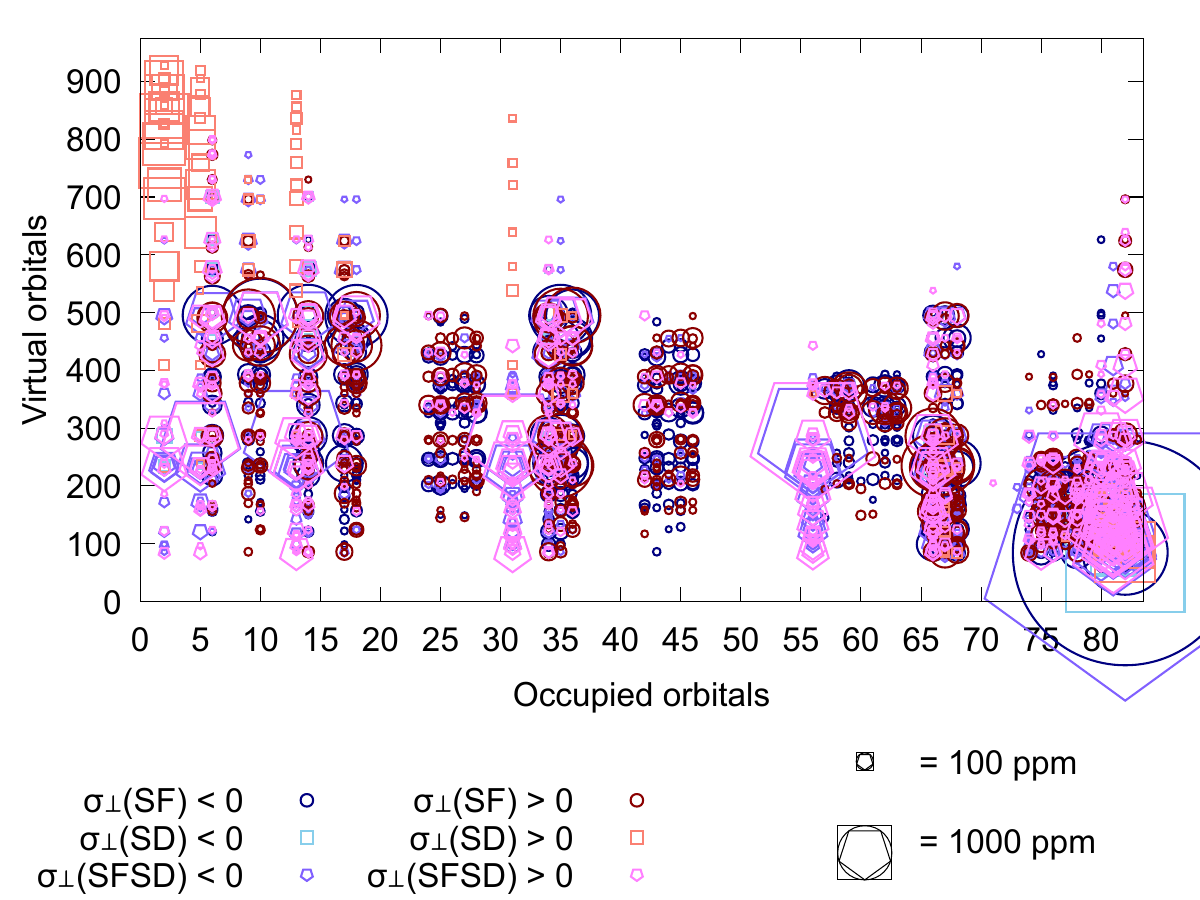}
      \end{minipage}
      \hfill
      \begin{minipage}{.48\textwidth}
      \includegraphics[width=\textwidth]{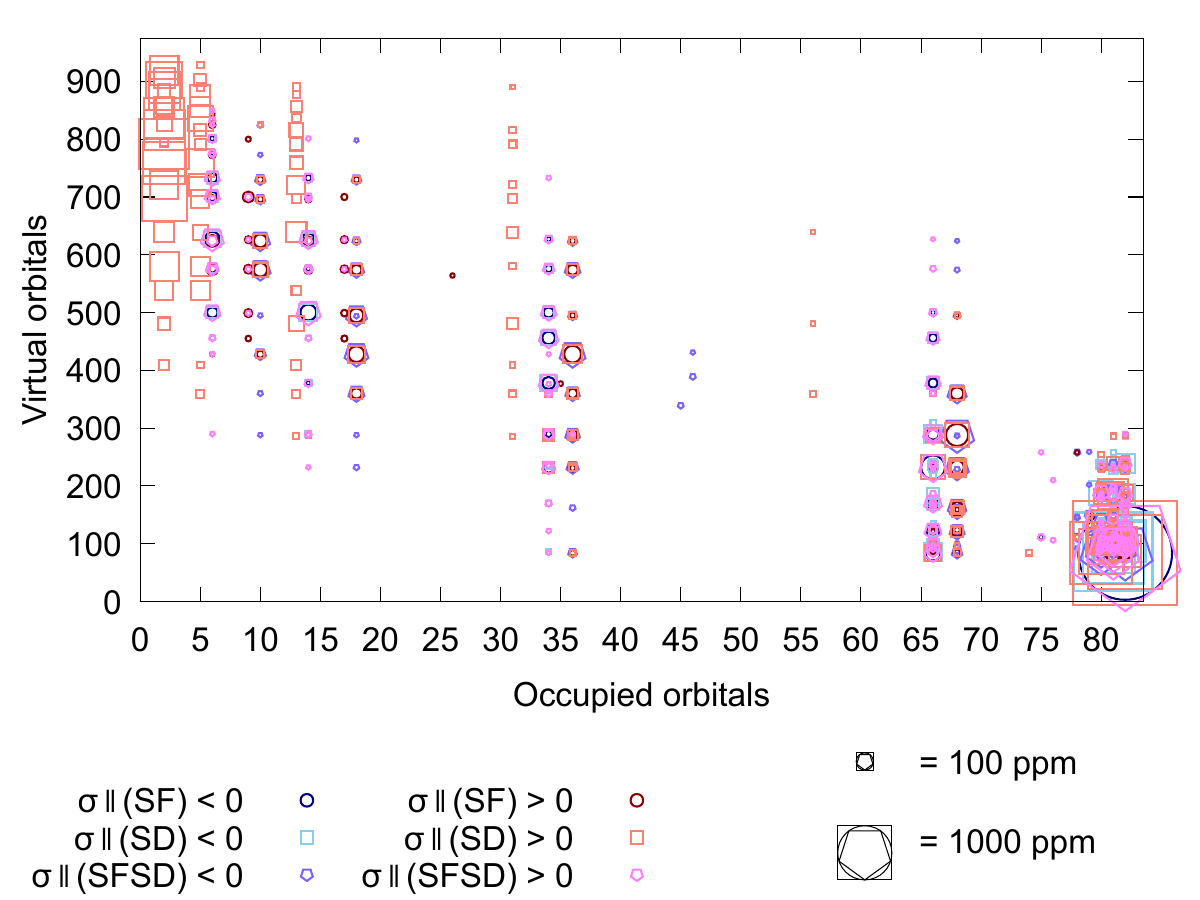}
      \end{minipage}%
      \caption{Pattern of excitations for $\sigma_\parallel$(Tl) (left)
                and $\sigma_\perp$(Tl) (right) for TlAt shown separately for the SD and SF parts of the NMR shielding propagator at the level of rZORA-LDA. The magnitude of the amplitude of each type is proportional to marked area. The numbers that label the occupied and virtual orbitals correspond to the Kramers pairs, such that 1 refers to the lowest energy occupied pair and 83 corresponds to the highest energy occupied pair.}
      \label{fig:propagator_rZORA-LDA_TlAt}
      \end{figure*}

      \newpage

      \begin{figure*}[h!]
      \begin{minipage}{.48\textwidth}
      \includegraphics[width=\textwidth]{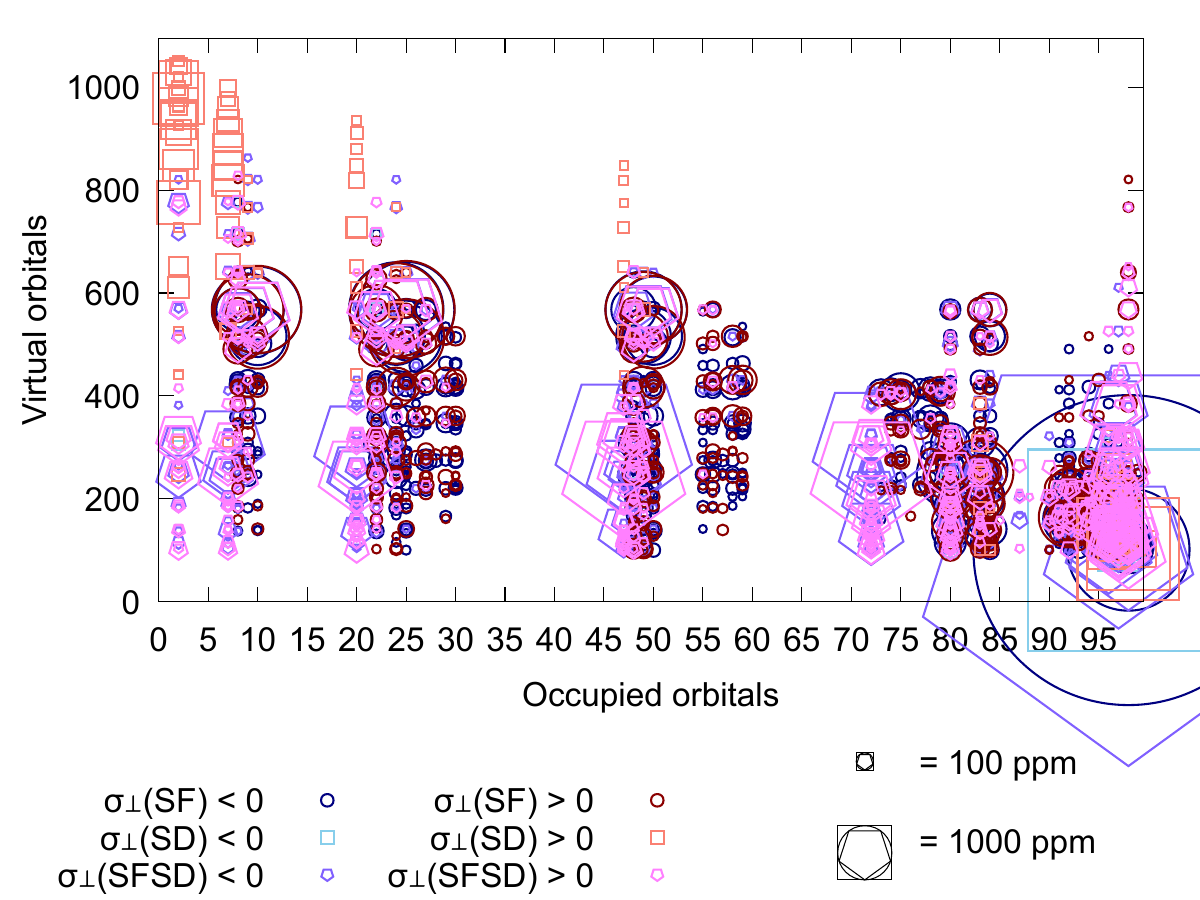}
      \end{minipage}
      \hfill
      \begin{minipage}{.48\textwidth}
      \includegraphics[width=\textwidth]{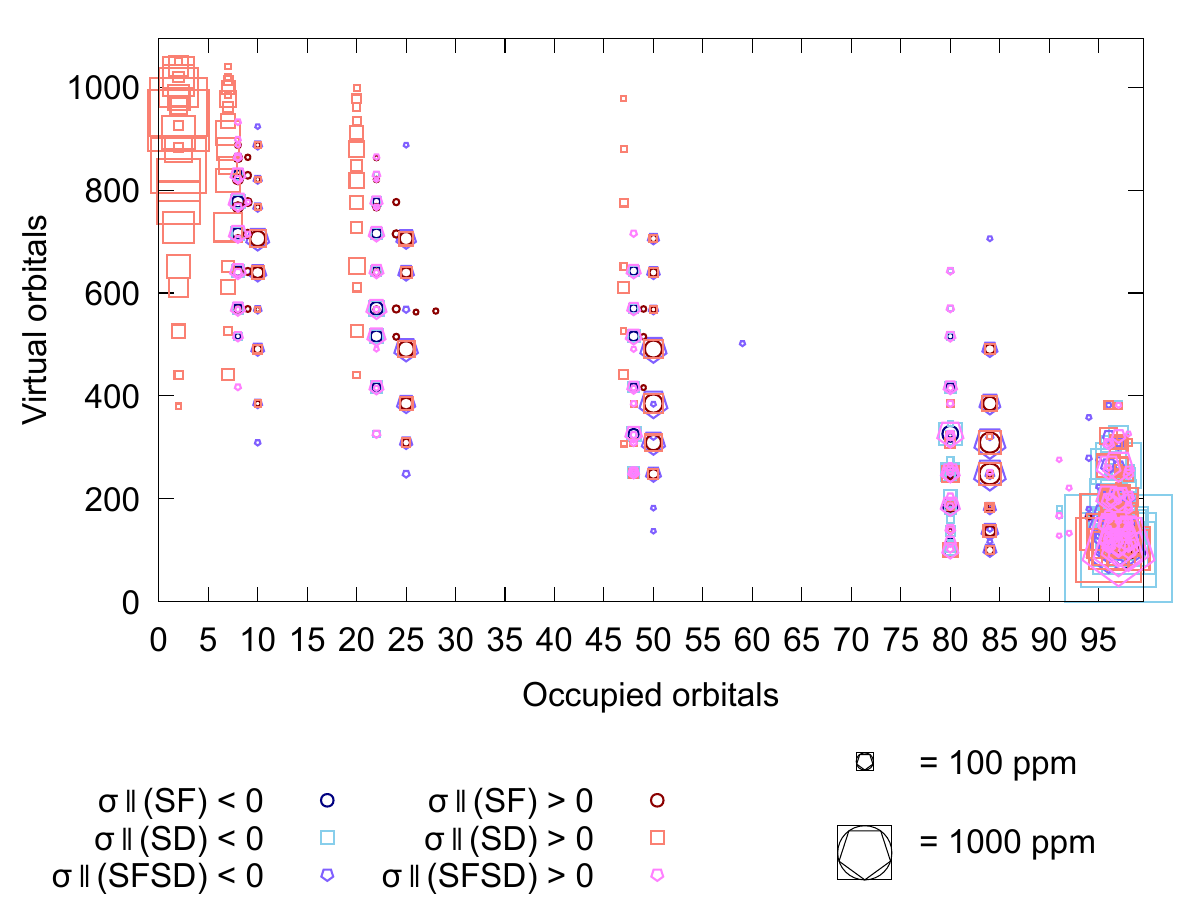}
      \end{minipage}%
      \caption{Pattern of excitations for $\sigma_\parallel$(Tl) (left)
                and $\sigma_\perp$(Tl) (right) for TlTs shown separately for the SD and SF parts of the NMR shielding propagator at the level of rZORA-LDA. The magnitude of the amplitude of each type is proportional to marked area. The numbers that label the occupied and virtual orbitals correspond to the Kramers pairs, such that 1 refers to the lowest energy occupied pair and 99 corresponds to the highest energy occupied pair.}
      \label{fig:propagator_rZORA-LDA_TlTs}
      \end{figure*}

      \newpage

\clearpage

\section{ZORA orbitals involved in HAVHA mechanism}

\begin{figure*}[h!]
\includegraphics[width=\textwidth]{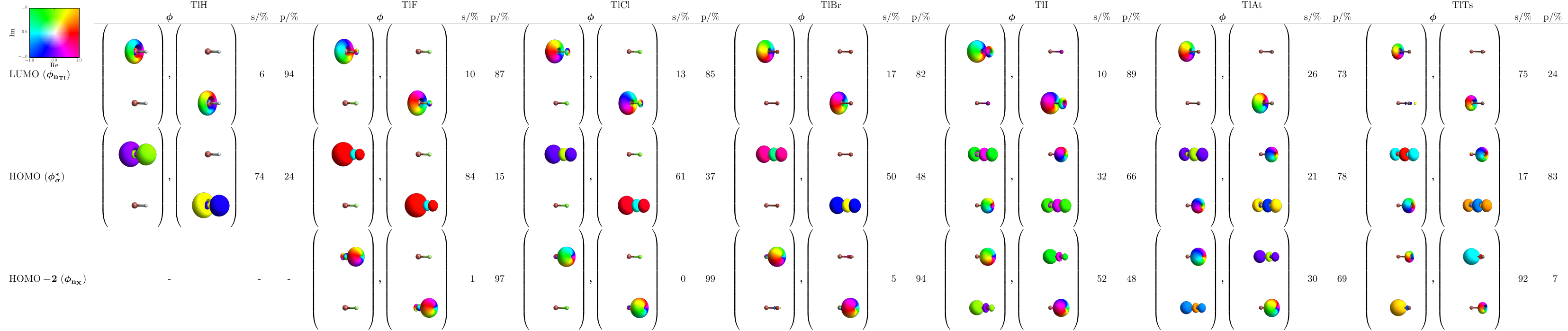}
\caption{Molecular two-component spinors at the level of ZORA HF which are relevant for the HAVHA mechanism. The complex two-component orbitals are visualized as both Kramers paired spinors by calculating orbital 
amplitudes on a three-dimensional grid and plotting them
with the help of \textsc{Mathematica} version 11 \cite{mathematica11} by mapping the phase in the complex plane via a color code on the contour surface with value 0.03 of the absolute value of each spinors. Note that the absolute phase of every single orbital is arbitrary. For each molecular orbital the contribution from s-type and p-type basis functions is given. In the first column in parenthesis the label of the orbital in the NR limit corresponding to Figure 8 in the main text is given.}
\label{fig:orbitalsHF}
\end{figure*}
\begin{figure*}[h!]
\includegraphics[width=\textwidth]{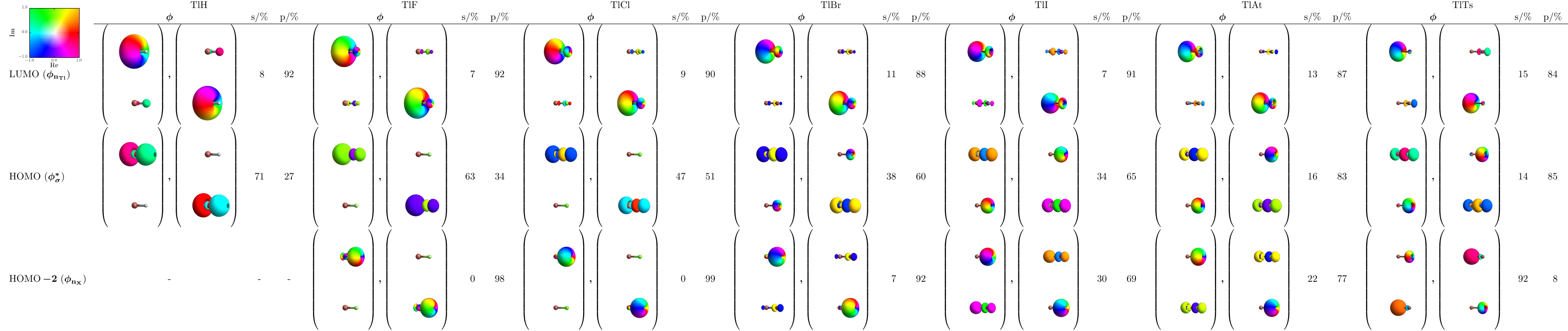}
\caption{Molecular two-component spinors at the level of ZORA PBE0 which are relevant for the HAVHA mechanism.
The complex two-component orbitals are visualized as both Kramers paired spinors by calculating orbital 
amplitudes on a three-dimensional grid and plotting them
with the help of \textsc{Mathematica} version 11 \cite{mathematica11} by mapping the phase in the complex plane via a color code on the contour surface with value 0.03 of the absolute value of each spinors. Note that the absolute phase of every single orbital is arbitrary. For each molecular orbital the contribution from s-type and p-type basis functions is given. In the first column in parenthesis the label of the orbital in the NR limit corresponding to Figure 8 in the main text is given. For TlAt and TlTs the orbital in the middle is not the HOMO but the HOMO$-1$.}
\label{fig:orbitalsPBE0}
\end{figure*}
\begin{figure*}[h!]
\includegraphics[width=\textwidth]{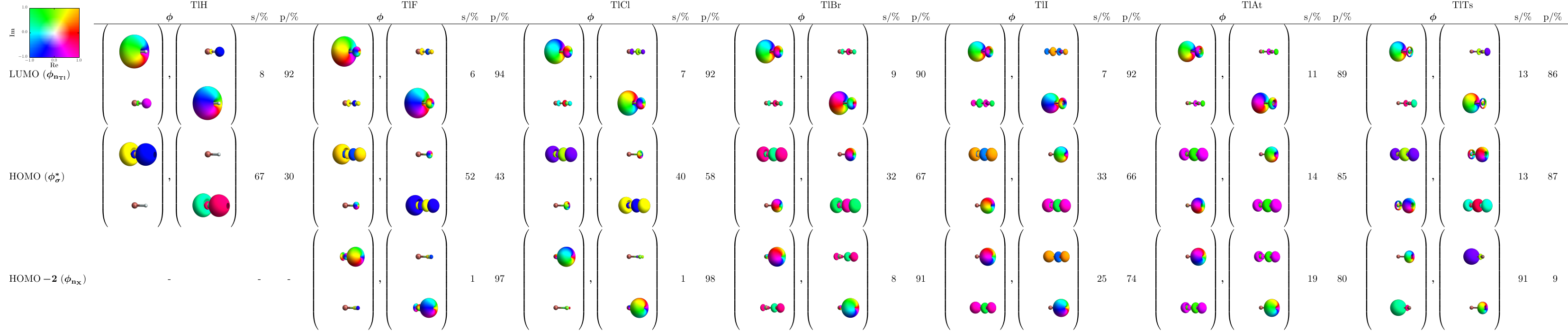}
\caption{Molecular two-component spinors at the level of ZORA LDA which are relevant for the HAVHA mechanism.
The complex two-component orbitals are visualized as both Kramers paired spinors by calculating orbital 
amplitudes on a three-dimensional grid and plotting them
with the help of \textsc{Mathematica} version 11 \cite{mathematica11} by mapping the phase in the complex plane via a color code on the contour surface with value 0.03 of the absolute value of each spinors. Note that the absolute phase of every single orbital is arbitrary. For each molecular orbital the contribution from s-type and p-type basis functions is given. In the first column in parenthesis the label of the orbital in the NR limit corresponding to Figure 8 in the main text is given. For TlBr, TlI, TlAt and TlTs the orbital in the middle is not the HOMO but the HOMO$-1$.}
\label{fig:orbitalsLDA}
\end{figure*}

\clearpage

\section{QED effects}

\begin{table}[!htb]
\caption{QED effects on ${\bm \sigma}$ for the Tl nucleus in Tl$^+$ ion and Tl$X$ ($X$ = H, F, Cl, Br, I, At, Ts) molecules, calculated using the DFT/PBE0 approach. Numbers in parentheses were obtained using the ``s-all'' approximation (see text for details).}
\tabcolsep=0.5\tabcolsep
\begin{tabular*}{\linewidth}{@{}l @{\extracolsep{\fill}} *{12}{S[parse-numbers=false]} @{}}
\toprule
 & {Tl$^+$} & \multicolumn{3}{c}{TlF} & \multicolumn{3}{c}{TlCl} & \multicolumn{3}{c}{TlBr} &&\\
\cmidrule{3-5}\cmidrule{6-8}\cmidrule{9-11}
 & {$\sigma_\text{iso}$} & {$\sigma_\perp$} &  {$\sigma_\parallel$} & {$\sigma_\text{iso}$} & {$\sigma_\perp$} &  {$\sigma_\parallel$} & {$\sigma_\text{iso}$} & {$\sigma_\perp$} &  {$\sigma_\parallel$} & {$\sigma_\text{iso}$} \\
\midrule
1s & -41.4 & -40.8 & -40.9 & -40.9 & -40.7 & -40.8 & -40.8 & -40.7 & -40.8 & -40.7 &  &  \\
 & (-42.8) & (-42.6) & (-42.7) & (-42.7) & (-42.5) & (-42.7) & (-42.6) & (-42.5) & (-42.7) & (-42.6) &  &  \\
2s & -16.1 & -15.4 & -15.6 & -15.4 & -15.3 & -15.4 & -15.3 & -15.2 & -15.4 & -15.3 &  &  \\
 & (-16.3) & (-16.3) & (-16.2) & (-16.3) & (-16.3) & (-16.2) & (-16.3) & (-16.4) & (-16.2) & (-16.3) &  &  \\
3s & -7.1 & -6.1 & -6.4 & -6.2 & -6.0 & -6.2 & -6.1 & -6.0 & -6.2 & -6.1 &  &  \\
 & (-7.1) & (-7.2) & (-7.0) & (-7.1) & (-7.2) & (-7.0) & (-7.2) & (-7.4) & (-7.0) & (-7.3) &  &  \\
4s & -3.1 & -2.1 & -2.4 & -2.2 & -2.2 & -2.2 & -2.2 & -2.1 & -2.2 & -2.2 &  &  \\
 & (-3.1) & (-3.3) & (-3.1) & (-3.2) & (-3.5) & (-3.1) & (-3.4) & (-3.9) & (-3.0) & (-3.6) &  &  \\
5s & -1.1 & -0.5 & -0.6 & -0.6 & -0.5 & -0.5 & -0.5 & -0.4 & -0.5 & -0.5 &  &  \\
 & (-1.1) & (-2.2) & (-1.0) & (-1.8) & (-2.8) & (-1.0) & (-2.2) & (-3.7) & (-0.9) & (-2.8) &  &  \\
Total & -68.8 & -65.0 & -65.9 & -65.3 & -64.7 & -65.1 & -64.8 & -64.5 & -65.0 & -64.6 &  &  \\
 & (-70.4) & (-71.6) & (-70.0) & (-71.1) & (-72.4) & (-70.1) & (-71.6) & (-74.0) & (-69.9) & (-72.6) &  &  \\
\midrule
 & \multicolumn{3}{c}{TlI} & \multicolumn{3}{c}{TlAt} & \multicolumn{3}{c}{TlH} & \multicolumn{3}{c}{TlTs} \\
\cmidrule{2-4}\cmidrule{5-7}\cmidrule{8-10}\cmidrule{11-13}
 & {$\sigma_\perp$} &  {$\sigma_\parallel$} & {$\sigma_\text{iso}$} & {$\sigma_\perp$} &  {$\sigma_\parallel$} & {$\sigma_\text{iso}$} & {$\sigma_\perp$} &  {$\sigma_\parallel$} & {$\sigma_\text{iso}$} & {$\sigma_\perp$} &  {$\sigma_\parallel$} & {$\sigma_\text{iso}$} \\
\midrule
1s & -40.4 & -40.8 & -40.5 & -40.2 & -40.9 & -40.5 & -39.8 & -40.2 & -39.9 & -40.0 & -41.1 & -40.4 \\
 & (-42.4) & (-42.7) & (-42.5) & (-42.1) & (-42.8) & (-42.3) & (-42.1) & (-42.8) & (-42.3) & (-41.9) & (-43.0) & (-42.3) \\
2s & -15.5 & -15.4 & -15.4 & -15.5 & -15.3 & -15.4 & -14.6 & -14.6 & -14.6 & -15.3 & -15.2 & -15.3 \\
 & (-16.5) & (-16.2) & (-16.4) & (-16.5) & (-16.2) & (-16.4) & (-16.0) & (-16.2) & (-16.1) & (-16.7) & (-16.0) & (-16.5) \\
3s & -6.4 & -6.2 & -6.3 & -6.3 & -6.1 & -6.3 & -5.6 & -5.5 & -5.5 & -6.1 & -6.0 & -6.1 \\
 & (-7.6) & (-7.0) & (-7.4) & (-7.8) & (-7.0) & (-7.5) & (-6.8) & (-7.0) & (-6.9) & (-8.0) & (-6.9) & (-7.6) \\
4s & -2.4 & -2.2 & -2.3 & -2.4 & -2.2 & -2.3 & -1.9 & -1.9 & -1.9 & -2.3 & -2.1 & -2.2 \\
 & (-4.3) & (-3.0) & (-3.9) & (-4.8) & (-3.0) & (-4.2) & (-2.8) & (-2.9) & (-2.9) & (-5.4) & (-2.8) & (-4.5) \\
5s & -0.5 & -0.5 & -0.5 & -0.5 & -0.5 & -0.5 & -0.9 & 0.0 & -0.6 & -0.4 & -0.5 & -0.4 \\
 & (-4.8) & (-0.8) & (-3.5) & (-6.0) & (-0.5) & (-4.2) & (-2.4) & (-0.5) & (-1.8) & (-7.3) & 0.5 & (-4.7) \\
Total & -65.1 & -65.0 & -65.1 & -64.8 & -65.0 & -64.9 & -62.7 & -62.2 & -62.6 & -64.1 & -65.0 & -64.4 \\
 & (-75.5) & (-69.8) & (-73.6) & (-77.2) & (-69.4) & (-74.6) & (-70.2) & (-69.4) & (-69.9) & (-79.3) & (-68.2) & (-75.6) \\
\bottomrule
\end{tabular*}
\end{table}

\sisetup{text-series-to-math=true,propagate-math-font=true}
\begin{table}[!htb]
\caption{$C_{QR}^\text{QED/DC}$ coefficients for atoms with $Z$ = 10--92. Only coefficients for $Z$ = 10, 18, 30, 36, 48, 54, 70, 80, 86, and 92 have been calculated directly, because only for these atomic numbers there are $D_\text{VP}(Z\alpha)$, $D_\text{SE}(Z\alpha)$, and $D_\text{VP,po}(Z\alpha)$ values available in the papers by Yerokhin \textit{et al.}~used in this work\cite{Yerokhin2011,Yerokhin2012}. Coefficients for the other $Z$ values have been estimated by using cubic interpolation. The highlighted values correspond to the Tl atom and are the coefficients that were used in this work.}
\begin{tabular}{@{}S[table-format=2.0] S[table-format=5.4e3]S[table-format=4.4e3] S[table-format=8.0] S[table-format=5.4e3]S[table-format=4.4e3] S[table-format=8.0] S[table-format=5.4e3]S[table-format=4.4e3]@{}}
\toprule
$Z$ & $C_{1s-ns}^\text{QED/DC}$ & $C_{2s-ns}^\text{QED/DC}$ & {\hspace{1cm}$Z$} & $C_{1s-ns}^\text{QED/DC}$ & $C_{2s-ns}^\text{QED/DC}$ & {\hspace{1cm}$Z$} & $C_{1s-ns}^\text{QED/DC}$ & $C_{2s-ns}^\text{QED/DC}$ \\
\midrule
10 & -5.2739E-04 & -5.2867E-04 & 38 & -3.8203E-03 & -3.7226E-03 & 66 & -6.4120E-03 & -6.3078E-03 \\[0.5ex]
11 & -6.5166E-04 & -6.4182E-04 & 39 & -3.9333E-03 & -3.8332E-03 & 67 & -6.5651E-03 & -6.4655E-03 \\[0.5ex]
12 & -7.8024E-04 & -7.6171E-04 & 40 & -4.0499E-03 & -3.9466E-03 & 68 & -6.7265E-03 & -6.6315E-03 \\[0.5ex]
13 & -9.1249E-04 & -8.8723E-04 & 41 & -4.1682E-03 & -4.0611E-03 & 69 & -6.8961E-03 & -6.8057E-03 \\[0.5ex]
14 & -1.0478E-03 & -1.0173E-03 & 42 & -4.2864E-03 & -4.1749E-03 & 70 & -7.0741E-03 & -6.9882E-03 \\[0.5ex]
15 & -1.1855E-03 & -1.1507E-03 & 43 & -4.4024E-03 & -4.2863E-03 & 71 & -7.2602E-03 & -7.1787E-03 \\[0.5ex]
16 & -1.3249E-03 & -1.2864E-03 & 44 & -4.5146E-03 & -4.3938E-03 & 72 & -7.4532E-03 & -7.3760E-03 \\[0.5ex]
17 & -1.4654E-03 & -1.4233E-03 & 45 & -4.6209E-03 & -4.4954E-03 & 73 & -7.6515E-03 & -7.5788E-03 \\[0.5ex]
18 & -1.6064E-03 & -1.5602E-03 & 46 & -4.7196E-03 & -4.5896E-03 & 74 & -7.8535E-03 & -7.7857E-03 \\[0.5ex]
19 & -1.7472E-03 & -1.6962E-03 & 47 & -4.8086E-03 & -4.6746E-03 & 75 & -8.0578E-03 & -7.9953E-03 \\[0.5ex]
20 & -1.8872E-03 & -1.8307E-03 & 48 & -4.8863E-03 & -4.7488E-03 & 76 & -8.2628E-03 & -8.2061E-03 \\[0.5ex]
21 & -2.0257E-03 & -1.9632E-03 & 49 & -4.9515E-03 & -4.8113E-03 & 77 & -8.4669E-03 & -8.4169E-03 \\[0.5ex]
22 & -2.1621E-03 & -2.0933E-03 & 50 & -5.0070E-03 & -4.8646E-03 & 78 & -8.6686E-03 & -8.6262E-03 \\[0.5ex]
23 & -2.2957E-03 & -2.2206E-03 & 51 & -5.0560E-03 & -4.9123E-03 & 79 & -8.8664E-03 & -8.8326E-03 \\[0.5ex]
24 & -2.4259E-03 & -2.3446E-03 & 52 & -5.1023E-03 & -4.9578E-03 & 80 & -9.0587E-03 & -9.0348E-03 \\[0.5ex]
25 & -2.5520E-03 & -2.4649E-03 & 53 & -5.1491E-03 & -5.0046E-03 & \hlt{81} & \hlt{-9.2462E-03} & \hlt{-9.2335E-03} \\[0.5ex]
26 & -2.6734E-03 & -2.5810E-03 & 54 & -5.2001E-03 & -5.0561E-03 & 82 & -9.4385E-03 & -9.4385E-03 \\[0.5ex]
27 & -2.7895E-03 & -2.6925E-03 & 55 & -5.2581E-03 & -5.1152E-03 & 83 & -9.6473E-03 & -9.6614E-03 \\[0.5ex]
28 & -2.8996E-03 & -2.7989E-03 & 56 & -5.3238E-03 & -5.1826E-03 & 84 & -9.8845E-03 & -9.9143E-03 \\[0.5ex]
29 & -3.0030E-03 & -2.8998E-03 & 57 & -5.3971E-03 & -5.2581E-03 & 85 & -1.0162E-02 & -1.0209E-02 \\[0.5ex]
30 & -3.0991E-03 & -2.9948E-03 & 58 & -5.4782E-03 & -5.3419E-03 & 86 & -1.0491E-02 & -1.0557E-02 \\[0.5ex]
31 & -3.1878E-03 & -3.0839E-03 & 59 & -5.5671E-03 & -5.4339E-03 & 87 & -1.0884E-02 & -1.0971E-02 \\[0.5ex]
32 & -3.2715E-03 & -3.1689E-03 & 60 & -5.6639E-03 & -5.5340E-03 & 88 & -1.1352E-02 & -1.1461E-02 \\[0.5ex]
33 & -3.3529E-03 & -3.2523E-03 & 61 & -5.7685E-03 & -5.6424E-03 & 89 & -1.1908E-02 & -1.2041E-02 \\[0.5ex]
34 & -3.4350E-03 & -3.3365E-03 & 62 & -5.8811E-03 & -5.7591E-03 & 90 & -1.2563E-02 & -1.2722E-02 \\[0.5ex]
35 & -3.5205E-03 & -3.4237E-03 & 63 & -6.0017E-03 & -5.8839E-03 & 91 & -1.3329E-02 & -1.3516E-02 \\[0.5ex]
36 & -3.6124E-03 & -3.5165E-03 & 64 & -6.1304E-03 & -6.0170E-03 & 92 & -1.4217E-02 & -1.4434E-02 \\[0.5ex]
37 & -3.7127E-03 & -3.6165E-03 & 65 & -6.2671E-03 & -6.1582E-03 &  &  &  \\
\bottomrule
\end{tabular}
\end{table}

\clearpage

\begin{figure*}[!htb]
\centering
\includegraphics[width=0.48\linewidth]{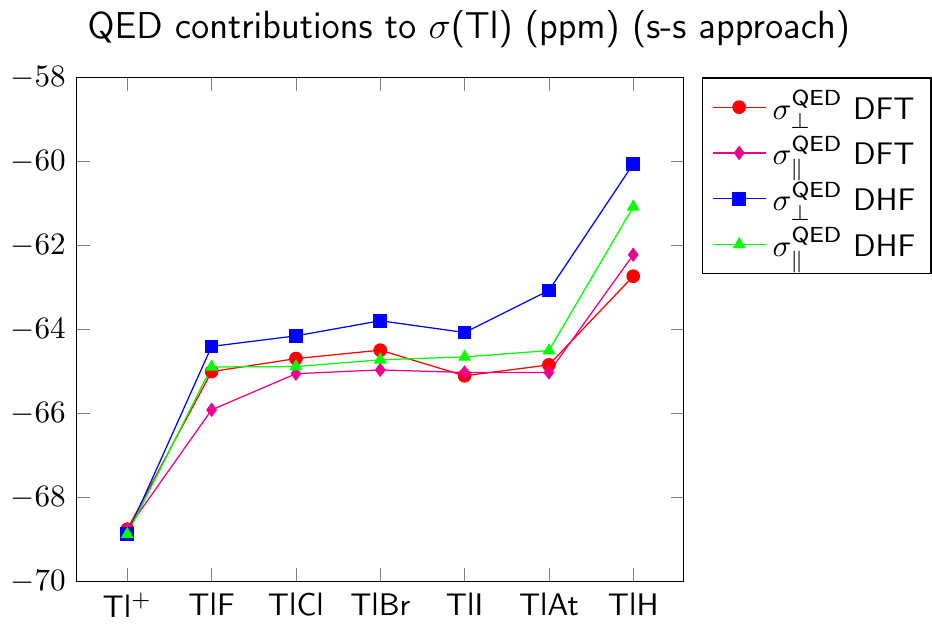}
\hfill
\includegraphics[width=0.48\linewidth]{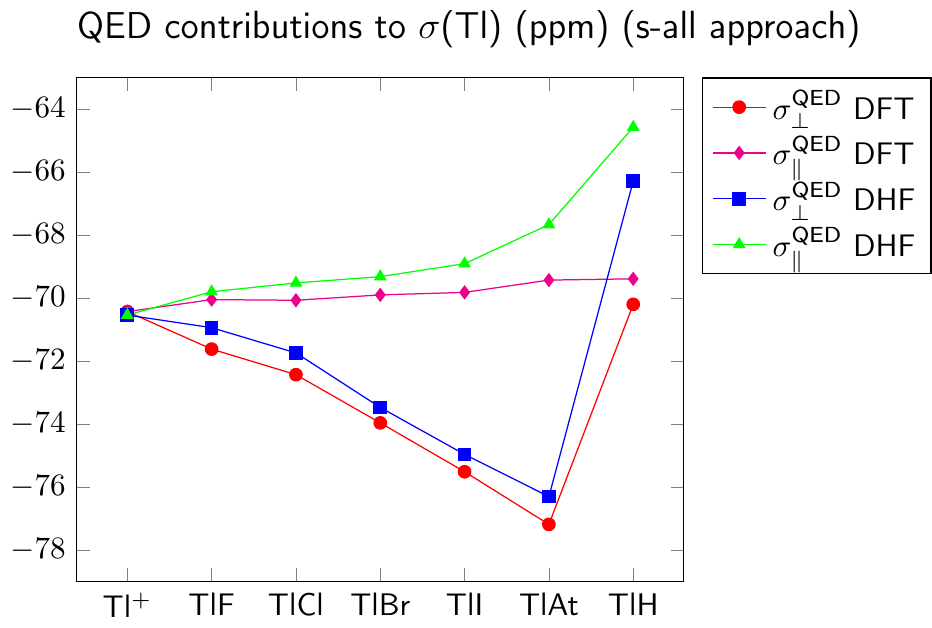}
\caption{QED effects on $\sigma$ for the Tl nucleus in Tl$^+$ ion and in Tl$X$ ($X$ = H, F, Cl, Br, I, At) molecules, calculated using the DHF and DFT approaches. Both ''s-s'' (left figure) and ''s-all'' (right figure) approches are presented.}
\label{fig:QED_contrib}
\end{figure*}

\bibliographystyle{aipnum4-2}
\bibliography{refs}